\pdfoutput=1 
\documentclass[11pt]{article}
\usepackage[utf8]{inputenc}
\usepackage{fullpage}

\usepackage{amssymb}
\usepackage{amsmath}
\usepackage{amsthm}
\usepackage{graphicx,color,colordvi}
\usepackage{bbm}
\usepackage{stmaryrd}
\usepackage[utf8]{inputenc}
\usepackage[blocks]{authblk}
\usepackage{dsfont}
\usepackage[showonlyrefs]{mathtools}
\usepackage{authblk}
\usepackage{wrapfig}
\usepackage[english]{babel}
\usepackage{physics}
\usepackage[table]{xcolor}
\usepackage[center]{caption}
\usepackage{tikz}
\usepackage{ifthen}
\usepackage{pgfplots}
\pgfplotsset{compat=1.9}
\usetikzlibrary{shapes,arrows.meta}
\usetikzlibrary{positioning}
\usetikzlibrary{shapes.geometric}
\RequirePackage[framemethod=default]{mdframed}
\usepackage[margin=1in]{geometry}
\usepackage{comment}
\usepackage{url}
\usepackage{ upgreek }
\usepackage{makecell}

\usepackage{nameref}
\usepackage{varioref}
\usepackage[colorlinks = true, linkcolor = blue, urlcolor  = blue, citecolor = red]{hyperref}
\usepackage[capitalise]{cleveref}

\usepackage{float}
\usepackage{enumitem}
\usepackage{subcaption}
\newfloat{algorithm}{t}{lop}
\usepackage{array,rotating ,makecell, multirow, tabularx}
\usepackage{scrextend}
\usepackage{mathrsfs}
\usepackage{enumitem}
\usepackage{soul}

\usepackage[normalem]{ulem}

\usepackage{blkarray}

\usepackage{cancel}
\usepackage{thmtools}

\usepackage{thm-restate} 

\usepackage{arydshln} 

\newtheorem{definition}{Definition}
\newtheorem{theorem}{Theorem}
\newtheorem{theorem*}{Theorem}
\newtheorem{lemma}[theorem]{Lemma}
\newtheorem{proposition}[theorem]{Proposition}
\newtheorem*{remark*}{Remark}
\newtheorem{corollary}[theorem]{Corollary}
\theoremstyle{definition}

\usepackage{tikz}
\usetikzlibrary{decorations.pathreplacing, calligraphy, decorations.markings}
\usetikzlibrary{arrows, automata, positioning}

\definecolor{purple}{HTML}{E5E3F5}
\definecolor{amaranth}{HTML}{F5CFD8}
\definecolor{yellow}{HTML}{FFEAB6}
\definecolor{pink}{HTML}{FFDBDA}
\definecolor{powderblue}{HTML}{B6CBE2}
\definecolor{melon}{HTML}{F8B4B4}
\definecolor{atomictangerine}{HTML}{FF8D5C}

\definecolor{darkpurple}{HTML}{3D348B}
\definecolor{darkamaranth}{HTML}{AB2346}
\definecolor{darkyellow}{HTML}{FFBA08}

\definecolor{tensor}{rgb}{0.5,0.8,0.5}
\definecolor{isometry}{rgb}{0.8,0.8,1}
\definecolor{unitary}{rgb}{0.8,0.5,.5}
\definecolor{gate}{rgb}{1.0,1.0,1.0}

\usepackage{tikz}
\usepackage{venndiagram}
\usetikzlibrary{decorations.pathreplacing, calligraphy, decorations.markings}
\usetikzlibrary{arrows, automata, positioning}

\newcommand{\MPSTensor}[3]{
	\begin{scope}[shift={(#1)}]
		\draw (-1,0) -- (1,0);
		\draw (0,1) -- (0,0);
		\filldraw[fill=#3] (-1/2,-1/2) -- (-1/2,1/2) -- (1/2,1/2) -- (1/2,-1/2) -- (-1/2,-1/2);
		\draw (0,0) node {\scriptsize #2};
	\end{scope}
}

\newcommand{\FullMPS}[3]{
	\begin{scope}[shift={(#1)}]
		\draw[shift={(0,0)},dotted] (0,0) -- (4.5,0);
        \MPSTensor{0,0}{#2}{#3}
        \MPSTensor{1.5,0}{#2}{#3}
        \MPSTensor{5,0}{#2}{#3}
	\end{scope}
}

\newcommand{\MatrixHorizontalIndices}[4]{
	\begin{scope}[shift={(#1)}]
		\draw (-1-#4/2,0) -- (1+#4/2,0);
		\filldraw[fill=#3] (-1/2-#4/2,-1/2) -- (-1/2-#4/2,1/2) -- (1/2+#4/2,1/2) -- (1/2+#4/2,-1/2) -- (-1/2-#4/2,-1/2);
		\draw (0,0) node {\scriptsize #2};
	\end{scope}
}

\newcommand{\FullMPSX}[5]{
	\begin{scope}[shift={(#1)}]
		\draw[shift={(0,0)},dotted] (0,0) -- (4.5,0);
        \MPSTensor{0,0}{#2}{#4}
        \MPSTensor{1.5,0}{#2}{#4}
        \MPSTensor{5,0}{#2}{#4}
        \MatrixHorizontalIndices{-1.5,0}{#3}{#5}{0}
        \draw (-2.5,0) -- (-2.5,-0.8) -- (6,-0.8) -- (6,0);
	\end{scope}
}

\newcommand{\highr}[1]{\colorbox{red!20}{$\displaystyle#1$}}
\newcommand{\highb}[1]{\colorbox{blue!20}{$\displaystyle#1$}}
\newcommand{\highg}[1]{\colorbox{green!20}{$\displaystyle#1$}}
\newcommand{\highy}[1]{\colorbox{yellow!99}{$\displaystyle#1$}}

\newcommand{\marta}[1]{{\color{teal}#1}}

\theoremstyle{plain}

\newtheorem{innercustomprop}{Proposition}
\newenvironment{myprop}[1]
  {\renewcommand\theinnercustomprop{#1}\innercustomprop}
  {\endinnercustomprop}

\begin{document}

\title{Uniform matrix product states with a boundary}

\author[1,2]{Marta Florido-Llin\`as\thanks{marta.florido.llinas@mpq.mpg.de}}
\author[3]{\'Alvaro M. Alhambra}
\author[4]{David P\'erez-Garc\'ia}
\author[1,2]{J. Ignacio Cirac}
\affil[1]{\footnotesize Max-Planck-Institut f{\"{u}}r Quantenoptik, Hans-Kopfermann-Str. 1, 85748 Garching, Germany}
\affil[2]{\footnotesize Munich Center for Quantum Science and Technology (MCQST), Schellingstr. 4, 80799 M{\"{u}}nchen, Germany}
\affil[3]{\footnotesize Instituto de F\'isica Te\'orica UAM/CSIC, C/ Nicol\'as Cabrera 13-15, Cantoblanco, 28049 Madrid, Spain}
\affil[4]{\footnotesize Departamento de An\'alisis Matem\'atico, Universidad Complutense de Madrid, 28040 Madrid, Spain}

\maketitle
\vspace{-1cm}
\begin{abstract}
    Canonical forms are central to the analytical understanding of tensor network states, underpinning key results such as the complete classification of one-dimensional symmetry-protected topological phases within the matrix product state (MPS) framework. Yet, the established theory applies only to uniform MPS with periodic boundary conditions, leaving many physically relevant states beyond its reach. Here we introduce a generalized canonical form for uniform MPS with a boundary matrix, thus extending the analytical MPS framework to a more general setting of wider physical significance. This canonical form reveals that any such MPS can be represented as a block-invertible matrix product operator acting on a structured class of algebraic regular language states that capture its essential long-range and scale-invariant features. Our construction builds on new algebraic results of independent interest that characterize the span and algebra generated by non-semisimple sets of matrices, including a generalized quantum Wielandt's inequality that gives an explicit upper bound on the blocking length at which the fixed-length span stabilizes to an algebra. Together, these results establish a unified theoretical foundation for uniform MPS with boundaries, bridging the gap between periodic and arbitrary-boundary settings, and providing the basis for extending key analytical and classification results of matrix product states to a much broader class of states and operators.
\end{abstract}

\newpage 
\setcounter{tocdepth}{2}        
\renewcommand{\contentsname}{Table of Contents}
\tableofcontents

\section{Introduction} 

Tensor networks (TNs) provide a powerful and unifying framework for the description, classification, and simulation of quantum many-body systems \cite{Cirac_review_2021, orus_2014_introTNs}. In one spatial dimension, matrix product states (MPS) \cite{fannes_finitely_1992, Perez-Garcia2007} have become particularly valuable, both from a numerical and a theoretical standpoint. Numerically, MPS allow for accurate approximations of low-energy properties in many-body systems, underpinning techniques such as the paradigmatic density matrix renormalization group (DMRG) \cite{DMRG92, Schollwock_2011},  as well as simulations of short-time dynamics or thermal equilibrium \cite{Vidal_2004, Osborne_2006, Verstraete_2004, White_2009, Kuwahara_2020}. On the theoretical side, tensor network techniques have enabled key analytical insights \cite{Cirac_review_2021}, such as the classification of gapped quantum phases in 1D systems \cite{Chen_2011_symm, Schuch_2011_symm, Pollmann_2010_phases}.

A central concept in the theoretical understanding and practical use of tensor networks is that of \textit{canonical forms}. Since multiple different sets of tensors can represent the same physical TN state, a natural question arises: how are these representations related? Specifically, can they be connected via a gauge transformation, i.e., inserting invertible matrices and their corresponding inverses along the internal bonds, which cancel upon contraction? For uniform MPS with periodic boundary conditions (PBC), consisting of a single repeated tensor, this question has a positive answer. A well-defined canonical form exists in this case, enabling us to extract structural and physical information about the state directly from the local tensors. This underlies applications such as classifying gapped topological phases and their symmetry-protected counterpart \cite{Chen_2011_symm, Schuch_2011_symm, Pollmann_2010_phases}, characterizing renormalization \cite{cirac_2017_mpdo} or time evolution through the classification of MPUs and quantum cellular automata \cite{cirac_2017_MPUs, styliaris_2024_MPUs}, understanding topological order and SPT phases in 2D PEPSs through matrix product operator (MPO) algebras \cite{bultinck_2017_anyons-MPOs, molnar_2022_matrixproductoperatoralgebras}, and proposing optimal state preparation protocols \cite{malz_2024_MPS-preparation}.

However, many physically relevant quantum states fall outside the scope of uniform PBC MPS. A paradigmatic example is the W-state \cite{Dur_2000_Wstate}, which cannot be represented as a uniform PBC MPS with constant bond dimension: the size of any such representation must scale as $\Omega(N^{1/(3+\delta)})$ for any $\delta>0$ \cite{Perez-Garcia2007, Michalek2018}. In contrast, it admits an exact description as a uniform MPS with a boundary matrix, a structure we refer to as an \textit{MPS-X}.

Analogous situations arise for many other important classes of states and operators where the MPS-X ansatz naturally appears, including Dicke states \cite{Dicke_1954}, domain wall states \cite{Haegeman_2012}, ansätze for elementary excitations \cite{ostlund_1995_dmrg-exc, haegeman_2013_elem-excitations, Osborne_2025_MPS-excitation, white_2025_sitebasisexcitationansatz, jha_2025_scattering-mpsx}, tangent-space methods \cite{haegeman_2013_postMPS-tangent-space, vanderstraeten_2019}, quantum many-body scars \cite{moudgalya_2018_scars-MPS, gioia_2025_distincttypesparenthamiltonians}, the algebraic Bethe ansatz \cite{murg_2012_bethe-TNs}, regular language states \cite{florido-llinas_2024_RLS}, MPO representations of local Hamiltonians \cite{Pirvu_2010_MPO-representations, McCulloch_2007}, and relevant classes of matrix product unitaries (MPUs) \cite{styliaris_2024_MPUs}. Despite their ubiquity, MPS-X have so far lacked a general theoretical framework as the one existing for TI PBC MPS. In particular, there is no canonical form or structural classification of the inherent freedom in these representations, limiting our ability to analyze or study them systematically.

In this work, we develop a canonical form for uniform MPS with a boundary. Our construction is based on new results on the structure of the algebra and span generated by sets of matrices, specifically in the case where off-diagonal blocks cannot be neglected anymore (\textit{non-semisimple} case), in contrast to the uniform PBC MPS ansatz. A central result of our paper is the formulation of a generalized canonical form (gCF) for TI MPS-X. This canonical form represents any MPS-X as a block-invertible uniform matrix product operator acting on a simpler class of states, which we call \textit{algebraic regular language states}. These serve as canonical representatives for MPS-X capturing their essential long-range structural features. 

We begin in Section \ref{sec:background-pbc} by reviewing the canonical form for uniform PBC MPS within the broader algebraic framework introduced in this work. In Section \ref{sec:mps-x}, we extend this framework by introducing the gCF for the full class of MPS-X. The structural results on the algebra and span of sets of matrices that underpin the gCF are developed in Section \ref{sec:algebra-span}. Finally, in Section \ref{sec:how-to}, we illustrate with an example how these results are combined to construct the gCF.

\section{Background on uniform PBC MPS}
\label{sec:background-pbc}

A matrix product state (MPS) tensor $A$ consists of a collection of matrices $\{A^i\}_{i=1}^d$ with $A^i \in \mathcal{M}_{D}(\mathbb{C})$, where $d$ and $D$ are the \textit{physical} and \textit{bond dimensions}, respectively. Graphically,
\begin{equation*}
    A_{\alpha\beta}^i \equiv
    \begin{tikzpicture}[scale=.45,thick,baseline={([yshift=-2ex]current bounding box.center)}]
        \MPSTensor{0,0}{$A$}{purple}
        \draw (-1.2,0) node {\scriptsize $\alpha$};
        \draw (1.2,0) node {\scriptsize $\beta$};
        \draw (0,1.3) node {\scriptsize $i$};
    \end{tikzpicture}.
\end{equation*}
A \emph{uniform MPS with periodic boundary conditions} (uniform PBC MPS) and tensor $A$ is the family $\{\ket{\psi_N(A)}\}_{N \in \mathbb{N}}$ defined by
\begin{equation*}
    \ket{\psi_N(A)} 
    =
    \sum_{i_1 \dots i_N = 1}^d \Tr[A^{i_1} \dots A^{i_N}] \ket{i_1 \dots i_N}
    \equiv \ 
    \begin{tikzpicture}[scale=.45,thick,baseline={([yshift=-1.15ex]current bounding box.center)}]
        \FullMPS{0,0}{$A$}{purple}
        \draw (-1,0) -- (-1,-0.75) -- (6,-0.75) -- (6,0);
    \end{tikzpicture} \ .
\end{equation*}
Uniform PBC MPS are by now well understood, and there is a complete theoretical framework describing them. This framework forms the basis for, among other results, the classification of gapped 1D phases of matter \cite{Chen_2011_symm, Schuch_2011_symm}, and naturally extends to matrix product operators (MPOs) after appropriately vectorizing them \cite{cirac_2017_mpdo, cirac_2017_MPUs} (see \cite{orus_2014_introTNs, Cirac_review_2021} for comprehensive reviews). Its two central ingredients are a \emph{canonical form}, and a \emph{fundamental theorem} describing the freedom in the representation \cite{Perez-Garcia2007, cirac_2017_mpdo}.

To bring any uniform PBC MPS to its canonical form, the first step consists of a simultaneous block-upper-triangularization procedure. This guarantees the existence of an invertible matrix $P$ \cite{radjavi_2000_simultaneous} such that
\begin{equation} \label{eq:block-upper-triangular-form}
    P A^i P^{-1} = {\scriptsize\begin{pmatrix}
        A_{11}^i & A_{12}^i & \dots & A_{1b}^i \\
        0 & A_{22}^i & \dots & A_{2b}^i \\
        \vdots & \vdots & \ddots & \vdots \\
        0 & 0 & \dots & A_{bb}^i
    \end{pmatrix} } ,
\end{equation}
for some number of blocks $b$. Each diagonal block is \emph{irreducible}, meaning that the corresponding matrices admit no common proper \textit{invariant subspace} $V \subseteq \mathbb{C}^D$ with $\mathrm{span}_i\{A_{jj}^i V\} = V$. However, such blocks may still exhibit additional invariant subspaces under blocking (so-called \emph{periodic subspaces}, i.e., $W \subseteq \mathbb{C}^D$ for which there exists a \textit{period} $r\geq 1$ such that $W_r = W$, where $W_{n+1} := \mathrm{span}_i\{A_{jj}^i W_n\}$ and $W_0 := W$). 

Blocking every $p$ sites, with $p$ being the least common multiple of the periods of the periodic subspaces of all the diagonal blocks, ensures that no further invariant subspaces appear after blocking. The resulting diagonal blocks give rise to \emph{normal} tensors \cite{cirac_2017_mpdo}. Setting the off-diagonal blocks in Eq. \eqref{eq:block-upper-triangular-form} to zero, which can be done without loss of generality for periodic boundary conditions, yields the canonical form of uniform PBC MPS.

We now reformulate this canonical form in a way we refer to as the \emph{pbcCF}, which is especially convenient for two reasons: 
\textit{(i)} it decomposes the MPS into its ``backbone'' (the lower part) that captures long-range features such as the phase and the renormalization fixed point, and an ``invertible'' upper part encoding only short-range entanglement; and 
\textit{(ii)} it provides the natural foundation for generalizing the canonical form to uniform MPS with a boundary matrix. 

\begin{definition}[Canonical form for uniform PBC MPS (pbcCF)] \label{def:pbcCF}
A family of quantum states $\{\ket{\psi_N}\}_N$ is in \textit{pbcCF} if, for each $N$,
\begin{equation*}
    \ket{\psi_N} = 
    \begin{tikzpicture}[scale=.45,thick,baseline={([yshift=-1.15ex]current bounding box.center)}]
        \begin{scope}[shift={(0,1)}]
            \draw[dotted] (1.5,0) -- (4.5,0);
            \MPSTensor{0,0}{$A$}{purple}
            \draw (0,-0.5) -- (0,-1);
            \MPSTensor{1.5,0}{$A$}{purple}
            \draw (1.5,-0.5) -- (1.5,-1);
            \MPSTensor{4.5,0}{$A$}{purple}
            \draw (4.5,-0.5) -- (4.5,-1);
        \end{scope}
        \filldraw[fill=amaranth] (-0.5,-1) -- (-0.5,0) -- (5,0) -- (5,-1) -- (-0.5,-1);
        \draw (2.25,-0.5) node {\scriptsize $\ket{L_N}$};
        \draw (-1,1) -- (-1,0.25) -- (-0.2,0.25);
        \draw (0.2,0.25) -- (1.3,0.25);
        \draw (1.7,0.25) -- (2.4,0.25);
        \draw[dotted] (2.5,0.25) -- (3.5,0.25);
        \draw (3.6,0.25) -- (4.3,0.25);
        \draw (4.7,0.25) -- (5.5,0.25) -- (5.5,1);
    \end{tikzpicture} \ , 
\end{equation*}
where
\begin{enumerate}[label = (\roman*)]
    \item $\{\ket{L_N}\}_N$ are a family of weighted GHZ-like states, with
    \begin{equation} \label{eq:PBC_lower_states}
        |L_N\rangle := \sum_{j\in \Sigma_\infty} \left( \sum_{k=1}^{r_j} \mu_{j,k}^N \right) \ket{j}^{\otimes N},
    \end{equation}
    where $\mu_{j,k} \in \mathbb{C}$, $r_j \in \mathbb{N}$, and $\{\ket{j}\}_{j\in \Sigma_\infty}$ is a computational basis of the local Hilbert space labeled by the symbols of some alphabet $\Sigma_\infty$. 

    \item $\exists L_{BI} \in \mathbb{N}$, called the \emph{block-injectivity length}, such that the set of tensors
    \begin{equation*}
    \left\{
    \begin{tikzpicture}[scale=.45,thick,baseline={([yshift=-0.6ex]current bounding box.center)}]
        \begin{scope}[shift={(0,0)}]
    		\draw (-1-0.3,0) -- (1+0.3,0);
    		\draw (-0.5-0.1,1) -- (-0.5-0.1,0);
            \draw (-0.35-0.1,1) -- (-0.35-0.1,0);
            \draw (-0.2-0.1,1) -- (-0.2-0.1,0);
            \draw[dotted] (-0.2,0.75) -- (0.4,0.75);
            \draw (0.5+0.1,1) -- (0.5+0.1,0);
            \draw (0.35+0.1,1) -- (0.35+0.1,0);
    		\filldraw[fill=purple] (-1/2-0.3,-1/2) -- (-1/2-0.3,1/2) -- (1/2+0.3,1/2) -- (1/2+0.3,-1/2) -- (-1/2-0.3,-1/2);
    		\draw (0,0) node {\scriptsize $B_j$};
            \draw [decorate,decoration = {calligraphic brace}] (-0.7,1.15) --  (0.7,1.15);
            \draw (0,1.75) node {\scriptsize $\ell$ sites};
    	\end{scope}
    \end{tikzpicture} \right\}_{j \in \Sigma_\infty} , \quad
    {}
    \text{where } \ 
    \begin{tikzpicture}[scale=.45,thick,baseline={([yshift=-1.3ex]current bounding box.center)}] 
        \begin{scope}[shift={(0,0)}]
    		\draw (-1-0.3,0) -- (1+0.3,0);
    		\draw (-0.5-0.1,1) -- (-0.5-0.1,0);
            \draw (-0.35-0.1,1) -- (-0.35-0.1,0);
            \draw (-0.2-0.1,1) -- (-0.2-0.1,0);
            \draw[dotted] (-0.2,0.75) -- (0.4,0.75);
            \draw (0.5+0.1,1) -- (0.5+0.1,0);
            \draw (0.35+0.1,1) -- (0.35+0.1,0);
    		\filldraw[fill=purple] (-1/2-0.3,-1/2) -- (-1/2-0.3,1/2) -- (1/2+0.3,1/2) -- (1/2+0.3,-1/2) -- (-1/2-0.3,-1/2);
    		\draw (0,0) node {\scriptsize $B_j$};
    	\end{scope}
    \end{tikzpicture} \ := \ 
    \begin{tikzpicture}[scale=.45,thick,baseline={([yshift=0.3ex]current bounding box.center)}] 
        \begin{scope}[shift={(0,0)}]
            \draw[dotted] (1.5,0) -- (4.5,0);
            \MPSTensor{0,0}{$A$}{purple}
            \draw (0,-0.5) -- (0,-1);
            \MPSTensor{1.5,0}{$A$}{purple}
            \draw (1.5,-0.5) -- (1.5,-1);
            \MPSTensor{4.5,0}{$A$}{purple}
            \draw (4.5,-0.5) -- (4.5,-1);
            \draw[fill=amaranth] (0,-1.4) circle (0.4);
            \draw (0,-1.4) node {\scriptsize $j$};
            \draw[fill=amaranth] (1.5,-1.4) circle (0.4);
            \draw (1.5,-1.4) node {\scriptsize $j$};
            \draw[fill=amaranth] (4.5,-1.4) circle (0.4);
            \draw (4.5,-1.4) node {\scriptsize $j$};
            \draw (3,0.5) node {\tiny $\ell$ times};
        \end{scope}
    \end{tikzpicture},
    \end{equation*}
    of bond dimension $D_j \times D_j$, is \emph{block-injective} for all $\ell \geq L_{BI}$, i.e. there exists a family of inverse tensors $\{B_k^{-1}\}_{k\in\Sigma_\infty}$ such that, for every $j \in \Sigma_\infty$ and every choice of indices $\alpha, \alpha', \beta, \beta'$,
    \begin{equation} \label{eq:def_block_injectivity}
        \begin{tikzpicture}[scale=.45,thick,baseline={([yshift=-0.5ex]current bounding box.center)}]
            \begin{scope}[shift={(0,0)}]
        		\draw (-1-0.3,0) -- (1+0.3,0);
        		\draw (-0.5-0.1,1) -- (-0.5-0.1,0);
                \draw (-0.35-0.1,1) -- (-0.35-0.1,0);
                \draw (-0.2-0.1,1) -- (-0.2-0.1,0);
                \draw[dotted] (-0.2,0.75) -- (0.4,0.75);
                \draw (0.5+0.1,1) -- (0.5+0.1,0);
                \draw (0.35+0.1,1) -- (0.35+0.1,0);
        		\filldraw[fill=purple] (-1/2-0.3,-1/2) -- (-1/2-0.3,1/2) -- (1/2+0.3,1/2) -- (1/2+0.3,-1/2) -- (-1/2-0.3,-1/2);
        		\draw (0,0) node {\scriptsize $B_j$};
                \draw[fill=amaranth] (-1.75,0) circle (0.45);
                \draw (-1.7,0) node {\scriptsize $\alpha$};
                \draw[fill=amaranth] (1.75,0) circle (0.45);
                \draw (1.7,0) node {\scriptsize $\beta$};
        	\end{scope}
            \begin{scope}[shift={(0,1.5)}]
        		\draw (-1-0.3,0) -- (1+0.3,0);
        		\filldraw[fill=purple] (-1/2-0.3,-1/2) -- (-1/2-0.3,1/2) -- (1/2+0.3,1/2) -- (1/2+0.3,-1/2) -- (-1/2-0.3,-1/2);
        		\draw (0,0) node {\scriptsize $B^{-1}_k$};
                \draw[fill=amaranth] (-1.75,0) circle (0.45);
                \draw (-1.7,0) node {\scriptsize $\alpha'$};
                \draw[fill=amaranth] (1.75,0) circle (0.45);
                \draw (1.7,0) node {\scriptsize $\beta'$};
        	\end{scope}
        \end{tikzpicture} = 
        \delta_{jk} \delta_{\alpha \alpha'} \delta_{\beta \beta'}.
        {}
    \end{equation}
    {}
\end{enumerate}
\end{definition}
The following two key results make this pbcCF particularly useful. Recall that $D$ is the bond dimension, and $p$ is the minimal blocking length that removes all periodic subspaces.
\begin{theorem}[\cite{Perez-Garcia2007, cirac_2017_mpdo}] \label{prop:generality_pbcCF}
    Every uniform MPS can be written in pbcCF after blocking every $p$ sites, with the block-injectivity length upper bounded by $L_{BI} \leq 3D^5$. 
\end{theorem}

\begin{theorem}[Fundamental theorem of uniform PBC MPS \cite{Perez-Garcia2007, cirac_2017_mpdo}] \label{prop:freedom_pbcCF}
    Given two uniform MPS in pbcCF denoted by 
    \begin{equation*}
        \left\{ 
        \begin{tikzpicture}[scale=.45,thick,baseline={([yshift=-1.15ex]current bounding box.center)}]
            \begin{scope}[shift={(0,0)}]
                \MPSTensor{0,0}{$A$}{powderblue!50}
                \draw (0,-0.5) -- (0,-1);
            \end{scope}
        \end{tikzpicture}, \ 
        \{ | L_N^A \rangle \}_N
        \right\}
        \quad \text{and} \quad
        \left\{
        \begin{tikzpicture}[scale=.45,thick,baseline={([yshift=-1.15ex]current bounding box.center)}]
            \begin{scope}[shift={(0,0)}]
                \MPSTensor{0,0}{$B$}{melon!50}
                \draw (0,-0.5) -- (0,-1);
            \end{scope}
        \end{tikzpicture}, \ 
        \{ | L_N^B \rangle \}_N
        \right\} \ , 
    \end{equation*}
    respectively, they generate the same family of states if and only if, upon some relabelling of the basis elements for $| L_N^B \rangle$, there exist invertible matrices $Z_j$ such that
    \begin{equation} \label{eq:freedom_pbc}
        \begin{tikzpicture}[scale=.45,thick,baseline={([yshift=-1.15ex]current bounding box.center)}]
            \begin{scope}[shift={(0,0)}]
                \MPSTensor{0,0}{$B$}{melon!50}
                \draw (0,-0.5) -- (0,-1);
            \end{scope}
        \end{tikzpicture}
        =
        \begin{tikzpicture}[scale=.45,thick,baseline={([yshift=-1.15ex]current bounding box.center)}]
            \MPSTensor{0,0}{$A$}{powderblue!50}
            \draw (0,-0.5) -- (0,-1.1);
            \draw (0,-0.5) -- (0,-1);
            \begin{scope}[shift={(1.5,0)}]
                \draw (-1,0) -- (1,0);
                \filldraw[fill=gray!10] (0.5,-0.5) -- (-0.5,-0.5) -- (-0.5,0.5) -- (0.5, 0.5) -- (0.5,-0.5);
                \draw (0,0) node {\scriptsize $Z$};
            \end{scope}
            \begin{scope}[shift={(-2,0)}]
                \draw (-1,0) -- (1.5,0);
                \filldraw[fill=gray!10] (1,-0.5) -- (-0.5,-0.5) -- (-0.5,0.5) -- (1, 0.5) -- (1,-0.5);
                \draw (0.3,0) node {\scriptsize $Z^{-1}$};
            \end{scope}
            \draw (-2,-0.5) -- (-2,-0.75) -- (1.5,-0.75) -- (1.5,-0.5);
            \fill (0,-0.75) circle (0.1);
        \end{tikzpicture}
        \quad \text{and} \quad
        | L_N^A \rangle = | L_N^B \rangle .
    \end{equation}
\end{theorem}
Note that the lines joined by a dot at the right-hand side of Eq.~\eqref{eq:freedom_pbc} represent a $\delta$ function:
\begin{equation*} 
    \begin{tikzpicture}[scale=.45,thick,baseline={([yshift=-1.15ex]current bounding box.center)}]
        \begin{scope}[shift={(0,0)}]
            \draw (0,-0.5) -- (0,0.5);
            \draw (-1,0) -- (1,0);
            \fill (0,0) circle (0.115);
            \draw[fill=amaranth] (0,-0.9) circle (0.4);
            \draw (0,-0.9) node {\scriptsize $l$};
            \draw[fill=amaranth] (-1.4,0) circle (0.4);
            \draw (-1.4,0) node {\scriptsize $i$};
            \draw[fill=amaranth] (0,0.9) circle (0.4);
            \draw (0,0.9) node {\scriptsize $j$};
            \draw[fill=amaranth] (1.4,0) circle (0.4);
            \draw (1.4,0) node {\scriptsize $k$};
        \end{scope}
    \end{tikzpicture}
    = \delta_{ijkl} \ .
\end{equation*}

{

The family of states $\{\ket{L_N}\}$ remains well-behaved in the thermodynamic limit, when the coefficients in Eq. \eqref{eq:PBC_lower_states} do not depend on $N$ (i.e. $\mu_{j,k} = 1$ for all $j,k$). In this case, it suffices to consider
\begin{equation} \label{eq:stable_PBC-MPS}
    \ket{L_N} := \sum_{j=1}^g c_j \ket{j}^{\otimes N}, \quad c_j \in \mathbb{N} \, .
\end{equation}
Any such family is \textit{invariant} under blocking: if we group every $m$ physical sites into a single block and define a new local basis via $\ket{\tilde{j}} := \ket{j}^{\otimes m}$ for each $j$, then the blocked family $\{\ket{L_{mk}}\}_{k \in \mathbb{N}}$, expressed in the $\{ \ket{\tilde{j}} \}$-basis, retains exactly the same form as the original.

\section{Uniform MPS with a boundary}
\label{sec:mps-x}

As mentioned in the introduction, many physically relevant states naturally admit an MPS representation of constant bond dimension if one allows for a boundary matrix, but not if one enforces periodic boundary conditions, even if the original state is translationally invariant. This motivates the following definition.

\begin{definition}[Uniform MPS with a boundary (MPS-X)]
    Given an MPS tensor $A$ and a boundary matrix $X$, with $A^i, X \in \mathcal{M}_{D\times D}(\mathbb{C})$, they define the \emph{MPS-X} family of states $\{\ket{\psi_N(X, A)}\}_N$ via
    \begin{equation*}
        \ket{\psi_N(X,a)} := \begin{tikzpicture}[scale=.45,thick,baseline={([yshift=-1.15ex]current bounding box.center)}]
            \FullMPSX{0,0}{$A$}{$X$}{purple}{yellow}
        \end{tikzpicture} \ .
    \end{equation*}
\end{definition}
Note that the PBC case is recovered choosing $X = \mathds{1}$. As an example, the W-state family, $\ket{W_N} := \ket{10 \dots 0} + \ket{01 \dots 0} + \dots + \ket{00 \dots 1}$, for which the bond dimension of any uniform PBC MPS representation must scale as $\Omega(N^{1/(3+\delta)})$ for any $\delta > 0$ \cite{Perez-Garcia2007, Michalek2018, klimov_2023_TI-MPS}, admits the following bond dimension 2 MPS-X representation:
\begin{equation*}
    X = {\scriptsize \begin{pmatrix}
        0 & 0 \\ 1 & 0
    \end{pmatrix}}, \quad A^0 = {\scriptsize \begin{pmatrix}
        1 & 0 \\ 0 & 1
    \end{pmatrix}}, \quad A^1 = {\scriptsize \begin{pmatrix}
        0 & 1 \\ 0 & 0
    \end{pmatrix}}.
\end{equation*}

The standard toolbox for uniform PBC MPS no longer applies in the MPS-X setting: $X = \mathds{1}$ allows to restrict the attention to block-diagonal MPS, but for arbitrary $X$, off-diagonal blocks introduce notable complications, rendering Theorems \ref{prop:generality_pbcCF} and \ref{prop:freedom_pbcCF} invalid. 


\subsection{Stable MPS-X} \label{sec:stableMPS-X}

Any TI MPS-X can be decomposed in an analogous way to the pbcCF, but with a more general class of structured backbone states $\{\ket{L_N}\}$. In what follows, we restrict to the class of \emph{stable MPS-X}. Their definition relies on two notions associated to the given MPS matrices $\{A^i\}$:
\begin{itemize}
    \item The \textit{length-$\ell$ span} $\mathcal{A}^{(\ell)} := \text{span}\{A^{i_1} A^{i_2} \dots A^{i_\ell} \mid i_1, \dots, i_\ell \in \{1, \dots, d\}\}$, containing all linear combinations of matrix products of fixed length $\ell$.  
    \item The \textit{algebra} $\mathcal{A} := \text{Alg}(\{A^i\}) = \cup_{\ell \geq 1} \mathcal{A}^{(\ell)}$, containing products of any arbitrary length.
\end{itemize}
\begin{definition}[Stable MPS-X] \label{def:stable_mps-x}
    An MPS-X with matrices $\{A^i\}$ is \emph{stable} if the span eventually stabilizes to an algebra upon blocking, i.e. if $\exists L_{\mathrm{stab}}$ such that $\mathcal{A}^{(L_{\mathrm{stab}})} = \mathrm{Alg}(\mathcal{A}^{(1)})$.
\end{definition}
This class of stable MPS-X has a direct physical motivation: we show that they are exactly those admitting a backbone family of states $\{\ket{L_N}\}$ whose coefficients are independent of the system size and which remain invariant under a self-consistent coarse-graining procedure. In fact, $\{\ket{L_N}\}$ are \emph{algebraic regular language states}, introduced in the next section. This ensures that $\{\ket{L_N}\}$ captures the essential long-range features of the MPS-X independently of the system size, analogous to how fixed points of renormalization-group transformations encode the relevant scale-invariant properties of a system \cite{Verstraete_2004, cirac_2017_mpdo}.

Moreover, stable MPS-X include several well-known families such as GHZ-, W-, Dicke-, domain-wall-states, and generalizations thereof. In particular, normal MPS (i.e. those that are short-range correlated and whose tensor consists of a single irreducible block without periodic subspaces) are always stable. More generally, stability is a standard assumption in the theory of uniform PBC MPS when the constants $\mu_{j,k}$ become simultaneously equal to one after blocking, resulting in the family $\{\ket{L_N}\}$ of Eq. \eqref{eq:stable_PBC-MPS}.

Non-stable MPS-X, on the other hand, do not contain any fundamentally distinct features with respect to stable MPS-X, beyond specific size-dependent polynomial or exponential coefficients described in Appendix \ref{app:spanRLS-spanCF}.

\subsection{Algebraic regular language states} \label{sec:algebraicRLS}

The class of algebraic regular language states (algebraic RLS) builds upon a subclass of the regular language states (RLS) introduced in \cite{florido-llinas_2024_RLS}, allowing for complex weights. 

Given an alphabet $\Sigma$, a \textit{word} is a finite string over $\Sigma$, and a \textit{language} is a set of such words. \textit{Regular languages} are those describable by \textit{regular expressions} built from concatenation, union, and Kleene star ($R^* := \varepsilon \cup R \cup RR \cup \dots$, where $\varepsilon$ denotes the empty word). For a regular language $L \subseteq \Sigma^*$, the corresponding RLS family is given by $\ket{L_N} := \sum_{w \in L \cap \Sigma^N} \ket{w}$. Some examples include: 
\begin{itemize}
    \item GHZ states, $\{\ket{0}^{\otimes N} + \ket{1}^{\otimes N}\}_N \equiv \ket{0^*} + \ket{1^*}$,
    \item W states, $\{\sum_{n=0}^N \ket{0^n 1 0^{N-n}}\}_N \equiv \ket{0^* 1 0^*}$, or
    \item 2-excitation Dicke states,  $\{\sum_{n_1+n_2=0}^{N-2} \ket{0^{n_1} 1 0^{n_2} 1 0^{N-n_1-n_2}}\}_N \equiv \ket{0^* 1 0^* 1 0^*}$.
\end{itemize}
We generalize them further by allowing substitutions with complex weights via an operator $\hat{S}^{(m)}$ analogous to the one used in \cite{florido-llinas_2024_RLS}. For instance, 
\begin{equation*}
    \hat{S}^{(2)} \ket{0^* f 1^* f 0^*} (\alpha \ket{23} + \beta \ket{45}) = \alpha \ket{0^* 2 1^* 3 0^*} + \beta \ket{0^* 4 1^* 5 0^*}.
\end{equation*}
The alphabet $\Sigma$ will be further partitioned as $\Sigma = \Sigma_\infty \cup \Sigma_f$, where $\Sigma_f$ contains symbols whose number of appearances in any ket is upper bounded by a constant, and $\Sigma_\infty$ contains the rest. Then, $\Sigma_f = \cup_{i,j \in \tilde{\Sigma}_\infty} \Sigma_f^{ij}$, where each of the disjoint subsets $\Sigma_f^{ij}$ has the $\Sigma_f$ symbols that can appear only between strings $i^*$ and $j^*$ of $\tilde{\Sigma}_\infty^*$. In the example above, $\Sigma_\infty = \{0,1\}$, $\Sigma_f^{01} = \{2,4\}$ and $\Sigma_f^{10} = \{3,5\}$. 

\begin{definition}[Algebraic RLS] \label{def:algebraicRLS}
    A family $\{\ket{L_N}\}_N$ is an \emph{algebraic RLS} on an alphabet $\Sigma$ partitioned as $\Sigma_\infty \cup (\cup_{i,j \in \tilde{\Sigma}_\infty} \Sigma_f^{ij})$ if 
    \begin{equation} \label{eq:def_algebraic_RLS_equation}
        \{\ket{L_N}\}_N = \sum_{m\leq M} \sum_{O \in \tilde{\Sigma}_\infty^{m+1}} \hat{S}^{(m)} \ket{O_0^\ast f O_1^\ast f O_2^\ast \dots O_{m-1}^\ast f O_m^\ast} | X_O \rangle,
    \end{equation}
    for some constant $M$, where $\tilde{\Sigma}_\infty := \Sigma_\infty \cup \{\varepsilon\}$ and $\ket{X_O}$ is a weighted superposition (with $N$-independent weights) of strings in $\Sigma_f^{O_0 O_1} \cdot \Sigma_f^{O_1 O_2} \cdot \ldots \cdot \Sigma_f^{O_{m-1}O_m}$. We refer to the set $\{\ket{X_O}\}_O$ as the \emph{defining states} of the algebraic RLS.
\end{definition} 
Note that any algebraic RLS admits an MPS-X representation (see Lemma \ref{lemma:bond_dimension_algebraic-RLS} in Appendix \ref{app:spanRLS-spanCF} for the explicit construction and bond dimension). The backbone states $\{\ket{L_N}\}$ for stable pbcCF in Eq. \eqref{eq:stable_PBC-MPS} are algebraic RLS with $\Sigma = \Sigma_\infty$, $\ket{X_O} = c_j \in \mathbb{N}$ for $O = j$, and $\ket{X_O} = 0$ for any other $O$.

\subsection{Blocking and $\Gamma$-invariance}

Algebraic RLS might only remain invariant under blocking upon the right choice of coarse-grained basis states. For example:
\begin{itemize}
    \item $\ket{0^* 1 0^*}$ is invariant with $\ket{\tilde{0}} := \ket{00}$, $\ket{\tilde{1}} := \ket{01} + \ket{10}$.
    \item $\ket{0^* 1 0^* 1 0^*}$ is not invariant under any blocking.
    \item $\ket{0^* 1 0^* 1 0^*} + \ket{0^* 2 0^*}$ is invariant with $\ket{\tilde{0}} := \ket{00}$, $\ket{\tilde{1}} := \ket{01} + \ket{10}$, $\ket{\tilde{2}} := \ket{11} + \ket{02} + \ket{20}$.
\end{itemize}
Therefore, ensuring the existence of a self-consistent coarse-graining procedure first requires a way to systematically specify how the blocking is performed. For this purpose, we introduce the $\Gamma$ and $\Gamma_\ell$ tensors, which prescribe how to block every two and every $\ell$ sites, respectively. $\Gamma_\ell$ is recursively obtained from $\Gamma$ as
\begin{equation*}
    \begin{tikzpicture}[scale=.45,thick,baseline={([yshift=-1ex]current bounding box.center)}]
        \begin{scope}[shift={(0,0)}]
            \draw (-0.35-0.1,1) -- (-0.35-0.1,0);
            \draw (-0.2-0.1,1) -- (-0.2-0.1,0);
            \draw[dotted] (-0.18,0.75) -- (0.4,0.75);
            \draw (0.35+0.1,1) -- (0.35+0.1,0);
            \filldraw[fill=gray!10] (-1/2-0.2,-1/2) -- (-1/2-0.2,1/2) -- (1/2+0.2,1/2) -- (1/2+0.2,-1/2) -- (-1/2-0.2,-1/2);
            \draw (0,0) node {\scriptsize $\Gamma_\ell$};
            \draw (0,-0.5) -- (0,-1);
        \end{scope}
    \end{tikzpicture} :=
    \begin{tikzpicture}[scale=.45,thick,baseline={([yshift=-1ex]current bounding box.center)}]
        \begin{scope}[shift={(0,0)}]
            \draw (-0.8,2.5) -- (-0.8,0);
            \draw (-0.6,2.5) -- (-0.6,0);
            \draw[dotted] (-0.4,1.5) -- (0.2,1.5);
            \draw (0.4,2.5) -- (0.4,0);
            \draw (0.8,1) -- (0.8,0);
            \filldraw[fill=gray!10] (-1/2-0.5,-1/2) -- (-1/2-0.5,1/2) -- (1/2+0.5,1/2) -- (1/2+0.5,-1/2) -- (-1/2-0.5,-1/2);
            \draw (0,0) node {\scriptsize $\Gamma_{\ell-1}$};
            \draw (0,-0.5) -- (0,-1);
        \end{scope}
        \draw [decorate,decoration = {calligraphic brace}] (-0.9,2.6) --  (0.5,2.6);
        \draw(-0.2,3.45) node {\tiny $\ell-2$};
        \draw(-0.2,3.1) node {\tiny sites};
        \begin{scope}[shift={(1.1,1.5)}]
            \filldraw[fill=gray!10] (-0.5,-0.5) -- (-0.5,0.5) -- (0.5,0.5) -- (0.5,-0.5) -- (-0.5,-0.5);
            \draw (0,0) node {\scriptsize $\Gamma$};
            \draw (-0.3,0.5) -- (-0.3,1);
            \draw (0.3,0.5) -- (0.3,1);
        \end{scope}
    \end{tikzpicture}
    \ , \quad 
    \begin{tikzpicture}[scale=.45, baseline={([yshift=-1ex]current bounding box.center)}, thick]
        \begin{scope}[shift={(0,0)}]
            \draw (0.3,0.5) -- (0.3,1);
            \draw (-0.3,0.5) -- (-0.3,1);
            \filldraw[fill=gray!10] (-0.5,-0.5) -- (-0.5,0.5) -- (0.5,0.5) -- (0.5,-0.5) -- (-0.5,-0.5);
            \draw (0,0) node {\scriptsize $\Gamma_2$};
            \draw (0,-0.5) -- (0,-1);
        \end{scope}
    \end{tikzpicture} 
    := 
    \begin{tikzpicture}[scale=.45, baseline={([yshift=-1ex]current bounding box.center)}, thick]
        \begin{scope}[shift={(0,0)}]
            \draw (0.3,0.5) -- (0.3,1);
            \draw (-0.3,0.5) -- (-0.3,1);
            \filldraw[fill=gray!10] (-0.5,-0.5) -- (-0.5,0.5) -- (0.5,0.5) -- (0.5,-0.5) -- (-0.5,-0.5);
            \draw (0,0) node {\scriptsize $\Gamma$};
            \draw (0,-0.5) -- (0,-1);
        \end{scope}
    \end{tikzpicture} \ .
\end{equation*}
Moreover, $\Gamma$ must be an associative tensor, since the blocking procedure should be independent of the order in which sites are grouped together, i.e.
\begin{equation*}
    \begin{tikzpicture}[scale=.45, baseline={([yshift=-1ex]current bounding box.center)}, thick]
        \begin{scope}[shift={(0,1.5)}]
            \filldraw[fill=gray!10] (-0.5,-0.5) -- (-0.5,0.5) -- (0.5,0.5) -- (0.5,-0.5) -- (-0.5,-0.5);
            \draw (0,0) node {\scriptsize $\Gamma$};
            \draw (-0.3,0.5) -- (-0.3,1);
            \draw (0.3,0.5) -- (0.3,1);
        \end{scope}
        \draw (0.3,0.5) -- (0.3,1);
        \draw (0.9,0.5) -- (0.9,2.5);
        \begin{scope}[shift={(0.6,0)}]
            \filldraw[fill=gray!10] (-0.5,-0.5) -- (-0.5,0.5) -- (0.5,0.5) -- (0.5,-0.5) -- (-0.5,-0.5);
            \draw (0,0) node {\scriptsize $\Gamma$};
            \draw (0,-0.5) -- (0,-1);
        \end{scope}
    \end{tikzpicture}
    \ = \ 
    \begin{tikzpicture}[scale=.45, baseline={([yshift=-1ex]current bounding box.center)}, thick]
        \begin{scope}[shift={(0.65,1.5)}]
            \filldraw[fill=gray!10] (-0.5,-0.5) -- (-0.5,0.5) -- (0.5,0.5) -- (0.5,-0.5) -- (-0.5,-0.5);
            \draw (0,0) node {\scriptsize $\Gamma$};
            \draw (-0.3,0.5) -- (-0.3,1);
            \draw (0.3,0.5) -- (0.3,1);
        \end{scope}
        \draw (0.3,0.5) -- (0.3,1);
        \draw (-0.3,0.5) -- (-0.3,2.5);
        \begin{scope}[shift={(0,0)}]
            \filldraw[fill=gray!10] (-0.5,-0.5) -- (-0.5,0.5) -- (0.5,0.5) -- (0.5,-0.5) -- (-0.5,-0.5);
            \draw (0,0) node {\scriptsize $\Gamma$};
            \draw (0,-0.5) -- (0,-1);
        \end{scope}
    \end{tikzpicture}.
\end{equation*}
\begin{definition}[$\Gamma$-blocking]
    Given an associative tensor $\Gamma$, state $\ket{\psi_{k}} \in \mathbb{C}^{\otimes k}$ is the result of \textit{$\Gamma$-blocking} every $\ell$ sites of another state $\ket{\psi_{\ell k}} \in \mathbb{C}^{\otimes \ell k}$ if
    \begin{equation*}
        \begin{tikzpicture}[scale=.45,thick,baseline={([yshift=-2.8ex]current bounding box.center)}]
            \begin{scope}[shift={(0,0)}]
                \draw (-0.35-0.1,1) -- (-0.35-0.1,0);
                \draw (-0.2-0.1,1) -- (-0.2-0.1,0);
                \draw[dotted] (-0.18,0.75) -- (0.4,0.75);
                \draw (0.35+0.1,1) -- (0.35+0.1,0);
        		\filldraw[fill=gray!10] (-1/2-0.2,-1/2) -- (-1/2-0.2,1/2) -- (1/2+0.2,1/2) -- (1/2+0.2,-1/2) -- (-1/2-0.2,-1/2);
        		\draw (0,0) node {\scriptsize $\Gamma_\ell$};
                \draw (0,-0.5) -- (0,-1);
        	\end{scope}
            \draw [decorate,decoration = {calligraphic brace}] (-0.5,1.1) --  (0.5,1.1);
            \draw (0,1.7) node {\tiny $\ell$ sites};
            \begin{scope}[shift={(1.6,0)}]
                \draw (-0.35-0.1,1) -- (-0.35-0.1,0);
                \draw (-0.2-0.1,1) -- (-0.2-0.1,0);
                \draw[dotted] (-0.18,0.75) -- (0.4,0.75);
                \draw (0.35+0.1,1) -- (0.35+0.1,0);
        		\filldraw[fill=gray!10] (-1/2-0.2,-1/2) -- (-1/2-0.2,1/2) -- (1/2+0.2,1/2) -- (1/2+0.2,-1/2) -- (-1/2-0.2,-1/2);
        		\draw (0,0) node {\scriptsize $\Gamma_\ell$};
                \draw (0,-0.5) -- (0,-1);
        	\end{scope}
            \begin{scope}[shift={(4.2,0)}]
                \draw (-0.35-0.1,1) -- (-0.35-0.1,0);
                \draw (-0.2-0.1,1) -- (-0.2-0.1,0);
                \draw[dotted] (-0.18,0.75) -- (0.4,0.75);
                \draw (0.35+0.1,1) -- (0.35+0.1,0);
        		\filldraw[fill=gray!10] (-1/2-0.2,-1/2) -- (-1/2-0.2,1/2) -- (1/2+0.2,1/2) -- (1/2+0.2,-1/2) -- (-1/2-0.2,-1/2);
        		\draw (0,0) node {\scriptsize $\Gamma_\ell$};
                \draw (0,-0.5) -- (0,-1);
        	\end{scope}
            \draw[dotted] (2.5,0) -- (3.3,0);
            \begin{scope}[shift={(0,-1.5)}]
                \filldraw[fill=amaranth] (-0.7,0.5) -- (-0.7,-0.5) -- (4.9,-0.5) -- (4.9,0.5) -- (-0.7,0.5);
        		\draw (2.3,0) node {\scriptsize $\ket{\psi_k}$};
        	\end{scope}
            \end{tikzpicture}
            \ = \ 
            \begin{tikzpicture}[scale=.45,thick,baseline={([yshift=-1ex]current bounding box.center)}]
                \filldraw[fill=amaranth] (0,0) -- (3.4,0) -- (3.4,1) -- (0,1) -- (0,0);
                \draw (1.7,0.5) node {\scriptsize $\ket{\psi_{\ell k}}$};
                \draw (0.2,1) -- (0.2,1.6);
                \draw (0.9,1) -- (0.9,1.6);
                \draw (3.2,1) -- (3.2,1.6);
                \draw[dotted] (1.2,1.3) -- (2.9,1.3);
            \end{tikzpicture} \ .
        \end{equation*}
\end{definition}
\begin{definition}[Invariance under $\Gamma$-blocking] 
    A family of states $\{\ket{\psi_N}\}_N$ is \textit{invariant under $\Gamma$-blocking} if, for every choice of $\alpha, \beta \in \mathbb{N}$, we have that $\ket{\psi_{\beta}}$ is the result of $\Gamma$-blocking every $\alpha$ sites of $\ket{\psi_{\alpha\beta}}$.
\end{definition}
For instance, stable PBC MPS are $\Gamma$-invariant for the blocking tensor $\Gamma_\ell$ such that $(\Gamma_\ell)_i^{ii\dots i} = 1$ for each $i \in \Sigma_\infty$ and 0 otherwise, for all $\ell$; while for the family $\ket{0^* 1 0^* 1 0^*} + \ket{0^* 2 0^*}$, the blocking specified above that leaves it invariant corresponds to the following associative $\Gamma$-tensor:
\begin{equation*}
    \begin{tikzpicture}[scale=.45, baseline={([yshift=-1ex]current bounding box.center)}, thick]
        \begin{scope}[shift={(0,0)}]
            \draw (0.3,0.5) -- (0.3,1);
            \draw (-0.3,0.5) -- (-0.3,1);
            \filldraw[fill=gray!10] (-0.5,-0.5) -- (-0.5,0.5) -- (0.5,0.5) -- (0.5,-0.5) -- (-0.5,-0.5);
            \draw (0,0) node {\scriptsize $\Gamma$};
            \draw (0,-0.5) -- (0,-1);
        \end{scope}
        \draw (0,-1.3) node {\scriptsize $0$};
    \end{tikzpicture}
    = \ket{00}, \quad 
    \begin{tikzpicture}[scale=.45, baseline={([yshift=-1ex]current bounding box.center)}, thick]
        \begin{scope}[shift={(0,0)}]
            \draw (0.3,0.5) -- (0.3,1);
            \draw (-0.3,0.5) -- (-0.3,1);
            \filldraw[fill=gray!10] (-0.5,-0.5) -- (-0.5,0.5) -- (0.5,0.5) -- (0.5,-0.5) -- (-0.5,-0.5);
            \draw (0,0) node {\scriptsize $\Gamma$};
            \draw (0,-0.5) -- (0,-1);
        \end{scope}
        \draw (0,-1.3) node {\scriptsize $1$};
    \end{tikzpicture}
    = \ket{01} + \ket{10}, \quad
    \begin{tikzpicture}[scale=.45, baseline={([yshift=-1ex]current bounding box.center)}, thick]
        \begin{scope}[shift={(0,0)}]
            \draw (0.3,0.5) -- (0.3,1);
            \draw (-0.3,0.5) -- (-0.3,1);
            \filldraw[fill=gray!10] (-0.5,-0.5) -- (-0.5,0.5) -- (0.5,0.5) -- (0.5,-0.5) -- (-0.5,-0.5);
            \draw (0,0) node {\scriptsize $\Gamma$};
            \draw (0,-0.5) -- (0,-1);
        \end{scope}
        \draw (0,-1.3) node {\scriptsize $2$};
    \end{tikzpicture}
    = \ket{11} + \ket{02} + \ket{20} .
\end{equation*}

Stable MPS-X always admit a $\Gamma$-tensor under which their backbone family of algebraic RLS remains invariant, determined by the algebra of the MPS matrices, as we will see in Section \ref{sec:how-to}. 

{}

\subsection{The generalized canonical form for stable MPS-X}

We are now ready to introduce the gCF for stable MPS-X, effectively generalizing the pbcCF of Definition \ref{def:pbcCF}.

\begin{definition}[Generalized canonical form for MPS-X (gCF)]  \label{def:gCF}
    A family of quantum states $\{\ket{\psi_N}\}$ is in gCF if there exists an associative tensor $\Gamma$ such that, for each $N$,
    \begin{equation*}
        \ket{\psi_N} = 
        \begin{tikzpicture}[scale=.45,thick,baseline={([yshift=-1.15ex]current bounding box.center)}]
            \begin{scope}[shift={(0,1)}]
                \draw[dotted] (1.5,0) -- (4.5,0);
                \MPSTensor{0,0}{$A$}{purple}
                \draw (0,-0.5) -- (0,-1);
                \MPSTensor{1.5,0}{$A$}{purple}
                \draw (1.5,-0.5) -- (1.5,-1);
                \MPSTensor{4.5,0}{$A$}{purple}
                \draw (4.5,-0.5) -- (4.5,-1);
            \end{scope}
            \filldraw[fill=amaranth] (-0.5,-1) -- (-0.5,0) -- (5,0) -- (5,-1) -- (-0.5,-1);
            \draw (2.25,-0.5) node {\scriptsize $\ket{L_N}$};
            \draw (-1,1) -- (-1,0.25) -- (-0.2,0.25);
            \draw (0.2,0.25) -- (1.3,0.25);
            \draw (1.7,0.25) -- (2.4,0.25);
            \draw[dotted] (2.5,0.25) -- (3.5,0.25);
            \draw (3.6,0.25) -- (4.3,0.25);
            \draw (4.7,0.25) -- (5.5,0.25) -- (5.5,1);
        \end{tikzpicture} \ ,
    \end{equation*}
    where 
    \begin{enumerate}[label=(\roman*)]
        \item $\{\ket{L_N}\}$ is a $\Gamma$-invariant family of algebraic RLS.
        \item $\exists L_{BI} \in \mathbb{N}$ such that the set of tensors 
            \begin{equation*}
    \left\{
    \begin{tikzpicture}[scale=.45,thick,baseline={([yshift=-0.6ex]current bounding box.center)}]
        \begin{scope}[shift={(0,0)}]
    		\draw (-1-0.3,0) -- (1+0.3,0);
    		\draw (-0.5-0.1,1) -- (-0.5-0.1,0);
            \draw (-0.35-0.1,1) -- (-0.35-0.1,0);
            \draw (-0.2-0.1,1) -- (-0.2-0.1,0);
            \draw[dotted] (-0.2,0.75) -- (0.4,0.75);
            \draw (0.5+0.1,1) -- (0.5+0.1,0);
            \draw (0.35+0.1,1) -- (0.35+0.1,0);
    		\filldraw[fill=purple] (-1/2-0.3,-1/2) -- (-1/2-0.3,1/2) -- (1/2+0.3,1/2) -- (1/2+0.3,-1/2) -- (-1/2-0.3,-1/2);
    		\draw (0,0) node {\scriptsize $B_j$};
            \draw [decorate,decoration = {calligraphic brace}] (-0.7,1.15) --  (0.7,1.15);
            \draw (0,1.75) node {\scriptsize $\ell$ sites};
    	\end{scope}
    \end{tikzpicture} \right\}_{j \in \Sigma} , \quad
    \text{where } \ 
    \begin{tikzpicture}[scale=.45,thick,baseline={([yshift=-1ex]current bounding box.center)}]
        \begin{scope}[shift={(0,0)}]
    		\draw (-1-0.3,0) -- (1+0.3,0);
    		\draw (-0.5-0.1,1) -- (-0.5-0.1,0);
            \draw (-0.35-0.1,1) -- (-0.35-0.1,0);
            \draw (-0.2-0.1,1) -- (-0.2-0.1,0);
            \draw[dotted] (-0.2,0.75) -- (0.4,0.75);
            \draw (0.5+0.1,1) -- (0.5+0.1,0);
            \draw (0.35+0.1,1) -- (0.35+0.1,0);
    		\filldraw[fill=purple] (-1/2-0.3,-1/2) -- (-1/2-0.3,1/2) -- (1/2+0.3,1/2) -- (1/2+0.3,-1/2) -- (-1/2-0.3,-1/2);
    		\draw (0,0) node {\scriptsize $B_j$};
    	\end{scope}
    \end{tikzpicture} \ :=
    \begin{tikzpicture}[scale=.45,thick,baseline={([yshift=-0ex]current bounding box.center)}]
        \begin{scope}[shift={(0,1)}]
            \draw[dotted] (1.5,0) -- (4.5,0);
            \MPSTensor{0,0}{$A$}{purple}
            \draw (0,-0.5) -- (0,-1);
            \MPSTensor{1.5,0}{$A$}{purple}
            \draw (1.5,-0.5) -- (1.5,-1);
            \MPSTensor{4.5,0}{$A$}{purple}
            \draw (4.5,-0.5) -- (4.5,-1);
        \end{scope}
        \filldraw[fill=gray!10] (-0.5,-1) -- (-0.5,0) -- (5,0) -- (5,-1) -- (-0.5,-1);
        \draw (2.25,-0.5) node {\scriptsize $\Gamma_\ell$};
        \draw (2.25,-1) -- (2.25,-1.5);
        \draw[fill=amaranth] (2.25,-1.9) circle (0.4);
        \draw (2.25,-1.9) node {\scriptsize $j$};
    \end{tikzpicture}
    \end{equation*}
        is block-injective for all $\ell \geq L_{BI}$.
    \end{enumerate}
\end{definition}
In analogy with Theorem \ref{prop:generality_pbcCF} for uniform PBC MPS, we show that any TI stable MPS-X can be brought into generalized canonical form after blocking. Recall from Eq.~\eqref{eq:block-upper-triangular-form} that the MPS matrices $\{A^i\}$ admit a block-upper-triangular form of $b$ blocks by $b$ blocks; $p$ denotes the minimal blocking length that removes all periodic subspaces; and $q$ is the additional blocking needed so that the constants $\mu_{j,k}$ relating equivalent diagonal blocks satisfy $\mu_{j,k}^{pq} = 1$ for all $j,k$, which necessarily exists for stable MPS-X. 
\begin{restatable}{theorem}{generalitygCF} \label{thm:generality_gCF}
    Every stable TI MPS-X can be written in gCF, upon blocking every $pqL_{\mathrm{span}}b2^b$ sites, with $L_{\mathrm{span}} \leq 45b^2 D^3 2^{b^2}$. Additionally, the block-injectivity length is upper bounded by $L_{BI} \leq D^2$.
\end{restatable} 
The proof of the theorem is given in Appendix \ref{app:proof_thm_generality-gCF}. A key ingredient is that translational invariance allows to write the boundary matrix $X$ in a simplified form $\tilde{X}$:
\begin{proposition}[Translational invariance, informal]
\label{prop:check_TI_informal}
    The boundary matrix of any TI MPS-X can be replaced by a simpler one whose blocks are each either zero or proportional to the identity.
\end{proposition}
The full formal version of this statement, together with the precise necessary and sufficient conditions for any MPS-X to be translational invariant, are proven in Appendix \ref{app:TI_proofs}. Theorem \ref{thm:generality_gCF} also applies to non-TI stable MPS-X, as long as the boundary matrix satisfies this blockwise constraint.

For completeness, we develop in Appendix \ref{app:spanRLS-spanCF} a more general canonical form, the \emph{spanCF}, which applies to all TI MPS-X, stable or not. This generality, however, comes at a cost: the structure that makes backbone states of stable MPS-X capture their essential, scale-invariant features is no longer present. In particular, non-stable MPS-X \textit{(i)} necessarily have coefficients in their backbone states that depend on the system size, or \textit{(ii)} they are not invariant under any self-consistent coarse-graining procedure.

Finally, in analogy with the fundamental theorem for uniform PBC MPS (Theorem \ref{prop:freedom_pbcCF}), we characterize the freedom of the gCF. This requires the pair of MPS-X tensors to be in \emph{reduced form}, informally meaning that they contain no negligible components and span the same physical subspace. The precise definition is given in Appendix \ref{app:fundthm_proof}. There, we show that any equivalent pair of general MPS-X can always be brought to a reduced pair by an explicit gauge transformation, and that any equivalent pair of uniform PBC MPS is automatically reduced, requiring no additional preprocessing. Let $\Sigma$ be the labels of the basis states for the backbone algebraic RLS, $\tilde{\Sigma}_f^{ij} := \Sigma_f^{ij}$ if $i \neq j$, or $\Sigma_f^{ii} \cup \{i\}$ if $i = j$, and $\langle \Sigma \rangle := \mathrm{span}\{ \ket{x} \mid x \in \Sigma \}$.


\begin{restatable}{theorem}{propfreedomgCF}
    \label{prop:freedom_gCF}
    Given a reduced pair of MPS-X with gCF representations
    \begin{equation*}
        \left\{ 
        \begin{tikzpicture}[scale=.45,thick,baseline={([yshift=-1.15ex]current bounding box.center)}]
            \begin{scope}[shift={(0,0)}]
                \MPSTensor{0,0}{$A$}{powderblue!50}
                \draw (0,-0.5) -- (0,-1);
            \end{scope}
        \end{tikzpicture}, \ 
        \{ \ket{X_O^A} \}
        \right\}
        \quad \text{and} \quad
        \left\{
        \begin{tikzpicture}[scale=.45,thick,baseline={([yshift=-1.15ex]current bounding box.center)}]
            \begin{scope}[shift={(0,0)}]
                \MPSTensor{0,0}{$B$}{melon!50}
                \draw (0,-0.5) -- (0,-1);
            \end{scope}
        \end{tikzpicture}, \ 
        \{ \ket{X_O^B} \}
        \right\} \ ,
    \end{equation*}
    where $\{\ket{X_O^A}\}$ and $\{\ket{X_O^B}\}$ are the defining states of the respective algebraic RLS on alphabets $\Sigma^A = \Sigma_\infty^A \cup (\cup_{i,j \in \tilde{\Sigma}_\infty^A} \Sigma_f^{A,ij})$ and $\Sigma^B = \Sigma_\infty^B \cup (\cup_{i,j \in \tilde{\Sigma}_\infty^B} \Sigma_f^{B,ij})$ (cf. Definition \eqref{def:algebraicRLS}), they generate the same family of MPS-X states if and only if, after sufficiently blocking and relabeling the $\Sigma_\infty^B$ symbols, the following relations hold:
    \begin{equation} \label{eq:freedom_gCF_relation_X_O} 
        \begin{tikzpicture}[scale=.45, baseline={([yshift=-0.5ex]current bounding box.center)}, thick]
            \begin{scope}[shift={(0,0)}]
                \draw (-1.2,0) -- (1.2,0);
                \draw (0,1) -- (0,-1);
                \filldraw[fill=melon!50] (-1/2-0.0,-1/2) -- (-1/2-0.0,1/2) -- (1/2+0.0,1/2) -- (1/2+0.0,-1/2) -- (-1/2-0.0,-1/2);
                \draw (0,0) node {\scriptsize $B$};
            \end{scope}
        \end{tikzpicture} =
        \begin{tikzpicture}[scale=.45, baseline={([yshift=2.2ex]current bounding box.center)}, thick]
            \begin{scope}[shift={(0,0)}]
                \draw (-1.2,0) -- (1.2,0);
                \draw (0,1) -- (0,-3);
                \filldraw[fill=powderblue!50] (-1/2-0.0,-1/2) -- (-1/2-0.0,1/2) -- (1/2+0.0,1/2) -- (1/2+0.0,-1/2) -- (-1/2-0.0,-1/2);
                \draw (0,0) node {\scriptsize $A$};
            \end{scope}
            \begin{scope}[shift={(1.5,0)}]
                \draw (-1,0) -- (1,0);
                \filldraw[fill=gray!10] (0.5,-0.5) -- (-0.5,-0.5) -- (-0.5,0.5) -- (0.5, 0.5) -- (0.5,-0.5);
                \draw (0,0) node {\scriptsize $Z$};
                \draw (0,-0.5) -- (0,-1);
            \end{scope}
            \begin{scope}[shift={(-2,0)}]
                \draw (-1,0) -- (1.5,0);
                \draw (0.25,-1) -- (0.25,-0.5);
                \filldraw[fill=gray!10] (1,-0.5) -- (-0.5,-0.5) -- (-0.5,0.5) -- (1, 0.5) -- (1,-0.5);
                \draw (0.3,0) node {\scriptsize $Z^{-1}$};
            \end{scope}
            \draw (-1.75,-0.5) -- (-1.75,-1) -- (1.5,-1) -- (1.5,-0.5);
            \fill (0,-1) circle (0.1);
            \begin{scope}[shift={(-0.875,-1)}]
                \filldraw[fill=yellow!75] (0.35,-0.35) -- (-0.35,-0.35) -- (-0.35,0.35) -- (0.35, 0.35) -- (0.35,-0.35);
                \draw (0,0) node {\scriptsize $r^1$};
            \end{scope} 
            \begin{scope}[shift={(0.75,-1)}]
                \filldraw[fill=yellow!75] (0.35,-0.35) -- (-0.35,-0.35) -- (-0.35,0.35) -- (0.35, 0.35) -- (0.35,-0.35);
                \draw (0,0) node {\scriptsize $r^2$};
            \end{scope}
            \begin{scope}[shift={(0,-2)}]
                \filldraw[fill=gray!10] (0.5,-0.5) -- (-0.5,-0.5) -- (-0.5,0.5) -- (0.5, 0.5) -- (0.5,-0.5);
                \draw (0,0) node {\scriptsize $P_B$};
            \end{scope}
        \end{tikzpicture}
        \quad \text{and} \quad
        \ket{X^A_{O}} = P_B^{\otimes m} \ket{X^B_{O}} , \quad \forall O \in \Sigma_\infty^{m+1},
    \end{equation}
    where $\Sigma_\infty^A = \Sigma_\infty^B =: \Sigma_\infty$ and $\Sigma_f^{A,ij} = \Sigma_f^{B,ij} =: \Sigma_f^{ij}$; $P_B, Z_j$ are invertible matrices; and functions $r^1, r^2: \Sigma \to \Sigma_\infty$ are defined such that, for all $t \in \Sigma$, it holds that $t \in \tilde{\Sigma}_f^{r^1_t r^2_t}$. Moreover, $P_B$ is block-diagonal and acts as
    \begin{equation*}
        \begin{cases}
            P_B \ket{x} = \ket{x} & \text{if } x \in \Sigma_\infty, \\
            P_B ( \langle \Sigma_f^{ij} \rangle ) \subseteq \langle \Sigma_f^{ij}\rangle  & \text{for } i,j \in \tilde{\Sigma}_\infty.
        \end{cases}
    \end{equation*}
\end{restatable}
Therefore, after sufficient blocking, the freedom of the gCF representation is fully characterized by the orbits of the defining states $\{\ket{X_O}\}$ of the algebraic RLS, under a specific class of block-diagonal SLOCC operations. The proof is given in Appendix \ref{app:sec_gCF_freedom}. The fundamental theorem for uniform PBC MPS (Theorem \ref{prop:freedom_pbcCF}) is recovered as a special case when the MPS-X tensors are block-diagonal and the boundary matrix is the identity. 

As an illustration, consider the following cases: block-diagonal TI MPS-X with an arbitrary boundary (for which $\Sigma = \Sigma_\infty$); W-like MPS-X with $\{\ket{L_N}\} = \ket{0^* 1 0^*}$; Dicke-like MPS-X with $\{\ket{L_N}\} = \ket{0^* (10^*)^k}$ for some $k \in \mathbb{N}$ (for which $|\Sigma_f^{00}| = 1$); and domain-wall-like MPS-X with $\{\ket{L_N}\} = \ket{0^* 1 2^*}$ (for which $|\Sigma_f^{02}| = 1$). In all these examples, Theorem \ref{prop:freedom_gCF} guarantees that any other equivalent MPS-X representation must have a gCF with
\begin{itemize}
    \item[\textit{(i)}] the same algebraic RLS as the backbone family $\{\ket{L_N}\}$, and
    \item[\textit{(ii)}] the MPS tensors of each free block within the matrix structure related by a gauge transformation.
\end{itemize}

\section{The algebraic structure of non-semisimple sets of matrices}
\label{sec:algebra-span}

Our generalized canonical form for MPS-X builds upon three key independent results: a structural characterization of matrix algebras, $\mathcal{A}$, a characterization of the length-$\ell$ span of a set of matrices, $\mathcal{A}^{(\ell)}$, and a practical criterion for the \textit{stability} of a set of matrices upon blocking. We elaborate on them in this section. 

    

We will follow two guiding questions to understand the structure of both $\mathcal{A}$ and $\mathcal{A}^{(\ell)}$ for a given block-upper-triangularized MPS tensor $\{A^i\}$: \textit{(1)} how does each block (denoted as $A_{mn}^i$ in Eq. \eqref{eq:block-upper-triangular-form}) look like, and how are different blocks related to each other?, and \textit{(2)} how do $\mathcal{A}, \mathcal{A}^{(\ell)}$ differ from each other? The first question determines the gCF of any MPS-X, while the second distinguishes stable from non-stable cases.

\subsection{The structure of the algebra and the span of a set of matrices} \label{sec:algebra-span-thmbasis}

To address the first question, we introduce three notions:
\begin{itemize}
    \item \textbf{The $\preceq$-order on blocks:} $(1,1) \preceq (2,2) \preceq \dots \preceq (b,b) \preceq (1,2) \preceq (2,3) \preceq \dots \preceq (1,b)$.
    
    \item \textbf{Free blocks:} Block $(i,j)$ of $\mathcal{A}$ or $\mathcal{A}^{(\ell)}$ is \textit{free} if, for any matrix $A \in \mathcal{M}_{D_i \times D_j}(\mathbb{C})$, there exists $a \in \mathcal{A}$ such that $a_{ij} = A$, where $a_{ij}$ denotes the $(i,j)$-th block of $a$, and $a_{mn} = 0$ for all $(m,n) \prec (i,j)$. Freeness is key for defining the upper MPO of the gCF and ensuring the block-injectivity of the associated tensors, as it guarantees that the subspaces spanned by different free blocks in the physical Hilbert space are linearly independent.

    \item \textbf{Sectors:} Owing to the theory of semisimple algebras and uniform PBC MPS, the diagonal blocks in $\mathcal{A}$ (or $\mathcal{A}^{(\ell)}$) are either free, or equal (or proportional) to another after a suitable change of basis. Grouping equivalent diagonal blocks together defines equivalence classes labeled by symbols in $\Sigma_\infty$. A block $(m,n)$ \textit{belongs in sector $[i,j]$} for $i,j \in \Sigma_\infty$ if block $(m,m)$ is in equivalence class $i$, and block $(n,n)$ is in equivalence class $j$. 
\end{itemize}
As an illustrative example, consider
\begin{equation} \label{eq:example_nonsemisimple_part}
    \mathcal{A} = \left\{ {\scriptsize\begin{pmatrix}
        \highg{A} & \highb{C} & \highg{D} & \highg{E} \\
        0 & \highy{B} & \highr{0} & \highr{0} \\
        0 & 0 & \highg{A} & \highg{D} \\
        0 & 0 & 0 & \highg{A}
    \end{pmatrix}} \mid A,\dots, E \right\}.
\end{equation}
Blocks $(1,1), (2,2), (1,2), (3,4), (1,4)$ are free, while $(3,3), (4,4), (1,3)$ are not. Thus, there are two equivalence classes of diagonal blocks labeled by $\Sigma_\infty = \{0,1\}$: $(2,2)$ belongs to 0, while the rest belong to 1.  The sectors are indicated with colors: green for $[0,0]$, blue for $[0,1]$, red for $[1,0]$ and yellow for $[1,1]$.

We now state the main results describing the structure of general (not necessarily semisimple) matrix sets. For the algebra characterization, similar statements appear in \cite{kostov_1995_burnside, Sergeichuk_2000}. In this work, we provide our own independent proof based on elementary algebra, explicitly constructing the change of basis matrix $P$ that transforms any block-upper-triangular matrix set into one with the desired structure.

\begin{restatable}[Algebra Structure]{proposition}{propstructurealgebra} \label{prop:structure_subalgebra}
Given any matrix algebra $\mathcal{A}$, there exists an invertible matrix $P$ such that $P\mathcal{A}P^{-1}$ is block-upper-triangular with the following properties:
\begin{enumerate}
    \item Each diagonal block is either free or equal to another diagonal block.
    \item Each off-diagonal block is either zero, free, or a linear combination of other off-diagonal free blocks in the same sector.
\end{enumerate}
\end{restatable}
The proof is given in Appendix \ref{app:structurealgebraproofs}, along with some examples. 
Unlike algebras, the length-$\ell$ span of a set of matrices has not yet been systematically studied in the literature. Nevertheless, it plays a central role in the analysis of MPS. In the following, we provide a general characterization of the span associated with an arbitrary set of matrices. We assume, without loss of generality, that they have been sufficiently blocked to remove periodic subspaces.

\begin{restatable}[Span Structure]{theorem}{propstructurespan} \label{prop:structure_span}
    Given the length-$\ell$ span of a set of matrices, $\mathcal{A}^{(\ell)}$, with $\ell \geq L_{\mathrm{span}}$ for some $L_{\mathrm{span}} \in \mathbb{N}$, there exists an invertible matrix $P$ such that $P\mathcal{A}^{(\ell)}P^{-1}$ is block-upper-triangular with the following properties:
    \begin{enumerate}
        \item Each diagonal block is either free or proportional to another diagonal block.
        \item Each off-diagonal block is either zero, free, or a linear combination of free blocks within the same sector (the diagonal free block $i \in \Sigma_\infty$ is included in sector $[i,i]$).
    \end{enumerate}
    Moreover, $L_{\mathrm{span}}$ is upper bounded as
    \begin{equation} \label{eq:upper_bound_difficult_Lspan}
        L_{\mathrm{span}} \leq \left(L_{BI}^{diag} + \frac{2}{3} L_0^{diag}\right) 2^{b(b-1)} - \frac{2}{3}L_0^{diag} , 
    \end{equation}
    where $b$ denotes that the block-upper-triangularized matrices have $b$ blocks by $b$ blocks, $L_{BI}^{diag} \leq 3(b-1)(L_0^{diag}+1)$ and $L_0^{diag} \leq \max_{j \in \Sigma_\infty} 2 D_j^2 (6 + \log_2(D_j))$, $D_j$ being the size of the $j$-th diagonal block.
\end{restatable}
We provide the proof in Appendix \ref{app:span_structure}. Thus, after sufficient blocking, the basis for the span can have the following additional properties with respect to the basis for the algebra:
\begin{itemize}
    \item Diagonal blocks can be proportional, rather than equal. For instance,
    \begin{equation}  \label{eq:ex_differences_alg-span_1}
        \mathcal{A}^{(1)} = \text{span}\left\{{
        \scriptsize
        \begin{pmatrix}
            1 & 0 \\
            0 & e^{i \sqrt{2}\pi}
        \end{pmatrix} 
        }\right\} \to 
        \mathcal{A}^{(\ell)} = \text{span}\left\{{
        \scriptsize
        \begin{pmatrix}
            1 & 0 \\
            0 & e^{i \ell\sqrt{2}\pi}
        \end{pmatrix} 
        }\right\} \neq \mathrm{Alg}(\mathcal{A}^{(\ell)})
        = \left\{ {\scriptsize \begin{pmatrix}
            a & 0 \\ 0 & b
        \end{pmatrix}} \mid a,b \in \mathbb{C} \right\} .
    \end{equation}
    
    \item Jordan-type structures can appear, i.e. the diagonal and off-diagonal blocks are no longer independent, but instead off-diagonal blocks might be proportional to the diagonal block, or linear combinations including them. For instance,  
    \begin{equation} \label{eq:ex_differences_alg-span_2}
        \mathcal{A}^{(1)} = \text{span} \left\{ {\scriptsize \begin{pmatrix}
        1 & 1 \\ 0 & 1
    \end{pmatrix}} \right\} \implies
    \mathcal{A}^{(\ell)} = \text{span}\left\{ {\scriptsize \begin{pmatrix}
        1 & \ell \\ 0 & 1
    \end{pmatrix}} \right\} \neq \text{Alg}(\mathcal{A}^{(\ell)}) = \left\{ {\scriptsize \begin{pmatrix}
        a & b \\ 0 & a
    \end{pmatrix}} \mid a,b \in \mathbb{C} \right\} .
    \end{equation}
\end{itemize}

Although the gCF for stable MPS-X can be obtained directly from the algebra structure of Proposition \ref{prop:structure_subalgebra}, Theorem \ref{prop:structure_span} for the span structure remains essential for several key reasons:
\textit{(i)} it provides an explicit upper bound on the block-injectivity length $L_{BI}$ and the amount of blocking required for the gCF,
\textit{(ii)} it underlies the proof of the generalized quantum Wielandt's inequality, introduced in the next section, and
\textit{(iii)} it characterizes the form of non-stable MPS-X (see the \emph{spanCF} in Appendix \ref{app:spanRLS-spanCF}).

\subsection{The generalized quantum Wielandt's inequality} \label{subsec:stable_sets_matrices}

We now present a theorem giving a general criterion to determine when an MPS-X tensor becomes stable (Def. \ref{def:stable_mps-x}) under blocking, together with explicit upper bounds on the required blocking length and on $L_{\mathrm{stab}}$.

For normal MPS, the so-called quantum Wielandt's inequality \cite{Sanz2010, Michalek2018} guarantees stability of the MPS tensor with $L_{\mathrm{stab}} \leq 2D^2(6+\log_2 D)$ (or even a tighter bound, $L_{\mathrm{stab}} \leq D^2 + 2D - 4$, according to recent work \cite{shitov_2023_growth}). Our result can thus be seen as a generalized quantum Wielandt's inequality, valid for any arbitrary set of matrices.


Recall that $p$ is the minimal blocking that removes all periodic subspaces, and $q$ is the smallest integer such that the proportionality constants $\mu_{j,k}$ between diagonal blocks satisfy $\mu_{j,k}^{pq} = 1$ for all $j,k$ (if no such $q$ exists, as in Eq. \eqref{eq:ex_differences_alg-span_1}, we set $q = \infty$). Let $\mathds{1}_0$ denote the identity matrix with zeros in the positions corresponding to vanishing diagonal blocks of the algebra, and let $r_{\mathrm{alg}}$ be the \textit{algebra length} (i.e. the minimal $r \in \mathbb{N}$ such that $\mathrm{Alg}(\mathcal{A}^{(1)}) =  \mathrm{span}\{\cup_{n=1}^{r}\mathcal{A}^{(n)}\}$, which satisfies $r_{\mathrm{alg}} \leq D^2$). 

\begin{restatable}[Generalized quantum Wielandt's inequality]{theorem}{genWielandtt} \label{prop:stability_criterion}
    A set of matrices $\{A^i\}$ becomes stable upon blocking if and only if $q < \infty$ and $\mathds{1}_0 \in \mathcal{A}^{(pq L_{\mathrm{span}} 2^b)}$. In this case, $\mathcal{A}^{(pqL_{\mathrm{span}}b2^b)}$ is stable with $L_{\mathrm{stab}} = r_{\mathrm{alg}}$. In particular, \begin{equation} \label{eq:main_stability_result}
        \mathcal{A}^{(pqL_{\mathrm{span}} 2^b (r_{\mathrm{alg}}b + s))} = \mathrm{Alg}(\mathcal{A}^{(pqL_{\mathrm{span}} b 2^b (1+t))}), \quad \forall s,t \geq 0.
    \end{equation}
    Otherwise, the set $\{A^i\}$ never becomes stable under blocking, and $\mathcal{A}^{(n)} \nsubseteq \mathcal{A}^{(n+m)}$ for all $m,n \in \mathbb{N}$.
\end{restatable}


\section{Bringing an MPS-X to the generalized canonical form}
\label{sec:how-to}

This section outlines how to bring any MPS-X to its gCF. The procedure mirrors the proof strategy of Theorem \ref{thm:generality_gCF} (Appendix \ref{app:proof_thm_generality-gCF}) and we illustrate it with a physically relevant class of MPS-X that we call \textit{W-like MPS-X}. These have the following form,
\begin{align*}
    \ket{\psi_N(X,A^i)} := \sum_{i_1 \dots i_N} \Tr[X A^{i_1} \dots A^{i_N}] \ket{i_1 \dots i_N}, \quad 
    \text{with }
    \
    X = 
    {\scriptsize \begin{pmatrix}
        X_{11} & X_{12} \\ X_{21} & X_{22}
    \end{pmatrix} }, 
    \
    A^i = 
    {\scriptsize \begin{pmatrix}
        B^i & C^i \\ 0 & B^i
    \end{pmatrix} }
    .
\end{align*}
Even though we consider arbitrary boundary matrices $X$ in what follows, note that choosing $X_{11} = X_{22} = X_{12} = 0$ and $X_{21} = \mathds{1}$ gives a W-like superposition of MPS,
\begin{equation*}
    \ket{\psi_N(X,A^i)} = \sum_n \sum_{i_1 \dots i_N}
    \Tr[B^{i_1} \dots B^{i_{n-1}} C^{i_n} B^{i_{n+1}} \dots B^{i_N}] \ket{i_1 \dots i_N},
\end{equation*}
which is the ansatz commonly used to represent low-lying excited states above an MPS ground state \cite{haegeman_2013_elem-excitations, Haegeman_2012, Osborne_2025_MPS-excitation, white_2025_sitebasisexcitationansatz} and tangent vectors to the manifold of uniform PBC MPS \cite{haegeman_2013_postMPS-tangent-space, vanderstraeten_2019}.

\paragraph{Preliminary step.} Use the stability criterion of Theorem \ref{prop:stability_criterion} to verify that the set of MPS matrices becomes stable after blocking. More specifically, block every $L$ sites such that $\mathcal{A}^{(\ell L)} = \mathrm{Alg}(\mathcal{A}^{(L)})$ for $\ell \geq r_{\mathrm{alg}}$, where $L$ is upper bounded as $L \leq pq L_{\mathrm{span}} b 2^b$. 

In our example, we assume for simplicity that
\begin{equation} \label{eq:example_wlike_1}
    \mathcal{A}^{(1)} = 
    \left\{
        {\scriptsize \begin{pmatrix}
            B & C \\ 0 & B
        \end{pmatrix} }
        \mid B, C\in \mathcal{M}_{D_B \times D_B}(\mathbb{C})
    \right\} ,
\end{equation}
which satisfies $\mathcal{A}^{(\ell)} = \mathrm{Alg}(\mathcal{A}^{(1)})$ for all $\ell \geq 1$, so no blocking is required.

\paragraph{Step 1.} Choose a gauge $P$ such that $P\mathrm{Alg}(\mathcal{A}^{(1)}) P^{-1}$ is in the block-upper-triangular form with the properties described in Proposition \ref{prop:structure_subalgebra}. The tensors of our W-like example already have this structure by the assumption on $\mathcal{A}^{(1)}$ shown in Eq. \eqref{eq:example_wlike_1}, so no additional gauge transformation is required.

\paragraph{Step 2.} In the new gauge, decompose the tensor according to a basis of $P \mathrm{Alg}(\mathcal{A}^{(1)})P^{-1}$ as
\begin{equation} \label{eq:find_gCF_matrix-CF_maintext}
    \begin{tikzpicture}[scale=.45, baseline={([yshift=-1ex]current bounding box.center)}, thick]
        \MPSTensor{0,0}{$A$}{purple}
        \begin{scope}[shift={(-1.5,0)}]
            \draw (-1,0) -- (1,0);
            \filldraw[fill=yellow] (0.5,-0.5) -- (-0.5,-0.5) -- (-0.5,0.5) -- (0.5, 0.5) -- (0.5,-0.5);
            \draw (0,0) node {\scriptsize $P$};
        \end{scope}
        \begin{scope}[shift={(1.5,0)}]
            \draw (-1,0) -- (1.5,0);
            \filldraw[fill=yellow] (1,-0.5) -- (-0.5,-0.5) -- (-0.5,0.5) -- (1, 0.5) -- (1,-0.5);
            \draw (0.3,0) node {\scriptsize $P^{-1}$};
        \end{scope}
    \end{tikzpicture}
    =
    \begin{tikzpicture}[scale=.45, baseline={([yshift=-1ex]current bounding box.center)}, thick]
        \begin{scope}[shift={(0,0)}]
            \draw (-1.2,0) -- (1.2,0);
            \draw (0,1) -- (0,0);
            \filldraw[fill=purple] (-1/2-0.2,-1/2) -- (-1/2-0.2,1/2) -- (1/2+0.2,1/2) -- (1/2+0.2,-1/2) -- (-1/2-0.2,-1/2);
            \draw (0,0) node {\scriptsize $A_{\text{\normalfont{low}}}$};
        \end{scope}
        \begin{scope}[shift={(0,1.5)}]
            \draw (-1.2,0) -- (1.2,0);
            \draw (0,1) -- (0,0);
            \filldraw[fill=purple] (-1/2-0.2,-1/2) -- (-1/2-0.2,1/2) -- (1/2+0.2,1/2) -- (1/2+0.2,-1/2) -- (-1/2-0.2,-1/2);
            \draw (0,0) node {\scriptsize $A_{\text{\normalfont{up}}}$};
        \end{scope}
    \end{tikzpicture},
\end{equation}
where the lower part $A_{\mathrm{low}}$ captures the algebraic relations between blocks in the algebra basis, and the upper part $A_{\mathrm{up}}$ contains the free blocks. We formalize this decomposition as the \emph{matrix-CF} in Appendix \ref{app:matrix-CF}, which underlies the proofs for Theorems \ref{thm:generality_gCF} and \ref{prop:freedom_gCF}. In our example,
\begin{equation} \label{eq:matrix-CF-Wstate-example}
    \begin{tikzpicture}[scale=.45, baseline={([yshift=-1.6ex]current bounding box.center)}, thick]
        \draw (-1.2,0) -- (1.2,0);
        \draw (0,1) -- (0,0);
        \filldraw[fill=purple] (-1/2-0.2,-1/2) -- (-1/2-0.2,1/2) -- (1/2+0.2,1/2) -- (1/2+0.2,-1/2) -- (-1/2-0.2,-1/2);
        \draw (0,0) node {\scriptsize $A_{\text{low}}$};
        \draw (0,1.3) node {\scriptsize $0$};
    \end{tikzpicture}
    = 
    {\scriptsize
    \begin{pmatrix}
        1 & 0 \\
        0 & 1 
    \end{pmatrix}}
    ,
    \
    \begin{tikzpicture}[scale=.45, baseline={([yshift=-1.6ex]current bounding box.center)}, thick]
        \draw (-1.2,0) -- (1.2,0);
        \draw (0,1) -- (0,0);
        \filldraw[fill=purple] (-1/2-0.2,-1/2) -- (-1/2-0.2,1/2) -- (1/2+0.2,1/2) -- (1/2+0.2,-1/2) -- (-1/2-0.2,-1/2);
        \draw (0,0) node {\scriptsize $A_{\text{low}}$};
        \draw (0,1.3) node {\scriptsize $1$};
    \end{tikzpicture}
    = 
    {\scriptsize
    \begin{pmatrix}
        0 & 1 \\
        0 & 0
    \end{pmatrix}}
    , 
    \ 
    \begin{tikzpicture}[scale=.45, baseline={([yshift=-0.5ex]current bounding box.center)}, thick]
        \begin{scope}[shift={(0,0)}]
            \draw (-1.2,0) -- (1.2,0);
            \draw (0,1) -- (0,-1);
            \filldraw[fill=purple] (-1/2-0.2,-1/2) -- (-1/2-0.2,1/2) -- (1/2+0.2,1/2) -- (1/2+0.2,-1/2) -- (-1/2-0.2,-1/2);
            \draw (0,0) node {\scriptsize $A_{\text{up}}$};
            \draw (0,1.3) node {\scriptsize $i$};
            \draw (0,-1.3) node {\scriptsize $0$};
        \end{scope}
    \end{tikzpicture}
    = 
    B^i , 
    \
    \begin{tikzpicture}[scale=.45, baseline={([yshift=-0.5ex]current bounding box.center)}, thick]
        \begin{scope}[shift={(0,0)}]
            \draw (-1.2,0) -- (1.2,0);
            \draw (0,1) -- (0,-1);
            \filldraw[fill=purple] (-1/2-0.2,-1/2) -- (-1/2-0.2,1/2) -- (1/2+0.2,1/2) -- (1/2+0.2,-1/2) -- (-1/2-0.2,-1/2);
            \draw (0,0) node {\scriptsize $A_{\text{up}}$};
            \draw (0,1.3) node {\scriptsize $i$};
            \draw (0,-1.3) node {\scriptsize $1$};
        \end{scope}
    \end{tikzpicture}
    = 
    C^i  .
\end{equation}
Since we choose $\{A_{\mathrm{low}}^e\}_{e \in \Sigma}$ to be a basis of the algebra $\mathcal{A}_{\mathrm{low}}$ of lower tensors, we can define $\Gamma$ as the associative tensor of structure constants of $\mathcal{A}_{\mathrm{low}}$ with respect to this basis. In our example, the tensor $\Gamma$ with respect to the basis $\{A_{\mathrm{low}}^0, A_{\mathrm{low}}^1\}$ is
\begin{equation} \label{eq:Gamma_tensor_W-state}
    \begin{tikzpicture}[scale=.45, baseline={([yshift=-1ex]current bounding box.center)}, thick]
        \begin{scope}[shift={(0,0)}]
            \draw (0.3,0.5) -- (0.3,1);
            \draw (-0.3,0.5) -- (-0.3,1);
            \filldraw[fill=gray!10] (-0.5,-0.5) -- (-0.5,0.5) -- (0.5,0.5) -- (0.5,-0.5) -- (-0.5,-0.5);
            \draw (0,0) node {\scriptsize $\Gamma$};
            \draw (0,-0.5) -- (0,-1);
        \end{scope}
        \draw (0,-1.3) node {\scriptsize $0$};
    \end{tikzpicture}
    = \ket{00}, \quad 
    \begin{tikzpicture}[scale=.45, baseline={([yshift=-1ex]current bounding box.center)}, thick]
        \begin{scope}[shift={(0,0)}]
            \draw (0.3,0.5) -- (0.3,1);
            \draw (-0.3,0.5) -- (-0.3,1);
            \filldraw[fill=gray!10] (-0.5,-0.5) -- (-0.5,0.5) -- (0.5,0.5) -- (0.5,-0.5) -- (-0.5,-0.5);
            \draw (0,0) node {\scriptsize $\Gamma$};
            \draw (0,-0.5) -- (0,-1);
        \end{scope}
        \draw (0,-1.3) node {\scriptsize $1$};
    \end{tikzpicture}
    = \ket{01} + \ket{10} .
\end{equation}

\paragraph{Step 3.} Apply Proposition \ref{prop:check_TI_informal} to simplify the boundary matrix of the MPS-X. For the W-like case, the formal version of the proposition, included in Appendix \ref{app:TI_proofs}, shows that translational invariance holds if and only if the boundary matrix blocks satisfy
\begin{equation*}
    X_{11} + X_{22} = \beta_0 \mathds{1}, \quad X_{21} = \beta_1 \mathds{1} .
\end{equation*}
Hence we may replace $X$ with a simplified matrix $\tilde{X}$ yielding the same state $|\psi_N(\tilde{X}, A^i) \rangle = |\psi_N(X,A^i)\rangle$, and introduce a $b \times b$ matrix $Y$ containing only the relevant proportionality constants, as follows:
\begin{equation*}
    \tilde{X} = 
    {\scriptsize \begin{pmatrix}
        \beta_0 \mathds{1} & 0 \\ \beta_1 \mathds{1} & 0
    \end{pmatrix} } , \quad 
    Y = 
    {\scriptsize \begin{pmatrix}
        \beta_0 & 0 \\ \beta_1 & 0
    \end{pmatrix} }.
\end{equation*}

\paragraph{Step 4.} The lower part of the gCF is the algebraic RLS family
\begin{equation*}
    \{\ket{L_N}\} := \{\ket{\psi_N(Y, A_{\mathrm{low}})}\} 
    = \beta_0 \ket{0^*} + \beta_1 \ket{0^* 1 0^*} ,
\end{equation*}
which is $\Gamma$-invariant with respect to the associative $\Gamma$-tensor defined in Eq. \eqref{eq:Gamma_tensor_W-state}. Using the notation of Eq. \eqref{eq:def_algebraic_RLS_equation}, this corresponds to \begin{equation*}
    \{\ket{L_N}\} = \ket{0^*} X_0 + \hat{S}^{(1)} \ket{0^* f 0^*} \ket{X_{00}}, 
    \quad \text{with } 
    \begin{cases}
        X_0 = \beta_0 \, ,
        \\ 
        \ket{X_{00}} = \beta_1 \ket{1} \, .
    \end{cases}
\end{equation*}

\paragraph{Step 5.} The upper tensor $A_{\mathrm{up}}$ in the gCF is given by the upper tensor $A_{\mathrm{up}}$ obtained in Step 2 (Eq. \eqref{eq:find_gCF_matrix-CF_maintext}). By Theorem \ref{thm:generality_gCF}, this tensor becomes block-injective after blocking at most $D^2$ sites, which completes the construction of the gCF.

\section{Conclusions}
 \label{sec:conclusions}

In this work, we have introduced a generalized canonical form (gCF) for uniform matrix product states with a boundary matrix, thereby extending the canonical framework previously available only for translationally invariant MPS with periodic boundary conditions. The gCF shows that any stable MPS-X can be expressed as the composition of two conceptually distinct layers: an upper block-invertible matrix product operator capturing local and short-range correlations, and a lower algebraic regular language state encoding the essential long-range and scale-invariant structure of the state. This decomposition also clarifies how structural properties in the theory of MPS such as injectivity, block-injectivity, basis of normal tensors, or gauge freedom, naturally generalize beyond the periodic boundary setting.

Our construction relies on new results concerning the algebraic and span structure of sets of matrices in the non-semisimple regime, where off-diagonal blocks cannot be neglected. We explicitly construct a structured basis for both the algebra and the finite-length span, together with the corresponding gauge transformations and an explicit upper bound on the blocking length required for these properties to hold. This extends the semisimple results that underpin the standard theory of uniform PBC MPS. Furthermore, we characterize stable and non-stable sets of matrices, providing rigorous statements connecting the algebra to the length-$\ell$ span of a given matrix set, along with a practical criterion to discern stability from non-stability that effectively generalizes the standard version of the quantum Wielandt's inequality. Building on these results, we formulate the gCF for stable MPS-X and fully characterize the freedom in the representation.

Our findings open several avenues for future research. Natural next steps include characterizing the freedom structure beyond stable MPS-X, developing an extension of the gCF that applies to MPS-X without requiring blocking, and analyzing the spectral properties of the transfer matrix in this setting and their connection to the gCF, questions that are well-understood in the PBC case but still unexplored for MPS-X. 

Beyond these directions, our framework invites to a re-examination of the analytical results in the MPS literature that rely on canonical forms and fundamental theorems, to understand how they extend to the broader class of MPS-X with the tools developed here. Examples include the parent-Hamiltonian construction for MPS-X ground states through their relation to the stability notion of MPS tensors introduced in \cite{garre-rubio_2025_mpsstabilityintersectionproperty}, or for MPS-X eigenstates that arise as quantum many-body scars \cite{gioia_2025_distincttypesparenthamiltonians}; the classification of matrix product unitaries (MPUs) beyond the known cases of quantum cellular automata \cite{styliaris_2024_MPUs}; the generalization of efficient MPS preparation protocols for translationally invariant systems \cite{malz_2024_MPS-preparation}; and potential extensions of finite-size phase classification, including in the presence of symmetries.

Altogether, this work establishes a unified algebraic and structural framework for uniform MPS with boundaries, bridging the existing gap between the theory of uniform PBC MPS and the broader landscape of uniform MPS with boundaries. It provides a foundation for a systematic understanding of a significantly larger class of tensor-network states, while offering results of independent interest: most notably, the explicit characterization of the span structure of non-semisimple matrix sets and its connection to the underlying algebra via the generalized Wielandt's inequality.

\section*{Acknowledgements}

We are grateful to Y. Liu and A. Molnar for valuable discussions, and to G. Styliaris for many insightful interactions and feedback on the manuscript. 
M.F.L. acknowledges support from the International Max Planck Research School for Quantum Science and Technology (IMPRS-QST).
The work at MPQ is partly funded by THEQUCO as part of the Munich Quantum Valley, which is supported by the Bavarian state government with funds from the Hightech Agenda Bayern Plus, as well as to the Klaus Tschira Foundation (project Decoding the Quanta of Space and Time).
A.M.A. acknowledges support from the Spanish Agencia Estatal de Investigación through the grants “IFT Centro de Excelencia Severo Ochoa CEX2020-001007-S", “PCI2024-153448" and “Ramón y Cajal RyC2021-031610-I", financed by MCIN/AEI/10.13039/501100011033 and the European Union NextGenerationEU/PRTR. This project was funded within the QuantERA II Programme that has received funding from the EU’s H2020 research and innovation programme under the GA No 101017733. 
D.P.G. acknowledges support from the Spanish Ministry of Science and Innovation MCIN/AEI/10.13039/501100011033 (CEX2023-001347-S and PID2023-146758NB-I00), and Comunidad Autonoma de Madrid (TEC-2024/COM-84-QUITEMAD-CM).

\bibliographystyle{abbrv}
\bibliography{references}

\newpage 

\appendix

\section{Technical notation} \label{app:notation_and_basis}

We consider sets of matrices of size $D \times D$ that are block-upper-triangular with $b$ blocks $\times \ b$ blocks, the $(i,j)$-th block having size $D_i \times D_j$. This entails no loss of generality: any collection of matrices $\{A^i\}$ can be simultaneously brought into a block-upper-triangular form by an invertible change of basis $P$, i.e., by replacing $A^i$ with $P A^i P^{-1}$, as shown in Eq. \eqref{eq:block-upper-triangular-form} of the main text. In addition to the notation already introduced in Section \ref{sec:algebra-span} of the main text, we will also use the following notions throughout the appendix:
\begin{itemize}
    \item $L_{BI}^\text{diag}$ denotes the block-injectivity length over the diagonal blocks, in the sense of the PBC MPS canonical form \cite{Perez-Garcia2007} (upper bounded by $3 D^5$).  
    
    \item $L_0^\text{diag}$ is defined as the maximum Wielandt length among all diagonal blocks, i.e. $L_0^{\text{diag}} := \max_{i \in \{1, \dots, b\}} L_0^{(i)}$, where $L_0^{(i)}$ is the Wielandt length of the normal block at $(i,i)$ \cite{Sanz2010, Michalek2018}.
    
    Note that $L_0^\text{diag} \leq L_{BI}^\text{diag}$, since block-injectivity requires not only injectivity of each diagonal block, but also linear independence between the physical subspaces corresponding to each of the free blocks indexed by $\Sigma_\infty$. 
    
    \item Block $(i,j)$ is \textit{isolatable} in $\mathcal{A}^{(\ell)}$ if $\exists a \in \mathcal{A}^{(\ell)}$ s.t. $a\mid_{\prec (i,j)} = 0$ and $a_{ij} \neq 0$. 
    
    \item $m_{ij}$ denotes the minimum length at which block $(i,j)$ becomes isolatable in the fixed-length span, i.e. $m_{ij} := \min \{ \ell \mid \text{block } (i,j) \text{ is isolatable in } \mathcal{A}^{(\ell)}\}$. If the block never becomes isolatable (because it is either 0 or dependent on other free blocks), then $m_{ij} = \infty$.

    \item Given the alphabet $\Sigma = \Sigma_\infty \cup \Sigma_f$ denoting the diagonal and off-diagonal free blocks, respectively, we relabel them for convenience as
    \begin{equation*}
        \Sigma_f = \{\{1\},\{2\}, \dots, \{|\Sigma_f|\}\}, \quad 
        \Sigma_\infty = \{\{0,1\},\{0,2\}, \dots, \{0,|\Sigma_\infty|\}\}.
    \end{equation*}
    %
    %
\end{itemize}
For clarity, let us revisit the example of Eq. \eqref{eq:example_nonsemisimple_part} in the main text:
\begin{equation} \label{eq:example_nonsemisimple_part_appendix}
    \mathcal{A} = \left\{ {\scriptsize\begin{pmatrix}
        \highg{A} & \highb{C} & \highg{D} & \highg{E} \\
        0 & \highy{B} & \highr{0} & \highr{0} \\
        0 & 0 & \highg{A} & \highg{D} \\
        0 & 0 & 0 & \highg{A}
    \end{pmatrix}} \mid A,\dots, E \right\}.
\end{equation}
Thus, there are two equivalence classes of diagonal blocks labeled by $\Sigma_\infty = \{\{0,1\},\{0,2\}\}$, denoted by $A, B$ in Eq. \eqref{eq:example_nonsemisimple_part_appendix}: $(2,2)$ belongs to $\{0,2\}$, while the rest belong to $\{0,1\}$. The three free off-diagonal blocks, denoted by $C, D, E$ in Eq.~\eqref{eq:example_nonsemisimple_part_appendix}, are labeled by $\{1\},\{2\},\{3\}$, respectively. The sectors are indicated with colors: green for $[\{0,1\},\{0,1\}]$, blue for $[\{0,1\},\{0,2\}]$, red for $[\{0,2\},\{0,1\}]$ and yellow for $[\{0,2\},\{0,2\}]$.

The structure described in Prop.~\ref{prop:structure_subalgebra} and Thm.~\ref{prop:structure_span} for the algebra $\mathcal{A}$ and the span $\mathcal{A}^{(\ell)}$, respectively, ensures the existence of constants $\{k_{ij;e}\}$ and $\{k_{ij;e}^{(\ell)}\}$, respectively, for each $i,j \in \{1, \dots, b\}$ and $e \in \Sigma$, such that the matrices
\begin{equation} \label{eq:def_basis_elements}
    [A]_e := \sum_{i\leq j} [k_{ij;e} A]_{ij}, \quad [A]_e^{(\ell)} := \sum_{i\leq j} [k_{ij;e}^{(\ell)} A]_{ij}, \quad A \in \mathcal{M}_{D_i \times D_j}(\mathbb{C})
\end{equation}
form a basis of $P\mathcal{A}P^{-1}$ or $P\mathcal{A}^{(\ell)}P^{-1}$, where $P$ is the invertible matrix prescribed by Prop. \ref{prop:structure_subalgebra} or Thm. \ref{prop:structure_span}. $[A]_{ij}$ denotes a matrix that equals $A$ in block $(i,j)$, and zero everywhere else. These basis elements satisfy additional properties, which are summarized in Table \ref{tab:basis_summary}:
\begin{itemize}
    \item \textbf{Basis elements $e \in \Sigma_\infty$:}
    \begin{itemize}
        \item \textit{For $\mathcal{A}$:} 
        \textit{(a)} $[A]_e$ is strictly block-diagonal (i.e. $k_{ij;e} = 0$ for $i \neq j$), and
        \textit{(b)} diagonal blocks are either free or equal to another diagonal block (i.e. $k_{ii;e} = 1$ only when $e = r_i$, and otherwise vanish).
        
        \item \textit{For $\mathcal{A}^{(\ell)}$:}         
        \textit{(a)} $[A]_e^{(\ell)}$ may contain off-diagonal contributions in sector $[e,e]$ (i.e. $k_{ij;e}^{(\ell)} \neq 0$ for $i \neq j$ only when $i< j$ and $r_i = r_j = e$), and        
        \textit{(b)} diagonal blocks are either free or proportional to another one (i.e. $k_{ii;e}^{(\ell)} \neq 0$ only when $e = r_i$, and otherwise vanish).
    \end{itemize}
    \item \textbf{Basis elements $e \in \Sigma_f^{pq}$ for some $p,q \in \Sigma_\infty$:}
    \begin{itemize}
        \item \textit{For both $\mathcal{A}$ and $\mathcal{A}^{(\ell)}$:}
        \textit{(a)} $[A]_e$ and $[A]_e^{(\ell)}$ are strictly block-upper-triangular (i.e. $k_{ij;e}, k_{ij;e}^{(\ell)} \neq 0$ only if $i < j$), and
        \textit{(b)} nonzero coefficients $k_{ij;e}$ or $k_{ij;e}^{(\ell)}$ can only occur when $e \in \Sigma_f^{r_i r_j}$ (i.e. $k_{ij;e}, k_{ij;e}^{(\ell)} \neq 0 \iff r_i = r_e^1$ and $r_j = r_e^2$).
    \end{itemize}
\end{itemize}

\begin{table}[h!]
\centering
\renewcommand{\arraystretch}{1.25}
\begin{tabular}{c||c|c|c}
 & \multicolumn{2}{c|}{$e\in\Sigma_\infty$} & $e\in\Sigma_f^{pq}$ \\ \hline
 & $\mathcal{A}$ & $\mathcal{A}^{(\ell)}$ & $\mathcal{A}$ and $\mathcal{A}^{(\ell)}$ \\ \hline\hline
\makecell{Block-\\structure\\support}
& Block-diagonal
& \makecell{Block-diagonal, and \\ also off-diagonal \\ in sector $[e,e]$}
& \makecell{Strictly \\ block-upper-\\triangular} \\ \hline
\makecell{Non-zero\\elements}
& $k_{ii;e}=\begin{cases} 1 &\text{for } e = r_i \\ 0 &\text{otherwise} \end{cases}$
& $k_{ii;e}^{(\ell)} \begin{cases}
    \in\mathbb{C}\setminus\{0\} &\text{for } e=r_i \\
    =0 & \text{otherwise}
\end{cases}$
& \makecell{$k_{ij;e}, \, k_{ij;e}^{(\ell)}\neq 0$ \\ only if $e\in\Sigma_f^{r_i r_j}$} \\ 
\end{tabular}
\caption{Properties of the basis elements $[A]_e$ and $[A]_e^{(\ell)}$. We let $r: \{1, \dots, b\} \to \Sigma_\infty$ assign to each diagonal block $(i,i)$ the symbol $r_i$ of its equivalence class in $\Sigma_\infty$.}
\label{tab:basis_summary}
\end{table}

We consider now the matrices $A_{\text{low}}^e$ (or $A_{\text{low}}^{(\ell),e}$) defined as
\begin{equation*}
    (A_{\text{low}}^e)_{mn} := k_{mn;e} \quad \text{or} \quad (A_{\text{low}}^{(\ell),e})_{mn} := k_{mn;e}^{(\ell)}.
\end{equation*} 
By construction, they form a basis of the algebra $\mathcal{A}_{\text{low}}$ and the span $\mathcal{A}_{\text{low}}^{(\ell)}$, respectively. For the algebra $\mathcal{A}_{\text{low}}$, we defined the corresponding structure constants tensor $\Gamma$ in Eq. \eqref{eq:Gamma_prescription}. Even though for the span $\mathcal{A}_{\text{low}}^{(\ell)}$ no such tensor exists, as it is not necessarily closed under product, we can still define a \textit{generalized structure constants tensor} $\Gamma^{(\ell_1, \ell_2)}$ with respect to a basis of $\mathcal{A}^{(\ell_1)}, \mathcal{A}^{(\ell_2)}$ and $\mathcal{A}^{(\ell_1+\ell_2)}$, as the values $\{(\Gamma^{(\ell_1, \ell_2)})^{ij}_k \}\subseteq \mathbb{C}$ such that for each $i,j \in \Sigma$,
\begin{equation*}
    A_{\text{low}}^{(\ell_1),i} \cdot A_{\text{low}}^{(\ell_2),j} = \sum_{k \in \Sigma} \left(\Gamma^{(\ell_1, \ell_2)}\right)^{ij}_k A_{\text{low}}^{(\ell_1+\ell_2),k} \ .
\end{equation*}
Therefore, in terms of the original matrices, for any $e,f \in \Sigma$,
\begin{equation*}
    [A]^{(\ell_1)}_e \cdot [B]^{(\ell_2)}_f = \sum_{k \in \Sigma} \left(\Gamma^{(\ell_1, \ell_2)}\right)^{ef}_g [AB]^{(\ell_1+\ell_2)}_g \ .
\end{equation*}
Prop. \ref{prop:structure_subalgebra} and Thm. \ref{prop:structure_span} ensure that the tensors $\Gamma$ and $\Gamma^{(\ell_1, \ell_2)}$ associated to the structured bases they prescribe satisfy the properties:
\begin{itemize}
    \item[(\textbf{P1})] 
    $\Gamma^{\{0,t\},\{0,s\}}_u =
        \left\{\def\arraystretch{1.2}\begin{tabular}{@{}l@{\quad}l@{}}
      $1$ & if $u = \{0,t\}$ and $\{0,t\} = \{0,s\}$, \\
      0 & otherwise,
    \end{tabular}
    \right.$    
    \item[(\textbf{P2})] 
    $\Gamma^{\{p\}, \{q\}}_{\{0,t\}} = 0, \ \forall \{p\}, \{q\} \in \Sigma_f$, $\forall \{0,t\} \in \Sigma_\infty$,
    \item[(\textbf{P3})] 
    $\Gamma^{uv}_{w} \neq 0 \implies \exists s_1, s_2, s_3 \in \Sigma_\infty$ such that $u \in \tilde{\Sigma}_f^{s_1 s_2}$, $v \in \tilde{\Sigma}_f^{s_2 s_3}$ and $w \in \tilde{\Sigma}_f^{s_1 s_3}$, where $\tilde{\Sigma}_f^{st} = \Sigma_f^{st} \cup \{s\}$ if $s= t$, and $\Sigma_f^{st}$ otherwise.
\end{itemize}

A feature of the span that is absent in the algebra is that new free blocks can emerge or disappear under blocking. For instance, consider:
\begin{equation*}
    \mathcal{A}^{(1)} = \left\{{\scriptsize \begin{pmatrix}
        A & B & 0 \\ & A & B \\ & & A 
    \end{pmatrix}
    } \mid A, B\right\} 
    \longrightarrow
    \mathcal{A}^{(2)} = \left\{{\scriptsize \begin{pmatrix}
        A & B & C \\ & A & B \\ & & A 
    \end{pmatrix}
    } \mid A, B, C\right\} . 
\end{equation*}
Here the new free block $C$ appears only after blocking twice. Naively, one might conclude that the set of free‐block labels $\Sigma_f$ depends on the blocking length $\ell$. However, since there are at most $b(b-1)/2$ potential free off‐diagonal blocks in a $b\times b$ block‐upper‐triangular form, we may fix a single global index set $\Sigma_f$ for all blocking lengths. Whenever a free block $e\in\Sigma_f$ does not appear yet at $\mathcal{A}^{(\ell)}$, we simply set its coefficient $k_{ij;e}^{(\ell)}=0$, for all $i,j$. In the example above, one takes $\Sigma_f = \{\{1\},\{2\}\}$ corresponding to the $B$ and $C$ free blocks, respectively, with $k_{ij;\{2\}}^{(2)} = 1$ but and $k_{ij;\{2\}}^{(1)} = 0$ for all $i,j$, since the second free block in $\Sigma_f$ only arises in $\mathcal{A}^{(2)}$.

\paragraph{Example 1.} Given
$\mathcal{A} = \left\{ {\scriptsize \begin{pmatrix} A & B \\ & A \end{pmatrix} } \mid A, B \right\}$, the notation we introduced above looks as follows:
\begin{itemize}
    \item $\Sigma$ can be partitioned into $\Sigma_\infty = \{\{0,1\}\}$ and $\Sigma_f = \Sigma_f^{11} = \{\{1\}\}$.
    \item The constants $\{k_{ij;s}\}$ are:
    \begin{itemize}
        \item $s \in \Sigma_\infty$: $k_{11;\{0,1\}} =  k_{22;\{0,1\}} = 1$, $k_{12;\{0,1\}} = 0$.
        \item $s \in \Sigma_f$: $k_{12;\{1\}} = 1$, $k_{11;\{1\}} =  k_{22;\{1\}} = 0$.
    \end{itemize}
    \item The structure constants tensor for this basis consists of all zeros except for:
    \begin{equation*}
        \begin{tikzpicture}[scale=.45, baseline={([yshift=-1ex]current bounding box.center)}, thick]
            \begin{scope}[shift={(0,0)}]
                \draw (0.3,0.5) -- (0.3,1);
                \draw (-0.3,0.5) -- (-0.3,1);
                \filldraw[fill=gray!10] (-0.5,-0.5) -- (-0.5,0.5) -- (0.5,0.5) -- (0.5,-0.5) -- (-0.5,-0.5);
                \draw (0,0) node {\scriptsize $\Gamma$};
                \draw (0,-0.5) -- (0,-1);
            \end{scope}
            \draw (0,-1.3) node {\scriptsize $\{0,1\}$};
            \draw (-0.8,1.4) node {\scriptsize $\{0,1\}$};
            \draw (0.8,1.4) node {\scriptsize $\{0,1\}$};
        \end{tikzpicture}
        =
        \begin{tikzpicture}[scale=.45, baseline={([yshift=-1ex]current bounding box.center)}, thick]
            \begin{scope}[shift={(0,0)}]
                \draw (0.3,0.5) -- (0.3,1);
                \draw (-0.3,0.5) -- (-0.3,1);
                \filldraw[fill=gray!10] (-0.5,-0.5) -- (-0.5,0.5) -- (0.5,0.5) -- (0.5,-0.5) -- (-0.5,-0.5);
                \draw (0,0) node {\scriptsize $\Gamma$};
                \draw (0,-0.5) -- (0,-1);
            \end{scope}
            \draw (0,-1.3) node {\scriptsize $\{1\}$};
            \draw (-0.8,1.4) node {\scriptsize $\{0,1\}$};
            \draw (0.5,1.4) node {\scriptsize $\{1\}$};
        \end{tikzpicture}
        =  
        \begin{tikzpicture}[scale=.45, baseline={([yshift=-1ex]current bounding box.center)}, thick]
            \begin{scope}[shift={(0,0)}]
                \draw (0.3,0.5) -- (0.3,1);
                \draw (-0.3,0.5) -- (-0.3,1);
                \filldraw[fill=gray!10] (-0.5,-0.5) -- (-0.5,0.5) -- (0.5,0.5) -- (0.5,-0.5) -- (-0.5,-0.5);
                \draw (0,0) node {\scriptsize $\Gamma$};
                \draw (0,-0.5) -- (0,-1);
            \end{scope}
            \draw (0,-1.3) node {\scriptsize $\{1\}$};
            \draw (-0.5,1.4) node {\scriptsize $\{1\}$};
            \draw (0.8,1.4) node {\scriptsize $\{0,1\}$};
        \end{tikzpicture}
        =
        1 . 
    \end{equation*}
\end{itemize}

\paragraph{Example 2.} Consider 
\begin{equation*}
    \mathcal{A}^{(1)} = 
    \left\{ {\scriptsize 
    \begin{pmatrix} 
    A & B & C \\ & A & A + \eta B \\ & & A 
    \end{pmatrix} } 
    \mid A, B, C \right\} \longrightarrow
    \mathcal{A}^{(\ell)} = 
    \left\{ {\scriptsize 
    \begin{pmatrix} 
    A & B & C \\ & A & \ell A + \eta B \\ & & A 
    \end{pmatrix} } 
    \mid A, B, C \right\}.
\end{equation*}
\begin{itemize}
    \item $\Sigma$ can be partitioned into $\Sigma_\infty = \{\{0,1\}\}$, and $\Sigma_f = \Sigma_f^{11} = \{\{1\}, \{2\}\}$ which denote free blocks $B$ and $C$, respectively.
    \item Then, the generalized structure constants tensor is all zeros except for:
    \begin{equation*}
        \begin{tikzpicture}[scale=.45, baseline={([yshift=-1ex]current bounding box.center)}, thick]
            \begin{scope}[shift={(0,0)}]
                \draw (0.7,0.5) -- (0.7,1);
                \draw (-0.7,0.5) -- (-0.7,1);
                \filldraw[fill=gray!10] (-1.2,-0.5) -- (-1.2,0.5) -- (1.2,0.5) -- (1.2,-0.5) -- (-1.2,-0.5);
                \draw (0,0) node {\scriptsize $\Gamma^{(\ell_1, \ell_2)}$};
                \draw (0,-0.5) -- (0,-1);
            \end{scope}
            \draw (-0.8,1.4) node {\scriptsize $\{0,1\}$};
            \draw (0.8,1.4) node {\scriptsize $\{0,1\}$};
            \draw (0,-1.3) node {\scriptsize $\{0,1\}$};
        \end{tikzpicture}
        =
        \begin{tikzpicture}[scale=.45, baseline={([yshift=-1ex]current bounding box.center)}, thick]
            \begin{scope}[shift={(0,0)}]
                \draw (0.7,0.5) -- (0.7,1);
                \draw (-0.7,0.5) -- (-0.7,1);
                \filldraw[fill=gray!10] (-1.2,-0.5) -- (-1.2,0.5) -- (1.2,0.5) -- (1.2,-0.5) -- (-1.2,-0.5);
                \draw (0,0) node {\scriptsize $\Gamma^{(\ell_1, \ell_2)}$};
                \draw (0,-0.5) -- (0,-1);
            \end{scope}
            \draw (-0.8,1.4) node {\scriptsize $\{0,1\}$};
            \draw (0.7,1.4) node {\scriptsize $\{1\}$};
            \draw (0,-1.3) node {\scriptsize $\{1\}$};
        \end{tikzpicture}
        =
        \begin{tikzpicture}[scale=.45, baseline={([yshift=-1ex]current bounding box.center)}, thick]
            \begin{scope}[shift={(0,0)}]
                \draw (0.7,0.5) -- (0.7,1);
                \draw (-0.7,0.5) -- (-0.7,1);
                \filldraw[fill=gray!10] (-1.2,-0.5) -- (-1.2,0.5) -- (1.2,0.5) -- (1.2,-0.5) -- (-1.2,-0.5);
                \draw (0,0) node {\scriptsize $\Gamma^{(\ell_1, \ell_2)}$};
                \draw (0,-0.5) -- (0,-1);
            \end{scope}
            \draw (-0.8,1.4) node {\scriptsize $\{0,1\}$};
            \draw (0.7,1.4) node {\scriptsize $\{2\}$};
            \draw (0,-1.3) node {\scriptsize $\{2\}$};
        \end{tikzpicture}
        = 
        \begin{tikzpicture}[scale=.45, baseline={([yshift=-1ex]current bounding box.center)}, thick]
            \begin{scope}[shift={(0,0)}]
                \draw (0.7,0.5) -- (0.7,1);
                \draw (-0.7,0.5) -- (-0.7,1);
                \filldraw[fill=gray!10] (-1.2,-0.5) -- (-1.2,0.5) -- (1.2,0.5) -- (1.2,-0.5) -- (-1.2,-0.5);
                \draw (0,0) node {\scriptsize $\Gamma^{(\ell_1, \ell_2)}$};
                \draw (0,-0.5) -- (0,-1);
            \end{scope}
            \draw (-0.7,1.4) node {\scriptsize $\{1\}$};
            \draw (0.8,1.4) node {\scriptsize $\{0,1\}$};
            \draw (0,-1.3) node {\scriptsize $\{1\}$};
        \end{tikzpicture}
        =
        \begin{tikzpicture}[scale=.45, baseline={([yshift=-1ex]current bounding box.center)}, thick]
            \begin{scope}[shift={(0,0)}]
                \draw (0.7,0.5) -- (0.7,1);
                \draw (-0.7,0.5) -- (-0.7,1);
                \filldraw[fill=gray!10] (-1.2,-0.5) -- (-1.2,0.5) -- (1.2,0.5) -- (1.2,-0.5) -- (-1.2,-0.5);
                \draw (0,0) node {\scriptsize $\Gamma^{(\ell_1, \ell_2)}$};
                \draw (0,-0.5) -- (0,-1);
            \end{scope}
            \draw (-0.7,1.4) node {\scriptsize $\{2\}$};
            \draw (0.8,1.4) node {\scriptsize $\{0,1\}$};
            \draw (0,-1.3) node {\scriptsize $\{2\}$};
        \end{tikzpicture}
        = 1,
    \ \
        \begin{tikzpicture}[scale=.45, baseline={([yshift=-1ex]current bounding box.center)}, thick]
            \begin{scope}[shift={(0,0)}]
                \draw (0.7,0.5) -- (0.7,1);
                \draw (-0.7,0.5) -- (-0.7,1);
                \filldraw[fill=gray!10] (-1.2,-0.5) -- (-1.2,0.5) -- (1.2,0.5) -- (1.2,-0.5) -- (-1.2,-0.5);
                \draw (0,0) node {\scriptsize $\Gamma^{(\ell_1, \ell_2)}$};
                \draw (0,-0.5) -- (0,-1);
            \end{scope}
            \draw (-0.7,1.4) node {\scriptsize $\{1\}$};
            \draw (0.8,1.4) node {\scriptsize $\{0,1\}$};
            \draw (0,-1.3) node {\scriptsize $\{2\}$};
        \end{tikzpicture}
        =
        \ell_2, \ \ 
        \begin{tikzpicture}[scale=.45, baseline={([yshift=-1ex]current bounding box.center)}, thick]
            \begin{scope}[shift={(0,0)}]
                \draw (0.7,0.5) -- (0.7,1);
                \draw (-0.7,0.5) -- (-0.7,1);
                \filldraw[fill=gray!10] (-1.2,-0.5) -- (-1.2,0.5) -- (1.2,0.5) -- (1.2,-0.5) -- (-1.2,-0.5);
                \draw (0,0) node {\scriptsize $\Gamma^{(\ell_1, \ell_2)}$};
                \draw (0,-0.5) -- (0,-1);
            \end{scope}
            \draw (-0.7,1.4) node {\scriptsize $\{1\}$};
            \draw (0.7,1.4) node {\scriptsize $\{1\}$};
            \draw (0,-1.3) node {\scriptsize $\{2\}$};
        \end{tikzpicture}
        =
        \eta.
    \end{equation*}
\end{itemize}
 
\newpage

\section{Proofs for the algebra structure in Proposition \ref{prop:structure_subalgebra}}
\label{app:structurealgebraproofs}

To show Prop.~\ref{prop:structure_subalgebra} in Section~\ref{sec:algebra-span} of the main text, we start by considering the specific case of block-upper-triangular algebras of 2 blocks by 2 blocks, and we will build upon them the general case later. Due to the theory of semi-simple algebras \cite{pierce_1982_associative-algebras, Farenick_2001}, only five different structures are possible based on the form of the diagonal blocks:
\begin{align}
    \text{Type } AA: \ 
    \mathcal{A} &= {\scriptsize \left\{ \begin{pmatrix} A & \ast \\ & A \end{pmatrix} \mid A \right\} }, \quad 
    \text{Type } AB: \
    \mathcal{A} = {\scriptsize \left\{ \begin{pmatrix} A & \ast \\ & B \end{pmatrix} \mid A, B \right\} }, \nonumber \\
    \text{Type } A0: \ 
    \mathcal{A} &= {\scriptsize \left\{ \begin{pmatrix} A & \ast \\ & 0 \end{pmatrix} \mid A \right\} }, \quad
    \text{Type } 0A: \
    \mathcal{A} = {\scriptsize \left\{ \begin{pmatrix} 0 & \ast \\ & A \end{pmatrix} \mid A \right\} }, \nonumber \\
    \text{Type } 00: \ 
    \mathcal{A} &= {\scriptsize \left\{ \begin{pmatrix} 0 & \ast \\ & 0 \end{pmatrix} \mid A \right\} }, \label{eq:def_types_algebras}
\end{align}
where ``$0$'' denotes a $1 \times 1$ block. Note that for the type 00 algebra, we have that $\mathcal{A}^{(s)} = 0$ for all $s \geq 2$. The following lemma tells us how to find a suitable change of basis that simplifies the structure of the off-diagonal block.

\begin{lemma} \label{lemma:isolate_subalgebra}
    Given any algebra $\mathcal{A}$ of block-upper-triangular matrices of 2 blocks by 2 blocks, we can explicitly construct a basis transformation such that the off-diagonal block in $\mathcal{A}$ is either
    \begin{enumerate}[label = (\roman*)]
        \item Zero.
        \item A free block independent of the diagonal. 
    \end{enumerate}
\end{lemma}
\begin{proof} 
    We divide the proof in two different cases depending on whether an element of the form ${\tiny \begin{pmatrix} 0 & \highg{\neq 0} \\ & 0 \end{pmatrix}}$ is in $\mathcal{A}$ or not.
    
    \paragraph{Case A.} First, assume that no element of the form ${\tiny \begin{pmatrix} 0 & \highg{\neq 0} \\ & 0 \end{pmatrix}}$ is in $\mathcal{A}$. For each possible type of algebra, this assumption implies that
    \begin{align*}
        \text{Type } AA: \ 
    \mathcal{A} &= {\scriptsize \left\{ \begin{pmatrix} A & f(A) \\ & A \end{pmatrix} \mid A \right\} }, \quad 
    \text{Type } AB: \
    \mathcal{A} = {\scriptsize \left\{ \begin{pmatrix} A & f(A)+g(B) \\ & B \end{pmatrix} \mid A, B \right\} }, \nonumber \\
    \text{Type } A0: \ 
    \mathcal{A} &= {\scriptsize \left\{ \begin{pmatrix} A & f(A) \\ & 0 \end{pmatrix} \mid A \right\} }, \quad
    \text{Type } 0A: \
    \mathcal{A} = {\scriptsize \left\{ \begin{pmatrix} 0 & g(A) \\ & A \end{pmatrix} \mid A \right\} }, \nonumber \\
    \text{Type } 00: \ 
    \mathcal{A} &= {\scriptsize \left\{ \begin{pmatrix} 0 & 0 \\ & 0 \end{pmatrix} \right\} },
    \end{align*}
    for some functions $f, g$ that have to be linear due to closedness of $\mathcal{A}$ under linear combinations. Closedness under multiplication implies that 
    \begin{equation*}
        {\scriptsize \begin{pmatrix} A & f(A) \\ & A \end{pmatrix}
        \begin{pmatrix} B & f(B) \\ & B \end{pmatrix} =
        \begin{pmatrix} AB & Af(B) + f(A) B \\ & AB \end{pmatrix}} \in \mathcal{A} , 
    \end{equation*}
    so it must necessarily hold that $A f(B) + f(A) B = f(AB), \ \forall A, B$. That is, for each algebra type, the following relations must necessarily hold for all $A, B$,
    \begin{align*}
        \text{Type } AA: \ 
        f(AB) &= A f(B) + f(A) B, \\
        \text{Type } AB: \
        f(AB) &= Af(B), \ g(AB) = g(A)B, \ 0 = Ag(B) + f(A)B,  \\
        \text{Type } A0: \ 
        f(AB) &= Af(B), \\
        \text{Type } 0A: \
        g(AB) &= g(A)B .
    \end{align*}
    We now employ a series of technical lemmas to construct the required basis transformation that eliminates the off-diagonal block. In particular, we apply Lemma \ref{lemma:technical2} for type AA, Lemma \ref{lemma:technical1} for type AB, Corollary \ref{cor:technical1} for type A0, and Corollary \ref{cor:technical2} for type 0A. In each case, these results provide an explicit invertible matrix $P$ (depending on $f,g$) that get rid of the corresponding off-diagonal block, so that
    \begin{equation*}
        P\mathcal{A} P^{-1}  = \left\{
        {\scriptsize \begin{pmatrix} A & 0 \\ & A \end{pmatrix}} \mid A \right\}, \ 
        \left\{
        {\scriptsize \begin{pmatrix} A & 0 \\ & B \end{pmatrix}} \mid A,B \right\}, \
        \left\{
        {\scriptsize \begin{pmatrix} A & 0 \\ & 0 \end{pmatrix}} \mid A \right\} \ \text{ or } \
        \left\{
        {\scriptsize \begin{pmatrix} 0 & 0 \\ & A \end{pmatrix}} \mid A \right\} ,
    \end{equation*}
    respectively.
    
    \paragraph{Case B.} Now, assume that there is an element of the form ${\tiny \begin{pmatrix} 0 & \highg{Z} \\ & 0 \end{pmatrix}}$ in $\mathcal{A}$ with $Z \neq 0$, which wlog can be taken to be $Z = \dyad{j}{k} + \sum_{(\alpha,\beta)\neq(j,k)} z_{\alpha\beta} \dyad{\alpha}{\beta}$. 
    
    For type AA and AB algebras, since $\mathcal{A}$ is closed under multiplication, for all values of $i,l$ we have that the following element should be contained in $\mathcal{A}$:
    \begin{equation*}
        {\footnotesize \begin{pmatrix} \dyad{i}{j} & \ast \\ & \ast \end{pmatrix} 
        \begin{pmatrix} 0 & \dyad{j}{k} + \dots \\ & 0 \end{pmatrix}
        \begin{pmatrix} \ast & \ast \\ & \dyad{k}{l} \end{pmatrix} = 
        \begin{pmatrix} 0 & \dyad{i}{l} \\ & 0 \end{pmatrix}}.
    \end{equation*}
    This implies that the full block $(1, 2)$ belongs in $\mathcal{A}$ and is independent of the diagonal. Note that, since the elements denoted by an asterisk are not involved in the computation, the same argument is valid both for type AA and AB algebras.

    For type A0, the 0 block has size $1 \times 1$, and thus $Z = \dyad{j}{1} + \dots$. In that case,
    \begin{equation*}
        {\small \begin{pmatrix} \dyad{i}{j} & \ast \\ & 0 \end{pmatrix} 
        \begin{pmatrix} 0 & \dyad{j}{1} + \dots \\ & 0 \end{pmatrix} = 
        \begin{pmatrix} 0 & \dyad{i}{1} \\ & 0 \end{pmatrix}} \in \mathcal{A},
    \end{equation*}
    and hence block $(1,2)$ is the full free block, independent of the diagonal. In the same way, for type 0A, we have $Z = \dyad{1}{k} + \dots$, and then
    \begin{equation*}
        {\small 
        \begin{pmatrix} 0 & \dyad{1}{k} + \dots \\ & 0 \end{pmatrix} 
        \begin{pmatrix} 0 & \ast \\ & \dyad{k}{l} \end{pmatrix}
        = 
        \begin{pmatrix} 0 & \dyad{1}{l} \\ & 0 \end{pmatrix}} \in \mathcal{A},
    \end{equation*}
    so block $(1,2)$ is again the full free block, independent of the diagonal. For type 00, the off-diagonal block has size $1 \times 1$ by definition, and thus it is also a full free block. This concludes the proof of the claim.
\end{proof}

\paragraph{Example.} Let us illustrate the lemma above by considering
\begin{align*}
    \mathcal{A}^{(1)} &= \text{span}\left\{  
    {\footnotesize \begin{pmatrix}
        1 & 0 & 0 & b \\ 
        0 & 0 & -c & 0 \\
        & & 1 & 0 \\
        & & 0 & 0
    \end{pmatrix}, 
    \begin{pmatrix}
        0 & 1 & c & a \\ 
        0 & 0 & 0 & -c \\
        & & 0 & 1 \\
        & & 0 & 0
    \end{pmatrix},
    \begin{pmatrix}
        0 & 0 & -b & 0 \\ 
        1 & 0 & -a & b \\
        & & 0 & 0 \\
        & & 1 & 0
    \end{pmatrix},
    \begin{pmatrix}
        0 & 0 & 0 & -b \\ 
        0 & 1 & c & 0 \\
        & & 0 & 0 \\
        & & 0 & 1
    \end{pmatrix}}
    \right\} \\
    &=: \{ e_{11}, e_{12}, e_{21}, e_{22} \}
\end{align*}
for some $a, b, c \in \mathbb{C}$.
One can readily check that $\mathcal{A}^{(1)}$ is closed under multiplication, so $\text{Alg}(\mathcal{A}^{(1)}) = \mathcal{A}$ and $\mathcal{A}$ is a type AA algebra. Now, using the explicit construction of $P$ of Lemma \ref{lemma:technical2}, we have
\begin{equation*}
    P = {\footnotesize \begin{pmatrix}
        1 & 0 & -f^{11}_{11} & -f^{2\ast}_{1\ast} \\ 
        0 & 1 & -f^{1\ast}_{2\ast} & -f^{21}_{21} \\
        & & 1 & 0 \\
        & & 0 & 1
    \end{pmatrix} = 
    \begin{pmatrix}
        1 & 0 & 0 & b \\ 
        0 & 1 & c & a \\
        & & 1 & 0 \\
        & & 0 & 1
    \end{pmatrix}} \implies 
    P\mathcal{A} P^{-1}  = \left\{ {\scriptsize \begin{pmatrix} A & 0 \\ & A \end{pmatrix}} \mid A \right\} . 
\end{equation*}

The previous example shows how the change of basis can be constructed for algebras with a structure of 2 blocks by 2 blocks. This serves as the fundamental step that can be iterated to treat arbitrarily complicated algebras, as stated in the following proposition. Recall that a block at position $(m,n)$ belongs to sector $[i,j]$ for some $i,j \in \Sigma_\infty$ if $(m,m)$ belongs to the equivalence class of diagonal free blocks labeled by $i$, and $(n,n)$ to the equivalence class labeled by $j$.

\propstructurealgebra*
\begin{proof}
    Part 1 of the statement follows directly from block-upper-triangularizing the set of matrices generating the algebra and applying the standard theory of semisimple algebras \cite{Farenick_2001, pierce_1982_associative-algebras}, as discussed in Section \ref{sec:algebra-span-thmbasis} of the main text.
    
    For part 2, we proceed in two steps. First, we construct a change-of-basis matrix $P$ such that in $P \mathcal{A}P^{-1}$, all off-diagonal blocks are either zero or fully free with respect to the diagonal blocks. We refer to this structural feature as the \textit{dichotomy property} in the rest of the proof. Then, we show that dependencies among off-diagonal blocks can only arise through linear combinations.

    \paragraph{Step 1 (Construction of $P$ to achieve the dichotomy property).}
    Assume that $\mathcal{A}$ already has the block-upper-triangular form described by part 1 of the statement. Our goal is to construct $P$ iteratively as
    \begin{equation*}
        P = P^{(b)} \dots P^{(2)} P^{(1)},
    \end{equation*}
    where each $P^{(n)}$ acts to modify the $n$-th diagonal so that it satisfies the dichotomy property, without affecting the lower diagonals.

    We begin with $P^{(1)} := \mathds{1} + \tilde{P}^{(1)}$, where $\tilde{P}^{(1)}$ is a nilpotent, block-upper-triangular matrix with nonzero blocks only at positions $(i,i+1)$, for $i \in \{1, \dots, b-1\}$. Each such block, $\tilde{P}^{(1)}_{i,i+1} \in \mathcal{M}_{D_i \times D_{i+1}}(\mathbb{C})$, is chosen as the gauge prescribed by Lemma \ref{lemma:isolate_subalgebra} when applied to the submatrix 
    \begin{equation*}
        {\footnotesize \begin{pmatrix} \mathcal{A}_{ii} & \mathcal{A}_{i,i+1} \\ 0 & \mathcal{A}_{i+1,i+1} \end{pmatrix}} .
    \end{equation*}
    This choice ensures that each block at $(i,i+1)$ is transformed into either zero or a full block independent of the diagonal. Therefore, if we let $\mathcal{A}_{(1)} := P^{(1)} \mathcal{A} (P^{(1)})^{-1}$, we have by construction that the blocks in the first diagonal of $\mathcal{A}_{(1)}$ satisfy the dichotomy property. 

    We now proceed inductively. Suppose that, for some $n \geq 2$, the first $n-1$ diagonals of blocks in $\mathcal{A}_{(n-1)}$ already satisfy the dichotomy property. We define $P^{(n)} := \mathds{1} + \tilde{P}^{(n)}$, where $\tilde{P}^{(n)}$ is zero everywhere except for blocks at positions $(i,i+n)$, which are again chosen using Lemma \ref{lemma:isolate_subalgebra} applied to the submatrix
    \begin{equation*}
        {\footnotesize \begin{pmatrix} \mathcal{A}_{ii} & \mathcal{A}_{i,i+n} \\ 0 & \mathcal{A}_{i+n,i+n} \end{pmatrix}} .
    \end{equation*}
    Then, performing the transformation
    \begin{equation*}
        \mathcal{A}_{(n)} := P^{(n)} \mathcal{A}_{(n-1)} (P^{(n)})^{-1}
    \end{equation*}
    modifies the blocks at the $n$-th diagonal to ensure the dichotomy property. All previously processed diagonals remain unchanged due to the structure of $P^{(n)}$, while higher diagonals may change but will be handled in subsequent iterations. 

    Continuing this process up to $n = b-1$, we obtain the final change of basis $P = P^{(b-1)} \dots P^{(2)} P^{(1)}$ such that the transformed algebra $\mathcal{A}_{(b-1)} := P\mathcal{A}P^{-1}$ satisfies the dichotomy property on all off-diagonal blocks.

    \paragraph{Step 2 (Relations between off-diagonal blocks).} Assume that the procedure in Step 1 has already been carried out, so that all off-diagonal blocks in $\mathcal{A}$ satisfy the dichotomy property. 

    Let us consider the off-diagonal block at position $(i,j)$, which can either be free or not. If it is not free, it means that there must exist some set of free off-diagonal blocks at $(m_1, n_1), \dots, (m_l, n_l) \preceq (i,j)$ such that
    \begin{equation*}
        \mathcal{A}_{ij} = f(\mathcal{A}_{m_1 n_1}, \dots, \mathcal{A}_{m_l n_l}),
    \end{equation*}
    meaning that all elements in the algebra $\mathcal{A}$ should satisfy this relation. Because $\mathcal{A}$ is closed under linear combinations, we can rewrite the expression above as
    \begin{equation*}
        \mathcal{A}_{ij} = \sum_{\gamma=1}^l f_\gamma (\mathcal{A}_{m_\gamma n_\gamma}),
    \end{equation*}
    for some linear maps $f_\gamma$.

    Fix one of the free blocks, say $\mathcal{A}_{m_p n_p}$, and consider the corresponding map $f_p$. We will show that $f_p(A) = k_p A$ for some scalar $k_p \in \mathbb{C}$, and that $k_p \neq 0$ implies $(i,j)\sim (m_p, n_p)$, i.e. both blocks belong to the same sector. Let $a, b \in \mathcal{A}$ where:
    \begin{itemize}
        \item $a$ is a purely block-diagonal element,
        \item $b$ satisfies $b_{m_p n_p} \neq 0$ and $b_{m_\gamma n_\gamma} = 0$ for all $\gamma \neq p$, meaning that $b_{ij} = f_p(b_{m_p n_p})$.
    \end{itemize}
    Now consider the product $ab$, and compute the $(i,j)$-block in two different ways.
    \begin{enumerate}
        \item Using block-matrix multiplication:
        \begin{equation*}
            (ab)_{ij} = \sum_{\alpha = i}^j a_{i\alpha} b_{\alpha j} = a_{ii} b_{ij} = a_{ii} f_p(b_{m_p n_p}).
        \end{equation*}

        \item Using closedness under multiplication and the assumed dependency of $\mathcal{A}_{ij}$:
        \begin{align*}
            (ab)_{ij} = \sum_{\gamma=1}^l f_\gamma ((ab)_{m_\gamma n_\gamma}) = \sum_{\gamma=1}^l f_\gamma (a_{m_\gamma m_\gamma} b_{m_\gamma n_\gamma}) = f_p (a_{m_p m_p} b_{m_p n_p}).
        \end{align*}
    \end{enumerate}
    Equating both expressions gives
    \begin{equation} \label{eq:structurealgebra_aux1}
        a_{ii} f_p(b_{m_p n_p}) = f_p (a_{m_p m_p} b_{m_p n_p}).
    \end{equation}
    Two cases can occur:
    \begin{itemize}
        \item \textbf{Case 1:}  $\mathcal{A}_{ii}$ and $\mathcal{A}_{m_p m_p}$ are independent, i.e. $(i,i) \nsim (m_p,m_p)$. Then, we can choose $a$ and $b$ with the additional properties that $a_{ii} = 0$, $a_{m_p m_p} = \mathds{1}$, and $b_{m_p n_p} = B$ for any arbitrary $B$. Eq. \eqref{eq:structurealgebra_aux1} then becomes 
        \begin{equation*}
            f_p(B) = 0 \ , \quad \forall B .
        \end{equation*}

        \item \textbf{Case 2:} $\mathcal{A}_{ii} = \mathcal{A}_{m_p m_p}$. Then, for arbitrary $A, B$, we can take $a_{ii} = a_{m_p m_p} = A$, and $b_{m_p n_p} = B$, and the relation becomes
        \begin{equation} \label{eq:structurealgebra_aux2}
            A f_p(B) = f_p(AB) \ , \quad \forall A, B. 
        \end{equation}
    \end{itemize}

    Similarly, if we reverse the roles of $a$ and $b$, we would conclude that $f_p = 0$ when $\mathcal{A}_{jj}$ and $\mathcal{A}_{n_p n_p}$ are independent, and otherwise the relation
    \begin{equation} \label{eq:structurealgebra_aux3}
        f_p(A)B = f_p(AB) \ , \quad \forall A, B. 
    \end{equation}
    
    The two identities in Eq. \eqref{eq:structurealgebra_aux2} and Eq. \eqref{eq:structurealgebra_aux3} fit into the conditions of Lemma \ref{lemma:technical3}, which implies that $f_p(A) = k_pA$ for some scalar $k_p \in \mathbb{C}$. Moreover, as we just explained, $k_p \neq 0$ requires that $\mathcal{A}_{ii} = \mathcal{A}_{m_p m_p}$ and $\mathcal{A}_{jj} = \mathcal{A}_{n_p n_p}$, i.e. $(i,j) \sim (m_p, n_p)$. In other words, $(i,j)$ must be in the same sector as $(m_p, n_p)$. This concludes the proof of Step 2, and therefore of the proposition.
\end{proof}

\paragraph{Example.} We now illustrate the procedure described in the proof above to bring a block-upper-triangular matrix algebra into a form with the properties described in Proposition \ref{prop:structure_subalgebra}.

Consider the algebra
\begin{equation*} 
\mathcal{A} = { \scriptsize \left\{
    \begin{pmatrix}
    A & f(A) & h(A) &   C  \\
      &   A  & g(A) & i(A,B) \\
      &      &   A  &   0  \\
      &      &      &   B 
    \end{pmatrix} \mid A, B, C \right\} }
\end{equation*}

Our goal is to apply the iterative procedure from Step 1 of the proof, transforming $\mathcal{A}$ via the change of basis $P = P^{(3)} P^{(2)} P^{(1)}$ such that in $P \mathcal{A} P^{-1}$, all off-diagonal blocks are either zero or independent of the diagonal. 
\begin{itemize}
\item \textbf{Processing the blocks of the 1st diagonal.}
Due to the closedness of $\mathcal{A}$ under product and linear combinations, the maps $f, g$ satisfy the assumptions of Lemma \ref{lemma:technical2}, which gives us the explicit $P_f$ and $P_g$ such that conjugation by
\begin{equation*}
    P^{(1)} := {\scriptsize
    \begin{pmatrix}
        \mathds{1} & P_f & 0 & 0 \\
          & \mathds{1} & P_g & 0 \\
          &      &   \mathds{1} &  0 \\
          &      &      &   \mathds{1} 
    \end{pmatrix}}
\end{equation*}
eliminates $\mathcal{A}_{12}$ and $\mathcal{A}_{23}$. That is, after conjugation, we obtain 
\begin{equation*}
    \mathcal{A}_{(1)} 
    = P^{(1)} \mathcal{A} (P^{(1)})^{-1}
    = \left\{ {\scriptsize\begin{pmatrix}
    A & 0 & \tilde{h}(A) & C  \\
      &   A  &  0 & i(A,B) \\
      &      &   A  &   0  \\
      &      &      &   B \end{pmatrix}} \mid A, B, C \right\},
\end{equation*}
where
\begin{equation*}
    \tilde{h}(A) := h(A) + P_f g(A) - f(A) P_g + [A,P_f] P_g = h(A) + P_f[A,P_g] \ .
\end{equation*}

\item \textbf{Processing the blocks of the 2nd diagonal.}
We now eliminate the remaining second-diagonal blocks. Again, maps $\tilde{h}$ and $i$ satisfy the assumptions of Lemma \ref{lemma:technical2} and Lemma \ref{lemma:technical1}, respectively. Therefore, we have the explicit form of $P_{\tilde{h}}$ and $P_i$ such that conjugation by
\begin{equation*}
    P^{(2)} := {\scriptsize
    \begin{pmatrix}
        \mathds{1} & 0 & P_{\tilde{h}} & 0 \\
          & \mathds{1} & 0 & P_i \\
          &      &   \mathds{1} &  0 \\
          &      &      &   \mathds{1} 
    \end{pmatrix}}
\end{equation*}
removes the blocks at $(1,3)$ and $(2,4)$. After conjugation, we obtain 
\begin{equation*}
    \mathcal{A}_{(2)} 
    = P^{(2)} \mathcal{A}_{(1)} (P^{(2)})^{-1}
    = \left\{ {\scriptsize\begin{pmatrix}
    A & 0 & 0 & C  \\
      &   A  &  0 & 0 \\
      &      &   A  &   0  \\
      &      &      &   B \end{pmatrix}} \mid A, B, C \right\}.
\end{equation*}

\item \textbf{Processing the blocks of the 3rd diagonal.}
Block $(1,4)$ is already independent of the diagonal, so it satisfies the dichotomy property. No further transformation is needed here and we just take $P^{(3)} = \mathds{1}$. 
\end{itemize}
This completes the construction of a change of basis $P = P^{(3)} P^{(2)} P^{(1)}$ that brings $\mathcal{A}$ into the standard form for matrix algebras shown in Proposition \ref{prop:structure_subalgebra}. 

\paragraph{Example.} This example illustrates that the converse of Proposition \ref{prop:structure_subalgebra} is not true. In particular, even if a linear subspace of matrices $\mathcal{A}^{(1)}$ admits a basis with the properties stated in Proposition \ref{prop:structure_subalgebra}, $\mathcal{A}^{(1)}$ is not necessarily an algebra. Indeed, it may happen that $\mathcal{A}^{(1)} \subsetneq \mathrm{Alg}(\mathcal{A}^{(1)})$, since not every arbitrary choice of linear dependencies between off-diagonal blocks is closed under multiplication, as an algebra requires. Consider the following set of matrices,
\begin{equation*} 
\mathcal{A}^{(1)} = { \scriptsize \left\{
    \begin{pmatrix}
    A & B & B+C \\
      &   A  & C\\
      &      &   A
    \end{pmatrix} \mid A, B, C \right\} }
\end{equation*}
Take the following two elements:
\begin{equation*} 
    a := { \scriptsize \begin{pmatrix}
        0 & \dyad{1}{1} & \dyad{1}{1} \\
          &   0  & 0\\
          &      &   0
    \end{pmatrix} } , \
    b := { \scriptsize \begin{pmatrix}
        0 & 0 & \dyad{1}{1} \\
          &   0  & \dyad{1}{1} \\
          &      &   0
    \end{pmatrix} }    
\end{equation*}
When we compute their product, we obtain 
\begin{equation*}
    ab = { \scriptsize \begin{pmatrix}
        0 & 0  & \dyad{1}{1} \\
          &   0  & 0\\
          &      &   0
    \end{pmatrix} }
\end{equation*}
and therefore we see that, even though $a, b \in \mathcal{A}^{(1)}$, we have $ab \notin \mathcal{A}^{(1)}$. This means that $\mathcal{A}^{(1)}$ is not closed under multiplication, and thus it is not an algebra even though it satisfies the properties of Proposition \ref{prop:structure_subalgebra}. In fact, 
\begin{equation*}
    \mathrm{Alg}(\mathcal{A}^{(1)}) = 
    { \scriptsize \left\{
    \begin{pmatrix}
    A & B & D \\
      &   A  & C\\
      &      &   A
    \end{pmatrix} \mid A, B, C, D \right\} }
    \supsetneq \mathcal{A}^{(1)}
\end{equation*}

\newpage

\section{Proofs for the span structure in Theorem \ref{prop:structure_span}}
\label{app:span_structure}

In this section, we provide a proof of Theorem \ref{prop:structure_span} from Section \ref{sec:algebra-span} of the main text. This result establishes that, after sufficient blocking, the span $\mathcal{A}^{(\ell)}$ of any given set of matrices admits a simplified structure in a suitable basis. As in the case of algebras, we begin by analyzing matrices with a block-upper-triangular structure consisting of $2$ blocks by $2$ blocks, and then extend the result to the general case via an inductive argument.

\subsection{Base case of the induction ($b = 2)$}

We assume that the initial set of matrices, whose span is denoted by $\mathcal{A}^{(1)}$, has been \textit{(i)} blocked sufficiently many times to eliminate any periodic subspaces in its diagonal part, and \textit{(ii)} expressed in a basis where the diagonal blocks are either proportional or independent from each other.
As guaranteed by the standard theory of PBC MPS \cite{Perez-Garcia2007, cirac_2017_mpdo}, such a reduction can always be achieved by blocking every $p\ell$ sites, where $p$ is the least common multiple of the periods of all diagonal periodic subspaces, and for any $\ell$ larger than a certain threshold $L_{\mathrm{BI}}^{\mathrm{diag}}$, where $L_{BI}^{\mathrm{diag}} \leq 3D^5$.

Therefore, the structure of the span $\mathcal{A}^{(\ell)}$ for $\ell \geq L_{\mathrm{BI}}^{\mathrm{diag}}$ falls into one of the following five types, depending on the form of the diagonal blocks:
\begin{align}
    \text{Type } AA: \ 
    \mathcal{A}^{(\ell)} &= {\scriptsize \left\{ \begin{pmatrix} A & \ast \\ & \omega^\ell A \end{pmatrix} \mid A \right\} }, \quad 
    \text{Type } AB: \
    \mathcal{A}^{(\ell)} = {\scriptsize \left\{ \begin{pmatrix} A & \ast \\ & B \end{pmatrix} \mid A, B \right\} }, \label{eq:def_types_span} \\
    \text{Type } A0: \ 
    \mathcal{A}^{(\ell)} &= {\scriptsize \left\{ \begin{pmatrix} A & \ast \\ & 0 \end{pmatrix} \mid A \right\} }, \quad
    \text{Type } 0A: \
    \mathcal{A}^{(\ell)} = {\scriptsize \left\{ \begin{pmatrix} 0 & \ast \\ & A \end{pmatrix} \mid A \right\} }, \nonumber \\
    \text{Type } 00: \ 
    \mathcal{A}^{(\ell)} &= {\scriptsize \left\{ \begin{pmatrix} 0 & \ast \\ & 0 \end{pmatrix} \mid A \right\} }, \nonumber
\end{align}
where $0$ denotes a $1 \times 1$ block and $\omega \in \mathbb{C} \setminus \{ 0 \}$. Note that, for type 00, $\mathcal{A}^{(s)} = 0$ for all $s \geq 2$.

With the following lemma, we establish Theorem \ref{prop:structure_subalgebra} in the special case of 2 blocks by 2 blocks. This case will serve as the foundation to address the general setting in the next subsection.
\begin{lemma} \label{lemma:basis_case_induction}
    Assume, without loss of generality, that after blocking $\mathcal{A}^{(1)}$ can be written in a basis without periodic subspaces and with diagonal blocks that are either proportional or independent. Then, we can explicitly construct an invertible $P$ such that, for all $\ell \geq 4L_{BI}^{\text{diag}} + 2L_0^{\text{diag}}$, the off-diagonal block in $P \mathcal{A}^{(\ell)} P^{-1}$ takes one of the following forms:
    \begin{enumerate}[label = (\roman*)]
        \item Zero.
        \item A free block, independent of the diagonal.
        \item A block proportional to the diagonal one, $\mathcal{A}_{12}^{(\ell)} = k_{\ell} \mathcal{A}_{11}^{(\ell)}$, where $k_{\ell} := k \left( \sum_{i = 0}^{\ell-1} \omega^i \right)$ for some $k, \omega \in \mathbb{C}$. We refer to this as a \emph{generalized Jordan block}. 
    \end{enumerate}
\end{lemma}
\begin{proof}
    We divide the proof into two cases, depending on whether $m_{12} \leq 4 L_{BI}^{\text{diag}}$ or $m_{12} > 4L_{BI}^{\text{diag}}$. Recall that $m_{12}$ was defined in Appendix \ref{app:notation_and_basis} as the minimal blocking length $\ell$ for which a matrix of the form ${\tiny \begin{pmatrix} 0 & \highg{\neq 0} \\ & 0 \end{pmatrix}}$ appears in $\mathcal{A}^{(\ell)}$, i.e. such an element only belongs in $\mathcal{A}^{(\ell)}$ for $\ell \geq m_{12}$.
    
    \paragraph{Step A.} Assume that $m_{12} > 4L_{BI}^{\text{diag}}$. This would necessarily mean that, for each $s$ with $L_{BI}^{\text{diag}} \leq s < m_{12}$, we have
    \begin{align*}
        \text{Type } AA: \ 
        \mathcal{A}^{(s)} &= {\scriptsize \left\{ \begin{pmatrix} A & f^{(s)}(A) \\ & \omega^s A \end{pmatrix} \mid A \right\} }, \quad 
        \text{Type } AB: \
        \mathcal{A}^{(s)} = {\scriptsize \left\{ \begin{pmatrix} A & f^{(s)}(A) + g^{(s)}(B) \\ & B \end{pmatrix} \mid A, B \right\} }, \nonumber \\
        \text{Type } A0: \ 
        \mathcal{A}^{(s)} &= {\scriptsize \left\{ \begin{pmatrix} A & f^{(s)}(A) \\ & 0 \end{pmatrix} \mid A \right\} }, \quad
        \text{Type } 0A: \
        \mathcal{A}^{(s)} = {\scriptsize \left\{ \begin{pmatrix} 0 & g^{(s)}(A) \\ & A \end{pmatrix} \mid A \right\} }, \nonumber \\
        \text{Type } 00: \ 
        \mathcal{A}^{(s)} &= {\scriptsize \left\{ \begin{pmatrix} 0 & 0 \\ & 0 \end{pmatrix} \right\} }, \label{eq:def_types_span}
    \end{align*}
    for some functions $f^{(s)}, g^{(s)}$ which have to be linear due to closeness of $\mathcal{A}^{(s)}$ under linear combinations. 
    
    Even though these sets are not necessarily algebras, and therefore they are not closed under multiplication, the facts that $\mathcal{A}^{(2L_{BI}^{\text{diag}})} = \text{span}\{a b \mid a, b \in \mathcal{A}^{(L_{BI}^{\text{diag}})}\}$ and $\mathcal{A}^{(4L_{BI}^{\text{diag}})} = \text{span}\{a b \mid a, b \in \mathcal{A}^{(2L_{BI}^{\text{diag}})}\}$ imply that the following relations must necessarily hold for all $A, B$:
    \begin{align*}
        \text{Type } AA :& \ 
        {\small \begin{cases}
            f^{(2L_{BI}^{\text{diag}})}(AB) = A f^{(L_{BI}^{\text{diag}})}(B) + \omega^{L_{BI}^{\text{diag}}} f^{(L_{BI}^{\text{diag}})}(A) B, \\
            f^{(4L_{BI}^{\text{diag}})}(AB) = A f^{(2L_{BI}^{\text{diag}})}(B) + \omega^{2L_{BI}^{\text{diag}}} f^{(2L_{BI}^{\text{diag}})}(A) B,
        \end{cases}}  \\ 
        \text{Type } AB :& \ 
        {\small \begin{cases}
            f^{(2L_{BI}^{\text{diag}})}(AB) = A f^{(L_{BI}^{\text{diag}})}(B), \\
            g^{(2L_{BI}^{\text{diag}})}(AB) = g^{(L_{BI}^{\text{diag}})}(A) B,  \\
            A g^{(L_{BI}^{\text{diag}})}(AB) + f^{(L_{BI}^{\text{diag}})}(A) B = 0, 
        \end{cases}} \\
        \text{Type } A0 :& \
        f^{(2L_{BI}^{\text{diag}})}(AB) = A f^{(L_{BI}^{\text{diag}})}(B), \\
        \text{Type } 0A :& \
        g^{(2L_{BI}^{\text{diag}})}(AB) = g^{(L_{BI}^{\text{diag}})}(A) B .
    \end{align*}
    Then, we can apply Lemma \ref{lemma:technical2} for type AA, Lemma \ref{lemma:technical1} for type AB, Corollary \ref{cor:technical1} for type A0, and Corollary \ref{cor:technical2} for type 0A, to explicitly construct an invertible $P$ that eliminates the off-diagonal block, meaning that
    \begin{align*}
        \text{Type } AA :& \ 
        P\mathcal{A}^{(2L_{BI}^{\text{diag}})}P^{-1} =
        \left\{  {\scriptsize
        \begin{pmatrix}
            A & k^{(2L_{BI}^{\text{diag}})} A \\ & \omega^{2L_{BI}^{\text{diag}}} A
        \end{pmatrix}} \mid A \right\}, \\
        \text{Type } AB, \ A0 \text{ or } 0A :& \ 
        P\mathcal{A}^{(L_{BI}^{\text{diag}})}P^{-1} = \left\{  {\scriptsize
        \begin{pmatrix}
            A & 0 \\ & B
        \end{pmatrix}} \mid A, B \right\}, \ 
        \left\{  {\scriptsize
        \begin{pmatrix}
            A & 0 \\ & 0
        \end{pmatrix}} \mid A \right\} 
        \ \text{ or } \ 
        \left\{  {\scriptsize
        \begin{pmatrix}
            0 & 0 \\ & A
        \end{pmatrix}} \mid A \right\}.
    \end{align*}
    Having established the structure for the span for a single blocking length, we can already extend the statement to all blocking lengths beyond the threshold $L_{BI}^{\mathrm{diag}}$, by invoking Lemmas \ref{lemma:same_form_AA_2x2} and \ref{lemma:same_form_AB_2x2}. Specifically, Lemma \ref{lemma:same_form_AA_2x2} shows that in Type AA the span always contains a single generalized Jordan block in the off-diagonal sector, for every blocking length. In contrast, Lemma \ref{lemma:same_form_AB_2x2} ensures that for Types AB, A0, and 0A no off-diagonal block ever appears. Consequently, we obtain $m_{12} = \infty$, and for every $\ell \geq L_{BI}^{\text{diag}}$ we have
    \begin{equation*}
        P\mathcal{A}^{(\ell)}P^{-1} = \left\{
        {\scriptsize
        \begin{pmatrix}
            A & k^{(\ell)} A \\ & \omega^{\ell} A
        \end{pmatrix}} \mid A \right\}, \
        \left\{  {\scriptsize
        \begin{pmatrix}
            A & 0 \\ & B
        \end{pmatrix}} \mid A, B \right\}, \ 
        \left\{  {\scriptsize
        \begin{pmatrix}
            A & 0 \\ & 0
        \end{pmatrix}} \mid A \right\} 
        \ \text{ or } \ 
        \left\{  {\scriptsize
        \begin{pmatrix}
            0 & 0 \\ & A
        \end{pmatrix}} \mid A \right\},
    \end{equation*}
    respectively, where $k^{(\ell)} = k \left( \sum_{i=0}^{\ell-1} \omega^i \right)$ for some $k, \omega \in \mathbb{C}$.

    \paragraph{Step B.} If $m_{12} \leq 4L_{BI}^{\text{diag}}$, meaning that there exists an element ${\tiny \begin{pmatrix} 0 & \highg{\neq 0} \\ & 0 \end{pmatrix}}$ in $\mathcal{A}^{(m_{12})}$, then by Lemma \ref{lemma:nonzero_elements} we know that block $(1,2)$ is a full free block independent of the diagonal in $\mathcal{A}^{(\ell)}$, for all $\ell \geq m_{12} + 2L_{0}^{\text{diag}}$. In particular, this is also true for all $\ell \geq 4L_{BI}^{\text{diag}} + 2L_0^{\text{diag}}$. Note that, even though Lemma \ref{lemma:nonzero_elements} does not apply for type 00 structure, the claim still holds: since $\mathcal{A}^{(1)}$ would only consist of nilpotent $2 \times 2$ matrices, block $(1,2)$ would vanish under blocking. Therefore, the proof is complete. 
\end{proof}

\subsection{Proof for the general case}

We now turn to the general setting of Theorem \ref{prop:structure_span}, where the matrices have a block-upper-triangular structure with $b$ blocks by $b$ blocks. The proof proceeds by induction, relying on Lemma \ref{lemma:basis_case_induction} established in the previous subsection.
\propstructurespan*
\begin{proof} 
    We proceed by induction, building a basis with the desired properties, one block at a time, in accordance with the $\preceq$–order. The base case for block $(1,2)$ was established in Lemma \ref{lemma:basis_case_induction}, upon blocking every $\ell \ge L_{1,2}$ sites with $L_{1,2} := 4L_{BI}^{\text{diag}} + 2L_0^{\text{diag}}$.

    For the inductive step, we assume as our \emph{inductive hypothesis} that a basis with the stated properties has already been constructed for all blocks strictly preceding $(i,j)$, i.e. for $\mathcal{A}^{(\ell)}|_{\prec (i,j)}$ with $\ell \geq L_{i-1,j-1}$. This means that we can employ the notation introduced in Appendix \ref{app:notation_and_basis} to characterize the existing basis elements up to block $\prec (i,j)$. Then, we show how to extend this basis to also describe block $(i,j)$, provided blocking $\ell \geq L_{i,j}$. 

    As in the proof of Lemma \ref{lemma:basis_case_induction}, we split the arguments according to the structure of the restricted span 
    \begin{equation*}
        \mathcal{A}^{(\ell)}|_{[i,j]} := 
        \left\{
            {\scriptsize \begin{pmatrix}
                \mathcal{A}_{ii}^{(\ell)} & \mathcal{A}_{ij}^{(\ell)} \\
                0 & \mathcal{A}_{jj}^{(\ell)}
            \end{pmatrix}}
        \right\},
    \end{equation*} 
    which admits five possible configurations depending on the form of the diagonal blocks: types AA, AB, A0, 0A and 00, as defined in Eq. \eqref{eq:def_types_span}. To simplify the notation, we will denote $L := L_{i-1,j-1}$.

    \paragraph{Roadmap.}
    To prove the statement by induction, we will proceed according to the following steps:
    \begin{enumerate}[label = \textbf{\Alph*.}]
        \item Assume that $m_{ij} > 4L$.
        \begin{enumerate}[label = \textbf{\arabic*}.]
            \item[\textbf{A.1}] Find the dependence of block $(i,j)$ on the \textit{diagonal} free blocks in $\Sigma_\infty$.
            
            \item[\textbf{A.2}] Find the dependence of block $(i,j)$ on the \textit{off-diagonal} free blocks in $\Sigma_f$.
        \end{enumerate}
        
        \item Assume that $m_{ij} \leq 4L$. Then, show that $(i,j)$ is a free block in $\mathcal{A}^{(\ell)}$ for all $\ell \geq 4L + 2L_0^{\text{diag}} =: L_{i,j}$.
    
        \item Argue that the constructed basis is still valid in $\mathcal{A}^{(\ell)}$, given $m_{ij} \leq \ell < m_{ij}+2L_0^{\text{diag}}$.

        \item Update the construction to extend the basis from $\mathcal{A}^{(\ell)}\mid_{\prec (i,j)}$ to $\mathcal{A}^{(\ell)}\mid_{\preceq (i,j)}$ for all $\ell > L_{i,j}$.
    \end{enumerate}

    \paragraph{Step A.} Assume that $m_{ij} > 4L$. Then, for every $\ell$ with $L \leq \ell < m_{ij}$, there must exist linear functions $f_e^{(\ell)}$ indexed by $e \in \Sigma$, such that for any $a \in \mathcal{A}^{(\ell)}$, expressed in the basis provided by the inductive hypothesis up to block $\prec (i,j)$ as $a |_{\prec (i,j)} = \sum_{e \in \Sigma} [A_e]_e^{(\ell)}$, the $(i,j)$-th block of $a$ satisfies
    \begin{equation} \label{eq:step-A_form_ij}
        a_{ij} = \sum_{\{0,s\} \in \Sigma_\infty} f^{(\ell)}_{\{0,s\}}(A_{\{0,s\}}) + \sum_{\{t\} \in \Sigma_f} f^{(\ell)}_{\{t\}}(A_{\{t\}}).
    \end{equation}

    \paragraph{Step A.1.} \textit{Find the dependence of block $(i,j)$ on the \textit{diagonal} free blocks in $\Sigma_\infty$.}
    
    The idea here mirrors that of the base case in Lemma \ref{lemma:basis_case_induction}. Although the spans $\mathcal{A}^{(\ell)}$ are not closed under multiplication, the facts that $\mathcal{A}^{(2L)} = \text{span}\{a b \mid a, b \in \mathcal{A}^{(L)}\}$ and $\mathcal{A}^{(4L)} = \text{span}\{a b \mid a, b \in \mathcal{A}^{(2L)}\}$ still impose constraints that we will exploit to derive the desired conclusions.

    Note that the $(i,j)$-th block of the product of any two basis elements of $\Sigma_\infty$, 
    \begin{equation*}
        a = [A]^{(\ell_1)}_{\{0,T_1\}}, \quad 
        b = [B]^{(\ell_2)}_{\{0,T_2\}},
    \end{equation*}
    with $\ell_1, \ell_2 \geq L$ and $\ell_1 + \ell_2 < m_{ij}$, can be written in two different ways:
    \begin{enumerate}
        \item Using block-matrix multiplication:
        \begin{align}
            (ab)_{ij} &= \sum_{\gamma=i}^{j} a_{i \gamma} b_{\gamma j}
            = a_{ii} b_{ij} + a_{ij} b_{jj} + \sum_{\gamma=i+1}^{j-1} a_{i \gamma} b_{\gamma j} \nonumber \\
            &= k_{ii;\{0,T_1\}}^{(\ell_1)} A f_{\{0,T_2\}}^{(\ell_2)}(B)
            + k_{jj;\{0,T_2\}}^{(\ell_2)} f_{\{0,T_1\}}^{(\ell_1)}(A) B 
            + \left( \sum_{\gamma=i+1}^{j-1} k_{i \gamma; \{0,T_1\}}^{(\ell_1)} k_{\gamma j;\{0,T_2\}}^{(\ell_2)} \right) AB ,
            \label{eq:blockinj_aux1_1} 
        \end{align}
        where we have used the inductive hypothesis for $a_{i\gamma}$ and $b_{\gamma j}$, since $(i,\gamma), (\gamma, j) \prec (i,j)$.

        \item Using the fact that $ab \in \mathcal{A}^{(\ell_1 + \ell_2)}$ with $\ell_1 + \ell_2 < m_{ij}$, which means block $(i,j)$ should still have the form of Eq. \eqref{eq:step-A_form_ij}:
        \begin{align}
            (ab)_{ij} &=
            \left[ \sum_{e \in \Sigma} f_{e}^{(\ell_1+\ell_2)} \left( \left( \Gamma^{(\ell_1, \ell_2)}\right)^{\{0,T_1\},\{0,T_2\}}_{e} AB \right) \right]_{ij} 
            = 
            \delta_{T_1,T_2} f_{\{0,T_1\}}^{(\ell_1+\ell_2)}\left( \underbrace{\left(\Gamma^{(\ell_1, \ell_2)}\right)^{\{0,T_1\},\{0,T_1\}}_{\{0,T_1\}}}_{=1} AB \right) \nonumber \\
            &= \begin{cases}
                f_{\{0,T_1\}}^{(\ell_1+\ell_2)}( AB ), & \text{if } \{0,T_1\} = \{0,T_2\}, \\
                0, & \text{otherwise} .
            \end{cases}
            \label{eq:blockinj_aux1_2}
        \end{align}
        Here we used property (P1) of $\Gamma^{(\ell_1,\ell_2)}$ introduced in Section \ref{app:notation_and_basis}.
    \end{enumerate}

    We now subdivide the proof of Step A.1 according to the type of structure of the diagonal blocks $(i,i)$ and $(j,j)$:
    \begin{itemize}
        \item \textbf{A.1.1.} For type AA structure, we show $f_{\{0,T\}}^{(\ell)}(A) = k_{ij;\{0,T\}}^{(\ell)} A$ for all $T$, given $2L \leq \ell < m_{ij}$, where $k_{ij;\{0,T\}}^{(\ell)} = 0$ if $\{0,T\} \neq \{0,r_i\}$. We do it in two steps:
        \begin{itemize}
            \item \textbf{A.1.1.i.} Block $(i,j)$ can only depend on free block $\{0,r_i\}$ by being proportional to it. Equivalently, $f_{\{0,r_i\}}^{(\ell)}(A) = k_{ij;\{0,r_i\}}^{(\ell)} A$ for some $k_{ij;\{0,r_i\}}^{(\ell)}\in\mathbb{C}$, given $2L \leq \ell < m_{ij}$.
            
            \item \textbf{A.1.1.ii.} Block $(i,j)$ cannot depend on any other diagonal free block apart from $\{0,r_i\}$. Equivalently, $f_{\{0,T\}}^{(\ell)}(A) = 0$ for all $\{0,T\} \neq \{0,r_i\}$, given $2L \leq \ell < m_{ij}$.
        \end{itemize}
        
        \item \textbf{A.1.2.} For type AB structure, we show $f_{\{0,T\}}^{(\ell)}(A) = 0$ for all $T$, given $L \leq \ell < m_{ij}$. We do it in two steps:
        \begin{itemize}
            \item \textbf{A.1.2.i.} Block $(i,j)$ cannot depend on free blocks $\{0,r_i\}$ nor $\{0,r_j\}$ in $\mathcal{A}^{(\ell)}$, given $L \leq \ell < m_{ij}$.
            
            \item \textbf{A.1.2.ii.} Block $(i,j)$ cannot depend on free block $\{0,T\}$ in $\mathcal{A}^{(\ell)}$, for any $\{0,T\} \neq \{0,r_i\}, \{0,r_j\}$, given $L \leq \ell < m_{ij}$.
        \end{itemize}

        \item \textbf{A.1.3.} For type A0 and 0A structures, we show $f_{\{0,T\}}^{(\ell)}(A) = 0$ for all $T$, given  $L \leq \ell < m_{ij}$.
    \end{itemize}
    
    \paragraph{Step A.1.1.} We now focus on the type AA structure, meaning that $\{0,r_i\} = \{0,r_j\}$, and therefore $(i,i)$ and $(j,j)$ are necessarily proportional to each other.

    \paragraph{Step A.1.1.i.} \textit{Block $(i,j)$ can only depend on free block $\{0,r_i\}$ by being proportional to it. Equivalently, $f_{\{0,r_i\}}^{(\ell)}(A) = k_{ij;\{0,r_i\}}^{(\ell)} A$ for some $k_{ij;\{0,r_i\}}^{(\ell)}\in\mathbb{C}$, given $2L \leq \ell < m_{ij}$.}

    Take the basis elements
    \begin{equation*}
        a = [A]^{(L)}_{\{0,r_i\}}, \ b = [B]^{(L)}_{\{0,r_i\}}, \ 
        \tilde{a} = [A]^{(2L)}_{\{0,r_i\}}, \ \tilde{b} = [B]^{(2L)}_{\{0,r_i\}},
    \end{equation*} 
    for any matrices $A, B$. By applying Eq. \eqref{eq:blockinj_aux1_1} and \eqref{eq:blockinj_aux1_2} on the $(i,j)$-th block of the products $ab$ and $\tilde{a} \tilde{b}$, we obtain:
    \begin{align*}
        f_{\{0,r_i\}}^{(2L)}(AB) &=
        k_{ii;\{0,r_i\}}^{(L)} A f_{\{0,r_i\}}^{(L)}(B) +
        k_{jj;\{0,r_i\}}^{(L)} f_{\{0,r_i\}}^{(L)}(A) B + 
        \left( \sum_{\gamma=i+1}^{j-1} k_{i \gamma; \{0,r_i\}}^{(L)} k_{\gamma j;\{0,r_i\}}^{(L)} \right) AB , \\
        f_{\{0,r_i\}}^{(4L)}(AB) &=
        k_{ii;\{0,r_i\}}^{(2L)} A f_{\{0,r_i\}}^{(2L)}(B) +
        k_{jj;\{0,r_i\}}^{(2L)} f_{\{0,r_i\}}^{(2L)}(A) B + 
        \left( \sum_{\gamma=i+1}^{j-1} k_{i \gamma; \{0,r_i\}}^{(2L)} k_{\gamma j;\{0,r_i\}}^{(2L)} \right) AB .
    \end{align*}
    These two expressions allow us to apply Lemma \ref{lemma:technical1}, which explicitly constructs an invertible matrix $P$ of the form $P = \mathds{1} + [\tilde{P}]_{ij}$ (so that $P^{-1} = \mathds{1} - [\tilde{P}]_{ij}$), such that for any $a \in P\mathcal{A}^{(2L)}P^{-1}$ written as $a |_{\prec (i,j)} = \sum_{r\in \Sigma} [A_r]_r^{(2L)}$, the $(i,j)$-th block satisfies
    \begin{equation} \label{eq:processed_1}
        a_{ij} = k_{ij; \{0,r_i\}}^{(2L)} A_{\{0,r_i\}} + \sum_{e \neq \{0,r_i\}} f_{e}^{(2L)}(A_{e}) .
    \end{equation}
    The structure of $P$ ensures that all blocks $\prec (i,j)$ remain unchanged under this transformation.
    
    Finally, by Lemma \ref{lemma:same_form_AA_2x2_extension}, this form persists under blocking: for all $\ell$ with $2L \leq \ell < m_{ij}$, the same Eq. \eqref{eq:processed_1} holds with updated constants $k_{ij;\{0,r_i\}}^{(\ell)} \in \mathbb{C}$.

    \paragraph{Step A.1.1.ii.} \textit{Block $(i,j)$ cannot depend on any other diagonal free block apart from $\{0,r_i\}$, that is, $f_{\{0,T\}}^{(\ell)}(A) = 0$ for all $\{0,T\} \neq \{0,r_i\}$, for $\ell$ such that $2L \leq \ell < m_{ij}$.}
    
    Let $\{0,T\} \neq \{0,r_i\}$ and take the basis elements
    \begin{equation*}
        a = [A]^{(L)}_{\{0,T\}}, \ b = [B]^{(L+s)}_{\{0,T\}}
    \end{equation*}
    for any matrices $A, B$, and $s$ such that $0 \leq s < m_{ij} - 2L$. Applying Eq. \eqref{eq:blockinj_aux1_1} and \eqref{eq:blockinj_aux1_2} on the $(i,j)$-th block of the product $ab$, we find
    \begin{equation} \label{eq:step_A.1.1.ii_aux1}
        f_{\{0,T\}}^{(2L+s)} (AB) 
        = \cancel{k_{ii;\{0,T\}}^{(L)}} A f_{\{0,T\}}^{(L+s)}(B) +
        \cancel{k_{jj;\{0,T\}}^{(L+s)}} f_{\{0,T\}}^{(L)}(A) B + 
        \left(\sum_{\gamma=i+1}^{j-1} k_{i \gamma; \{0,T\}}^{(L)} k_{\gamma j;\{0,T\}}^{(L+s)}\right) AB.
    \end{equation}
    The first two terms vanish because $\{0,T\} \neq \{0,r_i\}$, and hence $k_{ii;\{0,T\}}^{(L)} = k_{jj;\{0,T\}}^{(L+s)} = 0$. The third term is also zero, since $k_{i\gamma;\{0,T\}}^{(L)} = 0$ for all $\gamma$. The reason is that, by the inductive hypothesis, the non-zero blocks of the basis element $a\mid_{\prec(i,j)}$ can only appear in sector $[\{0,T\},\{0,T\}]$. Thus, 
    \begin{equation*}
        k_{i\gamma;\{0,T\}}^{(L)} \neq 0 \implies (i,\gamma) \in [\{0,T\}, \{0,T\}].
    \end{equation*}
    However, this contradicts the assumption $\{0,T\} \neq \{0,r_i\}$. In conclusion, given $2L \leq \ell < m_{ij}$, we obtain
    \begin{equation*}
        f_{\{0,T\}}^{(\ell)}(A) = 0, \quad \text{whenever } \{0,T\} \neq \{0,r_i\}. 
    \end{equation*}

    \paragraph{Step A.1.2.} We now turn to the case of a type AB structure, meaning that $\{0,r_i\} \neq \{0,r_j\}$. Consequently, blocks $(i,i)$ and $(j,j)$ can be chosen independently. We proceed with a similar strategy to that of Step A.1.1. 

    \paragraph{Step A.1.2.i.} \textit{Block $(i,j)$ cannot depend on free blocks $\{0,r_i\}$ nor $\{0,r_j\}$ in $\mathcal{A}^{(\ell)}$, given $L \leq \ell < m_{ij}$.}

    Let $s_1, s_2 \in \mathbb{N}$ such that $s_1 + s_2 < m_{ij} - 2L$. For any $A,B$, take $a = [A]^{(L+s_1)}_{\{0,r_i\}}, b = [B]^{(L+s_2)}_{\{0,r_i\}}$. Using Eq. \eqref{eq:blockinj_aux1_1} and \eqref{eq:blockinj_aux1_2}, we have
    \begin{equation} \label{eq:step_A.1.2_1}
        f^{(2L+s_1+s_2)}_{\{0,r_i\}}(AB) 
        = k_{ii;\{0,r_i\}}^{(L+s_1)} A f_{\{0,r_i\}}^{(L+s_2)}(B) +
        \cancel{k_{jj;\{0,r_i\}}^{(L+s_2)}} f_{\{0,r_i\}}^{(L+s_1)}(A) B + 
        \left(\sum_{\gamma=i+1}^{j-1} k_{i \gamma; \{0,r_i\}}^{(L+s_1)} \cancel{k_{\gamma j;\{0,r_i\}}^{(L+s_2)}} \right) AB .
    \end{equation}
    The assumption $\{0,r_i\} \neq \{0,r_j\}$ implies that $k_{jj;\{0,r_i\}}^{(L+s_2)} = 0$ and also $k_{\gamma j;\{0,r_i\}}^{(L+s_2)} = 0$. Indeed, $k_{\gamma j;\{0,r_i\}}^{(L+s_2)} \neq 0$ is only possible if $\{0,r_\gamma\} = \{0,r_j\} = \{0,r_i\}$ due to the inductive hypothesis, which contradicts the type AB assumption, $\{0,r_i\} \neq \{0,r_j\}$. 
    
    Now, let $a = [A]^{(L+s_1)}_{\{0,r_j\}}, b = [B]^{(L+s_2)}_{\{0,r_j\}}$. Similarly as before, we have
    \begin{equation} \label{eq:step_A.1.2_2}
        f^{(2L+s_1+s_2)}_{\{0,r_j\}}(AB) = \cancel{k_{ii;\{0,r_j\}}^{(L+s_1)}} A f_{\{0,r_j\}}^{(L+s_2)}(B) 
        + k_{jj;\{0,r_j\}}^{(L+s_2)} f_{\{0,r_j\}}^{(L+s_1)}(A) B 
        + \left(\sum_{\gamma=i+1}^{j-1} \cancel{k_{i \gamma; \{0,r_j\}}^{(L+s_1)}} k_{\gamma j;\{0,r_j\}}^{(L+s_2)} \right) AB.
    \end{equation}

    Finally, take $a = [A]^{(L)}_{\{0,r_i\}}, b = [B]^{(L)}_{\{0,r_j\}}$. Then, since $2L < m_{ij} - 2L$, we have
    \begin{align*}
        (ab)_{ij} &= \sum_{e \in \Sigma} f_{e}^{(2L)} \left( \left( \Gamma^{(L, L)} \right)^{\{0,r_i\}, \{0,r_j\}}_e AB \right) = 0\\
        &= k_{ii;\{0,r_i\}}^{(L)} A f_{\{0,r_j\}}^{(L)}(B) +
        k_{jj;\{0,r_j\}}^{(L)} f_{\{0,r_i\}}^{(L)}(A) B + 
        \left(\sum_{\gamma=i+1}^{j-1} \cancel{k_{i \gamma; \{0,r_i\}}^{(L)} k_{\gamma j;\{0,r_j\}}^{(L)}} \right) AB.
    \end{align*}
    The first line vanishes due to property (P1) of the $\Gamma$ tensor (see section \ref{app:notation_and_basis}). The last term in the second line also vanishes because for any of the terms in the sum to be non-zero, we should have 
    \begin{equation*} \begin{cases}
        k_{i\gamma;\{0,r_i\}}^{(L)} &\neq 0 
        \implies \{0,r_i\} = \{0,r_\gamma\}
         \\
        k_{\gamma j; \{0,r_j\}}^{(L)} &\neq 0 \implies \{0,r_\gamma\} = \{0,r_j\} ,
    \end{cases} \end{equation*}
    which contradicts the assumption $\{0,r_i\} \neq \{0,r_j\}$. This equation, together with Eq. \eqref{eq:step_A.1.2_1} and \eqref{eq:step_A.1.2_2} setting $s_1 = s_2 = 0$, gives the following relations:
    \begin{equation*}
        \begin{cases}
            Af(B) = \hat{f}(AB) \\
            g(A)B = \hat{g}(AB) \\
            Ag(B)+f(A)B = 0
        \end{cases} \text{ with } 
        \begin{cases}
            f(A) := k_{jj;\{0,r_j\}}^{(L)} f_{\{0,r_i\}}^{(L)}(A) \\
            g(A) := k_{ii;\{0,r_i\}}^{(L)} f_{\{0,r_j\}}^{(L)}(A) 
        \end{cases} 
        \text{ and }
        \begin{cases}
            \hat{f}(A) := \frac{k_{jj;\{0,r_j\}}^{(L)}}{k_{ii;\{0,r_i\}}^{(L)}} f_{\{0,r_i\}}^{(2L)}(A) \\
            \hat{g}(A) := \frac{k_{ii;\{0,r_i\}}^{(L)}}{k_{jj;\{0,r_j\}}^{(L)}} f_{\{0,r_j\}}^{(2L)}(A)
        \end{cases}.
    \end{equation*}
    These allow us to apply Lemma \ref{lemma:technical1} to explicitly construct an invertible matrix $P$ of the form $P = \mathds{1} + [\tilde{P}]_{ij}$, such that for any $a \in P \mathcal{A}^{(L)}P^{-1}$ written as $a |_{\prec (i,j)} = \sum_{r\in\Sigma} [A_r]_r^{(L)}$, the $(i,j)$-th block satisfies
    \begin{equation*}
        a_{ij} = \sum_{e \neq \{0,r_i\}, \{0,r_j\}} f_e^{(L)}(A_e),
    \end{equation*}
    meaning that in this new basis, $f_{\{0,r_i\}}^{(L)} = f_{\{0,r_j\}}^{(L)} = 0$. Again, all blocks $\prec(i,j)$ remain unchanged under this transformation. 
    
    To extend this property to all $\ell$ with $L \leq \ell < m_{ij}$, note that from Eq. \eqref{eq:step_A.1.2_1} (choosing $(s_1, s_2) = (s,0)$ and $(0,s)$) and the fact that $f_{\{0,r_i\}}^{(L)} = 0$, it follows that for any $s$ with $0 \leq s < m_{ij} - 2L$,
    \begin{equation*}
        f_{\{0,r_i\}}^{(2L+s)}(AB) = 
        k_{ii;\{0,r_i\}}^{(L+s)} A \cancel{f_{\{0,r_i\}}^{(L)}(B)}
        = k_{ii;\{0,r_i\}}^{(L)} A f_{\{0,r_i\}}^{(L+s)}(B).
    \end{equation*}
    This gives $f_{\{0,r_i\}}^{(\ell)} = 0$ for all $L \leq \ell < m_{ij}$ (holding for all such values of $\ell$ because $2L \leq m_{ij}-L$ when $m_{ij} > 4L$). An analogous argument using Eq. \eqref{eq:step_A.1.2_2} establishes the same conclusion for $f_{\{0,r_j\}}^{(\ell)}$.

    \paragraph{Step A.1.2.ii.} \textit{Block $(i,j)$ cannot depend on free block $\{0,T\}$ in $\mathcal{A}^{(\ell)}$, for any $\{0,T\} \neq \{0,r_i\}, \{0,r_j\}$, given $L \leq \ell < m_{ij}$.}

    Take $\{0,T\} \neq \{0,r_i\}, \{0,r_j\}$. For arbitrary $A, B$, consider $a = [A]^{(L+s)}_{\{0,T\}}, b = [B]^{(L)}_{\{0,r_j\}}$, with $0 \leq s < m_{ij} - 2L$. Then, using Eq. \eqref{eq:blockinj_aux1_1} and \eqref{eq:blockinj_aux1_2}, we have
    \begin{align}
        (ab)_{ij} &=     
        \sum_{e \in \Sigma} f_e^{(2L+s)} \left( \left(\Gamma^{(L+s, L)}\right)^{\{0,T\}, \{0,r_j\}}_{e} AB \right) = 0 \label{eq:step_A.1.2.iii_1} \\
        &= 
        \cancel{k_{ii;\{0,T\}}^{(L+s)}} A f_{\{0,r_j\}}^{(L)}(B) +
        k_{jj;\{0,r_j\}}^{(L)} f_{\{0,T\}}^{(L+s)}(A) B + 
        \left(\sum_{\gamma=i+1}^{j-1} \cancel{k_{i \gamma; \{0,T\}}^{(L+s)} k_{\gamma j;\{0,r_j\}}^{(L)}} \right) AB, \nonumber
    \end{align}
    where the last term in the second line vanishes because any of the summands is non-zero only if
    \begin{equation*} \begin{cases}
        k_{i\gamma;\{0,T\}}^{(L+s)} \neq 0 
        &\implies \{0,T\} = \{0,r_\gamma\}
         \\
        k_{\gamma j; \{0,r_j\}}^{(L)} \neq 0 &\implies \{0,r_\gamma\} = \{0,r_j\}
    \end{cases} \end{equation*}
    which contradicts the assumption $\{0,T\} \neq \{0,r_j\}$. Therefore, Eq.~\eqref{eq:step_A.1.2.iii_1} implies that $f_{\{0,T\}}^{(\ell)} = 0$, given $L \leq \ell < m_{ij} - L$.
    
    To extend this conclusion to all $\ell$ with $L \leq \ell < m_{ij}$, we take $a = [A]_{\{0,T\}}^{(L)}$ and $b = [B]_{\{0,T\}}^{(L+s)}$. Using again Eq. \eqref{eq:blockinj_aux1_1} and \eqref{eq:blockinj_aux1_2} for their product, we obtain
    \begin{align}
        (ab)_{ij} &= f^{(2L+s)}_{\{0,T\}}(AB) \\
        &= \cancel{k_{ii;\{0,T\}}^{(L)}} A f_{\{0,T\}}^{(L+s)}(B) +
        \cancel{k_{jj;\{0,T\}}^{(L+s)}} f_{\{0,T\}}^{(L)}(A) B + 
        \left(\sum_{\gamma=i+1}^{j-1} k_{i \gamma; \{0,T\}}^{(L+s)} \cancel{k_{\gamma j;\{0,T\}}^{(L)}} \right) AB .
    \end{align}
    Therefore, we have that $f^{(\ell)}_{\{0,T\}}(A) = 0$ for all $s: 2L \leq \ell < m_{ij}$. Combining this with the previous result, and given that $2L \leq m_{ij}-L$, we obtain as desired that
    \begin{equation}
        f^{(\ell)}_{\{0,T\}} = 0, \quad \forall \ell : L \leq \ell < m_{ij}. 
    \end{equation}  
    This completes the proof of Step A.1.2.

    \paragraph{Step A.1.3.} \textit{For type A0 and 0A structures, we show that $f_{\{0,T\}}^{(\ell)}(A) = 0$ for all $T$, and for all $\ell$ such that $L \leq \ell < m_{ij}$.}

    The proof for both A0 and 0A structures proceeds analogously to Step A.1.2, but the argument simplifies due to the structure-specific technical tools available.

    To begin, one can show that the $(i,j)$-th block does not depend on $\{0,r_i\}$ nor $\{0,r_j\}$. As in step A.1.2.i, a change of basis achieving this can be built using Corollary \ref{cor:technical1} for type A0, and Corollary \ref{cor:technical2} for type 0A. Each corollary requires only a single condition to be satisfied, in contrast to the three conditions required by Lemma \ref{lemma:technical1}, making the argument more direct.

    To establish that the $(i,j)$-th block does not depend on any other pair $\{0,T\}$, with $\{0,T\} \neq \{0,r_i\}, \{0,r_j\}$, we proceed as in Step A.1.2.ii. For the A0 case, we use the same basis elements as in there. For the 0A case, we would instead consider $a = [A]_{\{0,r_i\}}^{(L)}, \ b = [B]_{\{0,T\}}^{(L+s)}$ for type A0.

    \paragraph{Step A.1 (Conclusion).}
    So far, we have established that for any $a \in \mathcal{A}^{(\ell)}$ of the form $a \mid_{\prec (i,j)} = \sum_{e\in\Sigma}[A_e]_e^{(\ell)}$ with $2L \leq \ell < m_{ij}$, the $(i,j)$-th block satisfies
    \begin{equation}
        a_{ij} = 
        k_{ij; \{0,r_i\}}^{(\ell)} \delta_{r_i r_j} A_{\{0,r_i\}} + \sum_{\{t\} \in \Sigma_f} f_{\{t\}}^{(\ell)}(A_{\{t\}}).
    \end{equation}
    This expression fully characterizes the dependence of the $(i,j)$-th block on the diagonal free blocks, indexed by $\Sigma_\infty$.

    \paragraph{Step A.2.} Now we are going to characterize the dependencies of block $(i,j)$ on the off-diagonal free blocks in $\Sigma_f$, which in turn is partitioned as $\Sigma_f = \cup_{s,t\in \tilde{\Sigma}_\infty} \Sigma_f^{st}$, where $\Sigma_f^{st}$ contains the free blocks in sector $[\{0,s\},\{0,t\}]$ and $\tilde{\Sigma}_\infty = \Sigma_\infty \cup \varepsilon$ (note that some of the sets $\Sigma_f^{st}$ may be empty). We will divide the proof of this step in the following parts:
    \begin{itemize}
        \item \textbf{A.2.1.} Block $(i,j)$ can only depend on other off-diagonal free blocks of the same sector by being linear combinations of them. Equivalently, given $\ell$ with $L \leq \ell < m_{ij} - L$, we have $f_{\{t\}}^{(\ell)}(A) = k_{ij;\{t\}}^{(\ell)} A$ for $k_{ij;\{t\}}^{(\ell)} \in \mathbb{C}$, and $k_{ij;\{t\}}^{(\ell)} = 0$ if $\{t\} \notin \Sigma_f^{r_ir_j}$.

        \item \textbf{A.2.2.} Extend the previous claim to hold in $\mathcal{A}^{(\ell)}$, given $L \leq \ell < m_{ij}$.
    \end{itemize}

    \paragraph{Step A.2.1.} Let us start by choosing some $\{T\} \in \Sigma_f$ and consider
    \begin{equation*}
        a = [A]_{\{0,r_i\}}^{(L)}, \quad  
        b = [B]_{\{T\}}^{(L+s)},
    \end{equation*}
    given $0 \leq s < m_{ij} - 2L$. The $(i,j)$-th block of the product $ab$ can be written in two different ways:
    \begin{enumerate}
        \item Using block-matrix multiplication:
        \begin{equation*}
            (ab)_{ij} = k_{ii;\{0,r_i\}}^{(L)} A f_{\{T\}}^{(L+s)}(B)
            + \cancel{k_{jj;\{T\}}^{(L+s)}} f_{\{0,r_i\}}^{(L)}(A) B + 
            \left( \sum_{\gamma=i+1}^{j-1} k_{i \gamma; \{0,r_i\}}^{(L)} k_{\gamma j;\{T\}}^{(L+s)} \right) AB 
        \end{equation*}

        \item Using the fact that $ab \in \mathcal{A}^{(2L+s)}$ with $2L+s < m_{ij}$, which means block $(i,j)$ should still have the form of Eq. \eqref{eq:step-A_form_ij}:
        \begin{align*}
            (ab)_{ij} 
            &= \left[ \sum_{e\in \Sigma} f_{e}^{(2L+s)}\left(  \left(\Gamma^{(L, L+s)}\right)^{\{0,r_i\},\{T\}}_{e} AB\right) \right]_{ij} \\
            &= 
            \left( \sum_{\{t\}\in \Sigma_f} \left(\Gamma^{(L, L+s)}\right)^{\{0,r_i\},\{T\}}_{\{t\}} f_{\{t\}}^{(2L+s)} \right) (AB) =: g_1(AB).
        \end{align*}
    \end{enumerate}
    
    If we took instead
    \begin{equation*}
        a = [A]_{\{T\}}^{(L+s)}, \quad b = [B]_{\{0,r_j\}}^{(L)},
    \end{equation*}
    we would obtain a similar equation with a linear function $g_2$ instead of $g_1$, defined as
    \begin{equation*}
        g_2(A):= \left( \sum_{\{t\}\in \Sigma_f} \left(\Gamma^{(L+s, L)}\right)^{\{T\},\{0,r_j\}}_{\{t\}} f_{\{t\}}^{(2L+s)} \right) (A)  .
    \end{equation*}
    Putting everything together, we get:
    \begin{align}
        g_1 (AB) &= k_{ii;\{0,r_i\}}^{(L)} A f_{\{T\}}^{(L+s)}(B) + 
        \left( \sum_{\gamma=i+1}^{j-1} k_{i \gamma; \{0,r_i\}}^{(L)} k_{\gamma j;\{T_1\}}^{(L+s)} \right) AB ,
        \label{eq:step_A.2.1_1}
        \\
        g_2 (AB) &= k_{jj;\{0,r_j\}}^{(L)} f_{\{T\}}^{(L+s)}(A) B + 
        \left( \sum_{\gamma=i+1}^{j-1} k_{i \gamma; \{T\}}^{(L+s)} k_{\gamma j;\{0,r_j\}}^{(L)} \right) AB .
        \label{eq:step_A.2.1_2}
    \end{align}
    Given these equations, we can apply Lemma \ref{lemma:technical3} to assert that there exist constants $k_{ij;\{T\}}^{(L+s)} \in \mathbb{C}$ such that
    \begin{equation*}
        f_{\{T\}}^{(L+s)}(A) = k_{ij;\{T\}}^{(L+s)} A,
    \end{equation*}
    for $s$ such that $0 \leq s < m_{ij} - 2L$.

    Let us now show that $k_{ij;\{T\}}^{(L+s)}$ vanishes unless $\{T\} \in \Sigma_f^{r_i r_j}$. 
    Assume that $\{T\} \in \Sigma_f^{r^1_T r^2_T}$ with $\{0,r^1_T\} \neq \{0,r_i\}$. In this case, the linear function $g_1$ is identically zero. This follows from property (P3) of the $\Gamma$-tensor in Appendix \ref{app:notation_and_basis}, which holds by the inductive hypothesis. That is, if the tensor element $(\Gamma^{(\ell_1,\ell_2)})^{ef}_g \neq 0$, then there must exist labels $s_1, s_2, s_3 \in \tilde{\Sigma}_\infty$ such that $e \in \tilde{\Sigma}_f^{s_1 s_2}$, $f \in \tilde{\Sigma}_f^{s_2 s_3}$ and $g \in \tilde{\Sigma}_f^{s_1 s_3}$. In our case, since $\{0,r_i\} \in \tilde{\Sigma}_f^{r_i r_i}$ and $\{T\} \in \tilde{\Sigma}_f^{r_1^T r_2^T}$ with $r_i \neq r_1^T$, the corresponding structure constant element vanishes, implying $g_1(A) = 0$.

    Moreover, the sum appearing in Eq. \eqref{eq:step_A.2.1_1} must also vanish. Indeed, for any term in the sum to be nonzero, the following conditions must hold:
    \begin{equation*}
        \begin{cases}
            k_{i \gamma;\{0,r_i\}}^{(L)} \neq 0 \implies \{0,r_i\} = \{0,r_{\gamma}\},\\
            k_{\gamma j;\{T_1\}}^{(L+s)} \neq 0 \implies \{0,r_\gamma\} = \{0,r^1_{T}\}, \ \{0,r_j\} = \{0,r^2_{T}\} ,
        \end{cases}
    \end{equation*}
    which are incompatible with the assumption that $\{0,r_T^1\} \neq \{0,r_i\}$. Hence, all such terms vanish.
Combining these two facts, we conclude that $f_{\{T\}}^{(L+s)} = 0$, given $0 \leq s < m_{ij} - 2L$.

    A completely analogous argument applies if $\{0,r^2_T\} \neq \{0,r_j\}$: in that case, the linear function $g_2$ is identically zero by property (P3), and from Eq. \eqref{eq:step_A.2.1_2} it follows that $f_{\{T\}}^{(L+s)} = 0$ given $0 \leq s < m_{ij}-2L$.

    \paragraph{Step A.2.2.} Let us now extend the previous claims to hold in $\mathcal{A}^{(\ell)}$, given $L \leq \ell < m_{ij}$. For each $0 \leq \ell < L$, the $(i,j)$–block of $\mathcal{A}^{(m_{ij}-L+\ell)}$ can be written as 
    \begin{align}
        \mathcal{A}^{(m_{ij}-L+\ell)}_{ij} 
        &= \text{span}\{ ab \mid 
        a \in \mathcal{A}^{(L)}, \ 
        b \in \mathcal{A}^{(m_{ij}-2L+\ell)} \} \mid_{(i,j)} \nonumber 
        \\ &= 
        \text{span}\{ [A]^{(L)}_e [B]^{(m_{ij}-2L+\ell)}_f\}_{e,f,A,B} \mid_{(i,j)} \label{eq:step_A.2.3_aux1}
        \\ &= 
        \left\{ \left( \sum_{\gamma=i}^j k_{i\gamma;e}^{(L)} k_{\gamma j;f}^{(m_{ij}-2L+\ell)}\right)C\right\}_{e,f,C} \nonumber
    \end{align}
    where we used the inductive hypothesis for both $\mathcal{A}^{(L)}$ and $\mathcal{A}^{(m_{ij}-2L+\ell)}$, since $L \leq m_{ij}-2L+\ell < m_{ij}-L$ and $(i,\gamma), (\gamma, j) \preceq (i,j)$. Hence, the $(i,j)$-block is either zero, a free block, or a linear combination of previously identified free blocks. In other words, there exist scalars $k^{(m_{ij}-L+\ell)}_{ij;\{T\}}\in\mathbb{C}$ such that
    \begin{equation*}
        f^{(m_{ij}-L+\ell)}_{\{T\}}(A)
        = k^{(m_{ij}-L+\ell)}_{ij;\{T\}} A .
    \end{equation*}
    
    It remains to show that $k^{(m_{ij}-L+\ell)}_{ij;\{T\}}=0$ whenever $\{T\}\notin\Sigma_f^{r_i r_j}$.  Indeed, by the inductive hypothesis, any basis element $[A]^{(L)}_e$ (resp. $[B]^{(m_{ij}-2L+\ell)}_f$) appearing in Eq. \eqref{eq:step_A.2.3_aux1} has support only in sector $[\{0,r^1_e\},\{0,r^2_e\}]$ (resp. $[\{0,r^1_f\},\{0,r^2_f\}]$). Therefore, their product can be nonzero only if $r^2_e=r^1_f$, in which case it belongs to the sector $[\{0,r^1_e\},\{0,r^2_f\}]$. For such a product to contribute to the $(i,j)$-th block, we must have $r_i=r^1_e$ and $r_j=r^2_f$. This means any non-trivial dependence of $(i,j)$ on a free block $\{T\}$ requires that $\{T\}$  is supported in the same sector, i.e. $\{T\} \in \Sigma_f^{r_i r_j}$, as we wanted to show. This concludes the proof of step A.2.2. 

    \paragraph{Step A (Conclusion).} Under the assumption that $m_{ij} > 4L$, we have shown that, for any $a \in \mathcal{A}^{(\ell)}$ of the form $a \mid_{\prec (i,j)} = \sum_{r\in\Sigma}[A_r]_r^{(\ell)}$ given $2L \leq \ell < m_{ij}$, we have
    \begin{equation*}
        a_{ij} = \sum_{e \in \tilde{\Sigma}_f^{r_i r_j}} k_{ij;e}^{(\ell)} A_e \ .
    \end{equation*}

    \paragraph{Step B.} If $m_{ij} \leq 4L$, we have that $(i,j)$ is isolatable in $\mathcal{A}^{(\ell)}$ for all $\ell \geq 4L$, and therefore by Lemma \ref{lemma:nonzero_elements} it is also the full free block in $\mathcal{A}^{(\ell)}$ for all $\ell \geq 4L+ 2L_0 =: L_{ij}$. Although Lemma \ref{lemma:nonzero_elements} does not apply when $(i,j)$ has type 00 structure, the claim remains valid: since type 00 blocks are $1 \times 1$ by construction, they are necessarily either zero, free, or linear combinations of other type 00 blocks.

    \paragraph{Step C.} \textit{Argue that the constructed basis is still valid in $\mathcal{A}^{(\ell)}$, given $m_{ij} \leq \ell < m_{ij}+2L_0$.}
    
    There remains one more step to show that a basis with the desired structural properties up to block $\preceq (i,j)$ exists in $\mathcal{A}^{(\ell)}$ for all $\ell \geq 4L+2L_0$. Indeed, what we have established so far is the following:
    \begin{itemize}
        \item If $m_{ij} \leq 4L$, then block $(i,j)$ becomes a free block in $\mathcal{A}^{(\ell)}$ for all $\ell \geq 4L + 2L_0$ (Step B).
        \item If $m_{ij} > 4L$, then block $(i,j)$ is a linear combination of other free blocks in the same sector in $\mathcal{A}^{(\ell)}$ for $2L \leq \ell < m_{ij}$ (Step A). 
    \end{itemize}
    However, there may be a gap between the first appearance of a nonzero isolatable element in $\mathcal{A}^{(m_{ij})}$ and the point at which block $(i,j)$ becomes a full free block. While we know that it eventually becomes free after blocking to $\mathcal{A}^{(m_{ij} + 2L_0 + s)}$ for all $s \geq 0$, it remains unclear whether the structured basis is still valid in the intermediate regime $\mathcal{A}^{(\ell)}$ for $m_{ij} \leq \ell < m_{ij} + 2L_0$. That is, in this range, block $(i,j)$ might be neither free nor a linear combination of existing free blocks.
    
    Here we tackle this issue and show that a basis of the desired form is still valid for $\mathcal{A}^{(m_{ij} + \ell)}$, for all $\ell \geq 0$, whenever $m_{ij} > 4L$. Therefore, such a basis also holds for $\mathcal{A}^{(\ell)} \mid_{\preceq(i,j)}$, $\forall \ell \geq 4L + 2L_0$.
    
    To begin, we observe that
    \begin{equation} \label{eq:crisis_span_aux1}
        \mathcal{A}^{(m_{ij})} = \text{span}\{\mathcal{A}^{(2L)} \cdot \mathcal{A}^{(m_{ij} - 2L)}\} .
    \end{equation}
    We can apply the inductive hypothesis up to block $\preceq (i,j)$ to both $\mathcal{A}^{(2L)}$ and $\mathcal{A}^{(m_{ij} - 2L)}$ since, together with Steps A and B, it guarantees that a structured basis exists for all intermediate lengths $2L \leq \ell < m_{ij}$. What remains is to confirm that block $(i,j)$ indeed becomes a full free block already in $\mathcal{A}^{(m_{ij})}$ and remains so in all subsequent $\mathcal{A}^{(m_{ij} + s)}$ for $0 \leq s < 2L_0$.
    
    Due to the fact that $m_{ij} := \min\{ M \mid \text{block } (i,j) \text{ is isolatable in } \mathcal{A}^{(M)}\}$, we know that $\exists a \in \mathcal{A}^{(m_{ij})}$ such that $a_{ij} = Z \neq 0$, and $a\mid_{\prec (i,j)} = 0$. From Eq. \eqref{eq:crisis_span_aux1}, there must exist $\{b_u\} \subset \mathcal{A}^{(2L)}$ and $\{c_u\} \subset \mathcal{A}^{(m_{ij}-2L)}$ such that $a = \sum_u b_u c_u$. Let us write
    \begin{equation*}
        b_u \mid_{\preceq(i,j)} = \sum_{t\in \Sigma} [B_t^u]_t^{(2L)}, \quad c_u \mid_{\preceq(i,j)} = \sum_{s\in \Sigma} [C_s^u]_s^{(m_{ij}-2L)} .
    \end{equation*} 
    where the basis elements $[B_t^u]_t^{(2L)}$ and $[C_s^u]_s^{(m_{ij}-2L)}$ possess the structural properties specified in the claim of the theorem. Then we have
    \begin{align*}
        \text{At } &(i,j): \ Z = \sum_{u,t,s} \sum_{\gamma=i}^j \left( [B_t^u]_t^{(2L)} \right)_{i \gamma}
        \left( [C^u_s]_s^{(m_{ij}-2L)} \right)_{\gamma j} = 
        \sum_{u,t,s} \left( \sum_{\gamma=i}^j k_{i\gamma;t}^{(2L)} k_{\gamma j;s}^{(m_{ij}-2L)} \right) C_t^u D_s^u
        \\
        \text{At } &(m,n): \ 0 = \sum_{u,t,s} \left( \sum_{\gamma=m}^n k_{m\gamma;t}^{(2L)} k_{\gamma n;s}^{(m_{ij}-2L)} \right) C_t^u D_s^u \ , \quad \forall (m,n) \prec (i,j).  
    \end{align*}
    
    Since $Z \neq 0$, we can write (after rescaling) $Z = \dyad{\alpha}{\beta} + \sum_{(\gamma, \delta) \neq (\alpha, \beta)} z_{\gamma\delta} \dyad{\gamma}{\delta}$. Now, for arbitrary $p,q$, we multiply the two expressions above on the left by $\dyad{p}{\alpha}$ and on the right by $\dyad{\beta}{q}$, obtaining:
    \begin{align*}
        \dyad{p}{q} &= 
        \sum_{u,t,s} \left( \sum_{\gamma=i}^j k_{i\gamma;t}^{(2L)} k_{\gamma j;s}^{(m_{ij}-2L)} \right)
        \left( \dyad{p}{\alpha} C_t^u \right)
        \left( D_s^u \dyad{\beta}{q} \right)
        \\
        0 &= \sum_{u,t,s} \left( \sum_{\gamma=m}^n k_{m\gamma;t}^{(2L)} k_{\gamma n;s}^{(m_{ij}-2L)} \right)
        \left( \dyad{p}{\alpha} C_t^u \right)
        \left( D_s^u \dyad{\beta}{q} \right) \ , \quad \forall (m,n) \prec (i,j).  
    \end{align*}
    Define $\tilde{b} \in \mathcal{A}^{(2L)}$ and $\tilde{c} \in \mathcal{A}^{(m_{ij} - 2L)}$ as 
    \begin{equation*}
        \tilde{b}\mid_{\preceq(i,j)} = \sum_u \sum_t [\dyad{p}{\alpha} B_t^u]_t^{(2L)}, \quad \tilde{c}\mid_{\preceq(i,j)} = \sum_u \sum_s [C_s^u \dyad{\beta}{q}]_s^{(m_{ij}-2L)}, 
    \end{equation*}
    where we have used that we can freely choose each of the free blocks. Then the product $\tilde{b} \tilde{c}$ satisfies:
    \begin{equation*}
        \tilde{b}\tilde{c} \in \mathcal{A}^{(m_{ij})}, \quad (\tilde{b} \tilde{c})_{ij} = \dyad{p}{q}, \quad \text{and} \quad (\tilde{b} \tilde{c})\mid_{\prec(i,j)} = 0 .
    \end{equation*}
    This means that block $(i,j)$ is not only isolatable, but also it is fully spanned and therefore a free block in $\mathcal{A}^{(m_{ij})}$.
    
    The same argument applies to each $\mathcal{A}^{(m_{ij}+s)}$ for $0 \leq s < 2L_0$ using the decomposition
    \begin{equation*}
        \mathcal{A}^{(m_{ij}+s)} = \text{span}\{\mathcal{A}^{(2L)} \cdot \mathcal{A}^{(m_{ij} - 2L + s)}\} .
    \end{equation*}
    The inductive hypothesis guarantees that a structured basis exists for both $\mathcal{A}^{(2L)}|_{\preceq (i,j)}$ and $\mathcal{A}^{(m_{ij}-2L+s)}|_{\preceq (i,j)}$, since
    \begin{equation*}
        2L \leq m_{ij} - 2L + s < m_{ij}, \quad \forall s \in \{0, \dots, 2L_0 - 1\},
    \end{equation*}
    where the first inequality follows from $m_{ij} > 4L$, and the second one from $2L > 2L_0-1$ (recall $L = L_{i-1,j-1} = 4L_{i-2,j-2} + 2L_0$). 
    
    Thus, the basis with the desired structure remains valid in $\mathcal{A}^{(\ell)}|_{\preceq (i,j)}$ for all $\ell \geq L_{ij} := 4L_{i-1,j-1} + 2L_0$, completing the proof of Step C.

    \paragraph{Step D.}
    We have now established that for any $a \in \mathcal{A}^{(\ell)}$ of the form $a \mid_{\prec (i,j)} = \sum_{r \in \Sigma} [A_r]_r^{(\ell)}$ with $\ell \geq 4L + 2L_0$, the $(i,j)$-th block can be written as
    \begin{equation*}
        a_{ij} = \sum_{e \in \tilde{\Sigma}_f^{r_i r_j}} k_{ij;e}^{(\ell)} A_e  .
    \end{equation*}
    This gives us a systematic way to extend the basis construction to block $(i,j)$ and thus proceed to the next one, $(i+1,j+1)$. The key updates are as follows:
    \begin{itemize}
        \item Set $L_{i,j} := 4L + 2L_0$. 
        
        \item Using the coefficients $k_{ij;e}^{(\ell)}$ established above for $\ell \geq L_{i,j}$, we extend the basis of $\mathcal{A}^{(\ell)}\mid_{\prec(i,j)}$ to include all blocks $\preceq (i,j)$, not just those strictly below $(i,j)$, whenever $L_{i,j} \leq \ell < m_{ij}$.

        \item For $\ell \geq m_{ij}$, the block $(i,j)$ becomes free. That is, while for $\ell < m_{ij}$ it could still be expressed as a linear combination of previously existing free blocks, from $\ell = m_{ij}$ onward this is no longer possible: $(i,j)$ must be regarded as an independent, isolatable block. Consequently, a new free block has to be introduced to the basis. As an illustrative example, consider:
        
        The explicit procedure for incorporating the new free block into the basis is as follows: 
        \begin{itemize}
            \item \textbf{Introduce a new symbol:} Add a symbol $\{T\}$ to the alphabet $\Sigma$, specifically to $\Sigma_f^{r_i r_j}$, representing the new free block. Set 
            \begin{equation*}
                k_{ij;\{T\}}^{(\ell)} =
                \begin{cases}
                    0 & \text{for } \ell < m_{ij}, \\
                    1 & \text{for } \ell \geq m_{ij}.
                \end{cases}
            \end{equation*}
            This is possible to do because block $(i,j)$ is isolatable in $\mathcal{A}^{(m_{ij})}$, and remains so in every $\mathcal{A}^{(\ell)}$ with $\ell \geq m_{ij}$ by Lemma \ref{lemma:nonzero_elements}, except for the special case where $(i,j)$ lies in the sector $[\{0,\varepsilon\},\{0,\varepsilon\}]$ (we return to this subtlety below).

            \item \textbf{Respect the $\preceq$-order:} Set $k_{mn;\{T\}}^{(\ell)} = 0$ for every $(m,n) \prec (i,j)$ and for all $\ell$, since the new free block $\{T\}$ does not appear at any position earlier than $(i,j)$ under the $\preceq$-ordering. 

            \item \textbf{Ensure uniqueness:} Impose that no other free block of $\Sigma_f$ appears at $(i,j)$ once $\ell \geq m_{ij}$. That is, set $k_{ij;e}^{(\ell)} = 0$ for all $e \in \Sigma_f \setminus \{T\}$ and all $\ell \geq m_{ij}$. Note that it is still possible that $k_{ij;e}^{(\ell)} \neq 0$ for $e \in \Sigma_\infty$ in order to preserve the desired behavior of the generalized structure constants for the $\Sigma_\infty$ blocks, that we specify below. 
        \end{itemize}
        As an example illustrating this procedure, consider:
        \begin{equation*}
            \mathcal{A}^{(1)} = 
            \left\{{\scriptsize
            \begin{pmatrix}
                A & B & A \\ & A & B \\ & & A
            \end{pmatrix}}\mid A, B\right\}
            \longrightarrow
            \mathcal{A}^{(2)} = \left\{
            {\scriptsize
            \begin{pmatrix}
                A & B & 2A + C \\ & A & B \\ & & A
            \end{pmatrix}} \mid A, B \right\} 
        \end{equation*}
        In $\mathcal{A}^{(1)}$ there are only two free blocks, $\Sigma_\infty = \{\{0,1\}\}$ and $\Sigma_f = \{\{1\}\}$ labeling $A$ and $B$, respectively. When blocking to $\mathcal{A}^{(2)}$, we observe that a new free block appears at $(1,3)$. Therefore, $m_{13} = 2$ and we label it introducing the new symbol $\{2\}$ to $\Sigma_f$. Thus,
        \begin{equation*}
            k_{13;\{2\}} = 
            \begin{cases}
                0 &\text{for } \ell < 2 \\
                1 &\text{for } \ell \geq 2
            \end{cases} \ , 
            \quad
            k_{mn;\{2\}}^{(\ell)} = 0 \ \text{ for } (m,n) \prec (1,3), 
            \quad 
            k_{13;\{1\}}^{(\ell)} = 0, \ \forall \ell \ .
        \end{equation*}
        Note that $k^{(\ell)}_{13;\{0,1\}} = \ell$, which does not contradict our prescription as $\{0,1\} \in \Sigma_\infty$.        

        \item Regarding the diagonal free blocks in $\Sigma_\infty$, we can define the corresponding coefficients for $s \geq 0$ without loss of generality as
        \begin{equation*}
            k_{ij;\{0,e\}}^{(L_{i,j}+s)} = 
            \begin{cases}
                \sum_{\gamma=i}^j 
                k_{i\gamma;\{0,e\}}^{(L_{i-1,j-1})}
                k_{\gamma j;\{0,e\}}^{(L_{i,j} + s - L_{i-1,j-1})}
                &\text{if } \{0,r_i\} = \{0,r_j\} = \{0,e\}, \\
                0 &\text{otherwise}.
            \end{cases}
        \end{equation*}
        This update guarantees the desired behavior under multiplication of the $\Sigma_\infty$ elements, namely:
        \begin{equation*}
            [A]_{\{0,e\}}^{(\ell_1)} [B]_{\{0,e\}}^{(\ell_2)} = [AB]_{\{0,e\}}^{(\ell_1+\ell_2)}
        \end{equation*}
        As an illustrative example, consider:
        \begin{equation*}
            \mathcal{A}^{(1)} = 
            \left\{{\scriptsize
            \begin{pmatrix}
                A & k_1 A & B \\ & A & k_2 A \\ & & A
            \end{pmatrix}}\mid A, B\right\}
            \longrightarrow
            \mathcal{A}^{(\ell)} = \left\{
            {\scriptsize
            \begin{pmatrix}
                A & \ell k_1 A & {\ell \choose 2} k_1 k_2 A + B \\ & A & \ell k_2 A \\ & & A
            \end{pmatrix}} \mid A, B \right\}, 
        \end{equation*}
        The basis elements defined upon blocking according to the rule above would be:
        \begin{equation*}
            [A]^{(\ell)}_{\{0,1\}} := {\scriptsize
            \begin{pmatrix}
                A & \ell k_1 A & {\ell \choose 2} k_1 k_2 A \\ & A & \ell k_2 A \\ & & A
            \end{pmatrix}}, \quad
            [B]^{(\ell)}_{\{1\}} := {\scriptsize
            \begin{pmatrix}
                0 & 0 & B \\ & 0 & 0 \\ & & 0
            \end{pmatrix}}.
        \end{equation*}
    
        \item These updates ensure that we can accordingly extend the generalized structure constants tensors $\Gamma^{(\ell_1, \ell_2)}$ for all $\ell_1,\ell_2 \geq L_{i,j}$, 
        \begin{equation*}
            [A]^{(\ell_1)}_e \cdot [B]^{(\ell_2)}_{f}\mid_{\preceq (i,j)} = \sum_{g \in \Sigma} \left(\Gamma^{(\ell_1, \ell_2)}\right)^{ef}_g [AB]^{(\ell_1+\ell_2)}_{g} \mid_{\preceq (i,j)} \ ,
        \end{equation*}
        while still preserving properties (P1), (P2), (P3) defined in Appendix \ref{app:notation_and_basis}, as desired.
    \end{itemize}

    \paragraph{Remarks.}
    \begin{itemize}
        \item \textbf{Cascading effects}: It is possible that while explicitly processing $(i,j)$, the blocking procedure triggers the appearance of new free blocks in positions $(m,n) \prec (i,j)$. For instance, consider the following example:
        \begin{equation*}
            \mathcal{A}^{(L)} = \left\{
            {\scriptsize
            \begin{pmatrix}
                A & B & 0 & \ast \\
                & A & B & \highg{C} \\
                & & A & 0 \\
                & & & A
            \end{pmatrix}
            } \mid A, B, C
            \right\}
            \to
            \mathcal{A}^{(2L)} = \left\{
            {\scriptsize
            \begin{pmatrix}
                A & B & D & \ast \\
                & A & B & \highg{C} \\
                & & A & 0 \\
                & & & A
            \end{pmatrix}
            } \mid A, B, C, D
            \right\}.
        \end{equation*}
        Here, although we are at a step of the proof where we are explicitly processing $(2,4)$, blocking every $2L$ sites together causes a new free block to emerge at position $(1,3) \prec (2,4)$.

        This phenomenon does not affect the correctness of our construction: the newly appearing blocks are necessarily generated from the multiplication of already free blocks in the lower diagonals, through a sort of ``cascading effect''. As such, they will themselves be full blocks and inherit the desired properties.
        
        \item \textbf{Nilpotent blocks}: Some blocks may disappear upon blocking, particularly those with type 00 structure (i.e. in sector $[\{0,\varepsilon\}, \{0,\varepsilon\}]$). For example:
        \begin{equation*}
            \mathcal{A}^{(L)} = \left\{
            {\scriptsize
            \begin{pmatrix}
                0 & B\\
                & 0 
            \end{pmatrix}
            } \mid B
            \right\}
            \to
            \mathcal{A}^{(2L)} = \left\{
            {\scriptsize
            \begin{pmatrix}
                0 & 0\\
                & 0 
            \end{pmatrix}
            }
            \right\}
        \end{equation*}

        However, not all off-diagonal blocks in the $[\{0,\varepsilon\},\{0,\varepsilon\}]$ sector vanish upon blocking. Consider
        \begin{equation*}
            \mathcal{A}^{(L)} = \left\{
            {\scriptsize
            \begin{pmatrix}
                0 & B & D \\
                & A & C \\
                & & 0 
            \end{pmatrix}
            } \mid A, B, C, D
            \right\}
            \to
            \mathcal{A}^{(2L)} = \mathcal{A}^{(L)}.
        \end{equation*}
        Here, block $(1,3)$ is a type 00 block, but it does \textit{not} vanish upon blocking. This is because we can multiply two elements in $\mathcal{A}^{(L)}$ as follows:
        \begin{equation*}
            {\scriptsize
            \begin{pmatrix}
                0 & B & \ast \\
                & 0 & 0 \\
                & & 0 
            \end{pmatrix}
            \begin{pmatrix}
                0 & 0 & \ast \\
                & 0 & C \\
                & & 0 
            \end{pmatrix}
            }
            =
            {\scriptsize
            \begin{pmatrix}
                0 & 0 & BC \\
                & 0 & 0 \\
                & & 0 
            \end{pmatrix}
            } \in \mathcal{A}^{(2L)} .
        \end{equation*}
        Thus, type 00 blocks may persist under blocking when they arise from a cascade of non-vanishing blocks in lower diagonals that themselves survive the blocking process. Remarkably, the presence of non-zero type 00 blocks is consistent with the desired properties of our basis construction: since we treat type 00 blocks as being of size $1 \times 1$, such blocks are automatically free. Moreover, they may also be linear combinations of other $1 \times 1$ blocks within the same sector $[\{0,\varepsilon\}, \{0,\varepsilon\}]$. 
    \end{itemize}

    \paragraph{Final step.} Lastly, we give the explicit value for the block-injectivity length $L_{BI}$, once all the induction steps are completed and we have reached block $(1,b)$. This comes after processing each of the $b(b-1)/2$ off-diagonal blocks, and therefore it is enough to solve the following 1st order linear non-homogeneous recurrence relation,
    \begin{equation*}
        L_n = 4L_{n-1}+2L_0, \quad \text{with} \quad L_1 := 4L_{BI}^{\text{diag}} + 2L_0^{\text{diag}} ,
    \end{equation*}
    which gives
    \begin{equation*}
        L_{b(b-1)/2} = \left( L_{BI}^{\text{diag}} + \frac{2}{3} L_0^{\text{diag}} \right) 2^{b(b-1)} - \frac{2}{3} L_0^{\text{diag}} =: L_{BI}.
    \end{equation*}
    The proof is now complete. 
\end{proof}

\newpage

\section{The matrix canonical form (matrix-CF)}
\label{app:matrix-CF}

The structure described in Prop. \ref{prop:structure_subalgebra} for the matrix algebra $\mathcal{A}$, and in Thm. \ref{prop:structure_span} for the span $\mathcal{A}^{(\ell)}$, guarantees the existence of sets of constants $\{k_{ij;e}\}$ and $\{k_{ij;e}^{(\ell)}\}$ for each $i,j \in \{1,\dots,b\}$ and $e \in \Sigma$, such that the elements
\begin{equation*}
    [A]_e := \sum_{i\leq j} [k_{ij;e} A]_{ij}, \quad [A]_e^{(\ell)} := \sum_{i\leq j} [k_{ij;e}^{(\ell)} A]_{ij}, \quad A \in \mathcal{M}_{D_i \times D_j}(\mathbb{C})
\end{equation*}
form a basis of $P\mathcal{A}P^{-1}$ or $P\mathcal{A}^{(\ell)}P^{-1}$ by construction, where $P$ is the invertible matrix constructed in Prop. \ref{prop:structure_subalgebra} or Thm. \ref{prop:structure_span}. The matrices $[A]_e$ and $[A]_e^{(\ell)}$ satisfy the structural properties that were summarized in Table \ref{tab:basis_summary} of Appendix \ref{app:notation_and_basis}.

Consider an MPS-X tensor $A^x$ written in the suitable gauge, such that its span after blocking every $\ell$ sites admits the basis $\{[A]_e^{(\ell)}\}$. Equivalently, there exist some matrices $\{A^x_e\}$ such that the tensor can be written as
\begin{equation*}
    A^x = \sum_{e \in \Sigma} [A^x_e]^{(\ell)}_e \ , \quad \forall x \in \{1, \dots, d\}.
\end{equation*}
Then, defining the tensors $A_{\mathrm{low}}$, $A_{\mathrm{up}}$ as
\begin{equation} \label{eq:Alow-Aup_definition}
    \begin{tikzpicture}[scale=.45, baseline={([yshift=-1.6ex]current bounding box.center)}, thick]
        \begin{scope}[shift={(0,0)}]
            \draw (-1.2,0) -- (1.2,0);
            \draw (0,1) -- (0,0);
            \filldraw[fill=purple] (-1/2-0.2,-1/2) -- (-1/2-0.2,1/2) -- (1/2+0.2,1/2) -- (1/2+0.2,-1/2) -- (-1/2-0.2,-1/2);
            \draw (0,0) node {\scriptsize $A_{\text{low}}$};
            \draw (-1.5,0) node {\scriptsize $i$};
            \draw (1.5,0) node {\scriptsize $j$};
            \draw (0,1.2) node {\scriptsize $e$};
        \end{scope}
    \end{tikzpicture}
    = 
    k_{ij;e}^{(\ell)} \ , 
    \quad 
    \begin{tikzpicture}[scale=.45, baseline={([yshift=-0.5ex]current bounding box.center)}, thick]
        \begin{scope}[shift={(0,0)}]
            \draw (-1.2,0) -- (1.2,0);
            \draw (0,1) -- (0,-1);
            \filldraw[fill=purple] (-1/2-0.2,-1/2) -- (-1/2-0.2,1/2) -- (1/2+0.2,1/2) -- (1/2+0.2,-1/2) -- (-1/2-0.2,-1/2);
            \draw (0,0) node {\scriptsize $A_{\text{up}}$};
            \draw (0,1.2) node {\scriptsize $x$};
            \draw (0,-1.2) node {\scriptsize $e$};
        \end{scope}
    \end{tikzpicture}
    = 
    A_e^x \ . 
\end{equation}
the tensor $A$ can be rewritten as
\begin{equation} \label{eq:matrix-CF-decomposition-tensors}
    \begin{tikzpicture}[scale=.45, baseline={([yshift=-1ex]current bounding box.center)}, thick]
        \MPSTensor{0,0}{$A$}{purple}
    \end{tikzpicture}
    =
    \begin{tikzpicture}[scale=.45, baseline={([yshift=-1ex]current bounding box.center)}, thick]
        \begin{scope}[shift={(0,0)}]
            \draw (-1.2,0) -- (1.2,0);
            \draw (0,1) -- (0,0);
            \filldraw[fill=purple] (-1/2-0.2,-1/2) -- (-1/2-0.2,1/2) -- (1/2+0.2,1/2) -- (1/2+0.2,-1/2) -- (-1/2-0.2,-1/2);
            \draw (0,0) node {\scriptsize $A_{\text{\normalfont{low}}}$};
        \end{scope}
        \begin{scope}[shift={(0,1.5)}]
            \draw (-1.2,0) -- (1.2,0);
            \draw (0,1) -- (0,0);
            \filldraw[fill=purple] (-1/2-0.2,-1/2) -- (-1/2-0.2,1/2) -- (1/2+0.2,1/2) -- (1/2+0.2,-1/2) -- (-1/2-0.2,-1/2);
            \draw (0,0) node {\scriptsize $A_{\text{\normalfont{up}}}$};
        \end{scope}
    \end{tikzpicture}.
\end{equation}
Moreover, the set of tensors defined by $A_{\mathrm{up}}$ is block-injective, since the span structure ensures that each free block can be isolated from the others by acting on the physical index alone, equivalently expressed as in Eq. \eqref{eq:def_block_injectivity} in the main text. Note that, whenever $\mathcal{A}^{(\ell)}$ forms an algebra, one can use the basis elements $\{[A]_e\}$ instead of $\{[A]_e^{(\ell)}\}$, which have a simpler structure. All of this motivates the introduction of a canonical form for the MPS matrices as follows.

\begin{definition}[Matrix-CF] \label{def:matrix-CF}
    An MPS tensor $A$ is said to be in \emph{matrix canonical form} (matrix-CF) if it can be decomposed in terms of two tensors $A_{\mathrm{low}}, A_{\mathrm{up}}$ as shown in Eq. \eqref{eq:matrix-CF-decomposition-tensors}, where $A_{\mathrm{up}}$ is a block-injective MPO, and $A_{\mathrm{low}}$ is an MPS tensor whose matrix entries satisfy the structural constraints in Table \ref{tab:Alow_summary}.
\end{definition}

\begin{table}[h!]
\centering
\renewcommand{\arraystretch}{1.25}
\begin{tabular}{c||c|c|c}
 & \multicolumn{2}{c|}{$e\in\Sigma_\infty$} & $e\in\Sigma_f^{pq}$ \\ \hline
 & $\mathcal{A}$ & $\mathcal{A}^{(\ell)}$ & $\mathcal{A}$ and $\mathcal{A}^{(\ell)}$ \\ \hline\hline
\makecell{Matrix\\structure}
& Diagonal
& \makecell{Diagonal, and off-diagonal \\ entries only in sector $[e,e]$}
& \makecell{Strictly \\ upper-triangular} \\ \hline
\makecell{Non-zero\\elements}
& $(A_{\mathrm{low}})_{ii}^e =
\begin{cases} 1 &\text{for } e = r_i \\ 0 &\text{otherwise} \end{cases}$
& $(A_{\mathrm{low}})_{ii}^e  \begin{cases}
    \in\mathbb{C}\setminus\{0\} &\text{for } e=r_i \\
    =0 & \text{otherwise}
\end{cases}$
& \makecell{$(A_{\mathrm{low}})_{ij}^e \neq 0$ \\ only if $e\in\Sigma_f^{r_i r_j}$} \\ 
\end{tabular}
\caption{Structure of the lower MPS tensor $A_{\mathrm{low}}$ in the matrix-CF. The constraints listed under the columns labeled $\mathcal{A}^{(\ell)}$ hold in general. If the span $\mathcal{A}^{(\ell)}$ forms an algebra for the chosen blocking length, then the additional properties in the columns labeled $\mathcal{A}$ also apply.}
\label{tab:Alow_summary}
\end{table}

\begin{theorem}[Existence of Matrix-CF] \label{thm:existence_matrixCF}
    Every MPS tensor can be brought into matrix-CF by a suitable gauge transformation after sufficient blocking:
    \begin{itemize}
        \item[(i)] For stable tensors which have been blocked every $pq L_{\mathrm{span}} b 2^b$ sites together (so that $\mathcal{A}^{(\ell)} = \mathrm{Alg}(\mathcal{A}^{(1)})$ holds for all $\ell \ge r_{\mathrm{alg}}$), the tensor $A_{\mathrm{low}}$ upon further blocking any $\ell \geq r_{\mathrm{alg}}$ sites satisfies the properties listed under the columns labeled $\mathcal{A}$ in Table \ref{tab:Alow_summary}.
        
        \item[(ii)] For arbitrary tensors, after blocking every $p\ell$ sites together for any $\ell \ge L_{\mathrm{span}}$, the tensor $A_{\mathrm{low}}$ satisfies the properties listed under the columns labeled $\mathcal{A}^{(\ell)}$ in Table \ref{tab:Alow_summary}.
    \end{itemize}
\end{theorem}
This result follows directly from Theorem \ref{prop:structure_span}, together with the stabilization properties of tensor sets stated in Proposition \ref{prop:stability_criterion}.

The structure of the degrees of freedom in the upper tensor $A_{\mathrm{up}}$ of the matrix-CF is formalized in the following proposition, and proven in Appendix \ref{app:sec_matrix-CF_freedom}.
This proposition is key to the general characterization of the freedom in the gCF. Recall that $r_1, r_2$ are functions $\Sigma \to \Sigma_\infty$ defined such that, for each $t \in \Sigma$, it holds that $t \in \tilde{\Sigma}_f^{r_t^1 r_t^2}$. 
\begin{restatable}{proposition}{freedommatrixCF} \label{prop:freedom_upper_matrixCF}
    Let $(X_A, A^i)$ and $(X_B, B^i)$ be two equivalent reduced MPS-X in matrix-CF, with decompositions $A_{\text{up}}, A_{\text{low}}$ and $B_{\text{up}}, B_{\text{low}}$, and free blocks labeled by alphabets $\Sigma^A = \Sigma_\infty^A \cup (\cup_{i,j \in \tilde{\Sigma}^A_\infty} \Sigma_f^{A,ij})$ and $\Sigma^B = \Sigma_\infty^B \cup (\cup_{i,j \in \tilde{\Sigma}^B_\infty} \Sigma_f^{B,ij})$, respectively. After blocking enough and relabeling the $\Sigma_\infty^B$ symbols, we can express $B_{\text{up}}$ in terms of $A_{\text{up}}$ as
    \begin{equation} \label{eq:final_relation_Bup=Aup}
        \begin{tikzpicture}[scale=.45, baseline={([yshift=-0.5ex]current bounding box.center)}, thick]
            \begin{scope}[shift={(0,0)}]
                \draw (-1.2,0) -- (1.2,0);
                \draw (0,1) -- (0,-1);
                \filldraw[fill=melon!50] (-1/2-0.2,-1/2) -- (-1/2-0.2,1/2) -- (1/2+0.2,1/2) -- (1/2+0.2,-1/2) -- (-1/2-0.2,-1/2);
                \draw (0,0) node {\scriptsize $B_{\text{up}}$};
            \end{scope}
        \end{tikzpicture} =
        \begin{tikzpicture}[scale=.45, baseline={([yshift=2.2ex]current bounding box.center)}, thick]
            \begin{scope}[shift={(0,0)}]
                \draw (-1.2,0) -- (1.2,0);
                \draw (0,1) -- (0,-3);
                \filldraw[fill=powderblue!50] (-1/2-0.2,-1/2) -- (-1/2-0.2,1/2) -- (1/2+0.2,1/2) -- (1/2+0.2,-1/2) -- (-1/2-0.2,-1/2);
                \draw (0,0) node {\scriptsize $A_{\text{up}}$};
            \end{scope}
            \begin{scope}[shift={(1.7,0)}]
                \draw (-1,0) -- (1,0);
                \filldraw[fill=gray!10] (0.5,-0.5) -- (-0.5,-0.5) -- (-0.5,0.5) -- (0.5, 0.5) -- (0.5,-0.5);
                \draw (0,0) node {\scriptsize $Z$};
                \draw (0,-0.5) -- (0,-1);
            \end{scope}
            \begin{scope}[shift={(-2.2,0)}]
                \draw (-1,0) -- (1.5,0);
                \draw (0.25,-1) -- (0.25,-0.5);
                \filldraw[fill=gray!10] (1,-0.5) -- (-0.5,-0.5) -- (-0.5,0.5) -- (1, 0.5) -- (1,-0.5);
                \draw (0.3,0) node {\scriptsize $Z^{-1}$};
            \end{scope}
            \draw (-1.95,-0.5) -- (-1.95,-1) -- (1.7,-1) -- (1.7,-0.5);
            \fill (0,-1) circle (0.1);
            \begin{scope}[shift={(-0.975,-1)}]
                \filldraw[fill=yellow!75] (0.35,-0.35) -- (-0.35,-0.35) -- (-0.35,0.35) -- (0.35, 0.35) -- (0.35,-0.35);
                \draw (0,0) node {\scriptsize $r^1$};
            \end{scope} 
            \begin{scope}[shift={(0.85,-1)}]
                \filldraw[fill=yellow!75] (0.35,-0.35) -- (-0.35,-0.35) -- (-0.35,0.35) -- (0.35, 0.35) -- (0.35,-0.35);
                \draw (0,0) node {\scriptsize $r^2$};
            \end{scope}
            \begin{scope}[shift={(0,-2)}]
                \filldraw[fill=gray!10] (0.5,-0.5) -- (-0.5,-0.5) -- (-0.5,0.5) -- (0.5, 0.5) -- (0.5,-0.5);
                \draw (0,0) node {\scriptsize $P_B$};
            \end{scope}
        \end{tikzpicture}
    \end{equation}
    where $\Sigma_\infty^A = \Sigma_\infty^B =: \Sigma$ and $\Sigma_f^{A, ij} = \Sigma_f^{B, ij} =: \Sigma_f^{ij}$. Here $P_B$ and $Z_j$ are invertible matrices, with $P_B$ block-diagonal and acting as
    \begin{equation} \label{prop:P_B-properties-old}
        \begin{cases}
            P_B \ket{x} = \alpha_x \ket{x} & \text{if } x \in \Sigma_\infty , \\
            P_B ( \langle \Sigma_f^{ij} \rangle ) \subseteq \langle \tilde{\Sigma}_f^{ij}\rangle  & \text{for } i,j \in \tilde{\Sigma}_\infty ,
        \end{cases}
    \end{equation}
    for some scalars $\alpha_x \in \mathbb{C}$.
\end{restatable}

{}

\newpage

\section{Proofs for the translational invariance property} \label{app:TI_proofs}

We now present some technical lemmas that will allow us to prove how translational invariance enables us to write the boundary matrix of an MPS-X in a simplified form, as informally stated in Proposition \ref{prop:check_TI_informal}. Actually, we establish a stronger result, which provides necessary and sufficient conditions characterizing when an MPS-X is translationally invariant. We will use the following characterization of translational invariance in terms of the algebra $\mathcal{A}$ of the MPS matrices, extracted from Lemma 3 in \cite{florido-llinas_2024_RLS}.

\begin{restatable}{lemma}{lemmaTI}
\label{lemma:TI}
    A family of MPS-X states $\{\ket{\psi_N(X, A)}\}_N$, 
    \begin{equation} \label{eq:general_MPS-X}
        \ket{\psi_N(X, A)} := 
        \begin{tikzpicture}[scale=.45, baseline={([yshift=-1ex]current bounding box.center)}, thick]
            \FullMPSX{(0,0)}{$A$}{$X$}{purple}{yellow}
            \node at (3,0.5) {\tiny $N$ \text{\normalfont times}};
        \end{tikzpicture},
    \end{equation}
    where $X, A^i \in \mathcal{M}_D(\mathbb{C})$ is TI for all $N$, if and only if,
    \begin{equation}
        \Tr[X[a,b]] = 0, \ \forall a, b \in \text{\normalfont Alg}(\{A^i\}).
    \end{equation}
\end{restatable}

The following two technical lemmas will enable us to leverage the algebra structure of Proposition \ref{prop:structure_subalgebra} to study the TI condition.
\begin{lemma} \label{lemma:comm=0}
    If $X \in \mathcal{M}_{D_1 \times D_2}(\mathbb{C})$ satisfies $\Tr[XC] = 0, \forall C \in \mathcal{M}_{D_2 \times D_1}(\mathbb{C})$, then $X = 0$.
\end{lemma}
\begin{proof}
    Take $C = \dyad{j}{i}$ for any $j \in \{1, \dots, D_2\}$, $i \in \{1, \dots, D_1\}$. Then, $\Tr[XC] = \mel{i}{X}{j} = X_{ij} = 0$, implying that $X = 0$. 
\end{proof}

\begin{lemma} \label{lemma:Xis1}
    If $X \in \mathcal{M}_D(\mathbb{C})$ satisfies $\Tr[X[C,D]] = 0$, $\forall C, D \in \mathcal{M}_D(\mathbb{C})$, then $X \propto \mathds{1}$. 
\end{lemma}
\begin{proof}
    Note that $\Tr[X[C,D]] = \Tr[[X,C]D]$. Then, the fact that $\Tr[[X,C]D] = 0, \forall D$, necessarily implies $[X,C] = 0$ for any $C$. Since the only matrices that commute with all the elements in $\mathcal{M}_D(\mathbb{C})$ are proportional to the identity, we have that $X = \alpha \mathds{1}$ for some constant $\alpha \in \mathbb{C}$.
\end{proof}
\begin{corollary}
    Given a normal tensor $A$, $\{\ket{\psi_N(X,A)}\}_N$ is TI $\iff X \propto \mathds{1}$. 
\end{corollary}

We proceed to prove now the full formal version of Prop. \ref{prop:check_TI_informal} in the main text. We denote by $\{k_{ij;t}\}$ the set of constants defining a basis of the algebra, as prescribed by Prop. \ref{prop:structure_subalgebra} and made precise in Appendix \ref{app:notation_and_basis}. For stable MPS-X, these constants coincide by construction with the lower tensor of the matrix-CF, i.e. $k_{ij;t} = (A_{\mathrm{low}}^t)_{ij}$. 
\begin{myprop}{4}[Translational invariance, formal]
\label{prop:check_TI}
    Let $(X, A^i)$ be the tensors of an MPS-X where $\{A^i\}$ is in matrix-CF. Then, the MPS-X is translationally invariant if and only if the following two conditions hold:
    \begin{enumerate}[label=(\roman*)]
        \item There exist constants $\beta_t \in \mathbb{C}$ such that, for each $t \in \Sigma$,
        \begin{equation*}
            \sum_{i,j} k_{ij;t} X_{ji} 
            = 
            \begin{cases}
                \beta_t \mathds{1} & \text{if } t \in [p,p] \text{ for some } p \in \Sigma_\infty , \\
                0 & \text{otherwise.} 
            \end{cases}
        \end{equation*}
        If this is the case, the simplified boundary matrix $\tilde{X}$ that is zero except for the blocks $\tilde{X}_{j_t i_t} = \beta_t \mathds{1}$, where $(i_t,j_t)$ denotes the first occurrence of each free block $t \in \Sigma$ in the $\preceq$-ordered block structure, generates the same family of states, i.e. $         |\psi_N(X,A) \rangle = | \psi_N (\tilde{X}, A) \rangle, \ \forall N$.
        
        \item The constants $\{\beta_t\}$ satisfy a set of linear relations determined by the structure constants $\Gamma$ of $\mathcal{A}_{\text{low}} := \mathrm{Alg}(\{A_{\mathrm{low}}^i\})$, 
        \begin{equation*}
            \sum_{r \in \Sigma_f} \beta_r (\Gamma^{pq}_r - \Gamma^{qp}_r) = 0 \ , \quad \forall p,q \in \Sigma_f. 
        \end{equation*}
    \end{enumerate}
\end{myprop}
\begin{proof}
    We are going to study the TI condition, $\Tr[X[a,b]] = 0, \forall a, b \in \mathcal{A}$, in terms of the basis elements of $\mathcal{A}$ described in Appendix \ref{app:notation_and_basis}.

    First, for some $\{0,p\}, \{0,q\} \in \Sigma_\infty$, take $a = [A]_{\{0,p\}}, b = [B]_{\{0,q\}}$ with arbitrary $A, B$. 
    If $\{0,p\} \neq \{0,q\}$, then $ab = ba = 0$ due to property (P1) of $\Gamma$ as stated in Appendix \ref{app:notation_and_basis}.

    Let $\{0,p\} = \{0,q\}$. Again, property (P1) of $\Gamma$ tells us that $ab = [AB]_{\{0,p\}}$. Therefore, we can rewrite the TI condition $\Tr[X[a,b]] = 0$ as
    \begin{equation*}
        0 
        = \Tr\left[ \left( \sum_{ij} k_{ij;\{0,p\}} X_{ji} \right) [A,B]\right] 
        = \Tr\left[ \left( \sum_{i} k_{ii;\{0,p\}} X_{ii} \right) [A,B]\right], \quad \forall A, B
    \end{equation*}
    which by Lemma \ref{lemma:Xis1} implies that, for each $\{0,p\} \in \Sigma_\infty$, there exist constants $\beta_{\{0,p\}} \in \mathbb{C}$ such that
    \begin{equation*}
        \sum_{i} k_{ii;\{0,p\}} X_{ii} = \beta_{\{0,p\}} \mathds{1}.
    \end{equation*}
    Note that, for the algebra, the constants $k_{ii;\{0,p\}}$ can only be either 0 or 1. 

    Now, take $a = [A]_{\{0,p\}}$, $b = [B]_{\{q\}}$ for any arbitrary $A, B$. Using the properties of $\Gamma$, we know that
    \begin{equation*}
        [a,b] 
        = 
        \sum_{\gamma} \left( 
        \Gamma^{\{0,p\},\{q\}}_{\gamma} [AB]_{\gamma} - 
        \Gamma^{\{q\}, \{0,p\}}_{\gamma} [BA]_{\gamma}  \right).
    \end{equation*}
    Due to property (P3) of $\Gamma$, we know that $\Gamma^{uv}_w \neq 0$ necessarily implies that $\exists s_1, s_2, s_3 \in \tilde{\Sigma}_\infty$ such that $u \in \tilde{\Sigma}_f^{s_1 s_2}$, $v \in \tilde{\Sigma}_f^{s_2 s_3}$ and $w \in \tilde{\Sigma}_f^{s_1 s_3}$. Thus, we can rewrite the expression above as
    \begin{equation*}
        [a,b] = \delta_{p r_q^1} [AB]_{\{q\}} - \delta_{p r_q^2}[BA]_{\{q\}}
    \end{equation*}
    In order for the TI condition $\Tr[X[a,b]] = 0$ to lead to a non-trivial equation, we need $p = r_q^1$ or $p = r_q^2$. Recall that the functions $r^1, r^2: \Sigma \to \Sigma_\infty$, as introduced in Thm. \ref{prop:freedom_gCF} of Sec. \ref{sec:mps-x} in the main text, indicate for each $t \in \Sigma$ that it belongs to $\tilde{\Sigma}_f^{r_t^1 r_t^2}$.

    First, let us assume that $r_q^1 \neq r_q^2$, and choose $\{0,p\} := \{0,r_q^1\}$. Then, we obtain that
    \begin{align*}
        0 &= \Tr[X[a,b]] = 
        \sum_{ij} \Tr[ X [k_{ij;\{q\}} AB]_{ij} ] \\
        &= \Tr[ \left(\sum_{ij} k_{ij;\{q\}} X_{ji} \right) AB ], \ \forall A, B \xRightarrow{\text{Lemma } \ref{lemma:comm=0}}
        \sum_{ij} k_{ij;\{q\}} X_{ji} = 0.
    \end{align*}
    Note that, if $r_q^1 = \varepsilon$, we could have just taken instead $\{0,p\} := \{0,r_q^2\}$ and the same equation would follow. 

    Now, assume that $r_q^1 = r_q^2$, and take $\{0,p\} := \{0,r_q^1\}$. We obtain that
    \begin{align}
        0 &= \Tr[X[a,b]] = 
        \sum_{ij} \Tr[ X [k_{ij;\{q\}}(AB-BA)]_{ij} ] \nonumber \\
        &= \Tr[ \left(\sum_{ij} k_{ij;\{q\}} X_{ji} \right) [A,B] ], \ \forall A, B \xRightarrow{\text{Lemma } \ref{lemma:Xis1}}
        \exists \beta_{\{q\}} \in \mathbb{C} \text{ s.t. }
        \sum_{ij} k_{ij;\{q\}} X_{ji} = \beta_{\{q\}} \mathds{1}. \label{eq:aaaa1}
    \end{align}
    In the special case where $r_q^1 = r_q^2 = \varepsilon$, the procedure above would not lead to any non-trivial equation, as there is no non-zero basis element associated to $\{0,\varepsilon\}$ by definition. However, Eq. \eqref{eq:aaaa1} still holds because any block $(i,j)$ in sector $[\{0,\varepsilon\}, \{0,\varepsilon\}]$, and thus also the corresponding $X_{ji}$, have size $1 \times 1$ by construction. 
    
    To prove part \emph{(ii)} of the claim, it only remains to take $a = [A]_{\{p\}}$ and $b = [B]_{\{q\}}$ with arbitrary $A, B$, for any $\{p\}, \{q\} \in \Sigma_f$.
    \begin{align}
        0 &= \Tr[X[a,b]] = \Tr[X \left( [A]_{\{p\}} [B]_{\{q\}} - [B]_{\{q\}} [A]_{\{p\}} \right)] \nonumber \\
        &= \sum_{\{r\} \in \Sigma_f} \left( \Gamma^{\{p\},\{q\}}_{\{r\}} \Tr[X[AB]_{\{r\}}] -\Gamma^{\{q\},\{p\}}_{\{r\}} \Tr[X[BA]_{\{r\}}] \right) \nonumber \\
        &= \sum_{\{r\} \in \Sigma_f} \left( \Gamma^{\{p\},\{q\}}_{\{r\}} \Tr[\left( \sum_{ij}  k_{ij;\{r\}} X_{ji} \right) AB] - 
        \Gamma^{\{q\},\{p\}}_{\{r\}} \Tr[\left( \sum_{ij} k_{ij;\{r\}} X_{ji} \right) BA] \right) \nonumber \\
        &= \sum_{\{r\} \in \Sigma_f} \left( \Gamma^{\{p\},\{q\}}_{\{r\}} - \Gamma^{\{q\},\{p\}}_{\{r\}} \right) \beta_{\{r\}} \Tr[AB] = 0, \ \forall A, B \nonumber \\
        &\quad \implies \sum_{\{r\}\in\Sigma_f} \beta_{\{r\}} \left( \Gamma^{\{p\},\{q\}}_{\{r\}} - \Gamma^{\{q\},\{p\}}_{\{r\}} \right) = 0, \ \forall \{p\}, \{q\} \in \Sigma_f. \label{eq:TI_condition_on_beta}
    \end{align}
    These are the additional restrictions on the constants $\beta_i$ that only depend on the algebra structure and ensure translational invariance.
    
    Finally, note that the only relevant quantities about $X$ that have a role in the state defined by the MPS-X are actually the $\beta_i$'s. Indeed, the coefficients in front of each ket $\ket{i_1 i_2 \dots i_N}$ are $\Tr[X a^{i_1} a^{i_2} \dots a^{i_N}]$. Since $a^{i_1} a^{i_2} \dots a^{i_N} \in \mathcal{A}$, it can be expressed in terms of the chosen basis as
    \begin{equation*}
        a^{i_1} a^{i_2} \dots a^{i_N} = \sum_{e \in \Sigma} [A_e]_e, 
    \end{equation*}
    so we have that
    \begin{align*}
        \Tr[Xa^{i_1} a^{i_2} \dots a^{i_N}] &= \Tr[X \left( \sum_{\{0,r\} \in \Sigma_\infty} [A_{\{0,r\}}]_{\{0,r\}} + \sum_{\{s\} \in \Sigma_f} [A_{\{s\}}]_{\{s\}} \right)] \\
        &= \Tr[ \sum_{\{0,r\} \in \Sigma_\infty} \left( \sum_{i} k_{ii;\{0,r\}} X_{ii} \right) A_{\{0,r\}} + \sum_{\{s\} \in \Sigma_f} \left( \sum_{ij} k_{ij;\{s\}} X_{ji} \right) A_{\{s\}} ] = \\
        &= \sum_{\{0,r\} \in \Sigma_\infty} \beta_{\{0,r\}} \Tr[A_{\{0,r\}}] + \sum_{\{s\} \in \Sigma_f} \beta_{\{s\}} \Tr[A_{\{s\}}].
    \end{align*}
    This shows that we can modify $X$ as much as we want, as long as the values of $\beta_{\{0,r\}}$ and $\beta_{\{s\}}$ do not change. In particular, we can without loss of generality define a new boundary matrix $Y$ that is all zeros except for 
    \begin{equation} \label{eq:simplified_boundary_TI}
        \tilde{X}_{j_t i_t} = \beta_t \mathds{1}
    \end{equation}
    for each $t \in \Sigma$, where $(i_t,j_t)$ marks the first occurrence under the $\preceq$-order of the free block labeled by $t$. With this simplified choice, we have that $\ket{\psi_N(X,A)} = \ket{\psi_N(\tilde{X},A)}$ for all $N$.    
\end{proof}

\paragraph{Example.}
Given $\mathcal{A} = \left\{ {\scriptsize\begin{pmatrix} 
    A & B & D \\ & A & C \\ & & A
\end{pmatrix}} \mid A,B,C,D \right\}$, we have that:
\begin{itemize}
    \item $\Sigma$ can be partitioned into $\Sigma_\infty = \{\{0,1\}\}$ and $\Sigma_f = \Sigma_f^{11} = \{\{1\},\{2\},\{3\}\}$ representing free blocks $B,C$ and $D$, respectively.

    \item The structure constants tensor when acting on the $\Sigma_f$ symbols is all zeros except for:
    \begin{equation*}
        \begin{tikzpicture}[scale=.45, baseline={([yshift=-1ex]current bounding box.center)}, thick]
            \begin{scope}[shift={(0,0)}]
                \draw (0.3,0.5) -- (0.3,1);
                \draw (-0.3,0.5) -- (-0.3,1);
                \filldraw[fill=gray!10] (-0.5,-0.5) -- (-0.5,0.5) -- (0.5,0.5) -- (0.5,-0.5) -- (-0.5,-0.5);
                \draw (0,0) node {\scriptsize $\Gamma$};
                \draw (0,-0.5) -- (0,-1);
            \end{scope}
            \draw (-0.5,1.4) node {\scriptsize $\{1\}$};
            \draw (0.5,1.4) node {\scriptsize $\{2\}$};
            \draw (0,-1.3) node {\scriptsize $\{3\}$};
        \end{tikzpicture}
        = 1 . 
    \end{equation*}
    Then, Eq. \eqref{eq:TI_condition_on_beta} reads as:
    \begin{equation} \label{eq:condition_on_beta3}
        0 = \sum_{\{r\}\in\Sigma_f} \beta_{\{r\}} \left( \Gamma^{\{1\},\{2\}}_{\{r\}} - \Gamma^{\{2\},\{1\}}_{\{r\}} \right) = \beta_{\{3\}}. 
    \end{equation}
    \item Therefore, according to Proposition \ref{prop:check_TI}, the set of boundary conditions that guarantee translational invariance is
    \begin{equation*}
        \left\{
            {\scriptsize \begin{pmatrix}
                X_{11} & \ast & \ast \\
                \beta_{\{1\}}\mathds{1} & X_{22} & \ast \\
                0 & \beta_{\{2\}} \mathds{1} & \beta_{\{0,1\}}\mathds{1}-X_{11}-X_{22}
            \end{pmatrix}}
            \mid X_{11}, X_{22}, \beta_{\{0,1\}}, \beta_{\{1\}}, \beta_{\{2\}}
        \right\} ,
    \end{equation*}
    where $X_{31}$ is necessarily zero due to the fact that $\beta_{\{3\}} = 0$ from Eq. \eqref{eq:TI_condition_on_beta}. The set of simplified boundary conditions $\tilde{X}$, as defined in Eq. \eqref{eq:simplified_boundary_TI}, that generate the same family of MPS-X states for each choice of $\beta_i$'s is thus given by
    \begin{equation*}
        \left\{
            {\scriptsize \begin{pmatrix}
                \beta_{\{0,1\}} \mathds{1} & \ast & \ast \\
                \beta_{\{1\}}\mathds{1} & 0 & \ast \\
                0 & \beta_{\{2\}} \mathds{1} & 0
            \end{pmatrix}} \mid
            \beta_{\{0,1\}}, \beta_{\{1\}}, \beta_{\{2\}} \in \mathbb{C}
        \right\} .
    \end{equation*}
    
\end{itemize}

\newpage

\section{Proof that every stable MPS-X can be written in gCF} \label{app:proof_thm_generality-gCF}

In this section, using the matrix-CF introduced in Appendix \ref{app:matrix-CF}, we prove one of the main results of this work (Thm. \ref{thm:generality_gCF}), which claims that any stable TI MPS-X can be brought into the gCF of Def. \ref{def:gCF}. We restate the claim below for convenience. 
\generalitygCF*
\begin{proof}
    Stability of the MPS implies, by Theorem \ref{prop:stability_criterion}, that
    \begin{equation*} 
        \mathcal{A}^{(pqL_{\mathrm{span}} 2^b (r_{\mathrm{alg}}b + s))} = \mathrm{Alg}(\mathcal{A}^{(pqL_{\mathrm{span}} b 2^b (1+t))}), \quad \forall s,t \geq 0.
    \end{equation*}
    With $L := pq L_{\mathrm{span}} b 2^b$ and $s = b\tilde{s}$ for any $\tilde{s} \in \mathbb{N}$, this becomes 
    \begin{equation} \label{eq:stability_constructing_gCF_1}
        \mathcal{A}^{(\tilde{s}L)} = \mathrm{Alg}(\mathcal{A}^{(L)}), \quad \forall \tilde{s} \geq r_{\mathrm{alg}}.
    \end{equation}
    Denote the blocked matrices as $A^i$, with $i = i_1 i_2 \dots i_L \in \{1, \dots, d\}^L$. Eq. \eqref{eq:stability_constructing_gCF_1} then reads as
    \begin{equation} \label{eq:stability_constructing_gCF_2}
        \mathcal{A}^{(s)} = \mathrm{Alg}(\mathcal{A}^{(1)}), \quad \forall s \ge r_{\mathrm{alg}} .
    \end{equation}
    Proposition \ref{prop:structure_subalgebra} (and more specifically, Theorem \ref{thm:existence_matrixCF} in Appendix \ref{app:matrix-CF}) ensures the existence of an invertible matrix $P$ such that the stable MPS tensors can be decomposed in matrix-CF as
    \begin{equation} \label{eq:find_gCF_matrix-CF}
        \begin{tikzpicture}[scale=.45, baseline={([yshift=-1ex]current bounding box.center)}, thick]
            \MPSTensor{0,0}{$A$}{purple}
            \begin{scope}[shift={(-1.5,0)}]
                \draw (-1,0) -- (1,0);
                \filldraw[fill=yellow] (0.5,-0.5) -- (-0.5,-0.5) -- (-0.5,0.5) -- (0.5, 0.5) -- (0.5,-0.5);
                \draw (0,0) node {\scriptsize $P$};
            \end{scope}
            \begin{scope}[shift={(1.5,0)}]
                \draw (-1,0) -- (1.5,0);
                \filldraw[fill=yellow] (1,-0.5) -- (-0.5,-0.5) -- (-0.5,0.5) -- (1, 0.5) -- (1,-0.5);
                \draw (0.3,0) node {\scriptsize $P^{-1}$};
            \end{scope}
        \end{tikzpicture}
        =
        \begin{tikzpicture}[scale=.45, baseline={([yshift=-1ex]current bounding box.center)}, thick]
            \begin{scope}[shift={(0,0)}]
                \draw (-1.2,0) -- (1.2,0);
                \draw (0,1) -- (0,0);
                \filldraw[fill=purple] (-1/2-0.2,-1/2) -- (-1/2-0.2,1/2) -- (1/2+0.2,1/2) -- (1/2+0.2,-1/2) -- (-1/2-0.2,-1/2);
                \draw (0,0) node {\scriptsize $A_{\text{\normalfont{low}}}$};
            \end{scope}
            \begin{scope}[shift={(0,1.5)}]
                \draw (-1.2,0) -- (1.2,0);
                \draw (0,1) -- (0,0);
                \filldraw[fill=purple] (-1/2-0.2,-1/2) -- (-1/2-0.2,1/2) -- (1/2+0.2,1/2) -- (1/2+0.2,-1/2) -- (-1/2-0.2,-1/2);
                \draw (0,0) node {\scriptsize $A_{\text{\normalfont{up}}}$};
            \end{scope}
        \end{tikzpicture},    
    \end{equation}
    with $A_{\mathrm{low}}, A_{\mathrm{up}}$ defined in Eq. \eqref{eq:Alow-Aup_definition}, in terms of the set of constants $\{k_{ij;e}\}\subseteq \mathbb{C}$ that determine a basis for $\mathrm{Alg}(\mathcal{A}^{(1)})$. This decomposition is valid under further blocking since $\mathcal{A}^{(\ell)} \subseteq \mathrm{Alg}(\mathcal{A}^{(1)})$ for all $\ell \geq 1$: $A_{\mathrm{low}}$ stays invariant, encoding the algebraic relations between the free blocks of $\mathrm{Alg}(\mathcal{A}^{(1)})$, and $A_{\mathrm{up}}$ contains the free blocks themselves. 

    By Proposition \ref{prop:check_TI}, translational invariance enables us to transform the boundary matrix $X$ into a simplified form $\tilde{X}$ whose blocks are either zero or proportional to the identity, i.e. $X_{ij} = x_{ij}\mathds{1}$ for some $x_{ij} \in \mathbb{C}$, $\forall i,j \in \{1,\dots, b\}$. 

    Define a lower-triangular matrix $Y \in \mathcal{M}_{b \times b}(\mathbb{C})$ by setting $Y_{ij} = x_{ij}$. The backbone states of the gCF are then given by
    \begin{equation*}
        \ket{L_N} := \ket{\psi_N(Y, A_{\mathrm{low}})} .
    \end{equation*}
    Due to the properties of $A_{\mathrm{low}}$ listed under the $\mathcal{A}$-columns in Table \ref{tab:Alow_summary}, the family $\{\ket{L_N}\}$ is an algebraic RLS. To see this, first note that $A_{\mathrm{low}}^e$ for $e \in \Sigma_\infty$ is zero except for some 1s in the diagonal, and thus contribute no nontrivial coefficients to $\ket{L_N}$. This means that all such nontrivial coefficients of $\ket{L_N}$ arise from $e \in \Sigma_f$, and can be entirely absorbed into the defining states $\ket{X_O}$ (cf. Def. \ref{def:algebraicRLS}), consisting of weighted superpositions of strings in $\Sigma_f^m$ for each $O \in \tilde{\Sigma}_\infty^{m+1}$.
    
    Off-diagonal blocks of $\mathrm{Alg}(\mathcal{A}^{(1)})$ are either zero, free, or linear combinations of free blocks within the same sector, so $(A_{\mathrm{low}}^t)_{mn} \neq 0$ for $m < n$ only if $t \in \Sigma_f$ and $(m,n) \in [r_1^t, r_2^t]$. Consequently, any ket $\ket{t_1 t_2 \dots t_N}$ in $\ket{X_O}$ with a nonzero coefficient requires  $A_{\mathrm{low}}^{t_1} A_{\mathrm{low}}^{t_2} \dots A_{\mathrm{low}}^{t_N} \neq 0$, which in turn occurs only if $t_1 t_2 \dots t_N \in \Sigma_f^{O_0 O_1} \cdot \Sigma_f^{O_1 O_2} \cdot \ldots \cdot \Sigma_f^{O_{m-1} O_m}$ for some $O = O_0 O_1 \dots O_m \in \tilde{\Sigma}_\infty^{m+1}$, as required by Definition \ref{def:algebraicRLS}. Hence $\{\ket{L_N}\}$ forms an algebraic RLS.

    Let $\Gamma$ be the structure constants tensor of the algebra $\mathcal{A}_{\mathrm{low}} := \mathrm{Alg}(\{A_{\mathrm{low}}^e\})$ with respect to its basis elements $A_{\mathrm{low}}^e$, i.e. 
    \begin{equation} \label{eq:Gamma_prescription}
        A_{\text{low}}^i \cdot A_{\text{low}}^j = \sum_k \Gamma^{ij}_k A_{\text{low}}^k \iff
        \begin{tikzpicture}[scale=.45, baseline={([yshift=-1ex]current bounding box.center)}, thick]
            \begin{scope}[shift={(0,0)}]
                \draw (-1.2,0) -- (1.2,0);
                \draw (0,1) -- (0,0);
                \filldraw[fill=purple] (-1/2-0.2,-1/2) -- (-1/2-0.2,1/2) -- (1/2+0.2,1/2) -- (1/2+0.2,-1/2) -- (-1/2-0.2,-1/2);
                \draw (0,0) node {\scriptsize $A_{\text{low}}$};
            \end{scope}
            \begin{scope}[shift={(1.9,0)}]
                \draw (-1.2,0) -- (1.2,0);
                \draw (0,1) -- (0,0);
                \filldraw[fill=purple] (-1/2-0.2,-1/2) -- (-1/2-0.2,1/2) -- (1/2+0.2,1/2) -- (1/2+0.2,-1/2) -- (-1/2-0.2,-1/2);
                \draw (0,0) node {\scriptsize $A_{\text{low}}$};
            \end{scope}
        \end{tikzpicture}
        =
        \begin{tikzpicture}[scale=.45, baseline={([yshift=-1.25ex]current bounding box.center)}, thick]
            \begin{scope}[shift={(0,0)}]
                \draw (-1.2,0) -- (1.2,0);
                \draw (0,1) -- (0,0);
                \filldraw[fill=purple] (-1/2-0.2,-1/2) -- (-1/2-0.2,1/2) -- (1/2+0.2,1/2) -- (1/2+0.2,-1/2) -- (-1/2-0.2,-1/2);
                \draw (0,0) node {\scriptsize $A_{\text{low}}$};
            \end{scope}
            \begin{scope}[shift={(0,1.5)}]
                \draw (0.5,0.5) -- (0.5,1);
                \draw (-0.5,0.5) -- (-0.5,1);
                \filldraw[fill=gray!10] (-0.5-0.2,-0.5) -- (-0.5-0.2,0.5) -- (0.5+0.2,0.5) -- (0.5+0.2,-0.5) -- (-0.5-0.2,-0.5);
                \draw (0,0) node {\scriptsize $\Gamma$};
                \draw (0,-0.5) -- (0,-1);
            \end{scope}
        \end{tikzpicture}
    \end{equation}
    Then, the backbone states are $\Gamma$-invariant, as required in the definition of the gCF. The associativity of the $\Gamma$-tensor follows from the associativity of the algebra. 

    The upper tensor in the gCF is $A_{\mathrm{up}}$. Under blocking every $\ell$ sites, it transforms according to $\Gamma_\ell$ as shown in Def.~\ref{def:gCF}\textit{(ii)}. Moreover, Eq. \eqref{eq:stability_constructing_gCF_2} guarantees block-injectivity of $A_{\mathrm{up}}$ with block-injectivity length $L_{BI} \leq r_{\mathrm{alg}}$. Upper bounding $L_{\mathrm{span}}$ in Eq. \eqref{eq:upper_bound_difficult_Lspan} with a simpler expression, and using the fact that $r_{\mathrm{alg}} \leq D^2$, we arrive at Theorem \ref{thm:generality_gCF}. This completes the proof.
\end{proof}

The procedure used in the proof to bring any stable MPS-X into its gCF was illustrated with the W-like example in Sec. \ref{sec:how-to}. For completeness, and to showcase a case with a richer block structure, we present here a second example.

Consider an MPS-X with tensors
\begin{equation*}
    a^i = {\scriptsize
    \begin{pmatrix}
        A^i & C^i & D^i & E^i \\
        & B^i & 0 & 0 \\
        & & A^i & D^i \\
        & & & A^i
    \end{pmatrix}
    }, \quad 
    X = {\scriptsize
    \begin{pmatrix}
        X_{11} & & & \\
        X_{21} & X_{22} & & \\
        X_{31} & X_{32} & X_{33} & \\
        X_{41} & X_{42} & X_{43} & X_{44}
    \end{pmatrix}
    }
\end{equation*}
and assume that the algebra they generate coincides with the one in Eq. \eqref{eq:example_nonsemisimple_part}, revisited in Eq. \eqref{eq:example_nonsemisimple_part_appendix}. We further assume stability, in particular
\begin{equation} \label{eq:example_nonsemisimple_part_appendix_2}
    \mathcal{A}^{(\ell)} = \mathrm{Alg}(\mathcal{A}^{(1)}) = \left\{
    {\scriptsize
    \begin{pmatrix}
        A & C & D & E \\
        & B & 0 & 0 \\
        & & A & D \\
        & & & A
    \end{pmatrix}
    } \mid A, B, C, D, E\right\}
    , \quad \forall \ell .
\end{equation}
Since this algebra is already in the form with the properties described in Proposition \ref{prop:structure_subalgebra}, we can start directly from Step 2 of the recipe in Section \ref{sec:how-to}.

\paragraph{Step 2.} Using the basis in Eq. \eqref{eq:example_nonsemisimple_part_appendix_2}, we decompose each tensor according to the basis of $\mathrm{Alg}(\mathcal{A}^{(1)})$ in Eq.~\eqref{eq:example_nonsemisimple_part_appendix_2} as
\begin{align*}
    \begin{tikzpicture}[scale=.45, baseline={([yshift=-1ex]current bounding box.center)}, thick]
        \MPSTensor{0,0}{$A$}{purple}
    \end{tikzpicture} \, 
    =
    \begin{tikzpicture}[scale=.45, baseline={([yshift=-1ex]current bounding box.center)}, thick]
        \begin{scope}[shift={(0,0)}]
            \draw (-1.2,0) -- (1.2,0);
            \draw (0,1) -- (0,0);
            \filldraw[fill=purple] (-1/2-0.2,-1/2) -- (-1/2-0.2,1/2) -- (1/2+0.2,1/2) -- (1/2+0.2,-1/2) -- (-1/2-0.2,-1/2);
            \draw (0,0) node {\scriptsize $A_{\mathrm{low}}$};
        \end{scope}
        \begin{scope}[shift={(0,1.5)}]
            \draw (-1.2,0) -- (1.2,0);
            \draw (0,1) -- (0,0);
            \filldraw[fill=purple] (-1/2-0.2,-1/2) -- (-1/2-0.2,1/2) -- (1/2+0.2,1/2) -- (1/2+0.2,-1/2) -- (-1/2-0.2,-1/2);
            \draw (0,0) node {\scriptsize $A_{\text{\normalfont{up}}}$};
        \end{scope}
    \end{tikzpicture}, 
\end{align*}
with the components $A_{\mathrm{low}}, A_{\mathrm{up}}$ explicitly given by:
\begin{align*}
    &\begin{tikzpicture}[scale=.45, baseline={([yshift=-1.6ex]current bounding box.center)}, thick]
        \draw (-1.2,0) -- (1.2,0);
        \draw (0,1) -- (0,0);
        \filldraw[fill=purple] (-1/2-0.2,-1/2) -- (-1/2-0.2,1/2) -- (1/2+0.2,1/2) -- (1/2+0.2,-1/2) -- (-1/2-0.2,-1/2);
        \draw (0,0) node {\scriptsize $A_{\text{low}}$};
        \draw (0,1.3) node {\scriptsize $\{0,1\}$};
    \end{tikzpicture}
    = 
    {\scriptsize
    \begin{pmatrix}
        1 & & &  \\
         & 0 & & \\
         & & 1 & \\
         & & & 1
    \end{pmatrix}}
    ,
    \
    \begin{tikzpicture}[scale=.45, baseline={([yshift=-1.6ex]current bounding box.center)}, thick]
        \draw (-1.2,0) -- (1.2,0);
        \draw (0,1) -- (0,0);
        \filldraw[fill=purple] (-1/2-0.2,-1/2) -- (-1/2-0.2,1/2) -- (1/2+0.2,1/2) -- (1/2+0.2,-1/2) -- (-1/2-0.2,-1/2);
        \draw (0,0) node {\scriptsize $A_{\text{low}}$};
        \draw (0,1.3) node {\scriptsize $\{0,2\}$};
    \end{tikzpicture}
    = 
    {\scriptsize
    \begin{pmatrix}
        0 & & &  \\
         & 1 & & \\
         & & 0 & \\
         & & & 0
    \end{pmatrix}}
    ,
    \
    \begin{tikzpicture}[scale=.45, baseline={([yshift=-1.6ex]current bounding box.center)}, thick]
        \draw (-1.2,0) -- (1.2,0);
        \draw (0,1) -- (0,0);
        \filldraw[fill=purple] (-1/2-0.2,-1/2) -- (-1/2-0.2,1/2) -- (1/2+0.2,1/2) -- (1/2+0.2,-1/2) -- (-1/2-0.2,-1/2);
        \draw (0,0) node {\scriptsize $A_{\text{low}}$};
        \draw (0,1.3) node {\scriptsize $\{1\}$};
    \end{tikzpicture}
    = 
    {\scriptsize
    \begin{pmatrix}
        0 & 1 & &  \\
         & 0 & & \\
         & & 0 & \\
         & & & 0
    \end{pmatrix}}
    ,
    \\
    &
    \begin{tikzpicture}[scale=.45, baseline={([yshift=-1.6ex]current bounding box.center)}, thick]
        \draw (-1.2,0) -- (1.2,0);
        \draw (0,1) -- (0,0);
        \filldraw[fill=purple] (-1/2-0.2,-1/2) -- (-1/2-0.2,1/2) -- (1/2+0.2,1/2) -- (1/2+0.2,-1/2) -- (-1/2-0.2,-1/2);
        \draw (0,0) node {\scriptsize $A_{\text{low}}$};
        \draw (0,1.3) node {\scriptsize $\{2\}$};
    \end{tikzpicture}
    = 
    {\scriptsize
    \begin{pmatrix}
        0 & & 1 & \\
         & 0 & & \\
         & & 0 & 1 \\
         & & & 0
    \end{pmatrix}}
    ,
    \
    \begin{tikzpicture}[scale=.45, baseline={([yshift=-1.6ex]current bounding box.center)}, thick]
        \draw (-1.2,0) -- (1.2,0);
        \draw (0,1) -- (0,0);
        \filldraw[fill=purple] (-1/2-0.2,-1/2) -- (-1/2-0.2,1/2) -- (1/2+0.2,1/2) -- (1/2+0.2,-1/2) -- (-1/2-0.2,-1/2);
        \draw (0,0) node {\scriptsize $A_{\text{low}}$};
        \draw (0,1.3) node {\scriptsize $\{3\}$};
    \end{tikzpicture}
    = 
    {\scriptsize
    \begin{pmatrix}
        0 & & & 1 \\
         & 0 & & \\
         & & 0 & \\
         & & & 0
    \end{pmatrix}}
    ,
    \\
    & \begin{tikzpicture}[scale=.45, baseline={([yshift=-0.5ex]current bounding box.center)}, thick]
        \begin{scope}[shift={(0,0)}]
            \draw (-1.2,0) -- (1.2,0);
            \draw (0,1) -- (0,-1);
            \filldraw[fill=purple] (-1/2-0.2,-1/2) -- (-1/2-0.2,1/2) -- (1/2+0.2,1/2) -- (1/2+0.2,-1/2) -- (-1/2-0.2,-1/2);
            \draw (0,0) node {\scriptsize $A_{\text{up}}$};
            \draw (0,1.3) node {\scriptsize $i$};
            \draw (0,-1.3) node {\scriptsize $\{0,1\}$};
        \end{scope}
    \end{tikzpicture}
    = 
    A^i , \ 
    \begin{tikzpicture}[scale=.45, baseline={([yshift=-0.5ex]current bounding box.center)}, thick]
        \begin{scope}[shift={(0,0)}]
            \draw (-1.2,0) -- (1.2,0);
            \draw (0,1) -- (0,-1);
            \filldraw[fill=purple] (-1/2-0.2,-1/2) -- (-1/2-0.2,1/2) -- (1/2+0.2,1/2) -- (1/2+0.2,-1/2) -- (-1/2-0.2,-1/2);
            \draw (0,0) node {\scriptsize $A_{\text{up}}$};
            \draw (0,1.3) node {\scriptsize $i$};
            \draw (0,-1.3) node {\scriptsize $\{0,2\}$};
        \end{scope}
    \end{tikzpicture}
    = 
    B^i , \ 
    \begin{tikzpicture}[scale=.45, baseline={([yshift=-0.5ex]current bounding box.center)}, thick]
        \begin{scope}[shift={(0,0)}]
            \draw (-1.2,0) -- (1.2,0);
            \draw (0,1) -- (0,-1);
            \filldraw[fill=purple] (-1/2-0.2,-1/2) -- (-1/2-0.2,1/2) -- (1/2+0.2,1/2) -- (1/2+0.2,-1/2) -- (-1/2-0.2,-1/2);
            \draw (0,0) node {\scriptsize $A_{\text{up}}$};
            \draw (0,1.3) node {\scriptsize $i$};
            \draw (0,-1.3) node {\scriptsize $\{1\}$};
        \end{scope}
    \end{tikzpicture}
    = 
    C^i , \ 
    \begin{tikzpicture}[scale=.45, baseline={([yshift=-0.5ex]current bounding box.center)}, thick]
        \begin{scope}[shift={(0,0)}]
            \draw (-1.2,0) -- (1.2,0);
            \draw (0,1) -- (0,-1);
            \filldraw[fill=purple] (-1/2-0.2,-1/2) -- (-1/2-0.2,1/2) -- (1/2+0.2,1/2) -- (1/2+0.2,-1/2) -- (-1/2-0.2,-1/2);
            \draw (0,0) node {\scriptsize $A_{\text{up}}$};
            \draw (0,1.3) node {\scriptsize $i$};
            \draw (0,-1.3) node {\scriptsize $\{2\}$};
        \end{scope}
    \end{tikzpicture}
    = 
    D^i , \ 
    \begin{tikzpicture}[scale=.45, baseline={([yshift=-0.5ex]current bounding box.center)}, thick]
        \begin{scope}[shift={(0,0)}]
            \draw (-1.2,0) -- (1.2,0);
            \draw (0,1) -- (0,-1);
            \filldraw[fill=purple] (-1/2-0.2,-1/2) -- (-1/2-0.2,1/2) -- (1/2+0.2,1/2) -- (1/2+0.2,-1/2) -- (-1/2-0.2,-1/2);
            \draw (0,0) node {\scriptsize $A_{\text{up}}$};
            \draw (0,1.3) node {\scriptsize $i$};
            \draw (0,-1.3) node {\scriptsize $\{3\}$};
        \end{scope}
    \end{tikzpicture}
    = 
    E^i . 
\end{align*}
The associative tensor $\Gamma$ of structure constants of $\mathcal{A}_{\mathrm{low}}$ with respect to the basis $\{A_{\mathrm{low}}^e\}_{e \in \Sigma}$ has the following non-zero components:
\begin{align}
    \begin{tikzpicture}[scale=.45, baseline={([yshift=-1ex]current bounding box.center)}, thick]
        \begin{scope}[shift={(0,0)}]
            \draw (0.3,0.5) -- (0.3,1);
            \draw (-0.3,0.5) -- (-0.3,1);
            \filldraw[fill=gray!10] (-0.5,-0.5) -- (-0.5,0.5) -- (0.5,0.5) -- (0.5,-0.5) -- (-0.5,-0.5);
            \draw (0,0) node {\scriptsize $\Gamma$};
            \draw (0,-0.5) -- (0,-1);
        \end{scope}
        \draw (0,-1.3) node {\scriptsize $\{0,1\}$};
        \draw (-0.8,1.4) node {\scriptsize $\{0,1\}$};
        \draw (0.8,1.4) node {\scriptsize $\{0,1\}$};
    \end{tikzpicture}
    &=
    \begin{tikzpicture}[scale=.45, baseline={([yshift=-1ex]current bounding box.center)}, thick]
        \begin{scope}[shift={(0,0)}]
            \draw (0.3,0.5) -- (0.3,1);
            \draw (-0.3,0.5) -- (-0.3,1);
            \filldraw[fill=gray!10] (-0.5,-0.5) -- (-0.5,0.5) -- (0.5,0.5) -- (0.5,-0.5) -- (-0.5,-0.5);
            \draw (0,0) node {\scriptsize $\Gamma$};
            \draw (0,-0.5) -- (0,-1);
        \end{scope}
        \draw (0,-1.3) node {\scriptsize $\{0,2\}$};
        \draw (-0.8,1.4) node {\scriptsize $\{0,2\}$};
        \draw (0.8,1.4) node {\scriptsize $\{0,2\}$};
    \end{tikzpicture}
    =
    \begin{tikzpicture}[scale=.45, baseline={([yshift=-1ex]current bounding box.center)}, thick]
        \begin{scope}[shift={(0,0)}]
            \draw (0.3,0.5) -- (0.3,1);
            \draw (-0.3,0.5) -- (-0.3,1);
            \filldraw[fill=gray!10] (-0.5,-0.5) -- (-0.5,0.5) -- (0.5,0.5) -- (0.5,-0.5) -- (-0.5,-0.5);
            \draw (0,0) node {\scriptsize $\Gamma$};
            \draw (0,-0.5) -- (0,-1);
        \end{scope}
        \draw (0,-1.3) node {\scriptsize $\{1\}$};
        \draw (-0.8,1.4) node {\scriptsize $\{0,1\}$};
        \draw (0.5,1.4) node {\scriptsize $\{1\}$};
    \end{tikzpicture}
    =
    \begin{tikzpicture}[scale=.45, baseline={([yshift=-1ex]current bounding box.center)}, thick]
        \begin{scope}[shift={(0,0)}]
            \draw (0.3,0.5) -- (0.3,1);
            \draw (-0.3,0.5) -- (-0.3,1);
            \filldraw[fill=gray!10] (-0.5,-0.5) -- (-0.5,0.5) -- (0.5,0.5) -- (0.5,-0.5) -- (-0.5,-0.5);
            \draw (0,0) node {\scriptsize $\Gamma$};
            \draw (0,-0.5) -- (0,-1);
        \end{scope}
        \draw (0,-1.3) node {\scriptsize $\{1\}$};
        \draw (-0.5,1.4) node {\scriptsize $\{1\}$};
        \draw (0.8,1.4) node {\scriptsize $\{0,2\}$};
    \end{tikzpicture}
    =  
    \begin{tikzpicture}[scale=.45, baseline={([yshift=-1ex]current bounding box.center)}, thick]
        \begin{scope}[shift={(0,0)}]
            \draw (0.3,0.5) -- (0.3,1);
            \draw (-0.3,0.5) -- (-0.3,1);
            \filldraw[fill=gray!10] (-0.5,-0.5) -- (-0.5,0.5) -- (0.5,0.5) -- (0.5,-0.5) -- (-0.5,-0.5);
            \draw (0,0) node {\scriptsize $\Gamma$};
            \draw (0,-0.5) -- (0,-1);
        \end{scope}
        \draw (0,-1.3) node {\scriptsize $\{2\}$};
        \draw (-0.8,1.4) node {\scriptsize $\{0,1\}$};
        \draw (0.5,1.4) node {\scriptsize $\{2\}$};
    \end{tikzpicture}
    =
    \begin{tikzpicture}[scale=.45, baseline={([yshift=-1ex]current bounding box.center)}, thick]
        \begin{scope}[shift={(0,0)}]
            \draw (0.3,0.5) -- (0.3,1);
            \draw (-0.3,0.5) -- (-0.3,1);
            \filldraw[fill=gray!10] (-0.5,-0.5) -- (-0.5,0.5) -- (0.5,0.5) -- (0.5,-0.5) -- (-0.5,-0.5);
            \draw (0,0) node {\scriptsize $\Gamma$};
            \draw (0,-0.5) -- (0,-1);
        \end{scope}
        \draw (0,-1.3) node {\scriptsize $\{2\}$};
        \draw (-0.5,1.4) node {\scriptsize $\{2\}$};
        \draw (0.8,1.4) node {\scriptsize $\{0,1\}$};
    \end{tikzpicture}
    \nonumber \\
    &=
    \begin{tikzpicture}[scale=.45, baseline={([yshift=-1ex]current bounding box.center)}, thick]
        \begin{scope}[shift={(0,0)}]
            \draw (0.3,0.5) -- (0.3,1);
            \draw (-0.3,0.5) -- (-0.3,1);
            \filldraw[fill=gray!10] (-0.5,-0.5) -- (-0.5,0.5) -- (0.5,0.5) -- (0.5,-0.5) -- (-0.5,-0.5);
            \draw (0,0) node {\scriptsize $\Gamma$};
            \draw (0,-0.5) -- (0,-1);
        \end{scope}
        \draw (0,-1.3) node {\scriptsize $\{3\}$};
        \draw (-0.8,1.4) node {\scriptsize $\{0,1\}$};
        \draw (0.5,1.4) node {\scriptsize $\{3\}$};
    \end{tikzpicture}
    =
    \begin{tikzpicture}[scale=.45, baseline={([yshift=-1ex]current bounding box.center)}, thick]
        \begin{scope}[shift={(0,0)}]
            \draw (0.3,0.5) -- (0.3,1);
            \draw (-0.3,0.5) -- (-0.3,1);
            \filldraw[fill=gray!10] (-0.5,-0.5) -- (-0.5,0.5) -- (0.5,0.5) -- (0.5,-0.5) -- (-0.5,-0.5);
            \draw (0,0) node {\scriptsize $\Gamma$};
            \draw (0,-0.5) -- (0,-1);
        \end{scope}
        \draw (0,-1.3) node {\scriptsize $\{3\}$};
        \draw (-0.5,1.4) node {\scriptsize $\{3\}$};
        \draw (0.8,1.4) node {\scriptsize $\{0,1\}$};
    \end{tikzpicture}
    =
    \begin{tikzpicture}[scale=.45, baseline={([yshift=-1ex]current bounding box.center)}, thick]
        \begin{scope}[shift={(0,0)}]
            \draw (0.3,0.5) -- (0.3,1);
            \draw (-0.3,0.5) -- (-0.3,1);
            \filldraw[fill=gray!10] (-0.5,-0.5) -- (-0.5,0.5) -- (0.5,0.5) -- (0.5,-0.5) -- (-0.5,-0.5);
            \draw (0,0) node {\scriptsize $\Gamma$};
            \draw (0,-0.5) -- (0,-1);
        \end{scope}
        \draw (0,-1.3) node {\scriptsize $\{3\}$};
        \draw (-0.5,1.4) node {\scriptsize $\{2\}$};
        \draw (0.5,1.4) node {\scriptsize $\{2\}$};
    \end{tikzpicture} =
    1 \, . \label{eq:gamma_tensor_example_2_appendix}
\end{align}

\paragraph{Step 3.} Apply Proposition \ref{prop:check_TI} to simplify the boundary matrix of the MPS-X. Translational invariance holds if and only if both conditions \textit{(i)} and \textit{(ii)} of the proposition are satisfied. For the present example these conditions become:
\begin{enumerate}
    \item[\textit{(i)}] There exist constants $\beta_t \in \mathbb{C}$ such that, for each $t \in \Sigma$,
    \begin{equation*}
        \sum_{i,j} k_{ij;t} X_{ji} = \begin{cases}
            \beta_t \mathds{1} & \text{if } t \in [p,p] \text{ for some } p \in \Sigma_\infty,
            \\
            0 & \text{otherwise.}
        \end{cases}
    \end{equation*}
    In this example, the equality above translates into the set of contraints
    \begin{equation*}
        \begin{cases}
            X_{11} + X_{33} + X_{44} = \beta_{\{0,1\}} \mathds{1}, \\
            X_{22} = \beta_{\{0,2\}} \mathds{1}, \\
            X_{21} = 0, \\
            X_{31} + X_{43} = \beta_{\{2\}} \mathds{1}, \\
            X_{41} = \beta_{\{3\}} \mathds{1}.
        \end{cases}
    \end{equation*}
    where $X_{21} = 0$ follows from the fact that $\{1\} \in [\{0,1\},\{0,2\}]$. 

    \item[\textit{(ii)}] The proportionality constants further satisfy
    \begin{equation*}
        \sum_{r \in \Sigma_f} \beta_r (\Gamma_r^{pq} - \Gamma_r^{qp}) = 0, \quad \forall p,q \in \Sigma_f.
    \end{equation*}
    Since the only non-zero coefficient $\Gamma_r^{pq}$ with $p,q,r \in \Sigma_f$ that is non-zero in this example is $\Gamma_{\{3\}}^{\{2\}\{2\}} = 1$, the equation above trivializes and does not impose additional contraints on the $\beta_r$. 
\end{enumerate}
Thus the boundary matrix $X$ can be replaced by the simplified matrix $\tilde{X}$ written below, which yields the same family of states $|\psi_N(\tilde{X},A^i)\rangle = |\psi_N(X,A^i)\rangle$ for all $N$. It is also convenient to define a $b \times b$ matrix $Y$ containing only the relevant proportionality constants:
\begin{equation*}
    \tilde{X} = {\scriptsize
    \begin{pmatrix}
        \beta_{\{0,1\}}\mathds{1} & & & \\
        0 & \beta_{\{0,2\}}\mathds{1} & & \\
        \beta_{\{2\}} \mathds{1} & 0 & 0 & \\
        \beta_{\{3\}} \mathds{1} & 0 & 0 & 0
    \end{pmatrix}}
    , \quad
    Y = {\scriptsize
    \begin{pmatrix}
        \beta_{\{0,1\}} & & & \\
        0 & \beta_{\{0,2\}} & & \\
        \beta_{\{2\}} & 0 & 0 & \\
        \beta_{\{3\}} & 0 & 0 & 0
    \end{pmatrix}}
\end{equation*}

\paragraph{Step 4.} The lower part of the gCF is the algebraic RLS family $\{\ket{L_N}\} := \{\ket{\psi_N(Y,A_{\mathrm{low}})}\}$, which in this example is
\begin{align*}
    \{\ket{L_N}\} 
    &= \beta_{\{0,1\}} \ket{\{0,1\}^*} + \beta_{\{0,2\}} \ket{\{0,2\}^*} +
    \beta_{\{2\}} \ket{\{0,1\}^* \{2\} \{0,1\}^*}
    \\ &\quad + \beta_{\{3\}} \left(
        \ket{\{0,1\}^* \{3\} \{0,1\}^*} + \ket{\{0,1\}^* \{2\} \{0,1\}^* \{2\} \{0,1\}^* } 
    \right) ,
\end{align*}
and it is $\Gamma$-invariant with respect to the associative $\Gamma$-tensor in Eq.~\eqref{eq:gamma_tensor_example_2_appendix}. In the algebraic RLS notation, this becomes
\begin{align*}
    \{\ket{L_N}\} &= 
    \ket{\{0,1\}^*} X_{\{0,1\}}
    + \ket{\{0,2\}^*} X_{\{0,2\}} + \hat{S}^{(1)} \ket{\{0,1\}^* f \{0,1\}^*} \ket{X_{\{0,1\}\{0,1\}}} \\
    & \qquad + \hat{S}^{(2)} \ket{\{0,1\}^* f \{0,1\}^* f \{0,1\}^*} \ket{X_{\{0,1\}\{0,1\}\{0,1\}}}, \\
    \text{with }
    & \begin{cases}
        X_{\{0,1\}} = \beta_{\{0,1\}} , \\
        X_{\{0,2\}} = \beta_{\{0,2\}} , \\
        \ket{X_{\{0,1\}\{0,1\}}} = \beta_{\{2\}} \ket{\{2\}} + \beta_{\{3\}} \ket{\{3\}} ,  \\
        \ket{X_{\{0,1\}\{0,1\}\{0,1\}}} = \beta_{\{3\}} \ket{\{2\}\{2\}} .
    \end{cases}
\end{align*}

\paragraph{Step 5.}
The block-injective upper tensor $A_{\mathrm{up}}$ in the gCF is given by the tensor $A_{\mathrm{up}}$ of the matrix-CF in Step 2.

\newpage

\section{A canonical form for stable and non-stable MPS-X} \label{app:spanRLS-spanCF}


Here we introduce the \emph{spanCF}, an alternative to the gCF that is fully general: any MPS-X (whether translationally invariant, or non-TI with the boundary matrix blocks proportional to $\mathds{1}$) can be brought into this form after sufficient blocking.

In contrast to the gCF, which applies only to stable MPS-X, the spanCF encompasses arbitrary (possibly non-stable) tensors. The trade-off is that this form lacks some of the convenient structural properties of the gCF, such as self-consistency under coarse-graining and a clear characterization of the gauge freedom. Nevertheless, this generalization remains conceptually valuable: by revealing the structural features common to all MPS-X, using Theorem~\ref{prop:structure_span}, it shows that the cases excluded by the gCF do not display any qualitatively new behavior.  

At the end of the appendix, we also provide two lemmas with a (generally non-optimal) explicit MPS-X representation of any algebraic RLS and span RLS, providing an explicit bound for the bond dimension.

\subsection{Span regular language states and the span canonical form}

The first step toward constructing the spanCF is to define a generalized notion of algebraic regular language states, which we call \emph{span RLS}. This class captures the most general type of backbone state that any MPS-X (stable or not) can realize.

\begin{definition}[Span RLS] \label{def:spanRLS}
    A family $\{\ket{L_N}\}_N$ over alphabet $\Sigma = \Sigma_\infty \cup (\cup_{i,j \in \tilde{\Sigma}_\infty} \Sigma_f^{ij})$ is a \emph{span RLS} on $\Sigma$ if 
    \begin{equation} \label{eq:def_span_RLS_equation}
        \{\ket{L_N}\} = \sum_{m \leq M} \sum_{O \in \tilde{\Sigma}_\infty^{m+1}}\sum_{\substack{\sum_{i=0}^m n_i = N-m \\ (n_i = 0 \text{ if } O_i = \varepsilon)} } 
        \hat{S}^{(m)} \ket{O_0^{n_0} f O_1^{n_1} f O_2^{n_2} \dots O_{m-1}^{n_{m-1}} f O_m^{n_m}} | X_O^{(n_0, \dots, n_m)} \rangle,
    \end{equation}
    for some constant $M$, where $\tilde{\Sigma}_\infty := \Sigma_\infty \cup \{\varepsilon\}$. 
    
    Each $| X_O^{(n_0, \dots, n_m)} \rangle$ is a weighted superposition of length-$m$ strings $\mathbf{x} \in \tilde{\Sigma}_f^{O_0 O_1} \cdot \ldots \cdot \tilde{\Sigma}_f^{O_{m-1}O_m}$, where for every such $\mathbf{x}$, its amplitude is given by
    \begin{equation*}
        \langle \mathbf{x} | X_O^{(n_0,\dots,n_m)} \rangle
        = \sum_{j=1}^K \alpha_j \prod_{i=0}^m \lambda_{j,i}^{n_i},   \quad \forall \{n_i\} \subseteq \mathbb{N},     
    \end{equation*}
    for some constants $\alpha_j, \lambda_{j,i} \in \mathbb{C}$ that depend on $O$ and $\mathbf{x}$.
\end{definition} 

Algebraic RLS are special cases of span RLS with the additional constraint that $|X_O^{(n_0, \dots, n_m)}\rangle$ is a weighted superposition of strings in $\Sigma_f^{O_0 O_1} \cdot \ldots \cdot \Sigma_f^{O_{m-1} O_m} \subsetneq \tilde{\Sigma}_f^{O_0 O_1} \cdot \ldots \cdot \tilde{\Sigma}_f^{O_{m-1}O_m}$, 
with $K = 1$ and coefficients independent of $(n_0, n_1, \dots, n_m)$, allowing us to write simply $\ket{X_O}$.

\begin{definition}[Span canonical form for any MPS-X (spanCF)]  \label{def:spanCF}
    A family of quantum states $\{\ket{\psi_N}\}_N$ is in spanCF if, for each $N$,
    \begin{equation*}
        \ket{\psi_N} = 
        \begin{tikzpicture}[scale=.45,thick,baseline={([yshift=-1.15ex]current bounding box.center)}]
            \begin{scope}[shift={(0,1)}]
                \draw[dotted] (1.5,0) -- (4.5,0);
                \MPSTensor{0,0}{$A$}{purple}
                \draw (0,-0.5) -- (0,-1);
                \MPSTensor{1.5,0}{$A$}{purple}
                \draw (1.5,-0.5) -- (1.5,-1);
                \MPSTensor{4.5,0}{$A$}{purple}
                \draw (4.5,-0.5) -- (4.5,-1);
            \end{scope}
            \filldraw[fill=amaranth] (-0.5,-1) -- (-0.5,0) -- (5,0) -- (5,-1) -- (-0.5,-1);
            \draw (2.25,-0.5) node {\scriptsize $\ket{L_N}$};
            \draw (-1,1) -- (-1,0.25) -- (-0.2,0.25);
            \draw (0.2,0.25) -- (1.3,0.25);
            \draw (1.7,0.25) -- (2.4,0.25);
            \draw[dotted] (2.5,0.25) -- (3.5,0.25);
            \draw (3.6,0.25) -- (4.3,0.25);
            \draw (4.7,0.25) -- (5.5,0.25) -- (5.5,1);
        \end{tikzpicture} \ ,
    \end{equation*}
    where 
    \begin{enumerate}[label=(\roman*)]
        \item $\{\ket{L_N}\}$ is a family of span RLS.
        \item The set of tensors
            \begin{equation*}
    \left\{ \,
    \begin{tikzpicture}[scale=.45,thick,baseline={([yshift=-0.6ex]current bounding box.center)}]
        \begin{scope}[shift={(0,0)}]
    		\draw (0,0.5) -- (0,1);
            \draw (-1,0) -- (1,0);
            \filldraw[fill=purple] (-1/2-0.0,-1/2) -- (-1/2-0.0,1/2) -- (1/2+0.0,1/2) -- (1/2+0.0,-1/2) -- (-1/2-0.0,-1/2);
    		\draw (0,0) node {\scriptsize $A$};
            \draw (0,-0.5) -- (0,-1);
            \draw[fill=amaranth] (0,-1.4) circle (0.4);
            \draw (0,-1.4) node {\scriptsize $j$};
    	\end{scope}
    \end{tikzpicture} \, \right\}_{j \in \Sigma}
    \end{equation*}
    is block-injective.
    \end{enumerate}
\end{definition}
Using Theorem \ref{prop:structure_span}, we obtain the following result about the generality of the spanCF.
\begin{theorem}
    Every TI MPS-X can be written in spanCF after blocking every $p \ell$ sites with $\ell \geq L_{\mathrm{span}}$ sites. 
\end{theorem}
The result also holds for non-TI MPS-X whose boundary matrix has blocks that are proportional to the identity. Even though the spanCF is completely general in contrast to the gCF, it has the following limitations:
\begin{itemize}
    \item[(i)] It does not provide a consistent coarse-graining procedure, as the family of backbone states depends on the system size and thus changes under blocking.
    \item[(ii)] There is no straightforward characterization of the corresponding gauge freedom.  
\end{itemize}

Moreover, the spanCF does not give rise to fundamentally new families of physical states. The differences with respect to the gCF arise from the only two structural differences that exist between the bases for the span and the algebra discussed in Sec. \ref{sec:algebra-span-thmbasis}:
\begin{itemize}
    \item Diagonal blocks might be proportional rather than equal, implying that $| X_O^{(n_0, \dots, n_m)} \rangle$ can carry non-trivial weights raised to powers that depend on the number $n_i$ of occurrences of each $\Sigma_\infty$ symbol $O_i$. 
    
    For instance, consider the MPS-X defined by tensors
    \begin{equation*}
        X = {\scriptsize \begin{pmatrix}
            0 & 0 \\ 1 & 0
        \end{pmatrix}}, \quad 
        A^0 = {\scriptsize \begin{pmatrix}
            1 & 0 \\ 0 & e^{i\sqrt{2}\pi}
        \end{pmatrix}}, \quad
        A^1 = {\scriptsize \begin{pmatrix}
            0 & 1 \\ 0 & 0
        \end{pmatrix}} 
        \implies
        \mathcal{A}^{(\ell)} = \mathrm{span} \left\{ 
        {\scriptsize \begin{pmatrix}
            a & b \\ 0 & a e^{i\sqrt{2}\pi \ell}
        \end{pmatrix}} \mid a, b \in \mathbb{C}
        \right\} .
    \end{equation*}
    This MPS-X can already be written in spanCF without any further blocking, with a trivial upper tensor $A_{\mathrm{up}}$ (bond dimension one, acting as the identity on the virtual leg), and a spanRLS as follows:
    \begin{equation*}
        \ket{L_N} = \sum_{n_0 + n_1 = N-1} e^{i\sqrt{2}\pi n_1} \ket{0^{n_0} 1 0^{n_1}}
        = \sum_{n_0 + n_1 = N-1} \hat{S}^{(1)} \ket{0^{n_0} f 0^{n_1}} (\underbrace{e^{i\sqrt{2}\pi n_1} \ket{1}}_{=:|X_{00}^{(n_0, n_1)}\rangle} ) .
    \end{equation*}
    \item Jordan-type structures might appear in sectors of the form $[e,e]$ with $e \in \Sigma_\infty$, captured by the fact that $| X_O^{(n_0, \dots, n_m)} \rangle$ now belongs to $\tilde{\Sigma}_f^{O_0 O_1} \cdot \ldots \cdot \tilde{\Sigma}_f^{O_{m-1} O_m}$ rather than $\Sigma_f^{O_0 O_1} \cdot \dots \cdot \Sigma_f^{O_{m-1} O_m}$. 

    For instance, consider the MPS-X defined by tensors
    \begin{equation*}
        X = {\scriptsize \begin{pmatrix}
            0 & & \\ 0 & 0 & \\ 1 & 0 & 0
        \end{pmatrix}}, \quad 
        A^0 = {\scriptsize \begin{pmatrix}
            1 & 1 & 0 \\ & 1 & 0 \\ & & 1
        \end{pmatrix}}, \quad
        A^1 = {\scriptsize \begin{pmatrix}
            0 & 0 & 0 \\ & 0 & 1 \\ & & 0
        \end{pmatrix}} 
        \implies
        \mathcal{A}^{(\ell)} = \mathrm{span} \left\{ 
        {\scriptsize \begin{pmatrix}
            a & \ell a & c \\ & a & b \\ & & a
        \end{pmatrix}} \mid a, b,c \in \mathbb{C}
        \right\} .
    \end{equation*}
    This MPS-X can already be written in spanCF without any further blocking, with a trivial upper tensor $A_{\mathrm{up}}$ (bond dimension one, acting as the identity on the virtual leg), and a spanRLS as follows:
    \begin{equation*}
        \ket{L_N} = \sum_{n_0 + n_1 = N-1}\ket{0^{n_0} 0 0^{n_1}1 0^{n_2}}
        = \sum_{n_0 + n_1 = N-1} \hat{S}^{(2)} \ket{0^{n_0} f 0^{n_1}} \underbrace{(\ket{01})}_{=:|X_{00}^{(n_0, n_1)}\rangle} .
    \end{equation*}
\end{itemize}

\subsection{An MPS-X representation for algebraic and span RLS}

We close this appendix by proving that every span RLS admits an MPS-X representation. We first give an explicit construction for the special case of algebraic RLS and then show how to extend it to general span RLS.

\begin{lemma} \label{lemma:bond_dimension_algebraic-RLS}
    A family $\{\ket{L_N}\}$ of algebraic RLS on an alphabet $\Sigma$ partitioned as $\Sigma = \Sigma_\infty \cup (\cup_{i,j \in \tilde{\Sigma}_\infty} \Sigma_f^{ij})$, with defining states $\{\ket{X_O}\}_{O \in \tilde{\Sigma}_\infty^{m+1}, \, m \leq M}$, admits an MPS-X with bond dimension at most
    \begin{equation*}
        |\Sigma_\infty| + \begin{cases}
            M(M+3)/2 & \text{if } |\Sigma_f| = 1 \\
            (M+1) |\Sigma_f|^{M+1} & \text{otherwise.} \\
        \end{cases}
    \end{equation*}
\end{lemma}
\begin{proof}
    We proceed by constructing an explicit MPS-X representation.

    Let $m \geq 1$. For each string $\mathbf{x} := x_1 x_2 \dots x_m \in \Sigma_f^{m}$ with $m \leq M$, the partition of $\Sigma$ fixes a unique $O = O_0 O_1 \dots O_m \in \tilde{\Sigma}_\infty^{m+1}$ such that $x_i \in \Sigma_f^{O_{i-1} O_i}$ for all $i$. Let $\alpha_{\mathbf{x}} := \braket{x_1 \dots x_m}{X_O}$, and define the MPS tensor $B_{\mathbf{x}}$ and boundary matrix $X_\mathbf{x}$ by
    \begin{equation*}
        \begin{cases}
        B^y_{\mathbf{x}} = \sum_{i : \, O_i = y} \dyad{i}{i}
        & \text{if } y \in \Sigma_\infty , \\
        B^y_{\mathbf{x}} = \sum_{i: \, x_i = y} \dyad{i-1}{i}
        & \text{if } y \in \Sigma_f , \\
        X_{\mathbf{x}} = \alpha_{\mathbf{x}} \dyad{m}{0}. 
        \end{cases}
    \end{equation*}
    To account for the terms in the algebraic RLS with $m = 0$, i.e. $\sum_{O_0 \in \Sigma_\infty} \alpha_{O_0} \ket{O_0^*}$, we define $B_\infty$ and $X_\infty$ as
    \begin{equation*}
        \begin{cases}
        B^y_{\infty} = \dyad{y}{y} & \text{if } y \in \Sigma_\infty , \\ 
        B^y_{\infty} = 0 & \text{if } y \notin \Sigma_\infty , \\
        X_\infty = \sum_{x \in \Sigma_\infty} \alpha_{x} \dyad{x}{x}. 
        \end{cases}
    \end{equation*}
    We now form the block-diagonal MPS-X tensors
    \begin{equation*}
        A^y := B_\infty^y 
        \oplus
        \left(\bigoplus_{m=1}^M \bigoplus_{\mathbf{x} \in \Sigma_f^m} B_{\mathbf{x}}^y \right)
        , \quad 
        X := X_\infty \oplus \left(\bigoplus_{m=1}^M \bigoplus_{\mathbf{x} \in \Sigma_f^m} X_{\mathbf{x}}\right).
    \end{equation*}

    Let us first show the correctness of our construction. The MPS-X state given by tensor $A$ and boundary $X$ can be written in terms of each of the blocks as
    \begin{align*}
        \ket{\psi_N(X, A)} 
        &= 
        \sum_{i_1, \ldots, i_N \in \Sigma} \Tr[X A^{i_1} \dots A^{i_N}] \ket{i_1 \dots i_N}
        \\
        &= 
        \sum_{i_1, \ldots, i_N \in \Sigma} 
        \Big( 
        \Tr[X_\infty B_\infty^{i_1} \dots B_\infty^{i_N}] 
        +
        \sum_{m=1}^M \sum_{\mathbf{x} \in \Sigma_f^M}
        \Tr[X_{\mathbf{x}} B_{\mathbf{x}}^{i_1} \dots B_{\mathbf{x}}^{i_N}] 
        \Big)
        \ket{i_1 \dots i_N}.
    \end{align*}
    The first term produces
    \begin{align}
        &\hspace{-1cm} \sum_{i_1, \dots, i_N\in \Sigma} \Tr[X_\infty B^{i_1}_\infty \dots B^{i_N}_\infty] \ket{i_1 \dots i_N} \nonumber \\
        &=
         \sum_{i_1, \dots, i_N\in \Sigma} \sum_{O_0 \in \Sigma_\infty}
         \alpha_{O_0} \mel{O_0}{B^{i_1}_\infty \dots B^{i_N}_\infty}{O_0} \ket{i_1 \dots i_N} \nonumber \\
        &= \sum_{O_0 \in \Sigma_\infty} \alpha_{O_0} \ket{O_0^N} \equiv \Big( \sum_{O_0 \in \Sigma_\infty} \alpha_{O_0} \ket{O_0^*} \Big)_N \label{eq:fist_term_algebraicRLS_proof},
    \end{align}
    corresponding to the $m = 0$ contributions of the algebraic RLS. To tackle the other terms in the sum, choose any $\mathbf{x} = x_1 x_2 \dots x_m \in \Sigma_f^m$ with $m \geq 1$. Using the structure of $B_{\mathbf{x}}$, one obtains:
    \begin{align}
        &\hspace{-1cm} \sum_{i_1, \dots, i_N\in \Sigma} \Tr[X_{\mathbf{x}} B^{i_1}_{\mathbf{x}} \dots B^{i_N}_{\mathbf{x}}] \ket{i_1 \dots i_N} \nonumber\\
        &=
         \sum_{i_1, \dots, i_N\in \Sigma}
         \alpha_{\mathbf{x}} \mel{0}{B^{i_1}_{\mathbf{x}} \dots B^{i_N}_{\mathbf{x}}}{m} \ket{i_1 \dots i_N} \nonumber \\
        &= \sum_{i_1, \dots, i_N\in \Sigma} \sum_{j_1,\dots,j_N=0}^m
        \alpha_{\mathbf{x}} B^{i_1}_{{\mathbf{x}}, 0 j_1} 
        B^{i_2}_{{\mathbf{x}}, j_1 j_2} 
        \dots 
        B^{i_{N-1}}_{{\mathbf{x}}, j_{N-1} j_N}
        B^{i_N}_{{\mathbf{x}}, j_N m} \ket{i_1 \dots i_N} \nonumber \\
        &= 
        \sum_{\sum_i n_i = N-m} \alpha_{\mathbf{x}} (B^{O_0}_{\mathbf{x},00})^{n_0}
        B^{x_1}_{\mathbf{x},01} (B^{O_1}_{\mathbf{x},11})^{n_1}
        \dots 
        B^{x_m}_{\mathbf{x},m-1,m}
        (B^{O_m}_{\mathbf{x},m m})^{n_m} 
        \ket{O_0^{n_0} x_1 O_1^{n_1} x_2 \dots  x_m O_m^{n_m}}
        \nonumber \\
        &= \sum_{\sum_i n_i = N-m} \alpha_{\mathbf{x}}
        \ket{O_0^{n_0} x_1 O_1^{n_1} x_2 \dots  x_m O_m^{n_m}} \nonumber\\
        &\equiv \left(\hat{S}^{(m)} \ket{O_0^* f O_1^* f \dots f O_m^*} (\alpha_{\mathbf{x}} \ket{x_1 x_2 \dots x_m}) \right)_N . \label{eq:second_term_algebraicRLS_proof}
    \end{align}
    Adding Eq. \eqref{eq:fist_term_algebraicRLS_proof} and \eqref{eq:second_term_algebraicRLS_proof} together for all $\mathbf{x}$ and $m \leq M$ recovers the algebraic RLS expression, 
    \begin{equation*}
        \{\ket{L_N}\}_N = \sum_{m \leq M} \sum_{O \in \tilde{\Sigma}_\infty^{m+1}} \hat{S}^{(m)} \ket{O_0^\ast f O_1^\ast f O_2^\ast \dots O_{m-1}^\ast f O_m^\ast} | X_O \rangle,
    \end{equation*}
    since $\ket{X_O} = \sum_{\mathbf{x} \in \Sigma_f^m} \alpha_\mathbf{x} \ket{\mathbf{x}}$ for $m \geq 1$, and $\ket{X_O} = \alpha_O$ for $m = 0$ (i.e. for $O = x \in \Sigma_\infty$).

    Now, let us compute the bond dimension of this MPS-X representation. The block $B_\infty$ has bond dimension $|\Sigma_\infty|$. Each block $B_{\mathbf{x}}$ with $\mathbf{x} = x_1 \dots x_m \in \Sigma_f^m$ has bond dimension $m+1$, and there are $|\Sigma_f|^m$ such blocks. The total bond dimension is therefore 
    \begin{equation*}
        D = |\Sigma_\infty| + \sum_{m=1}^M (m+1) |\Sigma_f|^m \leq 
        |\Sigma_\infty| + \begin{cases}
            M(M+3)/2 & \text{if } |\Sigma_f| = 1, \\
            (M+1)|\Sigma_f|^{M+1} & \text{otherwise}.
        \end{cases}
    \end{equation*}
    Note that $|\Sigma_f| = 0$ in the case $M = 0$. This completes the proof.
\end{proof}

\paragraph{Example.} To illustrate the construction described in the proof above, let us consider the following algebraic RLS,
\begin{equation*}
    \{\ket{L_N}\} = \hat{S}^{(2)} \ket{0^* f 1^* f 0^*} (\alpha_{24} \ket{24} + \alpha_{32} \ket{32})
\end{equation*}
on an alphabet partitioned as $\Sigma_\infty = \{0,1\}$, $\Sigma_f^{01} = \{2, 3\}$, $\Sigma_f^{10} = \{4\}$, and $M = 2$. Then, the MPS tensor $A$ and boundary matrix $X$ constructed above look as follows:
\begin{align*}
    A^0 &= {\scriptsize
    \left(
    \begin{array}{ccc:ccc}
    1 &   &   &   &   &   \\
      & 0 &   &   &   &   \\
      &   & 1 &   &   &   \\ \hdashline
      &   &   & 1 &   &   \\
      &   &   &   & 0 &   \\
      &   &   &   &   & 1
    \end{array}
    \right)
    }, \ 
    A^1 = {\scriptsize
    \left(
    \begin{array}{ccc:ccc}
    0 &   &   &   &   &   \\
      & 1 &   &   &   &   \\
      &   & 0 &   &   &   \\ \hdashline
      &   &   & 0 &   &   \\
      &   &   &   & 1 &   \\
      &   &   &   &   & 0
    \end{array}
    \right)
    }, \ 
    A^2 = {\scriptsize
    \left(
    \begin{array}{ccc:ccc}
    0 & 1 &   &   &   &   \\
      & 0 &   &   &   &   \\
      &   & 0 &   &   &   \\ \hdashline
      &   &   & 0 &   &   \\
      &   &   &   & 0 &   \\
      &   &   &   &   & 0
    \end{array}
    \right)
    }, \\
    A^3 &= {\scriptsize
    \left(
    \begin{array}{ccc:ccc}
    0 &   &   &   &   &   \\
      & 0 &   &   &   &   \\
      &   & 0 &   &   &   \\ \hdashline
      &   &   & 0 & 1 &   \\
      &   &   &   & 0 &   \\
      &   &   &   &   & 0
    \end{array}
    \right)
    }, \
    A^4 = {\scriptsize
    \left(
    \begin{array}{ccc:ccc}
    0 &   &   &   &   &   \\
      & 0 & 1 &   &   &   \\
      &   & 0 &   &   &   \\ \hdashline
      &   &   & 0 &   &   \\
      &   &   &   & 0 & 1 \\
      &   &   &   &   & 0
    \end{array}
    \right)
    }, \
    X = {\scriptsize
    \left(
    \begin{array}{ccc:ccc}
     0 &   &   &   &   &   \\
      & 0  &   &   &   &   \\
     \alpha_{24} &  & 0  &   &   &   \\ \hdashline
      &   &   &  0&  &   \\
      &   &   &   & 0&   \\
      &   &   & \alpha_{34}  &   & 0
    \end{array}
    \right)
    }, \
\end{align*}

Let us now generalize the lemma above to span RLS, as defined in Eq. \eqref{eq:def_span_RLS_equation}. 
\begin{lemma} \label{lemma:bond_dimension_span-RLS}
    A span RLS $\{\ket{L_N}\}$ as defined in Def. \ref{def:spanRLS} admits an MPS-X representation with bond dimension at most
    \begin{equation*}
        K |\Sigma_\infty| + 
        \begin{cases}
            M(M+3) K / 2 & \text{if } |\Sigma| = 1 \\
            (M+1) K |\Sigma|^{M+1} & \text{otherwise.} \\
        \end{cases}
    \end{equation*}
\end{lemma}
\begin{proof}
    We construct an explicit block-diagonal MPS-X. The proof follows the same block-by-block idea used in Lemma \ref{lemma:bond_dimension_algebraic-RLS} for algebraic RLS. 

    Fix $m \geq 1$. For each string $\mathbf{x} = x_1 x_2 \dots x_m \in \Sigma^m$, there exists a unique $O = O_0 O_1 \dots O_m \in \Sigma_\infty$ such that $x_i \in \tilde{\Sigma}_f^{O_{i-1}O_i}$ for all $i$. Let us denote the weight of string $\mathbf{x}$ in $|X_O^{(n_0,\dots,n_m)}\rangle$ as
    \begin{equation*}
        \langle x_1 \dots x_m | X_O \rangle = \sum_{j=1}^{K}\alpha_{\mathbf{x},j} \prod_{i=0}^m \lambda_{(\mathbf{x},j),i}^{n_i}
    \end{equation*}
    Now, let us define tensors $B_{\mathbf{x},j}$ and boundary matrices $X_{\mathbf{x},j}$ as
    \begin{equation*}
        \begin{cases}
        B^y_{\mathbf{x},j} = \sum_{i : \, O_i = y} \lambda_{(\mathbf{x},j),i}\dyad{i}{i} + \sum_{k: \, x_k = y} \dyad{k-1}{k}
        & \text{if } y \in \Sigma_\infty , \\
        B^y_{\mathbf{x},j} = \sum_{i: \, x_i = y} \dyad{i-1}{i}
        & \text{if } y \in \Sigma_f , \\
        X_{\mathbf{x},j} = \alpha_{\mathbf{x},j} \dyad{m}{0}. 
        \end{cases}.
    \end{equation*}
    For the $m = 0$ terms, for each $O_0 \in \Sigma_\infty$ we have that $X_{O_0}\in \mathbb{C}$ has the form $\sum_{j=1}^K \alpha_{O_0,j} \lambda_{(O_0,j)}^N$. We accordingly define
    \begin{equation*}
        \begin{cases}
        B^y_{O_0} = \sum_{k=1}^{K} \lambda_{(O_0,k)} \dyad{k}{k} & \text{if } y \in \Sigma_\infty , \\ 
        B^y_{O_0} = 0 & \text{if } y \notin \Sigma_\infty , \\
        X_{O_0} = \sum_{k=1}^{K} \alpha_{O_0,k} \dyad{k}{k}. 
        \end{cases}
    \end{equation*}
    Then, take 
    \begin{equation*}
        A^y :=\left( \bigoplus_{O_0 \in \Sigma_\infty} B_{O_0}^y \right) \oplus \left( 
        \bigoplus_{m=0}^M
        \bigoplus_{\mathbf{x} \in \Sigma^m}
        \bigoplus_{k=1}^K B_{\mathbf{x}, k}^y
        \right), 
        \quad 
        X := 
        \left( \bigoplus_{O_0 \in \Sigma_\infty} X_{O_0} \right) \oplus \left( 
        \bigoplus_{m=1}^M
        \bigoplus_{\mathbf{x} \in \Sigma^m}
        \bigoplus_{k=1}^K X_{\mathbf{x}, k}
        \right).
    \end{equation*}
    A short contraction check, analogous to the algebraic RLS case, shows the correctness of the MPS-X constructed above as a representation of the span RLS. 

    Let us compute the bond dimension of this MPS-X construction. The $m = 0$ part contributes $K|\Sigma_\infty|$ to the bond dimension. For $m \geq 1$, each block has dimension $m+1$. For fixed $m$ there are at most $K |\Sigma|^m$ blocks, and therefore the total bond dimension satisfies
    \begin{equation*}
        D \leq K|\Sigma_\infty| + \sum_{m=1}^M (m+1) K |\Sigma|^m 
        \leq 
        K|\Sigma_\infty| + \begin{cases}
            M(M+3)K/2 & \text{if } |\Sigma| = 1, \\
            K(M+1)K |\Sigma|^{M+1} & \text{otherwise.}
        \end{cases}
    \end{equation*}
    This completes the proof. 
\end{proof} 
Note that the construction for algebraic RLS in Lemma \ref{lemma:bond_dimension_span-RLS} (which would correspond to $K = 1$) imposes a looser bound on the bond dimension than the one derived specifically for algebraic RLS in Lemma \ref{lemma:bond_dimension_algebraic-RLS}, since $|\Sigma_f| \leq |\Sigma|$.

\newpage

\section{Proofs for the freedom in the gCF} \label{app:fundthm_proof}

In this appendix we prove Thm. \ref{prop:freedom_gCF}, which characterizes the freedom in the gCF representation. We consider two MPS-X, with tensors $\{X_A, A^i\}$ and $\{X_B, B^i\}$, that generate the same family of states,
\begin{equation*}
    \ket{\psi_N(X_A, A^i)} = \ket{\psi_N(X_B, B^i)}, \quad \forall N.
\end{equation*}
Our goal is to determine the relation between their gCF representations.

We begin by introducing the notion of \emph{reduced pairs of equivalent MPS-X}:
\begin{definition}[Reduced MPS-X representations]
\label{def:reduced_MPS-X}
    A pair of equivalent MPS-X is \emph{reduced} if
    \begin{enumerate}
        \item[\textit{(i)}] neither contains negligible blocks,
        \item[\textit{(ii)}] they possess no additive gauge freedom relative to each other, and
        \item[\textit{(iii)}] their tensors span the same physical subspace.
    \end{enumerate}
\end{definition}
In the next sections (Sec. \ref{sec:freedom_condition(i)}-\ref{sec:freedom_condition(iii)}) we formalize these conditions and describe how any pair of equivalent MPS-X can be transformed to satisfy them, yielding the following lemma.
\begin{lemma} \label{lemma:freedom_always-possible-reduced}
    Any pair of equivalent MPS-X can be transformed into an equivalent reduced pair.
\end{lemma}
We further show that any two equivalent uniform PBC-MPS are already in reduced form, requiring no additional transformation, as stated next.
\begin{lemma} \label{lemma:pbc-mps_always_reduced}
    Any pair of equivalent uniform PBC MPS is reduced.
\end{lemma}

Finally, we combine these ingredients to establish the freedom in the matrix-CF (Sec. \ref{app:sec_matrix-CF_freedom}) and in the gCF (Sec. \ref{app:sec_gCF_freedom}).


\subsection{Condition \textit{(i)}: No negligible blocks} 
\label{sec:freedom_condition(i)}

A first natural question is whether an MPS-X in matrix-CF or gCF can be expressed in a more economical form. For instance, can certain blocks be safely removed? In this section we describe a straightforward procedure to achieve this.

\begin{definition}[Negligible block] \label{def:negligible-block}
    Let $(X_A, A^i)$ be an MPS-X in matrix-CF with tensors $A_{\mathrm{up}}, A_{\mathrm{low}}$. If the free blocks encoded in $A_{\mathrm{up}}$ are indexed by $\Sigma$, we say that block $t \in \Sigma$ is \emph{negligible} if replacing $A_{\text{up}}$ with
    \begin{equation*}
        \begin{tikzpicture}[scale=.45, baseline={([yshift=0.4ex]current bounding box.center)}, thick]
            \begin{scope}[shift={(0,0)}]
                \draw (-1.2,0) -- (1.2,0);
                \draw (0,1) -- (0,-1);
                \filldraw[fill=powderblue!50] (-1/2-0.2,-1/2) -- (-1/2-0.2,1/2) -- (1/2+0.2,1/2) -- (1/2+0.2,-1/2) -- (-1/2-0.2,-1/2);
                \draw (0,0) node {\scriptsize $\tilde{A}_{\text{up}}$};
            \end{scope}
            \draw (0,-1.3) node {\scriptsize $x$};
        \end{tikzpicture}
        := 
        \begin{cases}
            0 &\text{if } x = t , \\
            \begin{tikzpicture}[scale=.45, baseline={([yshift=0.4ex]current bounding box.center)}, thick]
                \begin{scope}[shift={(0,0)}]
                    \draw (-1.2,0) -- (1.2,0);
                    \draw (0,1) -- (0,-1);
                    \filldraw[fill=powderblue!50] (-1/2-0.2,-1/2) -- (-1/2-0.2,1/2) -- (1/2+0.2,1/2) -- (1/2+0.2,-1/2) -- (-1/2-0.2,-1/2);
                    \draw (0,0) node {\scriptsize $A_{\text{up}}$};
                \end{scope}
                \draw (0,-1.3) node {\scriptsize $x$};
            \end{tikzpicture}  &\text{otherwise,}
        \end{cases}
    \end{equation*}
    generates the same family of states as the original MPS-X.

    Equivalently, for an MPS-X with gCF representation $\{A_{\mathrm{up}}, \{\ket{L_N}\}\}$, it has no negligible non-zero blocks if every $j \in \Sigma$ appears in at least one ket of the family $\{\ket{L_N}\}$.
\end{definition}

Note that determining whether a block $t \in \Sigma$ is negligible is computationally efficient. For example, weighted finite automata methods allow one to check this with runtime $O(dD^3)$ (if $\mathrm{rank}(X) = 1$) \cite{kiefer_2020_WFA-notes} or $O(dD^6)$ (otherwise). Therefore, by setting all negligible blocks to zero, we obtain a more economical representation of the same MPS-X family.

\paragraph{Example 1.} Consider the following MPS-X $(X_D, D^i)$ in matrix-CF (with a trivial $D_{\text{up}}$, i.e. with bond dimension one, and acting as the identity on the physical leg):
\begin{align*}
    &D^i = \ \begin{tikzpicture}[scale=.45, baseline={([yshift=-0.6ex]current bounding box.center)}, thick]
    \begin{scope}[shift={(0,0)}]
        \draw (-1.25,0) -- (1.25,0);
        \draw (0,0) -- (0,1);
        \filldraw[fill=powderblue!50] (-0.75,-0.5) -- (-0.75,0.5) -- (0.75,0.5) -- (0.75,-0.5) -- (-0.75,-0.5);
        \draw (0,0) node {\scriptsize $D_{\text{low}}$};
    \end{scope}
    \end{tikzpicture} \ , \quad  
    \begin{cases}
        \begin{tikzpicture}[scale=.45, baseline={([yshift=-0.6ex]current bounding box.center)}, thick]
        \begin{scope}[shift={(0,0)}]
            \draw (0,0) -- (0,1);
            \draw (-1.25,0) -- (1.25,0);
            \filldraw[fill=powderblue!50] (-0.75,-0.5) -- (-0.75,0.5) -- (0.75,0.5) -- (0.75,-0.5) -- (-0.75,-0.5);
            \draw (0,0) node {\scriptsize $D_{\text{low}}$};
            \draw (0,1.3) node {\scriptsize $0$};
        \end{scope}
    \end{tikzpicture}
    = \mathds{1}_4 , \
    \begin{tikzpicture}[scale=.45, baseline={([yshift=-0.6ex]current bounding box.center)}, thick]
        \begin{scope}[shift={(0,0)}]
            \draw (0,0) -- (0,1);
            \draw (-1.25,0) -- (1.25,0);
            \filldraw[fill=powderblue!50] (-0.75,-0.5) -- (-0.75,0.5) -- (0.75,0.5) -- (0.75,-0.5) -- (-0.75,-0.5);
            \draw (0,0) node {\scriptsize $D_{\text{low}}$};
            \draw (0,1.3) node {\scriptsize $1$};
        \end{scope}
    \end{tikzpicture}
    = {\scriptsize \begin{pmatrix}
        0 & 1 & 0 & 0 \\ 
        & 0 & 0 & 0 \\
        & & 0 & 1 \\
        & & & 0
    \end{pmatrix}} , \ 
    \begin{tikzpicture}[scale=.45, baseline={([yshift=-0.6ex]current bounding box.center)}, thick]
        \begin{scope}[shift={(0,0)}]
            \draw (0,0) -- (0,1);
            \draw (-1.25,0) -- (1.25,0);
            \filldraw[fill=powderblue!50] (-0.75,-0.5) -- (-0.75,0.5) -- (0.75,0.5) -- (0.75,-0.5) -- (-0.75,-0.5);
            \draw (0,0) node {\scriptsize $D_{\text{low}}$};
            \draw (0,1.3) node {\scriptsize $2$};
        \end{scope}
    \end{tikzpicture}
    = {\scriptsize \begin{pmatrix}
        0 & 0 & 0 & 0 \\ 
        & 0 & 1 & 0 \\
        & & 0 & -1 \\
        & & & 0
    \end{pmatrix}} , \\
    \begin{tikzpicture}[scale=.45, baseline={([yshift=-0.6ex]current bounding box.center)}, thick]
        \begin{scope}[shift={(0,0)}]
            \draw (0,0) -- (0,1);
            \draw (-1.25,0) -- (1.25,0);
            \filldraw[fill=powderblue!50] (-0.75,-0.5) -- (-0.75,0.5) -- (0.75,0.5) -- (0.75,-0.5) -- (-0.75,-0.5);
            \draw (0,0) node {\scriptsize $D_{\text{low}}$};
            \draw (0,1.3) node {\scriptsize $3$};
        \end{scope}
    \end{tikzpicture}
    = {\scriptsize \begin{pmatrix}
        0 & 0 & 1 & 0 \\ 
        & 0 & 0 & 0 \\
        & & 0 & 0 \\
        & & & 0
    \end{pmatrix}} , \
    \begin{tikzpicture}[scale=.45, baseline={([yshift=-0.6ex]current bounding box.center)}, thick]
        \begin{scope}[shift={(0,0)}]
            \draw (0,0) -- (0,1);
            \draw (-1.25,0) -- (1.25,0);
            \filldraw[fill=powderblue!50] (-0.75,-0.5) -- (-0.75,0.5) -- (0.75,0.5) -- (0.75,-0.5) -- (-0.75,-0.5);
            \draw (0,0) node {\scriptsize $D_{\text{low}}$};
            \draw (0,1.3) node {\scriptsize $4$};
        \end{scope}
    \end{tikzpicture}
    = {\scriptsize \begin{pmatrix}
        0 & 0 & 0 & 0 \\ 
        & 0 & 0 & 1 \\
        & & 0 & 0 \\
        & & & 0
    \end{pmatrix}} , \ 
    \begin{tikzpicture}[scale=.45, baseline={([yshift=-0.6ex]current bounding box.center)}, thick]
        \begin{scope}[shift={(0,0)}]
            \draw (0,0) -- (0,1);
            \draw (-1.25,0) -- (1.25,0);
            \filldraw[fill=powderblue!50] (-0.75,-0.5) -- (-0.75,0.5) -- (0.75,0.5) -- (0.75,-0.5) -- (-0.75,-0.5);
            \draw (0,0) node {\scriptsize $D_{\text{low}}$};
            \draw (0,1.3) node {\scriptsize $5$};
        \end{scope}
    \end{tikzpicture}
    = {\scriptsize \begin{pmatrix}
        0 & 0 & 0 & 1 \\ 
        & 0 & 0 & 0 \\
        & & 0 & 0 \\
        & & & 0
    \end{pmatrix}} ,
    \end{cases} \\
    &X_D =  {\scriptsize \begin{pmatrix}
        0 & & & \\ 
        1 & 0 & & \\
        0 & 1 & 0 & \\
        0 & 0 & 1 & 0
    \end{pmatrix}} . 
\end{align*}
This matrix-CF decomposition of $D^i$ arises from expressing the span $\mathcal{D}^{(1)}$ in the basis
\begin{equation*}
    \mathcal{D}^{(1)} = 
    \left\{
    {\scriptsize \begin{pmatrix}
        a & b & d & f \\ 
        & a & c & e \\
        & & a & b-c \\
        & & & a
    \end{pmatrix}} \mid 
    a, b, \dots, f \in \mathbb{C}
    \right\} ,
\end{equation*}
which has the structural properties described in Theorem \ref{prop:structure_span}. The corresponding family of algebraic RLS in the gCF (which coincides with $\ket{\psi_N(X_D, D^i)}$ since $D_{\text{up}}$ is trivial), is given by
\begin{equation*}
    \{\ket{L_N}\} = \ket{0^* 1 0^*}.
\end{equation*} 
Therefore, blocks labeled by $\ket{2}, \ket{3}, \ket{4}$ and $\ket{5}$ are negligible, as they do not contribute to the final states under this choice of boundary matrix.

\begin{lemma} \label{lemma:pbc-mps_no_negligible_blocks}
    Uniform PBC MPS have no negligible blocks. 
\end{lemma}
\begin{proof}
    Let $(X_D, D^i)$ be the tensors defining a uniform PBC MPS, so $X_D = \mathds{1}$ and $D^i$ are block-diagonal. The free blocks are labeled by $\Sigma$. Since there are no off-diagonal blocks, we have $\Sigma = \Sigma_\infty$ (and hence $\Sigma_f = \emptyset$).

    By Theorem \ref{prop:structure_span}, each diagonal block is either free or proportional to another diagonal block. Accordingly, after decomposing the MPS-X in its matrix-CF with tensors $D_{\mathrm{low}}^i, D_{\mathrm{up}}^i$, the states in the lower part of the gCF take the form
    \begin{equation*}
        \ket{\psi_N(\mathds{1}, D_{\text{low}}^i)}
        =
        \sum_{j \in \Sigma_\infty} (1 + \sum_{k=1}^{r_j} \mu_{j,k}^N) \ket{x}^{\otimes N},
    \end{equation*}
    where $\mu_{j,k}$ denote the proportionality constants between diagonal blocks. We see that, for a block $j \in \Sigma_\infty$ to be negligible, one would need
    \begin{equation*}
        \sum_{k=1}^{r_j} \mu_{j,k}^N = -1, \quad \forall N,
    \end{equation*}
    which is not possible. Hence, uniform MPS with PBC (with initailly block-diagonal tensors) do not have negligible blocks as defined in Def. \ref{def:negligible-block}.
\end{proof}

\subsection{Condition \textit{(ii)}: No additive gauge freedom} 
\label{sec:freedom_condition(ii)}

The constructive proofs of Proposition \ref{prop:structure_subalgebra} and Theorem \ref{prop:structure_span} (Appendices \ref{app:structurealgebraproofs} and \ref{app:span_structure}) show that, once a set of matrices has been put in block-upper-triangular form, the gauge $P$ that brings $\mathcal{A}^{(\ell)}$ into matrix-CF can always be factored as $P = P_{\mathrm{off}} P_{\mathrm{diag}}$, where
\begin{itemize}
    \item $P_{\mathrm{diag}}$ is strictly block-diagonal and enforces diagonal blocks that are not independent from each other to be proportional, and
    \item $P_{\mathrm{off}}$ is the identity plus non-zero contributions supported on off-diagonal blocks that still require processing, ensuring that off-diagonal blocks are either zero, free blocks, or linear combinations of other free blocks in the same sector.
\end{itemize}
An example where we explicitly constructed $P = P_{\mathrm{off}} P_{\text{diag}}$ is provided at the end of Appendix \ref{app:structurealgebraproofs}.

Before moving on to discuss condition \textit{(ii)}, we introduce the notation of \emph{stacked MPS matrices} $C^i$, which result from stacking the $A^i$ and $B^i$ matrices, as 
\begin{equation*}
    C^i := A^i \oplus B^i \equiv 
    {\scriptsize \begin{pmatrix}
        A^i & 0 \\ 0 & B^i
    \end{pmatrix}}.
\end{equation*}
\begin{definition}[No additive gauge freedom] \label{def:no_additive_freedom_formal}
    An MPS tensor $B^i$ has \emph{no additive gauge freedom} with respect to $A^i$ if the stacked tensor $C^i := A^i \oplus B^i$ can be put into matrix-CF using a strictly block-diagonal gauge transformation. 
\end{definition}
Equivalently, for the gauge $P = P_{\mathrm{off}} P_{\mathrm{diag}}$ prescribed by Theorem \ref{prop:structure_span} to bring $\mathcal{C}^{(\ell)}$ into matrix-CF, the off-diagonal part is trivial: $P_{\mathrm{off}} = \mathds{1}$. In fact, a direct corollary of Theorem \ref{prop:structure_span} is that there always exists a gauge $P$ such that the transformed tensor $\tilde{B}^i = PB^iP^{-1}$ has no additive gauge freedom with respect to $A^i$. 
\begin{corollary} \label{cor:no_additive_freedom_is_general}
For any two MPS tensors, there exists a gauge transformation under which the resulting pair has no additive gauge freedom with respect to each other.
\end{corollary}

To build intuition on how the additive gauge freedom can manifest itself in the tensor, consider two equivalent W-like MPS-X representations:
\begin{equation*}
    A^i = 
    {\scriptsize \begin{pmatrix}
        B^i & C^i \\  & B^i
    \end{pmatrix} }
    , \quad 
    \tilde{A}^i =
    {\scriptsize \begin{pmatrix}
        B^i & \tilde{C}^i \\  & B^i
    \end{pmatrix} }, 
    \quad X = \tilde{X} = 
    {\scriptsize \begin{pmatrix}
        0 &  \\ \mathds{1} & 0
    \end{pmatrix} }
\end{equation*}
whose span has the form
\begin{equation*}
    \mathcal{A}^{(1)} = \tilde{\mathcal{A}}^{(1)} = \left\{ 
    {\scriptsize \begin{pmatrix}
        B & C \\ & B
    \end{pmatrix}} \mid B,C
    \right\}.
\end{equation*}
The MPS tensors $A^i$ and $\tilde{A}^i$ form the stacked tensor
$D^i$:
\begin{equation*}
    D^i = {\scriptsize \begin{pmatrix}
        B^i & C^i & & \\ 
         & B^i & & \\
        & & B^i & \tilde{C}^i \\
        & & & B^i
    \end{pmatrix} } .
\end{equation*}
Although $\tilde{C}^i$ is a free block in $\tilde{\mathcal{A}}^{(1)}$, it cannot define a new free block in $\mathcal{D}^{(1)} := \mathrm{span}\{D^i\}$; otherwise the two MPS-X would define linearly independent states and hence would not be equivalent. Thus $\tilde{C}^i$ must be expressible in terms of $B^i$ and $C^i$. This implies that
\begin{equation*}
    \mathcal{D}^{(1)} = \left\{
    {\scriptsize \begin{pmatrix}
        B & C & & \\ 
        & B & & \\
        & & B & f(B) + g(C) \\
        & & & B 
    \end{pmatrix} } \mid
    B,C \right\}
\end{equation*}
for some linear functions $f, g$. The same arguments used in the proof of Theorem \ref{prop:structure_span} imply $f(A) = \alpha A + [A, P]$ and $g(B) = \beta B$ for some $\alpha, \beta \in \mathbb{C}$ and some matrix $P$. Hence,
\begin{equation*}
    \tilde{C}^i = \alpha B^i + \beta C^i + [B^i, P].
\end{equation*}
The commutator term corresponds to the part of $\tilde{C}^i$ that cannot be written as a linear combination of existing free blocks; this is the \emph{additive gauge freedom} of the tensor. The gauge transformation prescribed in Thm. \ref{prop:structure_span}, $\tilde{P} = \tilde{P}_{\mathrm{off}} = {\scriptsize \begin{pmatrix}
    \mathds{1} & P \\ & \mathds{1}
\end{pmatrix}}$, eliminates this term:
\begin{equation*}
    \tilde{P} \tilde{A}^i \tilde{P}^{-1} =
    {\scriptsize \begin{pmatrix}
        B^i & \alpha B^i + \beta C^i \\ & B^i
    \end{pmatrix}},
\end{equation*}
so that $\tilde{C}^i = \alpha B^i + \beta C^i$ in the new basis. We refer to this as \textit{removing the additive gauge freedom} of the stacked tensor, a procedure that can always be performed. This example shows that the notion in Def. \ref{def:no_additive_freedom_formal} can be equivalently phrased in a way that highlights the structure of these degrees of freedom:
\begin{remark*}
    An MPS tensor has \emph{additive gauge freedom} with respect to another if some of its free blocks, although not independent from those of the other tensor, cannot be written as linear combinations of them (after a suitable strictly block-diagonal gauge transformation is performed).
\end{remark*}

Note that the additive gauge freedom can adopt more intricate forms than just a single commutator:
\begin{itemize}
    \item Consider two equivalent MPS-X representations of the form:
    \begin{equation*}
        A^i = 
        {\scriptsize \begin{pmatrix}
            B^i & C^i & D^i \\  & B^i & C^i \\  &  & B^i
        \end{pmatrix} }
        , \quad 
        \tilde{A}^i =
        {\scriptsize \begin{pmatrix}
            B^i & \tilde{C}^i & \tilde{D}^i \\  & B^i & \tilde{C}^i \\  &  & B^i
        \end{pmatrix} } 
        \quad \to \quad 
        E^i = {\scriptsize \begin{pmatrix}
            B^i & C^i & D^i & & & \\  & B^i & C^i & & & \\  &  & B^i & & & \\
            & & & B^i & \tilde{C}^i & \tilde{D}^i \\ & & &  & B^i & \tilde{C}^i \\ & & &  &  & B^i
        \end{pmatrix} }.
    \end{equation*}
    By the same reasoning as in the previous example, we obtain
    \begin{align*}
        \tilde{C}^i &= \alpha B^i + \beta C^i + \gamma D^i + [B^i, P_f ] \\
        \tilde{D}^i &= \hat{\alpha} B^i + \hat{\beta} C^i + \hat{\gamma} D^i - [P_f, \alpha B^i + \beta C^i + \gamma D^i] - P_f [B^i, P_f] - [P_{\hat{g}}, B^i],
    \end{align*}
    for some $\alpha, \beta, \ldots \in \mathbb{C}$, and some matrices $P_f, P_{\hat{g}}$. 

    \item Consider two equivalent MPS-X representations of the form:
    \begin{equation*}
        A^i = 
        {\scriptsize \begin{pmatrix}
            B^i & C^i & D^i \\  & B^i & E^i \\  &  & B^i
        \end{pmatrix} }
        , \quad 
        \tilde{A}^i =
        {\scriptsize \begin{pmatrix}
            B^i & \tilde{C}^i & \tilde{D}^i \\  & B^i & \tilde{E}^i \\  &  & B^i
        \end{pmatrix} } 
        \quad \to \quad 
        E^i = {\scriptsize \begin{pmatrix}
            B^i & C^i & D^i & & & \\  & B^i & E^i & & & \\  &  & B^i & & & \\
            & & & B^i & \tilde{C}^i & \tilde{D}^i \\ & & &  & B^i & \tilde{E}^i \\ & & &  &  & B^i
        \end{pmatrix} }.
    \end{equation*}
    By the same reasoning as in the previous example, we obtain
    \begin{align*}
        \tilde{C}^i &= \alpha B^i + \beta C^i + \gamma D^i + [B^i, P_f] \ , \\
        \tilde{E}^i &= \alpha' B^i + \beta' C^i + \gamma' D^i + [B^i, P_h] \ , \\
        \tilde{D}^i &= \hat{\alpha} B^i + \hat{\beta} C^i + \hat{\gamma} D^i + [B^i, P_g] - P_f ( \alpha' B^i + \beta' C^i + \gamma' D^i + [B^i, P_h]) \\
        &\quad \quad +
        ( \alpha B^i + \beta C^i + \gamma D^i + [B^i, P_h]) P_g
        - [A^i, P_f] P_g \ ,
    \end{align*}
    for some $\alpha, \beta, \ldots \in \mathbb{C}$, and some matrices $P_f, P_g, P_h$. 
\end{itemize}

It remains an open question whether one can fix the gauge of the free blocks in a way that automatically removes additive freedom, without the need for further transformations. However, since the above procedure always exists, and the required change of basis can be explicitly constructed by Theorem \ref{prop:structure_span}, we can safely impose the removal of the additive gauge freedom as an assumption in what follows, as was expressed in Corollary \ref{cor:no_additive_freedom_is_general} above.

Furthermore, any pair of equivalent uniform PBC-MPS automatically satisfies this property.
\begin{lemma} \label{lemma:pbc-mps_no_additive_freedom}
    Equivalent uniform PBC MPS have no additive gauge freedom with respect to each other.
\end{lemma}
\begin{proof}
    In the PBC case, the tensors contain no off-diagonal blocks to be processed. Hence, in the gauge $P = P_{\mathrm{off}} P_{\mathrm{diag}}$ prescribed by Theorem \ref{prop:structure_span}, we necessarily have $P_{\mathrm{off}} = \mathds{1}$, and hence no additive gauge degrees of freedom arise. 
\end{proof}

\subsection{Condition \textit{(iii)}: Tensors spanning the same physical subspace}
\label{sec:freedom_condition(iii)}

The natural assumption that the tensors $A^i$ and $B^i$ span the same physical subspace will be key for proving the freedom in both the matrix-CF and gCF representations. This condition ensures that the two tensors \textit{do not contain negligible blocks with respect to each other}.

To formalize this idea, we introduce a technical tool, the \textit{stacking trick}, which plays a central role in the proofs. The main idea is to embed the two MPS-X tensors into a larger common tensor and exploit the structural properties of its span or algebra, as characterized in Prop. \ref{prop:structure_subalgebra} and Thm. \ref{prop:structure_span}. This construction is independent of the boundary matrix. For a set of symbols $\Sigma$ labeling the computational basis elements, we will use the shorthand $\langle \Sigma \rangle := \text{span}\{\ket{x} \mid x \in \Sigma\}$.
\begin{lemma}[The stacking trick] \label{lemma:stack-trick}
    Let $A, B$ be two MPS tensors in matrix-CF with decompositions $A_{\text{up}}, A_{\text{low}}$ and $B_{\text{up}}, B_{\text{low}}$, and free blocks labeled by alphabets $\Sigma^A = \Sigma_\infty^A \cup (\cup_{i,j \in \tilde{\Sigma}^A_\infty} \Sigma_f^{A,ij})$ and $\Sigma^B = \Sigma_\infty^B \cup (\cup_{i,j \in \tilde{\Sigma}^B_\infty} \Sigma_f^{B,ij})$, with no additive gauge freedom with respect to each other. Then there exist
    \begin{itemize}
        \item a block-injective tensor $C_{\text{up}}$,
        \item invertible matrices $Z_j$, and
        \item a map $\pi : \Sigma_\infty^B \to \Sigma_\infty^C$,
    \end{itemize}
    such that, after blocking sufficiently many sites, the following hold:
    \begin{enumerate}
        \item The tensor $A_{\text{up}}$ coincides with the first blocks of $C_{\text{up}}$ indexed by $\Sigma_f^C$ and $\Sigma_\infty^C$, i.e.
        \begin{equation} \label{eq:choose_tensor_C}
            \begin{tikzpicture}[scale=.45, baseline={([yshift=0.4ex]current bounding box.center)}, thick]
                \begin{scope}[shift={(0,0)}]
                    \draw (-1.2,0) -- (1.2,0);
                    \draw (0,1) -- (0,-1);
                    \filldraw[fill=powderblue!50] (-0.7,-0.5) rectangle (0.7,0.5);
                    \draw (0,0) node {\scriptsize $A_{\text{up}}$};
                \end{scope}
                \draw (0,-1.3) node {\scriptsize $x$};
            \end{tikzpicture}
            \ = \
            \begin{tikzpicture}[scale=.45, baseline={([yshift=0.4ex]current bounding box.center)}, thick]
                \begin{scope}[shift={(0,0)}]
                    \draw (-1.2,0) -- (1.2,0);
                    \draw (0,1) -- (0,-1);
                    \filldraw[fill=atomictangerine!50] (-0.7,-0.5) rectangle (0.7,0.5);
                    \draw (0,0) node {\scriptsize $C_{\text{up}}$};
                \end{scope}
                \draw (0,-1.3) node {\scriptsize $x$};
            \end{tikzpicture}
            \quad \text{for } x \in \{ \{0,1\}, \dots, \{0,|\Sigma_\infty^A|\}\} \cup \{ \{1\}, \dots, \{|\Sigma_f^A|\}\}.
        \end{equation}

        \item The tensor $B_{\text{up}}$ can be expressed in terms of $C_{\text{up}}$ as
        \begin{equation}  \label{eq:B_in_terms_of_C}
        \begin{tikzpicture}[scale=.45, baseline={([yshift=0.4ex]current bounding box.center)}, thick]
            \begin{scope}[shift={(0,0)}]
                \draw (-1.2,0) -- (1.2,0);
                \draw (0,1) -- (0,-1);
                \filldraw[fill=melon!50] (-1/2-0.2,-1/2) -- (-1/2-0.2,1/2) -- (1/2+0.2,1/2) -- (1/2+0.2,-1/2) -- (-1/2-0.2,-1/2);
                \draw (0,0) node {\scriptsize $B_{\text{up}}$};
            \end{scope}
            \draw (0,-1.3) node {\scriptsize $t$};
        \end{tikzpicture} =
            \begin{tikzpicture}[scale=.45, baseline={([yshift=0.4ex]current bounding box.center)}, thick]
            \begin{scope}[shift={(0,0)}]
                \draw (-1.2,0) -- (1.2,0);
                \draw (0,1) -- (0,-1);
                \filldraw[fill=atomictangerine!50] (-1/2-0.2,-1/2) -- (-1/2-0.2,1/2) -- (1/2+0.2,1/2) -- (1/2+0.2,-1/2) -- (-1/2-0.2,-1/2);
                \draw (0,0) node {\scriptsize $C_{\text{up}}$};
            \end{scope}
            \begin{scope}[shift={(1.7,0)}]
                \draw (-1,0) -- (1,0);
                \filldraw[fill=gray!10] (0.5,-0.5) -- (-0.5,-0.5) -- (-0.5,0.5) -- (0.5, 0.5) -- (0.5,-0.5);
                \draw (0,0) node {\scriptsize $Z$};
                \draw (0,-0.5) -- (0,-1);
                \draw (0,-1.3) node {\scriptsize $\pi(j)$};
            \end{scope}
            \begin{scope}[shift={(-2.2,0)}]
                \draw (-1,0) -- (1.5,0);
                \draw (0.25,-1) -- (0.25,-0.5);
                \filldraw[fill=gray!10] (1,-0.5) -- (-0.5,-0.5) -- (-0.5,0.5) -- (1, 0.5) -- (1,-0.5);
                \draw (0.3,0) node {\scriptsize $Z^{-1}$};
                \draw (0.25,-1.3) node {\scriptsize $\pi(i)$};
            \end{scope}
            \begin{scope}[shift={(0,-1.5)}]
                \filldraw[fill=gray!10] (0.5,-0.5) -- (-0.5,-0.5) -- (-0.5,0.5) -- (0.5, 0.5) -- (0.5,-0.5);
                \draw (0,0) node {\scriptsize $P_B$};
                \draw (0,-0.5) -- (0,-1);
                \draw (0,-1.3) node {\scriptsize $t$};
            \end{scope}
        \end{tikzpicture}
        \quad \text{for } t \in \tilde{\Sigma}_f^{B,ij},  
    \end{equation}
        where $\alpha_{s} \in \mathbb{C}$ and $P_B : \langle\Sigma^B\rangle \to \langle\Sigma^C\rangle$ acts as 
        \begin{equation*}
            \begin{cases}
                P_B \ket{x} = \alpha_{\pi(x)} \ket{\pi(x)} & \text{if } x \in \Sigma_\infty, \\
                P_B ( \langle \Sigma_f^{B,ij} \rangle ) \subseteq \langle \tilde{\Sigma}_f^{C, \pi(i)\pi(j)}\rangle  & \text{for } i,j \in \tilde{\Sigma}_\infty.
            \end{cases}
        \end{equation*}
    \end{enumerate}
\end{lemma}
\begin{proof} 
Define the block-diagonal tensor $\tilde{C}^i$ that results from stacking $A^i$ and $B^i$ together,
\begin{equation*}
    \tilde{C}^i := A^i \oplus B^i = {\scriptsize \begin{pmatrix} A^i & 0 \\ 0 & B^i\end{pmatrix}}.
\end{equation*}
Find the matrix-CF of $\tilde{C}^i$, obtaining the new tensor $C^i := \tilde{Z} \tilde{C}^i \tilde{Z}^{-1}$ decomposed in $C_{\text{up}}, C_{\text{low}}$, upon some invertible $\tilde{Z}$ and enough blocking. Since $A^i$ was already in matrix-CF, we can choose without loss of generality $\tilde{Z} = \mathds{1} \oplus Z$, and thus $C_{\text{low}}^x |_{\text{A-part}} = A_{\text{low}}^x$. In this way, the first free blocks of $C_{\mathrm{up}}$ can be indexed so that
\begin{equation*}
    \begin{tikzpicture}[scale=.45, baseline={([yshift=0.4ex]current bounding box.center)}, thick]
            \begin{scope}[shift={(0,0)}]
                \draw (-1.2,0) -- (1.2,0);
                \draw (0,1) -- (0,-1);
                \filldraw[fill=powderblue!50] (-1/2-0.2,-1/2) -- (-1/2-0.2,1/2) -- (1/2+0.2,1/2) -- (1/2+0.2,-1/2) -- (-1/2-0.2,-1/2);
                \draw (0,0) node {\scriptsize $A_{\text{up}}$};
            \end{scope}
            \draw (0,-1.3) node {\scriptsize $x$};
        \end{tikzpicture} =
        \begin{tikzpicture}[scale=.45, baseline={([yshift=0.4ex]current bounding box.center)}, thick]
            \begin{scope}[shift={(0,0)}]
                \draw (-1.2,0) -- (1.2,0);
                \draw (0,1) -- (0,-1);
                \filldraw[fill=atomictangerine!50] (-1/2-0.2,-1/2) -- (-1/2-0.2,1/2) -- (1/2+0.2,1/2) -- (1/2+0.2,-1/2) -- (-1/2-0.2,-1/2);
                \draw (0,0) node {\scriptsize $C_{\text{up}}$};
            \end{scope}
            \draw (0,-1.3) node {\scriptsize $x$};
        \end{tikzpicture}         
        \quad \text{if } x \in \{\{0,1\}, \dots, \{0,|\Sigma_\infty^A|\}\} \cup \{\{1\}, \dots, \{|\Sigma_f^A|\}\}.
\end{equation*}
This proves the first part of the claim. For the second part, we invoke the properties of the span structure:
\begin{itemize}
    \item \textit{Relation between the $\Sigma_\infty^B$ and $\Sigma_\infty^C$ blocks:}  
    By Theorem \ref{prop:structure_span}, in matrix-CF each diagonal block is either free or proportional to another diagonal block. Hence, every block $s \in \Sigma_\infty^B$ is proportional to one of the blocks in $\Sigma_\infty^C$, up to the gauge transformation $Z_s$. More precisely, there exists a map $\pi:\Sigma_\infty^B \to \Sigma_\infty^C$ and scalars $\alpha_s \in \mathbb{C}$ such that
    \begin{equation}
        \begin{tikzpicture}[scale=.45, baseline={([yshift=0.4ex]current bounding box.center)}, thick]
            \begin{scope}[shift={(0,0)}]
                \draw (-1.2,0) -- (1.2,0);
                \draw (0,1) -- (0,-1);
                \filldraw[fill=melon!50] (-1/2-0.2,-1/2) -- (-1/2-0.2,1/2) -- (1/2+0.2,1/2) -- (1/2+0.2,-1/2) -- (-1/2-0.2,-1/2);
                \draw (0,0) node {\scriptsize $B_{\text{up}}$};
            \end{scope}
            \draw (0,-1.3) node {\scriptsize $s$};
        \end{tikzpicture} =
        \alpha_{\pi(s)} \cdot
        \begin{tikzpicture}[scale=.45, baseline={([yshift=0.4ex]current bounding box.center)}, thick]
            \begin{scope}[shift={(0,0)}]
                \draw (-1.2,0) -- (1.2,0);
                \draw (0,1) -- (0,-1);
                \filldraw[fill=atomictangerine!50] (-1/2-0.2,-1/2) -- (-1/2-0.2,1/2) -- (1/2+0.2,1/2) -- (1/2+0.2,-1/2) -- (-1/2-0.2,-1/2);
                \draw (0,0) node {\scriptsize $C_{\text{up}}$};
            \end{scope}
            \draw (0,-1.3) node {\scriptsize $\pi(s)$};
            \begin{scope}[shift={(1.7,0)}]
                \draw (-1,0) -- (1,0);
                \filldraw[fill=gray!10] (0.5,-0.5) -- (-0.5,-0.5) -- (-0.5,0.5) -- (0.5, 0.5) -- (0.5,-0.5);
                \draw (0,0) node {\scriptsize $Z$};
                \draw (0,-0.5) -- (0,-1);
                \draw (0,-1.3) node {\scriptsize $\pi(s)$};
            \end{scope}
            \begin{scope}[shift={(-2.2,0)}]
                \draw (-1,0) -- (1.5,0);
                \draw (0.25,-1) -- (0.25,-0.5);
                \filldraw[fill=gray!10] (1,-0.5) -- (-0.5,-0.5) -- (-0.5,0.5) -- (1, 0.5) -- (1,-0.5);
                \draw (0.3,0) node {\scriptsize $Z^{-1}$};
                \draw (0.25,-1.3) node {\scriptsize $\pi(s)$};
            \end{scope}
        \end{tikzpicture}
        \quad \text{for each } s \in \Sigma_\infty^B.
        \label{eq:Bup_Cup_relation_infty}
    \end{equation}

    \item \textit{Relation between the $\Sigma_f^B$ and $\Sigma_f^C$ blocks:}  
    By Theorem \ref{prop:structure_span} and the assumption that there is no additive gauge freedom between the two MPS-X, in matrix-CF each off-diagonal block is either zero, free, or a linear combination of free blocks \emph{within the same sector}, labeled by $\tilde{\Sigma}_f^{ij}$, where
    \begin{equation*}
        \tilde{\Sigma}_f^{ij} =
        \begin{cases}
            \{i\} \cup \Sigma_f^{ij} & \text{if } i=j, \\
            \Sigma_f^{ij} & \text{if } i \neq j.
        \end{cases}
    \end{equation*}
    Therefore, each block in $\Sigma_f^{B,ij}$ is a linear combination of blocks in $\tilde{\Sigma}_f^{C,\pi(i)\pi(j)}$, again up to the gauge $Z$. Equivalently, there exists a linear map $P_B : \langle \Sigma_f^B \rangle \longrightarrow \langle \Sigma^C \rangle$ such that
    \begin{equation} \label{eq:Bup_Cup_relation_f}
        \begin{tikzpicture}[scale=.45, baseline={([yshift=0.4ex]current bounding box.center)}, thick]
            \begin{scope}[shift={(0,0)}]
                \draw (-1.2,0) -- (1.2,0);
                \draw (0,1) -- (0,-1);
                \filldraw[fill=melon!50] (-1/2-0.2,-1/2) -- (-1/2-0.2,1/2) -- (1/2+0.2,1/2) -- (1/2+0.2,-1/2) -- (-1/2-0.2,-1/2);
                \draw (0,0) node {\scriptsize $B_{\text{up}}$};
            \end{scope}
            \draw (0,-1.3) node {\scriptsize $t$};
        \end{tikzpicture} =
        \begin{tikzpicture}[scale=.45, baseline={([yshift=0.4ex]current bounding box.center)}, thick]
            \begin{scope}[shift={(0,0)}]
                \draw (-1.2,0) -- (1.2,0);
                \draw (0,1) -- (0,-1);
                \filldraw[fill=atomictangerine!50] (-1/2-0.2,-1/2) -- (-1/2-0.2,1/2) -- (1/2+0.2,1/2) -- (1/2+0.2,-1/2) -- (-1/2-0.2,-1/2);
                \draw (0,0) node {\scriptsize $C_{\text{up}}$};
            \end{scope}
            \begin{scope}[shift={(1.7,0)}]
                \draw (-1,0) -- (1,0);
                \filldraw[fill=gray!10] (0.5,-0.5) -- (-0.5,-0.5) -- (-0.5,0.5) -- (0.5, 0.5) -- (0.5,-0.5);
                \draw (0,0) node {\scriptsize $Z$};
                \draw (0,-0.5) -- (0,-1);
                \draw (0,-1.3) node {\scriptsize $\pi(j)$};
            \end{scope}
            \begin{scope}[shift={(-2.2,0)}]
                \draw (-1,0) -- (1.5,0);
                \draw (0.25,-1) -- (0.25,-0.5);
                \filldraw[fill=gray!10] (1,-0.5) -- (-0.5,-0.5) -- (-0.5,0.5) -- (1, 0.5) -- (1,-0.5);
                \draw (0.3,0) node {\scriptsize $Z^{-1}$};
                \draw (0.25,-1.3) node {\scriptsize $\pi(i)$};
            \end{scope}
            \begin{scope}[shift={(0,-1.5)}]
                \filldraw[fill=gray!10] (0.5,-0.5) -- (-0.5,-0.5) -- (-0.5,0.5) -- (0.5, 0.5) -- (0.5,-0.5);
                \draw (0,0) node {\scriptsize $P_B$};
                \draw (0,-0.5) -- (0,-1);
                \draw (0,-1.3) node {\scriptsize $t$};
            \end{scope}
        \end{tikzpicture}
        \quad \text{for each } t \in \Sigma_f^{B,ij},
    \end{equation}
    where the restriction per sectors implies
    \begin{equation*}
        P_B \big( \langle \Sigma_f^{B,ij} \rangle \big) \subseteq \langle \tilde{\Sigma}_f^{C,\pi(i)\pi(j)} \rangle .
    \end{equation*}
    \end{itemize}
    By extending $P_B$ to all of $\langle \Sigma^B \rangle$ via $P_B \ket{x} = \alpha_{\pi(x)} \ket{\pi(x)}$ for each $x \in \Sigma_\infty^B$, such that both Eq. \eqref{eq:Bup_Cup_relation_infty} and \eqref{eq:Bup_Cup_relation_f} hold, the claim follows.
\end{proof}

\paragraph{Example 3.} This example illustrates the implications of Lemma \ref{lemma:stack-trick} in a concrete setting. Consider tensors $A$ and $B$ defined as
\begin{align*}
    A^0 = \mathds{1}_2, \  
    A^1 = {\scriptsize\begin{pmatrix}
        0 & 1 \\ & 0
    \end{pmatrix}}, \
    A^2 = {\scriptsize\begin{pmatrix}
        0 & 1 \\ & 0
    \end{pmatrix}} 
    &\implies 
    \mathcal{A}^{(1)} = \left\{ {\scriptsize\begin{pmatrix}
        a_0 & a_1 \\ & a_0
    \end{pmatrix}} \mid a_0, a_1 \in \mathbb{C}\right\}
    \\
    B^0 = \mathds{1}_3, \  
    B^1 = {\scriptsize\begin{pmatrix}
        0 & 1 & \\ & 0 & \\ & & 0
    \end{pmatrix}}, \
    B^2 = {\scriptsize\begin{pmatrix}
        0 & & \\ & 0 & 1 \\ & & 0
    \end{pmatrix}}
    &\implies 
    \mathcal{B}^{(1)} = \left\{ {\scriptsize\begin{pmatrix}
        b_0 & b_1 & \\ & b_0 & b_2 \\ & & b_0
    \end{pmatrix}} \mid b_0, b_1, b_2 \right\}
\end{align*}
According to the bases in which we have written $\mathcal{A}^{(1)}, \mathcal{B}^{(1)}$, both $A$ and $B$ can be expressed in matrix-CF form as
\begin{align*}
    A^i &= \ \begin{tikzpicture}[scale=.45, baseline={([yshift=-0.6ex]current bounding box.center)}, thick]
        \begin{scope}[shift={(0,1.5)}]
            \draw (0,-1) -- (0,1);
            \draw (-1.25,0) -- (1.25,0);
            \filldraw[fill=powderblue!50] (-0.75,-0.5) -- (-0.75,0.5) -- (0.75,0.5) -- (0.75,-0.5) -- (-0.75,-0.5);
            \draw (0,0) node {\scriptsize $A_{\text{up}}$};
        \end{scope}
        \begin{scope}[shift={(0,0)}]
            \draw (-1.25,0) -- (1.25,0);
            \filldraw[fill=powderblue!50] (-0.75,-0.5) -- (-0.75,0.5) -- (0.75,0.5) -- (0.75,-0.5) -- (-0.75,-0.5);
            \draw (0,0) node {\scriptsize $A_{\text{low}}$};
        \end{scope}
    \end{tikzpicture} \ , 
    \quad \text{with} \quad 
    \begin{tikzpicture}[scale=.45, baseline={([yshift=-0ex]current bounding box.center)}, thick]
        \begin{scope}[shift={(0,0)}]
            \draw (0,-1) -- (0,1);
            \filldraw[fill=powderblue!50] (-0.75,-0.5) -- (-0.75,0.5) -- (0.75,0.5) -- (0.75,-0.5) -- (-0.75,-0.5);
            \draw (0,0) node {\scriptsize $A_{\text{up}}$};
            \draw (0,-1.3) node {\scriptsize $\{0,1\}$};
        \end{scope}
    \end{tikzpicture}
    = \ket{0} , \
    \begin{tikzpicture}[scale=.45, baseline={([yshift=-0ex]current bounding box.center)}, thick]
        \begin{scope}[shift={(0,0)}]
            \draw (0,-1) -- (0,1);
            \filldraw[fill=powderblue!50] (-0.75,-0.5) -- (-0.75,0.5) -- (0.75,0.5) -- (0.75,-0.5) -- (-0.75,-0.5);
            \draw (0,0) node {\scriptsize $A_{\text{up}}$};
            \draw (0,-1.3) node {\scriptsize $\{1\}$};
        \end{scope}
    \end{tikzpicture}
    = \ket{1} + \ket{2} , \ 
    \begin{tikzpicture}[scale=.45, baseline={([yshift=-0.6ex]current bounding box.center)}, thick]
        \begin{scope}[shift={(0,0)}]
            \draw (0,0) -- (0,1);
            \draw (-1.25,0) -- (1.25,0);
            \filldraw[fill=powderblue!50] (-0.75,-0.5) -- (-0.75,0.5) -- (0.75,0.5) -- (0.75,-0.5) -- (-0.75,-0.5);
            \draw (0,0) node {\scriptsize $A_{\text{low}}$};
            \draw (0,1.3) node {\scriptsize $\{0,1\}$};
        \end{scope}
    \end{tikzpicture}
    = {\scriptsize \begin{pmatrix}
        1 & \\ & 1
    \end{pmatrix}} , \ 
    \begin{tikzpicture}[scale=.45, baseline={([yshift=-0.6ex]current bounding box.center)}, thick]
        \begin{scope}[shift={(0,0)}]
            \draw (0,0) -- (0,1);
            \draw (-1.25,0) -- (1.25,0);
            \filldraw[fill=powderblue!50] (-0.75,-0.5) -- (-0.75,0.5) -- (0.75,0.5) -- (0.75,-0.5) -- (-0.75,-0.5);
            \draw (0,0) node {\scriptsize $A_{\text{low}}$};
            \draw (0,1.3) node {\scriptsize $\{1\}$};
        \end{scope}
    \end{tikzpicture}
    = {\scriptsize \begin{pmatrix}
        0 & 1 \\ & 0
    \end{pmatrix}} , \\
    B^i &= \ \begin{tikzpicture}[scale=.45, baseline={([yshift=-0.6ex]current bounding box.center)}, thick]
            \begin{scope}[shift={(0,1.5)}]
                \draw (0,-1) -- (0,1);
                \draw (-1.25,0) -- (1.25,0);
                \filldraw[fill=melon!50] (-0.75,-0.5) -- (-0.75,0.5) -- (0.75,0.5) -- (0.75,-0.5) -- (-0.75,-0.5);
                \draw (0,0) node {\scriptsize $B_{\text{up}}$};
            \end{scope}
            \begin{scope}[shift={(0,0)}]
                \draw (-1.25,0) -- (1.25,0);
                \filldraw[fill=melon!50] (-0.75,-0.5) -- (-0.75,0.5) -- (0.75,0.5) -- (0.75,-0.5) -- (-0.75,-0.5);
                \draw (0,0) node {\scriptsize $B_{\text{low}}$};
            \end{scope}
        \end{tikzpicture} \ , \quad \text{with} \quad
        \begin{tikzpicture}[scale=.45, baseline={([yshift=-0ex]current bounding box.center)}, thick]
            \begin{scope}[shift={(0,0)}]
                \draw (0,-1) -- (0,1);
                \filldraw[fill=melon!50] (-0.75,-0.5) -- (-0.75,0.5) -- (0.75,0.5) -- (0.75,-0.5) -- (-0.75,-0.5);
                \draw (0,0) node {\scriptsize $B_{\text{up}}$};
                \draw (0,-1.3) node {\scriptsize $\{0,1\}$};
            \end{scope}
        \end{tikzpicture}
        = \ket{0} , \
        \begin{tikzpicture}[scale=.45, baseline={([yshift=-0ex]current bounding box.center)}, thick]
            \begin{scope}[shift={(0,0)}]
                \draw (0,-1) -- (0,1);
                \filldraw[fill=melon!50] (-0.75,-0.5) -- (-0.75,0.5) -- (0.75,0.5) -- (0.75,-0.5) -- (-0.75,-0.5);
                \draw (0,0) node {\scriptsize $B_{\text{up}}$};
                \draw (0,-1.3) node {\scriptsize $\{1\}$};
            \end{scope}
        \end{tikzpicture}
        = \ket{1} , \ 
        \begin{tikzpicture}[scale=.45, baseline={([yshift=-0ex]current bounding box.center)}, thick]
            \begin{scope}[shift={(0,0)}]
                \draw (0,-1) -- (0,1);
                \filldraw[fill=melon!50] (-0.75,-0.5) -- (-0.75,0.5) -- (0.75,0.5) -- (0.75,-0.5) -- (-0.75,-0.5);
                \draw (0,0) node {\scriptsize $B_{\text{up}}$};
                \draw (0,-1.3) node {\scriptsize $\{2\}$};
            \end{scope}
        \end{tikzpicture}
        = \ket{2} , \
        \begin{tikzpicture}[scale=.45, baseline={([yshift=-0.6ex]current bounding box.center)}, thick]
            \begin{scope}[shift={(0,0)}]
                \draw (0,0) -- (0,1);
                \draw (-1.25,0) -- (1.25,0);
                \filldraw[fill=melon!50] (-0.75,-0.5) -- (-0.75,0.5) -- (0.75,0.5) -- (0.75,-0.5) -- (-0.75,-0.5);
                \draw (0,0) node {\scriptsize $B_{\text{low}}$};
                \draw (0,1.3) node {\scriptsize $i$};
            \end{scope}
        \end{tikzpicture}
        = B^i .
\end{align*}
The $C$ tensor that results from stacking $A$ and $B$ together, $C^i = A^i \oplus B^i$, is
\begin{equation*}
    C^0 = \mathds{1}_5, \quad
    C^1 = {\scriptsize \begin{pmatrix}
        0 & 1 & & & \\
        & 0 & & & \\
        & & 0 & 1 & \\
        & & & 0 & \\
        & & & & 0
    \end{pmatrix}}, \quad
    C^2 = {\scriptsize \begin{pmatrix}
        0 & 1 & & & \\
        & 0 & & & \\
        & & 0 & & \\
        & & & 0 & 1 \\
        & & & & 0
    \end{pmatrix}}. 
\end{equation*}
Its span $\mathcal{C}^{(1)}$ can be written in a basis that that coincides with the basis of $\mathcal{A}^{(1)}$ above as follows, 
\begin{equation*}
    \mathcal{C}^{(1)} = \left\{ 
    {\scriptsize \begin{pmatrix}
        c_0 & c_1 & & & \\
        & c_0 & & & \\
        & & c_0 & c_2 & \\
        & & & c_0 & c_1 - c_2 \\
        & & & & c_0
    \end{pmatrix}} \mid c_0, c_1, c_2 \in \mathbb{C}
    \right\},
\end{equation*}
leading to the following matrix-CF decomposition:
\begin{align}
    C^i = \ \begin{tikzpicture}[scale=.45, baseline={([yshift=-0.6ex]current bounding box.center)}, thick]
        \begin{scope}[shift={(0,1.5)}]
            \draw (0,-1) -- (0,1);
            \draw (-1.25,0) -- (1.25,0);
            \filldraw[fill=atomictangerine!50] (-0.75,-0.5) -- (-0.75,0.5) -- (0.75,0.5) -- (0.75,-0.5) -- (-0.75,-0.5);
            \draw (0,0) node {\scriptsize $C_{\text{up}}$};
        \end{scope}
        \begin{scope}[shift={(0,0)}]
            \draw (-1.25,0) -- (1.25,0);
            \filldraw[fill=atomictangerine!50] (-0.75,-0.5) -- (-0.75,0.5) -- (0.75,0.5) -- (0.75,-0.5) -- (-0.75,-0.5);
            \draw (0,0) node {\scriptsize $C_{\text{low}}$};
        \end{scope}
    \end{tikzpicture}, 
    \quad \text{with} \quad
    \begin{tikzpicture}[scale=.45, baseline={([yshift=-0ex]current bounding box.center)}, thick]
        \begin{scope}[shift={(0,0)}]
            \draw (0,-1) -- (0,1);
            \filldraw[fill=atomictangerine!50] (-0.75,-0.5) -- (-0.75,0.5) -- (0.75,0.5) -- (0.75,-0.5) -- (-0.75,-0.5);
            \draw (0,0) node {\scriptsize $C_{\text{up}}$};
            \draw (0,-1.3) node {\scriptsize $\{0,1\}$};
        \end{scope}
    \end{tikzpicture}
    &= \ket{0} , \
    \begin{tikzpicture}[scale=.45, baseline={([yshift=-0ex]current bounding box.center)}, thick]
        \begin{scope}[shift={(0,0)}]
            \draw (0,-1) -- (0,1);
            \filldraw[fill=atomictangerine!50] (-0.75,-0.5) -- (-0.75,0.5) -- (0.75,0.5) -- (0.75,-0.5) -- (-0.75,-0.5);
            \draw (0,0) node {\scriptsize $C_{\text{up}}$};
            \draw (0,-1.3) node {\scriptsize $\{1\}$};
        \end{scope}
    \end{tikzpicture}
    = \ket{1} + \ket{2} , \
    \begin{tikzpicture}[scale=.45, baseline={([yshift=-0ex]current bounding box.center)}, thick]
        \begin{scope}[shift={(0,0)}]
            \draw (0,-1) -- (0,1);
            \filldraw[fill=atomictangerine!50] (-0.75,-0.5) -- (-0.75,0.5) -- (0.75,0.5) -- (0.75,-0.5) -- (-0.75,-0.5);
            \draw (0,0) node {\scriptsize $C_{\text{up}}$};
            \draw (0,-1.3) node {\scriptsize $\{2\}$};
        \end{scope}
    \end{tikzpicture}
    = \ket{1} , \
    \begin{tikzpicture}[scale=.45, baseline={([yshift=-0.6ex]current bounding box.center)}, thick]
        \begin{scope}[shift={(0,0)}]
            \draw (0,0) -- (0,1);
            \draw (-1.25,0) -- (1.25,0);
            \filldraw[fill=atomictangerine!50] (-0.75,-0.5) -- (-0.75,0.5) -- (0.75,0.5) -- (0.75,-0.5) -- (-0.75,-0.5);
            \draw (0,0) node {\scriptsize $C_{\text{low}}$};
            \draw (0,1.3) node {\scriptsize $\{0,1\}$};
        \end{scope}
    \end{tikzpicture}
    = \mathds{1}_5 , \nonumber \\ 
    \begin{tikzpicture}[scale=.45, baseline={([yshift=-0.6ex]current bounding box.center)}, thick]
        \begin{scope}[shift={(0,0)}]
            \draw (0,0) -- (0,1);
            \draw (-1.25,0) -- (1.25,0);
            \filldraw[fill=atomictangerine!50] (-0.75,-0.5) -- (-0.75,0.5) -- (0.75,0.5) -- (0.75,-0.5) -- (-0.75,-0.5);
            \draw (0,0) node {\scriptsize $C_{\text{low}}$};
            \draw (0,1.3) node {\scriptsize $\{1\}$};
        \end{scope}
    \end{tikzpicture}
    &= {\scriptsize \begin{pmatrix}
        0 & 1 & & & \\ & 0 & & & \\
        & & 0 & & \\ & & & 0 & 1 \\ & & & & 0
    \end{pmatrix}} , \
    \begin{tikzpicture}[scale=.45, baseline={([yshift=-0.6ex]current bounding box.center)}, thick]
        \begin{scope}[shift={(0,0)}]
            \draw (0,0) -- (0,1);
            \draw (-1.25,0) -- (1.25,0);
            \filldraw[fill=atomictangerine!50] (-0.75,-0.5) -- (-0.75,0.5) -- (0.75,0.5) -- (0.75,-0.5) -- (-0.75,-0.5);
            \draw (0,0) node {\scriptsize $C_{\text{low}}$};
            \draw (0,1.3) node {\scriptsize $\{2\}$};
        \end{scope}
    \end{tikzpicture}
    = {\scriptsize \begin{pmatrix}
        0 & & & & \\ & 0 & & & \\
        & & 0 & 1 & \\ & & & 0 & -1 \\ & & & & 0
    \end{pmatrix}} \ .
    \label{eq:example1_stacked_C}
\end{align}
We observe that Lemma \ref{lemma:stack-trick} holds as expected, since
\begin{equation} \label{eq:relations_Aup_Bup_Cup}
    \begin{tikzpicture}[scale=.45, baseline={([yshift=-0ex]current bounding box.center)}, thick]
        \begin{scope}[shift={(0,0)}]
            \draw (0,-1) -- (0,1);
            \filldraw[fill=powderblue!50] (-0.75,-0.5) -- (-0.75,0.5) -- (0.75,0.5) -- (0.75,-0.5) -- (-0.75,-0.5);
            \draw (0,0) node {\scriptsize $A_{up}$};
            \draw (0,-1.3) node {\scriptsize $\{1\}$};
        \end{scope}
    \end{tikzpicture}
    =
    \begin{tikzpicture}[scale=.45, baseline={([yshift=-0ex]current bounding box.center)}, thick]
        \begin{scope}[shift={(0,0)}]
            \draw (0,-1) -- (0,1);
            \filldraw[fill=atomictangerine!50] (-0.75,-0.5) -- (-0.75,0.5) -- (0.75,0.5) -- (0.75,-0.5) -- (-0.75,-0.5);
            \draw (0,0) node {\scriptsize $C_{up}$};
            \draw (0,-1.3) node {\scriptsize $\{1\}$};
        \end{scope}
    \end{tikzpicture} , 
    \quad
    \begin{tikzpicture}[scale=.45, baseline={([yshift=-0ex]current bounding box.center)}, thick]
        \begin{scope}[shift={(0,0)}]
            \draw (0,-1) -- (0,1);
            \filldraw[fill=melon!50] (-0.75,-0.5) -- (-0.75,0.5) -- (0.75,0.5) -- (0.75,-0.5) -- (-0.75,-0.5);
            \draw (0,0) node {\scriptsize $B_{up}$};
            \draw (0,-1.3) node {\scriptsize $\{1\}$};
        \end{scope}
    \end{tikzpicture}
    =
    \begin{tikzpicture}[scale=.45, baseline={([yshift=-0ex]current bounding box.center)}, thick]
        \begin{scope}[shift={(0,0)}]
            \draw (0,-1) -- (0,1);
            \filldraw[fill=atomictangerine!50] (-0.75,-0.5) -- (-0.75,0.5) -- (0.75,0.5) -- (0.75,-0.5) -- (-0.75,-0.5);
            \draw (0,0) node {\scriptsize $C_{up}$};
            \draw (0,-1.3) node {\scriptsize $\{2\}$};
        \end{scope}
    \end{tikzpicture} ,
    \quad
    \begin{tikzpicture}[scale=.45, baseline={([yshift=-0ex]current bounding box.center)}, thick]
        \begin{scope}[shift={(0,0)}]
            \draw (0,-1) -- (0,1);
            \filldraw[fill=melon!50] (-0.75,-0.5) -- (-0.75,0.5) -- (0.75,0.5) -- (0.75,-0.5) -- (-0.75,-0.5);
            \draw (0,0) node {\scriptsize $B_{up}$};
            \draw (0,-1.3) node {\scriptsize $\{2\}$};
        \end{scope}
    \end{tikzpicture}
    =
    \begin{tikzpicture}[scale=.45, baseline={([yshift=-0ex]current bounding box.center)}, thick]
        \begin{scope}[shift={(0,0)}]
            \draw (0,-1) -- (0,1);
            \filldraw[fill=atomictangerine!50] (-0.75,-0.5) -- (-0.75,0.5) -- (0.75,0.5) -- (0.75,-0.5) -- (-0.75,-0.5);
            \draw (0,0) node {\scriptsize $C_{up}$};
            \draw (0,-1.3) node {\scriptsize $\{1\}$};
        \end{scope}
    \end{tikzpicture} 
    -
    \begin{tikzpicture}[scale=.45, baseline={([yshift=-0ex]current bounding box.center)}, thick]
        \begin{scope}[shift={(0,0)}]
            \draw (0,-1) -- (0,1);
            \filldraw[fill=atomictangerine!50] (-0.75,-0.5) -- (-0.75,0.5) -- (0.75,0.5) -- (0.75,-0.5) -- (-0.75,-0.5);
            \draw (0,0) node {\scriptsize $C_{up}$};
            \draw (0,-1.3) node {\scriptsize $\{2\}$};
        \end{scope}
    \end{tikzpicture} \ .
\end{equation}
Hence, the $P_B$ matrix acts on $\langle \Sigma^B \rangle$ as
\begin{equation*}
    \begin{cases}
        P_B \ket{\{0,1\}} = \ket{\{0,1\}}, \\
        P_B \ket{\{1\}} = \ket{\{2\}}, \\
        P_B \ket{\{2\}} = \ket{\{1\}} - \ket{\{2\}}.
    \end{cases}
\end{equation*}
    

With the stacking trick, given two equivalent MPS-X representations in matrix-CF, $(X_A, A^i)$ and $(X_B, B^i)$, with no negligible blocks, we can obtain an even more compressed representation for each, exploiting the fact that they generate the same family of states. To formalize this, we define:
\begin{definition}[Negligible part of one MPS-X with respect to another] \label{def:negligible_part}
    Let $(X_A, A^i)$ and $(X_B, B^i)$ be two MPS-X representations with no negligible blocks. We say that $B$ has a negligible part with respect to $A$ if, after applying the stacking trick to form the tensor $C$ in matrix-CF with $C_{\text{low}}^i |_{\text{A-part}} = A^i_{\text{low}}$, there exists some $t \in \Sigma^C$ such that 
    \begin{equation} \label{eq:negligible_part}
        C_{\text{low}}^t |_{\text{A-part}} = 0 \quad \text{while} \quad C_{\text{low}}^t |_{\text{B-part}} \neq 0 \ . 
    \end{equation}
\end{definition}
The following lemma shows that Def. \ref{def:negligible_part} follows from the more natural assumption that tensors $A^i$ and $B^i$ should span the same physical subspace.
\begin{lemma} \label{lemma:freedom1}
    Let $(X_A, A^i)$ and $(X_B, B^i)$ be two MPS-X representations such that their physical subspaces coincide, i.e. $V_A = V_B$, where
    \begin{equation*}
        V_A := \left\{
        \begin{tikzpicture}[scale=.45,thick,baseline={([yshift=-1ex]current bounding box.center)}]
            \MPSTensor{0,0}{$A$}{powderblue!50}
            \draw (-2.5,0) -- (-2.5,-1) -- (1,-1) -- (1,0);
            \begin{scope}[shift={(-1.5,0)}]
                \draw (-1,0) -- (0,0);
                \filldraw[fill=gray!10] (0.5,-0.5) -- (-0.5,-0.5) -- (-0.5,0.5) -- (0.5, 0.5) -- (0.5,-0.5);
                \draw (0,0) node {\scriptsize $Y$};
            \end{scope}
        \end{tikzpicture} 
        \mid 
        \forall Y \in \mathcal{M}_{D_A\times D_A}(\mathbb{C})
        \right\},
        \quad 
        V_B := \left\{
        \begin{tikzpicture}[scale=.45,thick,baseline={([yshift=-1ex]current bounding box.center)}]
            \MPSTensor{0,0}{$B$}{melon!50}
            \draw (-2.5,0) -- (-2.5,-1) -- (1,-1) -- (1,0);
            \begin{scope}[shift={(-1.5,0)}]
                \draw (-1,0) -- (0,0);
                \filldraw[fill=gray!10] (0.5,-0.5) -- (-0.5,-0.5) -- (-0.5,0.5) -- (0.5, 0.5) -- (0.5,-0.5);
                \draw (0,0) node {\scriptsize $Z$};
            \end{scope}
        \end{tikzpicture} 
        \mid 
        \forall Z \in \mathcal{M}_{D_B\times D_B}(\mathbb{C})
        \right\}.
    \end{equation*}
    Then, neither $A$ nor $B$ have a negligible part with respect to the other.
\end{lemma}
\begin{proof}
    We proceed by contradiction. Assume that $B$ has a negligible part with respect to $A$. By applying the stacking construction from Lemma \ref{lemma:stack-trick}, this implies that in the stacked tensor $C$ obtained after sufficient blocking, with $C_{\mathrm{low}}^i\mid_{\mathrm{A-part}} = A_{\mathrm{low}}^i$, there exists a free block labeled by $t \in \Sigma_C$ that is fully supported in the B-part of tensor $C$.

    Let $V_t \neq \{0\}$ denote the physical subspace spanned by block $t$. Singe gauge transformations in the virtual space do not affect the physical subspace, we have that 
    \begin{equation*}
        V_B = V_{C |_{\mathrm{B-part}}} \supseteq V_t . 
    \end{equation*}

    On the other hand, because block $t$ is free and independent of the free blocks supported on the $A$-part of $C$ (as guaranteed by Eq. \eqref{eq:negligible_part}), it follows that $V_t \cap V_A = \{0\}.$

    Using the assumption that $V_A = V_B$, we thus obtain
    \begin{equation*}
        V_t \subseteq V_B = V_A ,
    \end{equation*}
    which contradicts $V_t \cap V_A = \{0\}$ when $V_t \neq \{0\}$. Hence, $B$ cannot have a negligible part with respect to $A$. By symmetry, the same conclusion holds with $A$ and $B$ exchanged.  
\end{proof}

Moreover, given any two equivalent MPS-X representations, one can always construct two other equivalent representations that generate the same family of states as the original ones, while ensuring that the physical subspaces spanned by the new tensors coincide, as established in the following lemma. 
\begin{lemma} \label{lemma:freedom2}
    Let $(X_A, A^i)$ and $(X_B, B^i)$ be two MPS-X representations of the same family of states $\{\ket{\psi_N}\}_N$. Define the modified tensors $\tilde{A}$ and $\tilde{B}$ by
    \begin{equation*}
        \begin{tikzpicture}[scale=.45,thick,baseline={([yshift=-1ex]current bounding box.center)}]
            \MPSTensor{0,0}{$\tilde{A}$}{powderblue!50}
        \end{tikzpicture} \
        := 
        \begin{tikzpicture}[scale=.45,thick,baseline={([yshift=-3ex]current bounding box.center)}]
            \MPSTensor{0,0}{$A$}{powderblue!50}
            \begin{scope}[shift={(0,1.5)}]
                \draw (0,0) -- (0,1);
                \filldraw[fill=gray!10] (0.5+0.8,-0.5) -- (-0.5-0.8,-0.5) -- (-0.5-0.8,0.5) -- (0.5+0.8, 0.5) -- (0.5+0.8,-0.5);
                \draw (0,0) node {\scriptsize $\mathbb{P}_{V_A \cap V_B}$};
            \end{scope}
        \end{tikzpicture}
        ,
        \quad
        \begin{tikzpicture}[scale=.45,thick,baseline={([yshift=-1ex]current bounding box.center)}]
            \MPSTensor{0,0}{$\tilde{B}$}{melon!50}
        \end{tikzpicture} \
        := 
        \begin{tikzpicture}[scale=.45,thick,baseline={([yshift=-3ex]current bounding box.center)}]
            \MPSTensor{0,0}{$B$}{melon!50}
            \begin{scope}[shift={(0,1.5)}]
                \draw (0,0) -- (0,1);
                \filldraw[fill=gray!10] (0.5+0.8,-0.5) -- (-0.5-0.8,-0.5) -- (-0.5-0.8,0.5) -- (0.5+0.8, 0.5) -- (0.5+0.8,-0.5);
                \draw (0,0) node {\scriptsize $\mathbb{P}_{V_A \cap V_B}$};
            \end{scope}
        \end{tikzpicture} ,
    \end{equation*}
    where $\mathbb{P}_{V_A \cap V_B}$ denotes the orthogonal projector onto the intersection $V_A \cap V_B$ of the corresponding physical subspaces.
    
    Then, the two MPS-X representations $(X_A, \tilde{A}^i)$ and $(X_B, \tilde{B}^i)$ generate the same family of states $\{\ket{\psi_N}\}_N$ as the original MPS-X.
\end{lemma}
\begin{proof}
    For each $N$, define
    \begin{equation*}
        V_A^{(N)} =
        \left\{
        \begin{tikzpicture}[scale=.45,thick,baseline={([yshift=-1.15ex]current bounding box.center)}]
            \FullMPSX{0,0}{$A$}{$Y$}{powderblue!50}{gray!10}
        \end{tikzpicture}
        \mid 
        \forall Y \in \mathcal{M}_{D_A \times D_A}(\mathbb{C})
        \right\}.
    \end{equation*}
    and define $V_B^{(N)}$ analogously. 
    
    For any $\ket{\eta} \in V_A^{(N)}$ and for any set of states $\{\ket{\phi_i}\}_i \subseteq \mathbb{C}^d$, projecting all but site $n$ onto them, for each $n \in \{1, \dots, N\}$, it holds that 
    \begin{equation*}
        \left[\left(\bigotimes_{j=1}^{n-1} \bra{\phi_j}\right) \otimes \mathds{1}_d \otimes \left(\bigotimes_{j=n+1}^N \bra{\phi_j}\right) \right] \ket{\eta} = 
        \begin{tikzpicture}[scale=.45,thick,baseline={([yshift=-1.15ex]current bounding box.center)}]
            \FullMPSX{0,0}{$A$}{$Y$}{powderblue!50}{gray!10}
            \begin{scope}[shift={(0,1.5)}]
                \draw[fill=yellow] (0,0) circle (0.5);
                \draw (0,0) node {\scriptsize $\phi_1$};
                \draw[fill=yellow] (1.5,0) circle (0.5);
                \draw (1.5,0) node {\scriptsize $\phi_2$};
                \draw[fill=yellow] (5,0.05) circle (0.55);
                \draw (5,0.025) node {\scriptsize $\phi_N$};
            \end{scope}
        \end{tikzpicture}
        = \
        \begin{tikzpicture}[scale=.45,thick,baseline={([yshift=-0.5ex]current bounding box.center)}]
            \MPSTensor{0,0}{$A$}{powderblue!50}
            \draw (-2.5,0) -- (-2.5,-0.8) -- (1,-0.8) -- (1,0);
            \begin{scope}[shift={(-1.5,0)}]
                \draw (-1,0) -- (0,0);
                \filldraw[fill=gray!10] (0.5,-0.5) -- (-0.5,-0.5) -- (-0.5,0.5) -- (0.5, 0.5) -- (0.5,-0.5);
                \draw (0,0) node {\scriptsize $Y'$};
            \end{scope}
        \end{tikzpicture} 
        \subseteq V_A^{(1)}
        .
    \end{equation*}
    Hence $V_A^{(N)} \subseteq (V_A^{(1)})^{\otimes N}$ for all $N$, and similarly $V_B^{(N)} \subseteq (V_B^{(1)})^{\otimes N}$.
    This immediately gives 
    \begin{equation} \label{eq:intersect-V_A^(N)}
        V_A^{(N)} \cap V_B^{(N)} \subseteq
        (V_A^{(1)})^{\otimes N} \cap (V_B^{(1)})^{\otimes N} 
        = (V_A^{(1)} \cap V_B^{(1)})^{\otimes N} ,
    \end{equation}
    where the last equality follows from the general identity $C^{\otimes N} \cap D^{\otimes N} = (C\cap D)^{\otimes N}$ for any linear subspaces $C, D$.

    Since both MPS-X representations generate the same states by assumption, i.e. $\ket{\psi_N} = | \psi_N(X_A, A^i) \rangle = | \psi_N(X_B, B^i) \rangle$, it holds that $\ket{\psi_N} \subseteq V_A^{(N)} \cap V_B^{(N)}$.
    
    Equation \eqref{eq:intersect-V_A^(N)} then implies $\ket{\psi_N} \subseteq (V_A^{(1)} \cap V_B^{(1)})^{\otimes N}$. Therefore,
    \begin{equation*}
        |\psi_N(X_A, \tilde{A}^i)\rangle = (\mathbb{P}_{V_A \cap V_B})^{\otimes N} \ket{\psi_N} = \ket{\psi_N},
    \end{equation*}
    and analogously for $|\psi_N(X_B, \tilde{B}^i)\rangle$. This proves the claim. 
\end{proof}

This concludes the proof of Lemma \ref{lemma:freedom_always-possible-reduced} showing that we can, without loss of generality, restrict to reduced pairs of MPS-X: Lemma \ref{lemma:freedom2} above guarantees that condition \textit{(iii)} of Def. \ref{def:reduced_MPS-X} can always be enforced, while conditions \textit{(i)} and \textit{(ii)} are ensured by the procedures described in Sections \ref{sec:freedom_condition(i)} and \ref{sec:freedom_condition(ii)}, respectively. 

Moreover, since condition \textit{(iii)} implies that $\tilde{A}$ and $\tilde{B}$ have no negligible parts with respect to each other (Def. \ref{def:negligible_part}), it follows that $|\Sigma^A| = |\Sigma^B|$. In particular, after enough blocking, $|\Sigma^A_\infty| = |\Sigma^B_\infty|$ and $|\Sigma^{A,ij}_f| = |\Sigma^{B,\pi(i)\pi(j)}_f|$ for some relabeling $\pi$ of the $\Sigma_\infty$ symbols, which establishes a one-to-one correspondence between the free blocks in the two MPS-X representations.

\paragraph{Example 3 (revisited).} Previously, we saw that block $\{2\}$ of the stacked tensor $C$ satisfies $C^{\{2\}}_{\text{low}} |_{\text{A-part}} = 0$. Thus, $B$ has a negligible part with respect to $A$, and the physical subspaces they generate are different, $V_A = \mathrm{span}\{\ket{0}, \ket{1}+\ket{2}\}\neq V_B = \mathrm{span}\{\ket{0}, \ket{1}, \ket{2}\}$.

Let us now apply $\mathbb{P}_{V_A\cap V_B}$ onto tensors $A, B$, where $V_A \cap V_B = V_A = \mathrm{span}\{\ket{0}, \ket{1}+\ket{2}\}$. Such orthogonal projector is 
\begin{equation*}
    \mathbb{P}_{V_A\cap V_B} = \dyad{0}{0} + \frac{1}{2}\left(\ket{1}+\ket{2}\right) \left(\bra{1}+\bra{2}\right).
\end{equation*}
Applying this onto $A$ does not modify it because $V_A = V_A \cap V_B$, so $\tilde{A} = A$. On the other hand, applying it onto $B$ gives
\begin{equation*}
    \begin{tikzpicture}[scale=.45, baseline={([yshift=-0ex]current bounding box.center)}, thick]
        \begin{scope}[shift={(0,0)}]
            \draw (0,-1) -- (0,1);
            \filldraw[fill=melon!50] (-0.75,-0.5) -- (-0.75,0.5) -- (0.75,0.5) -- (0.75,-0.5) -- (-0.75,-0.5);
            \draw (0,0) node {\scriptsize $\tilde{B}_{up}$};
            \draw (0,-1.3) node {\scriptsize $\{1\}$};
        \end{scope}
    \end{tikzpicture}
    :=
    \mathbb{P}_{V_A \cap V_B} \ket{1} = \frac{1}{2} (\ket{1} + \ket{2}),
    \quad
    \begin{tikzpicture}[scale=.45, baseline={([yshift=-0ex]current bounding box.center)}, thick]
        \begin{scope}[shift={(0,0)}]
            \draw (0,-1) -- (0,1);
            \filldraw[fill=melon!50] (-0.75,-0.5) -- (-0.75,0.5) -- (0.75,0.5) -- (0.75,-0.5) -- (-0.75,-0.5);
            \draw (0,0) node {\scriptsize $\tilde{B}_{up}$};
            \draw (0,-1.3) node {\scriptsize $\{2\}$};
        \end{scope}
    \end{tikzpicture}
    :=
    \mathbb{P}_{V_A \cap V_B} \ket{2} = \frac{1}{2} (\ket{1}+ \ket{2}),
    \end{equation*}
    which, combined with $B_{\text{low}}$, defines the MPS-X $(X_B, \tilde{B}^i)$ equivalent to the original one, $(X_B, B^i)$, where
    \begin{equation*}
        \tilde{B}^0 = \mathds{1}_3, \  
        \tilde{B}^1 = \tilde{B}^2 =  {\scriptsize \begin{pmatrix}
            0 & \frac{1}{2} & \\ & 0 & \frac{1}{2} \\ & & 0
        \end{pmatrix}} .
\end{equation*}
Note that one MPS-X representation may have negligible blocks relative to another, but not necessarily vice versa. In Example 3, $B$ contains negligible blocks with respect to $A$. Let us apply the stacking trick in the reverse order, that is, fixing $C_{\text{low}}^x \mid_{\text{B-block}} = B_{\text{low}}^x$. We can do so with the matrix-CF decomposition arising from writing $\mathcal{C}^{(1)}$ in the following basis,
\begin{equation*}
    \mathcal{C}^{(1)} = \left\{ 
    {\scriptsize \begin{pmatrix}
        c_0 & c_1+c_2 & & & \\
        & c_0 & & & \\
        & & c_0 & c_1 & \\
        & & & c_0 & c_2 \\
        & & & & c_0
    \end{pmatrix}} \mid c_0, c_1, c_2 \in \mathbb{C}
    \right\},
\end{equation*}
which leads to 
\begin{align}
    C^i = \ \begin{tikzpicture}[scale=.45, baseline={([yshift=-0.6ex]current bounding box.center)}, thick]
        \begin{scope}[shift={(0,1.5)}]
            \draw (0,-1) -- (0,1);
            \draw (-1.25,0) -- (1.25,0);
            \filldraw[fill=atomictangerine!50] (-0.75,-0.5) -- (-0.75,0.5) -- (0.75,0.5) -- (0.75,-0.5) -- (-0.75,-0.5);
            \draw (0,0) node {\scriptsize $C_{\text{up}}$};
        \end{scope}
        \begin{scope}[shift={(0,0)}]
            \draw (-1.25,0) -- (1.25,0);
            \filldraw[fill=atomictangerine!50] (-0.75,-0.5) -- (-0.75,0.5) -- (0.75,0.5) -- (0.75,-0.5) -- (-0.75,-0.5);
            \draw (0,0) node {\scriptsize $C_{\text{low}}$};
        \end{scope}
    \end{tikzpicture}, 
    \quad \text{with} \quad
    \begin{tikzpicture}[scale=.45, baseline={([yshift=-0ex]current bounding box.center)}, thick]
        \begin{scope}[shift={(0,0)}]
            \draw (0,-1) -- (0,1);
            \filldraw[fill=atomictangerine!50] (-0.75,-0.5) -- (-0.75,0.5) -- (0.75,0.5) -- (0.75,-0.5) -- (-0.75,-0.5);
            \draw (0,0) node {\scriptsize $C_{\text{up}}$};
            \draw (0,-1.3) node {\scriptsize $\{0,1\}$};
        \end{scope}
    \end{tikzpicture}
    &= \ket{0} , \
    \begin{tikzpicture}[scale=.45, baseline={([yshift=-0ex]current bounding box.center)}, thick]
        \begin{scope}[shift={(0,0)}]
            \draw (0,-1) -- (0,1);
            \filldraw[fill=atomictangerine!50] (-0.75,-0.5) -- (-0.75,0.5) -- (0.75,0.5) -- (0.75,-0.5) -- (-0.75,-0.5);
            \draw (0,0) node {\scriptsize $C_{\text{up}}$};
            \draw (0,-1.3) node {\scriptsize $\{1\}$};
        \end{scope}
    \end{tikzpicture}
    = \ket{1} , \
    \begin{tikzpicture}[scale=.45, baseline={([yshift=-0ex]current bounding box.center)}, thick]
        \begin{scope}[shift={(0,0)}]
            \draw (0,-1) -- (0,1);
            \filldraw[fill=atomictangerine!50] (-0.75,-0.5) -- (-0.75,0.5) -- (0.75,0.5) -- (0.75,-0.5) -- (-0.75,-0.5);
            \draw (0,0) node {\scriptsize $C_{\text{up}}$};
            \draw (0,-1.3) node {\scriptsize $\{2\}$};
        \end{scope}
    \end{tikzpicture}
    = \ket{2} , \
    \begin{tikzpicture}[scale=.45, baseline={([yshift=-0.6ex]current bounding box.center)}, thick]
        \begin{scope}[shift={(0,0)}]
            \draw (0,0) -- (0,1);
            \draw (-1.25,0) -- (1.25,0);
            \filldraw[fill=atomictangerine!50] (-0.75,-0.5) -- (-0.75,0.5) -- (0.75,0.5) -- (0.75,-0.5) -- (-0.75,-0.5);
            \draw (0,0) node {\scriptsize $C_{\text{low}}$};
            \draw (0,1.3) node {\scriptsize $\{0,1\}$};
        \end{scope}
    \end{tikzpicture}
    = \mathds{1}_5 , \nonumber \\ 
    \begin{tikzpicture}[scale=.45, baseline={([yshift=-0.6ex]current bounding box.center)}, thick]
        \begin{scope}[shift={(0,0)}]
            \draw (0,0) -- (0,1);
            \draw (-1.25,0) -- (1.25,0);
            \filldraw[fill=atomictangerine!50] (-0.75,-0.5) -- (-0.75,0.5) -- (0.75,0.5) -- (0.75,-0.5) -- (-0.75,-0.5);
            \draw (0,0) node {\scriptsize $C_{\text{low}}$};
            \draw (0,1.3) node {\scriptsize $\{1\}$};
        \end{scope}
    \end{tikzpicture}
    &= {\scriptsize \begin{pmatrix}
        0 & 1 & & & \\ & 0 & & & \\
        & & 0 & 1 & \\ & & & 0 &  \\ & & & & 0
    \end{pmatrix}} , \
    \begin{tikzpicture}[scale=.45, baseline={([yshift=-0.6ex]current bounding box.center)}, thick]
        \begin{scope}[shift={(0,0)}]
            \draw (0,0) -- (0,1);
            \draw (-1.25,0) -- (1.25,0);
            \filldraw[fill=atomictangerine!50] (-0.75,-0.5) -- (-0.75,0.5) -- (0.75,0.5) -- (0.75,-0.5) -- (-0.75,-0.5);
            \draw (0,0) node {\scriptsize $C_{\text{low}}$};
            \draw (0,1.3) node {\scriptsize $\{2\}$};
        \end{scope}
    \end{tikzpicture}
    = {\scriptsize \begin{pmatrix}
        0 & 1 & & & \\ & 0 & & & \\
        & & 0 & & \\ & & & 0 & 1 \\ & & & & 0
    \end{pmatrix}} \ .
    \label{eq:example1_stacked_C}
\end{align}
Lemma \ref{lemma:stack-trick} still holds reversing the roles of $A$ and $B$, since
\begin{equation} \label{eq:relations_Aup_Bup_Cup_2}
    \begin{tikzpicture}[scale=.45, baseline={([yshift=-0ex]current bounding box.center)}, thick]
        \begin{scope}[shift={(0,0)}]
            \draw (0,-1) -- (0,1);
            \filldraw[fill=powderblue!50] (-0.75,-0.5) -- (-0.75,0.5) -- (0.75,0.5) -- (0.75,-0.5) -- (-0.75,-0.5);
            \draw (0,0) node {\scriptsize $A_{up}$};
            \draw (0,-1.3) node {\scriptsize $\{1\}$};
        \end{scope}
    \end{tikzpicture}
    =
    \begin{tikzpicture}[scale=.45, baseline={([yshift=-0ex]current bounding box.center)}, thick]
        \begin{scope}[shift={(0,0)}]
            \draw (0,-1) -- (0,1);
            \filldraw[fill=atomictangerine!50] (-0.75,-0.5) -- (-0.75,0.5) -- (0.75,0.5) -- (0.75,-0.5) -- (-0.75,-0.5);
            \draw (0,0) node {\scriptsize $C_{up}$};
            \draw (0,-1.3) node {\scriptsize $\{1\}$};
        \end{scope}
    \end{tikzpicture} 
    +
    \begin{tikzpicture}[scale=.45, baseline={([yshift=-0ex]current bounding box.center)}, thick]
        \begin{scope}[shift={(0,0)}]
            \draw (0,-1) -- (0,1);
            \filldraw[fill=atomictangerine!50] (-0.75,-0.5) -- (-0.75,0.5) -- (0.75,0.5) -- (0.75,-0.5) -- (-0.75,-0.5);
            \draw (0,0) node {\scriptsize $C_{up}$};
            \draw (0,-1.3) node {\scriptsize $\{2\}$};
        \end{scope}
    \end{tikzpicture}
    \ , 
    \quad
    \begin{tikzpicture}[scale=.45, baseline={([yshift=-0ex]current bounding box.center)}, thick]
        \begin{scope}[shift={(0,0)}]
            \draw (0,-1) -- (0,1);
            \filldraw[fill=melon!50] (-0.75,-0.5) -- (-0.75,0.5) -- (0.75,0.5) -- (0.75,-0.5) -- (-0.75,-0.5);
            \draw (0,0) node {\scriptsize $B_{up}$};
            \draw (0,-1.3) node {\scriptsize $\{1\}$};
        \end{scope}
    \end{tikzpicture}
    =
    \begin{tikzpicture}[scale=.45, baseline={([yshift=-0ex]current bounding box.center)}, thick]
        \begin{scope}[shift={(0,0)}]
            \draw (0,-1) -- (0,1);
            \filldraw[fill=atomictangerine!50] (-0.75,-0.5) -- (-0.75,0.5) -- (0.75,0.5) -- (0.75,-0.5) -- (-0.75,-0.5);
            \draw (0,0) node {\scriptsize $C_{up}$};
            \draw (0,-1.3) node {\scriptsize $\{1\}$};
        \end{scope}
    \end{tikzpicture} 
    \ ,
    \quad
    \begin{tikzpicture}[scale=.45, baseline={([yshift=-0ex]current bounding box.center)}, thick]
        \begin{scope}[shift={(0,0)}]
            \draw (0,-1) -- (0,1);
            \filldraw[fill=melon!50] (-0.75,-0.5) -- (-0.75,0.5) -- (0.75,0.5) -- (0.75,-0.5) -- (-0.75,-0.5);
            \draw (0,0) node {\scriptsize $B_{up}$};
            \draw (0,-1.3) node {\scriptsize $\{2\}$};
        \end{scope}
    \end{tikzpicture}
    =
    \begin{tikzpicture}[scale=.45, baseline={([yshift=-0ex]current bounding box.center)}, thick]
        \begin{scope}[shift={(0,0)}]
            \draw (0,-1) -- (0,1);
            \filldraw[fill=atomictangerine!50] (-0.75,-0.5) -- (-0.75,0.5) -- (0.75,0.5) -- (0.75,-0.5) -- (-0.75,-0.5);
            \draw (0,0) node {\scriptsize $C_{up}$};
            \draw (0,-1.3) node {\scriptsize $\{2\}$};
        \end{scope}
    \end{tikzpicture} 
    \ .
\end{equation}
However, there is no free block in $C$ that satisfies Eq. \eqref{eq:negligible_part}, and hence $A$ does not have a negligible part with respect to $B$.

Finally, any two equivalent uniform PBC MPS automatically span the same physical subspace, in contrast to the general MPS-X case where a projection may be required. This is shown in the following lemma.
\begin{lemma} \label{lemma:pbc-mps_same_physical_subspace}
    The tensors of any two equivalent uniform PBC MPS span the same physical subspace, and therefore have no negligible part with respect to each other.
\end{lemma}
\begin{proof}
    For uniform PBC MPS, we take $X = \mathds{1}$ and the canonical decomposition of the tensor $A^i$ reads \cite[Eq. (20b)]{cirac_2017_mpdo}
    \begin{equation*}
        A^i = \bigoplus_{j\in \Sigma_\infty} \bigoplus_{k=1}^{r_j} \mu_{j,k} X_{j,k} A_j^i X_{j,k}^{-1} \ ,
    \end{equation*}
    where $\{A_j\}_{j\in \Sigma_\infty}$ is a basis of normal tensors (BNT) for $A^i$. The associated state is therefore
    \begin{equation*}
        \ket{\psi_N(\mathds{1},A)} = \sum_{j\in \Sigma_\infty} \left( \sum_{k=1}^{r_j} \mu_{j,k}^N \right) \ket{\psi_N(\mathds{1},A_j)}.
    \end{equation*}
    The fundamental theorem for PBC MPS \cite[Thm. 2.10]{cirac_2017_mpdo} tells us that, if two tensors $A$ and $B$ decomposed in terms of a BNT $\{A_j\}, \{B_j\}$  generate proportional states for all $N$, then (after possibly relabelling their BNT), each BNT element satisfies
    \begin{equation*}
        B_j^i = e^{i\phi_j} X_j A_j^{-1} X_j^{-1}
    \end{equation*}
    for some phases $\phi_j$ and invertible matrices $X_j$. Neither the phase nor the gauge affect the physical subspace spanned by the corresponding tensors, and hence $V_{B_j} = V_{A_j}$ for all $j \in \Sigma_\infty$. Thus, 
    \begin{equation*}
        V_A = \bigoplus_{j \in \Sigma_\infty^A} V_{A_j} = \bigoplus_{j \in \Sigma_\infty^B} V_{B_j} = V_B ,
    \end{equation*}
    so equivalent uniform PBC MPS necessarily span the same physical subspace. By Lemma \ref{lemma:freedom1}, this implies that neither $A$ nor $B$ have a negligible part with respect to each other.
\end{proof}

Together with Lemmas \ref{lemma:pbc-mps_no_negligible_blocks} and \ref{lemma:pbc-mps_no_additive_freedom}, Lemma \ref{lemma:pbc-mps_same_physical_subspace} completes the proof of Lemma \ref{lemma:pbc-mps_always_reduced}: any pair of equivalent uniform PBC MPS is already reduced, with no additional transformations required.

\subsection{Freedom in the matrix-CF} \label{app:sec_matrix-CF_freedom}

We now use the notions and technical tools introduced in the previous subsections to establish how the upper MPO of the matrix-CF associated with two equivalent MPS-X representations are related. This result will then serve as the starting point for analyzing the freedom in the gCF representation in the following subsection.
\freedommatrixCF*
\begin{proof}
Our goal is to express $B_{\text{up}}$ in terms of $A_{\text{up}}$. 
Let $C^i$ denote the stacked tensor with matrix-CF components $C_{\text{low}}, C_{\text{up}}$ obtained from Lemma \ref{lemma:stack-trick}, so that both $A_{\text{up}}$ and $B_{\text{up}}$ can be written in terms of $C_{\text{up}}$ as in Eqs. \eqref{eq:choose_tensor_C} and \eqref{eq:B_in_terms_of_C}.

Since the pair of MPS-X under consideration is reduced, by Lemma \ref{lemma:freedom1} we have that neither $A$ nor $B$ have a negligible part with respect to each other. That is, for each $t \in \Sigma^C$,
\begin{equation*}
    C_{\text{low}}^t|_{\text{A-part}} = A_{\text{low}}^t
    \quad \text{and} \quad 
    C_{\text{low}}^t|_{\text{B-part}} \neq 0,
\end{equation*}
which implies that there is no $t \in \Sigma^C$ that is solely supported on the $B$-part of tensor $C$, i.e. with $ C_{\text{low}}^t|_{\text{A-part}} = 0$ and $ C_{\text{low}}^t|_{\text{B-part}} \neq 0$. 
This means that all blocks appearing in the $B$-part must be proportional to, or linear combinations of, blocks in the $A$-part.  
But by construction, we chose the matrix-CF of $C$ to coincide with that of $A$ (without loss of generality), so we can replace $C_{\text{up}}$ with $A_{\text{up}}$ in Eq. \eqref{eq:B_in_terms_of_C}. 
This yields
\begin{equation*} 
        \begin{tikzpicture}[scale=.45, baseline={([yshift=0.4ex]current bounding box.center)}, thick]
            \begin{scope}[shift={(0,0)}]
                \draw (-1.2,0) -- (1.2,0);
                \draw (0,1) -- (0,-1);
                \filldraw[fill=melon!50] (-1/2-0.2,-1/2) -- (-1/2-0.2,1/2) -- (1/2+0.2,1/2) -- (1/2+0.2,-1/2) -- (-1/2-0.2,-1/2);
                \draw (0,0) node {\scriptsize $B_{\text{up}}$};
            \end{scope}
            \draw (0,-1.3) node {\scriptsize $t$};
        \end{tikzpicture} =
            \begin{tikzpicture}[scale=.45, baseline={([yshift=0.4ex]current bounding box.center)}, thick]
            \begin{scope}[shift={(0,0)}]
                \draw (-1.2,0) -- (1.2,0);
                \draw (0,1) -- (0,-1);
                \filldraw[fill=powderblue!50] (-1/2-0.2,-1/2) -- (-1/2-0.2,1/2) -- (1/2+0.2,1/2) -- (1/2+0.2,-1/2) -- (-1/2-0.2,-1/2);
                \draw (0,0) node {\scriptsize $A_{\text{up}}$};
            \end{scope}
            \begin{scope}[shift={(1.7,0)}]
                \draw (-1,0) -- (1,0);
                \filldraw[fill=gray!10] (0.5,-0.5) -- (-0.5,-0.5) -- (-0.5,0.5) -- (0.5, 0.5) -- (0.5,-0.5);
                \draw (0,0) node {\scriptsize $Z$};
                \draw (0,-0.5) -- (0,-1);
                \draw (0,-1.3) node {\scriptsize $\pi(j)$};
            \end{scope}
            \begin{scope}[shift={(-2.2,0)}]
                \draw (-1,0) -- (1.5,0);
                \draw (0.25,-1) -- (0.25,-0.5);
                \filldraw[fill=gray!10] (1,-0.5) -- (-0.5,-0.5) -- (-0.5,0.5) -- (1, 0.5) -- (1,-0.5);
                \draw (0.3,0) node {\scriptsize $Z^{-1}$};
                \draw (0.25,-1.3) node {\scriptsize $\pi(i)$};
            \end{scope}
            \begin{scope}[shift={(0,-1.5)}]
                \filldraw[fill=gray!10] (0.5,-0.5) -- (-0.5,-0.5) -- (-0.5,0.5) -- (0.5, 0.5) -- (0.5,-0.5);
                \draw (0,0) node {\scriptsize $P_B$};
                \draw (0,-0.5) -- (0,-1);
                \draw (0,-1.3) node {\scriptsize $t$};
            \end{scope}
        \end{tikzpicture}
        \quad \text{for } t \in \tilde{\Sigma}_f^{B,ij},  
    \end{equation*}
    where $\alpha_{s} \in \mathbb{C}$ and $P_B : \langle\Sigma^B\rangle \to \langle\Sigma^A\rangle$ acts as 
    \begin{equation*}
        \begin{cases}
            P_B \ket{x} = \alpha_{\pi(x)} \ket{\pi(x)} & \text{if } x \in \Sigma_\infty^B, \\
            P_B ( \langle \Sigma_f^{B,ij} \rangle ) \subseteq \langle \tilde{\Sigma}_f^{A, \pi(i)\pi(j)}\rangle  & \text{for } i,j \in \tilde{\Sigma}_\infty^B.
        \end{cases}
    \end{equation*}

Finally, let us verify how Eqs. \eqref{eq:final_relation_Bup=Aup} and \eqref{prop:P_B-properties-old} follow. Since the pair of MPS-X is reduced, we have $|\Sigma_\infty^A| = |\Sigma_\infty^B|$. Hence we may identify and replace both with a common alphabet $\Sigma_\infty$. By re-indexing the blocks in $B$ so that $\pi = \mathds{1}$ (which we may do without loss of generality), it follows that $|\Sigma_f^{A,ij}| = |\Sigma_f^{B,ij}|$, and hence we can also identify them with common alphabets $\Sigma_f^{ij}$. 

Under these identifications, the relations stated in 
Eqs. \eqref{eq:final_relation_Bup=Aup} and \eqref{prop:P_B-properties-old} hold. Recall that the functions $r^1$ and $r^2$ in Eq. \eqref{eq:final_relation_Bup=Aup} assign to each free block $t \in \Sigma$ its sector $[r_t^1, r_t^2]$, equivalently expressed as $t \in \Sigma_f^{r_t^1 r_t^2}$. This completes the proof.
\end{proof}

\subsection{Freedom in the gCF} \label{app:sec_gCF_freedom}

Finally, we bring together the results obtained so far to establish a relation between the gCF elements of two MPS-X representations, covering not only the MPO in the upper part, already addressed in Proposition \ref{prop:freedom_upper_matrixCF}, but also the algebraic RLS in the lower part. Up to this point, our results applied to arbitrary MPS-X, regardless of stability. We now impose stability, which ensures that the MPS-X can indeed be expressed in gCF with an algebraic RLS family of states in the lower part. In this setting, we show that the freedom is completely determined by the orbits of the algebraic RLS components $\ket{X_O}$ under a specific subset of TI SLOCC operations.

\propfreedomgCF*
\begin{proof} 
The implication ``$\Longleftarrow$'' is immediate: if the upper tensors and the algebraic RLS $\{\ket{X_O^A}\},\{\ket{X_O^B}\}$ are related as in Eq. \eqref{eq:freedom_gCF_relation_X_O}, then the two MPS-X generate the same family of states for all system sizes. We therefore focus on the converse direction ``$\Longrightarrow$''.

The relation between $B_{\text{up}}$ and $A_{\text{up}}$ follows directly from Proposition \ref{prop:freedom_upper_matrixCF}, together with the assumptions that each gCF representation contains no negligible blocks and that both tensor families span the same physical subspace. These conditions ensure that they are a pair of equivalent reduced MPS-X representations.
Our remaining task is to establish the corresponding relation between the sets $\{\ket{X_O^A}\}$ and $\{\ket{X_O^B}\}$ appearing in each gCF.

First, apply the left-inverse of the block-injective MPO $A_{\mathrm{up}}$ onto the identity $\ket{\psi_N(X_A, A^i)} = \ket{\psi_N(X_B, B^i)}$ for each $N$. Using the assumption that $B_{\text{up}}$ is expressed in terms of $A_{\text{up}}$ as in Eq. \eqref{eq:freedom_gCF_relation_X_O}, we obtain for every $N$ the identity
\begin{align}
    &\sum_{m \leq M_A} \sum_{O \in \tilde{\Sigma}_\infty^{m+1}} 
    \sum_{\substack{\sum_{i=1}^m n_i = N - m \\ n_i \geq 0}}
    \hat{S}^{(m)} \ket{O_0^{n_0} f O_1^{n_1} f \dots f O_m^{n_m}} \ket{X_O^A} \label{eq:general_LHS=RHS}
    \\ &\quad \quad 
    =\sum_{m \leq M_B} \sum_{O \in \tilde{\Sigma}_\infty^{m+1}} 
    \sum_{\substack{\sum_{i=1}^m n_i = N - m \\ n_i \geq 0}}
    \hat{S}^{(m)} \ket{O_0^{n_0} f O_1^{n_1} f \dots f O_m^{n_m}} \left( \alpha_{O_0}^{n_0} \alpha_{O_1}^{n_1} \dots \alpha_{O_m}^{n_m} P_B^{\otimes m} \ket{X_O^B} \right). \nonumber
\end{align}
From this equation it follows that $M_A \leq M_B$: otherwise, the non-vanishing terms on the LHS with $m > M_B$ would be orthogonal to the RHS. By symmetry, we conclude that $M_A = M_B := M$. 

Nevertheless, Eq. \eqref{eq:general_LHS=RHS} alone does not yet guarantee that 
\begin{equation} \label{eq:auxiliar_desired_freedom}
    \ket{X_O^A} = P_B^{\otimes m} \ket{X_O^B}, \quad \forall O \in \tilde{\Sigma}_\infty^{m+1}.
\end{equation} 
To enforce this stronger relation, we can either \textit{(i)} apply suitable projectors that isolate each term on both the LHS and the RHS, or \textit{(ii)} exploit the algebraic RLS structure of $\ket{X_O^A}$ and $\ket{X_O^B}$.

To ease the notation, we begin by assuming that whenever $\ket{X_O} \neq 0$ for $O = O_0 O_1 \dots O_m$, then all symbols satisfy $O_i \neq \varepsilon$. This restriction will be lifted later.

Let us first pursue approach \textit{(i)}. Fix $N = m \leq M$ and project both sides of Eq. \eqref{eq:general_LHS=RHS} with
\begin{equation*}
    \mathbb{P}_f^{O_0 O_1} \otimes \mathbb{P}_f^{O_1 O_2} \otimes \dots \otimes \mathbb{P}_f^{O_{m-1}O_m},
\end{equation*}
which single out the subspace $\langle\Sigma_f^{O_0 O_1} \rangle \otimes \langle\Sigma_f^{O_1 O_2} \rangle \otimes \dots \otimes \langle\Sigma_f^{O_{m-1} O_m} \rangle$. This yields 
\begin{equation} \label{eq:freedom_wrong_condition}
    \ket{X_O^A} = \bigotimes_{i=0}^{m-1} \left(  \mathbb{P}_{\Sigma_f^{O_i O_{i+1}}} P_B \right) \ket{X_O^B}
\end{equation}
for every $O \in \Sigma_\infty^{m+1}$. This equation is, however, weaker than the desired equation in Eq. \eqref{eq:auxiliar_desired_freedom}); in fact, it would not suffice to establish the ``$\Longleftarrow$'' direction of the statement. But we can still pose the following question: when does Eq. \eqref{eq:freedom_wrong_condition} coincide with Eq. \eqref{eq:auxiliar_desired_freedom}? Equivalently, when does it hold that 
\begin{equation} \label{eq:desired_wrong_freedom_becomes_right}
    \ket{X_O^A} = \bigotimes_{i=0}^{m-1} \left(  \mathbb{P}_{\Sigma_f^{O_i O_{i+1}}} P_B \right) \ket{X_O^B} = P_B^{\otimes m} \ket{X_O^B} \ \ ?
\end{equation}
Recall that Proposition \ref{prop:freedom_upper_matrixCF} guarantees that
\begin{equation*}
    P_B(\langle \Sigma_f^{ij} \rangle) \subseteq \langle \tilde{\Sigma}_f^{ij} \rangle \ , 
    \quad \text{with } \ 
    \tilde{\Sigma}_f^{ij} = 
    \begin{cases}
        \Sigma_f^{ii} \cup \{i\} &\text{if } i = j, \\
        \Sigma_f^{ij} &\text{if } i \neq j.
    \end{cases}
\end{equation*}
Thus, a sufficient condition for Eq. \eqref{eq:desired_wrong_freedom_becomes_right} to hold is that all non-zero $\ket{X_O}$ occur only for $O \in \Sigma_\infty^{m+1}$ with $O_i \neq O_{i+1}$ for every $i$, since in this case $P_B(\langle \Sigma_f^{ij} \rangle) \subseteq \langle \Sigma_f^{ij} \rangle$.

In general, however, this condition is not satisfied. To circumvent this, we turn to approach \textit{(ii)}, exploiting the algebraic RLS structure of $\{\ket{X_O^A}\}$ and $\{\ket{X_O^B}\}$ to show that the desired relation holds in full generality.

To illustrate the proof strategy, let us consider $\Sigma_\infty = \{0\}$. Then, Eq. \eqref{eq:general_LHS=RHS} becomes
\begin{align}
    &\sum_{m \leq M} \sum_{\substack{n_i \geq 0 \\ \sum_{i=0}^m n_i = N-m}} \hat{S}^{(m)} \ket{0^{n_0} f 0^{n_1} f \dots f 0^{n_m}} \ket{X_{0^{m+1}}^A} \label{eq:general_LHS=RHS_illustrative-example}
    \\ &\quad \quad 
    =\sum_{m \leq M} \sum_{\substack{n_i \geq 0 \\ \sum_{i=0}^m n_i = N - m}}
    \alpha_{0}^{N-m} \hat{S}^{(m)} \ket{0^{n_0} f 0^{n_1} f \dots f 0^{n_m}} \left(  P_B^{\otimes m} \ket{X_{0^{m+1}}^B} \right), \nonumber
\end{align}
Multiplying this equation by $\bra{0}^{\otimes N}$ on both sides yields
\begin{equation} \label{eq:illustrative-key1}
    X_0^A = \sum_{m \leq M} \alpha_0^{N-m} {N \choose m} (\bra{0} P_B)^{\otimes m} \ket{X_{0^{m+1}}^B} \ , \forall N \geq M ,
\end{equation}
where the term ${N \choose m}$ in the RHS reflects the number of ways to place the $m$ separators $f$ among the $N$ sites of Eq. \eqref{eq:general_LHS=RHS_illustrative-example}. Let us rewrite it as
\begin{equation*}\label{eq:Qdef}
    X_0^A \alpha_0^{-N} = Q(N)\ , \quad \text{where } 
    Q(N) := \sum_{m=0}^M (\bra{0} P_B)^{\otimes m} \ket{X_{0^{m+1}}^B} \alpha_0^{-m} \binom{N}{m} \ .
\end{equation*}
Observe that $Q(N)$ is a polynomial in $N$ of degree at most $M$, while the LHS scales exponentially with $N$. As long as the equation does not trivialize to zero, it forces $\alpha_0 = 1$ and $Q(N)$ to be a constant polynomial. Due to the fact that each term ${N \choose m}$ in $Q(N)$ is a polynomial in $N$ of degree $m$ with leading coefficient $1/m!$, the coefficients corresponding to ${N \choose m}$ must vanish for all $m \geq 1$, which necessarily implies $(\bra{0} P_B)^{\otimes m} \ket{X_{0^{m+1}}^B} = 0$ for all $m \geq 1$, and $X_0^B = X_0^A$.

We now multiply Eq. \eqref{eq:general_LHS=RHS_illustrative-example} by $\bra{0}^{\otimes p_0} \otimes \bra{y_1} \otimes \bra{0}^{\otimes p_1}$, denoted as $\bra{0^{p_0} y_1 0^{p_1}}$, for some $y_1 \in \Sigma_f^{00}$ and $N = p_0 + p_1 + 1$. This yields
\begin{align*}
    \braket{y_1}{X_{00}^A}
    &= 
    \sum_{1 \leq m \leq M}  \sum_{\substack{r_0+r_1=m-1 \\ r_0, r_1 \geq 0}} \alpha_0^{p_0+p_1+1-m}
    {p_0 \choose r_0} {p_1 \choose r_1} \mel{0^{r_0} y_1 0^{r_1}}{P_B^{\otimes m}}{X_{0^{m+1}}^B}
\end{align*}
Similarly as before, and using the fact that $\alpha_0 = 1$, we have
\begin{equation*}
    \braket{y_1}{X_{00}^A} \alpha_0^{-(p_0+p_1)}
    = Q(p_0, p_1), 
    \quad
    \text{where }
    Q(p_0, p_1) := 
    \sum_{1 \leq m \leq M}  \sum_{\substack{r_0+r_1=m-1 \\ r_0, r_1 \geq 0}} 
    \alpha_0^{1-m}
    d_{r_0, r_1, m}
    {p_0 \choose r_0} {p_1 \choose r_1} 
\end{equation*}
and $d_{r_0, r_1, m} := \mel{0^{r_0} y_1 0^{r_1}}{P_B^{\otimes m}}{X_{0^{m+1}}^B}$. Since $Q(p_0,p_1)$ is a multivariate polynomial in $p_0, p_1$, while the LHS scales exponentially with $p_0,p_1$, as long as the equation does not trivialize to zero, it follows that $\alpha_0 = 1$ and all coefficients $d_{r_0, r_1, m}$ must vanish except for $d_{0,0,1}$. Hence,
\begin{equation*}
    \braket{y_1}{X_{00}^A} = d_{0,0,1} = \mel{y_1}{P_B}{X_{00}^B} .
\end{equation*}

More generally, multiplying Eq. \eqref{eq:general_LHS=RHS_illustrative-example} with a bra of the form $\bra{0^{p_0} y_1 0^{p_1} y_2 \dots y_\gamma 0^{p_\gamma}}$ where $y_i \in \Sigma_f^{00}$, for any choice of $\gamma$ and $p_i$ satisfying $\gamma + \sum_i p_i = N$, we obtain the identity
\begin{equation*}
    \braket{y_1 y_2 \dots y_\gamma}{X_{0^{\gamma + 1}}^A} = 
    \sum_{\gamma \leq m \leq M}
    \sum_{\substack{0 \leq r_i \leq p_i \\ \sum_{i=0}^\gamma r_i = m-\gamma}}
    \alpha_0^{\sum_i p_i + \gamma - m}
    \prod_{j=0}^m
    {p_j \choose r_j} 
    \mel{0^{r_0} y_1 0^{r_1} y_2 \dots y_\gamma 0^{r_\gamma}}{P_B^{\otimes m}}{X_{0^{m+1}}^B}
\end{equation*}
Repeating the same argument as before, it follows that $\alpha_0 = 1$ and all terms vanish except for the constant one, leading to
\begin{equation*}
    \begin{cases}
        \mel{0^{r_0} y_1 0^{r_1} y_2 \dots y_\gamma 0^{r_\gamma}}{P_B^{\otimes m}}{X_{0^{m+1}}^B} = 0, 
        \quad
        \text{if } \sum_i r_i > 0 , \\
        \mel{y_1 y_2 \dots y_m}{P_B^{\otimes m}}{X_{0^{m+1}}^B} = \braket{y_1 y_2 \dots y_m}{X_{0^{m+1}}^A} , 
    \end{cases}
\end{equation*}
for all $y_1, \dots, y_m \in \Sigma_f^{00}$.

The same reasoning applies for arbitrary $\Sigma_\infty$, multiplying Eq. \eqref{eq:general_LHS=RHS} by bras of the form 
$$\bra{O_0^{p_0} y_1 O_1^{p_1} y_2 O_2^{p_2} \dots y_\gamma O_{\gamma}^{p_\gamma}},$$ 
with $O_i \in \Sigma_\infty$, $y_i \in \Sigma_f^{O_{i-1} O_i}$. We would express the identity including the right combinatorial prefactors, and this would lead to an equation that holds for any $p_0, p_1, \dots, p_\gamma$, implying again that 
\begin{equation*}
    \begin{cases}
        \mel{O_0^{r_0} y_1 O_1^{r_1} y_2 O_2^{r_2} \dots y_\gamma O_\gamma^{r_\gamma}}{P_B^{\otimes m}}{X_O^B} = 0, 
        \quad
        \text{if } \sum_i r_i > 0 , \\
        \mel{y_1 y_2 \dots y_m}{P_B^{\otimes m}}{X_O^B} = \braket{y_1 y_2 \dots y_m}{X_O^A} , 
    \end{cases}
\end{equation*}
for all $y_1, \dots, y_m \in \Sigma_f^{O_{i-1} O_i}$. Equivalently,
\begin{equation*}
    P_B^{\otimes m} \ket{X_O} \subseteq \bigotimes_{i=1}^m 
    \langle \Sigma_f^{O_{i-1}O_i} \rangle ,
    \quad
    \forall O = O_0 O_1 \dots O_m \in \Sigma_\infty^{m+1},
\end{equation*}
and thus Eq. \eqref{eq:desired_wrong_freedom_becomes_right} holds as desired.

We now extend the conclusion to general $O \in \tilde{\Sigma}_\infty^{m+1}$ possibly containing one or more $\varepsilon$ entries. We proceed by induction on the number $k$ of indices with $O_i = \varepsilon$, the goal being to prove $\ket{X_O^A} = P_B^{\otimes m} \ket{X_O^B}$ for all $O \in \tilde{\Sigma}_\infty^{m+1}$.

For the base case $k = 0$, one can proceed exactly as above. Indeed, when multiplying Eq. \eqref{eq:general_LHS=RHS} by a bra of the form
\begin{equation*}
    \bra{O_0^{p_0} y_1 O_1^{p_1} y_2 \dots y_\gamma O_\gamma^{p_\gamma}}, 
    \quad \text{with } p_i \geq 1, \ O_i \in \Sigma_\infty, \ y_i \in \Sigma_f^{O_{i-1}O_i},
\end{equation*}
the LHS still yields  $\braket{y_1 \dots y_\gamma}{X_O^A}$, since $\Sigma_f^{\varepsilon \ast}$ and $\Sigma_f^{\ast \varepsilon}$ are disjoint from $\tilde{\Sigma}_f^{ij}$ for any $i,j \in \Sigma_\infty$. On the RHS, nothing changes either, because we already know that 
\begin{equation*}
    P_B(\langle \Sigma_f^{\varepsilon \ast} \rangle) \subseteq \langle \tilde{\Sigma}_f^{\varepsilon \ast} \rangle = \langle \Sigma_f^{\varepsilon \ast} \rangle, 
    \quad 
    P_B(\langle \Sigma_f^{\ast \varepsilon} \rangle) \subseteq \langle \Sigma_f^{\ast \varepsilon } \rangle 
\end{equation*}
by the span-structure properties of Theorem \ref{prop:structure_span}. Hence the conclusion still holds for $k = 0$.

Assume now that the claim holds for all $O$ with fewer than $k$ occurrences of $\varepsilon$. Consider Eq. \eqref{eq:general_LHS=RHS}, and subtract from both sides all contributions corresponding to such $O$, which are equal by the induction hypothesis. The remaining terms thus correspond only to words with at least $k$ $\varepsilon$'s.

We now isolate the contributions with exactly $k$ occurrences of $\varepsilon$ in $O$, such that $O_i = \varepsilon$ for all $i$ at some fixed set of positions $I = \{i_1, \dots, i_k\}$. To this end, it is enough to multiply the equation by the bra $\bra{O_0^{p_0} y_1 O_1^{p_1} y_2 \dots y_\gamma O_\gamma^{p_\gamma}}$ enforcing $p_j > 0$ for $j \notin I$, and $p_j = 0$ for $j \in I$. This yields an identity in terms of the variables $\{p_j\}$ to which the same reasoning as before can be applied.

Hence, we conclude that $\ket{X_O^A} = P_B^{\otimes m} \ket{X_O^B}$ for all $O \in \tilde{\Sigma}_\infty^{m+1}$, as claimed.
\end{proof}

\noindent To illustrate Proposition \ref{prop:freedom_gCF}, consider the following representative cases of algebraic RLS:
\begin{enumerate}
    \item \textit{Block-diagonal MPS with a nontrivial boundary} with $\{\ket{L_N}\} = \sum_{j\in \Sigma} \alpha_j \ket{j^*}$ for some $\alpha_j \in \mathbb{C}$, for which $\Sigma = \Sigma_\infty$;
    \item \textit{W-like MPS-X} with backbone family $\{\ket{L_N}\} = \ket{0^* 1 0^*}$, satisfying $|\Sigma_f^{00}| = 1$;
    \item \textit{Dicke-like states:} with $\{\ket{L_N}\} = \ket{0^* (10^*)^k}$ for some $k \in \mathbb{N}$, also with $|\Sigma_f^{00}| = 1$;
    \item \textit{Domain-wall-like MPS-X:} with $\{\ket{L_N}\} = \ket{0^* 1 2^*}$, for which $|\Sigma_f^{02}| = |\Sigma_f^{20}| = 1$.
\end{enumerate}
In each case, Proposition \ref{prop:freedom_gCF} guarantees that any alternative gCF representation must share the same algebraic RLS as the backbone family ${\ket{L_N}}$, with the MPS tensors differing only by a gauge transformation.

The Dicke-like case illustrates an important subtlety. Although $\ket{0^* 1 0^* 1 0^*}$ is not strictly a valid backbone for the gCF since it is not closed under blocking (i.e. no blocking tensor $\Gamma$ exists to ensure $\Gamma$-invariance), the proof of Proposition \ref{prop:freedom_gCF} only uses the algebraic RLS structure, not the $\Gamma$-invariance.
Consequently, as long as the Dicke-like state is initially in matrix-CF (as introduced in Def. \ref{def:matrix-CF}), the proposition applies: the gCF of any alternative and reduced MPS-X representation must exhibit the same backbone family of states.

\newpage

\section{Proofs for the generalized quantum Wielandt's inequality}
\label{app:Wielandt}

In this section, we establish the precise conditions under which a set of matrices is stable, thereby deepening the connection between the algebra they generate and their length-$\ell$ span. These lead to the generalized quantum Wielandt's inequality of Theorem \ref{prop:stability_criterion} in the main text. For clarity, we formulate the definition of stable sets of matrices below, analogous to the one of stable MPS-X (Def. \ref{def:stable_mps-x}).
\begin{restatable}[Stable set of matrices]{definition}{defstableMPStensor} \label{def:stable_MPS_tensor}
    A set of matrices $\{A^i\}$ is \textit{stable} if the span eventually stabilizes to an algebra upon blocking, i.e. if $\exists L_{\mathrm{stab}}$ such that $\mathcal{A}^{(L_{\mathrm{stab}})} = \mathrm{Alg}(\mathcal{A}^{(1)})$.
\end{restatable}

We begin by proving the stability characterization for semisimple sets of matrices, which are those relevant to the standard theory of uniform MPS with PBC. We then extend the result to the more general case including non-semisimple sets of matrices.

\subsection{Stability in the uniform PBC MPS case} \label{app:sec:Wielandt-PBC}

We call a set of matrices $\{A^i\}$ \textit{semisimple} if the algebra they generate is semisimple. Equivalently, upon a suitable change of basis, we have
\begin{equation*}
    \mathrm{Alg}(\mathcal{A}^{(1)}) = 
    \bigoplus_{j=1}^{\tilde{g}} \left( 
    \mathds{1}_{\tilde{d}_j} \otimes \mathcal{M}_{D_j \times D_j} (\mathbb{C}) 
    \right) \oplus \mathbf{0}_{d_0}.
\end{equation*}
If we block-upper-triangularize $\{A^i\}$ such that the diagonal blocks are \textit{irreducible} (i.e. they admit no proper invariant subspace), then semisimplicity of $\{A^i\}$ is equivalent to the resulting form being strictly block-diagonal. By Burnside's theorem, the irreducibility of each of the diagonal blocks means that they generate the full matrix algebra, that is, $\mathrm{Alg}(\mathcal{A}_{jj}^{(1)}) = \mathcal{M}_{D_j\times D_j}(\mathbb{C})$ \cite{burnside_1905_condition, Lomonosov_2004_Burnsides-thm}. 

Not every set of semisimple matrices is stable. For example, consider the following one-dimensional set spanned by a single diagonal matrix:
\begin{equation*}
    \mathcal{A}^{(1)} = \mathrm{span} \left\{ 
    {\scriptsize \begin{pmatrix}
        1 & 0 \\ 0 & e^{i \sqrt{2\pi}}
    \end{pmatrix}}
    \right\}.
\end{equation*}
For any $\ell \in \mathbb{N}$, its length-$\ell$ span is again one-dimensional, while the algebra it generates is two-dimensional. Explicitly,
\begin{equation*}
    \mathcal{A}^{(\ell)} = \mathrm{span} \left\{ 
    {\scriptsize \begin{pmatrix}
        1 & 0 \\ 0 & e^{i \ell \sqrt{2\pi}}
    \end{pmatrix}}
    \right\}, \quad \mathrm{Alg}(\mathcal{A}^{(1)}) = \mathrm{Alg}(\mathcal{A}^{(\ell)}) = 
        \left\{
        {\scriptsize\begin{pmatrix}
            a & 0 \\ 0 & b
        \end{pmatrix}} \mid a, b \in \mathbb{C}
        \right\}
\end{equation*}
Therefore, no finite $L_{\mathrm{stab}}$ exists for which condition \textit{(i)} of the stability definition is satisfied.

However, a non-stable set of matrices can become stable after blocking. Consider, for instance,
\begin{equation*}
    \mathcal{A}^{(1)} = \mathrm{span} \left\{ 
    {\scriptsize \begin{pmatrix}
        1 & 0 \\ 0 & i
    \end{pmatrix}}
    \right\} \to \mathrm{Alg}(\mathcal{A}^{(1)}) = 
    \left\{
    {\scriptsize\begin{pmatrix}
        a & 0 \\ 0 & b
    \end{pmatrix}} \mid a, b \in \mathbb{C}
    \right\} \neq \mathcal{A}^{(\ell)} \text{ for any } \ell,
\end{equation*}
which implies that $\mathcal{A}^{(1)}$ is not stable. Yet, after blocking every four sites, $\mathcal{A}^{(4)}$ becomes stable because
\begin{equation*}
    \mathcal{A}^{(4\ell)} = \mathrm{span} \left\{
    {\scriptsize\begin{pmatrix}
        1 & 0 \\ 0 & 1
    \end{pmatrix}}
    \right\} = \mathrm{Alg}(\mathcal{A}^{(4)}) \ .
\end{equation*}

To proceed to formalize the conditions for stability, let us first recall some notions. An \textit{invariant subspace} of $\{A^i\}$ is a linear subspace $V \subseteq \mathbb{C}^D$ such that $\mathcal{A}^{(1)} V = V$. A \textit{periodic subspace} of $\mathcal{A}^{(1)}$ is a subspace $W \subseteq \mathbb{C}^D$ such that, defining $W_{i+1} := \mathcal{A}^{(1)} W_i$ with $W_0 := W$, there exists a smallest integer $p \geq 1$ for which $W_p = W_0$. We refer to $p$ as the \textit{period} of $W$ under $\mathcal{A}^{(1)}$. In this case, each $W_0,\dots,W_{p-1}$ is an invariant subspace of $\mathcal{A}^{(p)}$.

Moreover, the theory of PBC MPS tells us that, after blocking and in an appropriate basis, the span takes the form
\begin{equation} \label{eq:decomposition_span}
    \mathcal{A}^{(p\ell)} = 
    \bigoplus_{j=1}^g \left( 
    (\Lambda_j)^{p\ell} \otimes \mathcal{M}_{D_j \times D_j} (\mathbb{C})
    \right) \oplus \mathbf{0}_{d_0}, \quad \forall \ell \geq L_{BI}
\end{equation}
where each $\Lambda_j$ is a diagonal matrix of size $d_j \times d_j$ with non-zero entries, $L_{BI}$ is the block-injectivity length already introduced in the main text and upper bounded as $L_{BI} \leq 3D^5$, and $p$ is the least common multiple of the periods of all the periodic subspaces of $\mathcal{A}^{(1)}$. We also let $q$ denote the additional amount of blocking that needs to be done such that $(\Lambda_j)^{pq} = \mathds{1}_{d_j}$ for all $j$; if such $q$ does not exist, we set $q = \infty$.

\begin{proposition}[Stability of semisimple sets of matrices] \label{prop:stability_semisimple}
    Given a semisimple set of matrices $\{A^i\}$, it is stable (i.e. $\exists L_{\mathrm{stab}}$ such that $\mathcal{A}^{(L_{\mathrm{stab}})} = \mathrm{Alg}(\mathcal{A}^{(L_{\mathrm{stab}})}) = \mathrm{Alg}(\mathcal{A}^{(1)})$) if and only if 
    \begin{itemize}
        \item[(i)] $\mathcal{A}^{(1)}$ has no periodic subspaces, and
        \item[(ii)] the proportionality constants between equivalent diagonal blocks in $\mathcal{A}^{(1)}$ are all equal to 1.
    \end{itemize}
    In particular, it holds that $\mathcal{A}^{(L_{BI} + \alpha)} = \mathrm{Alg}(\mathcal{A}^{(1+\beta)})$ for any $\alpha, \beta \geq 0$. 
    
    More generally, as long as $q < \infty$, $\mathcal{A}^{(pq)}$ is stable with $L_{\mathrm{stab}} = L_{BI}$, such that $\mathcal{A}^{(pq(L_{BI} + \alpha))} = \mathrm{Alg}(\mathcal{A}^{(pq(1+\beta))}$ for any $\alpha, \beta \geq 0$.  
\end{proposition}
\begin{proof}
    We prove the forward direction $\Rightarrow$ of the claim by contradiction.
    
    Assume that $\mathcal{A}^{(1)}$ admits a periodic subspace $W$ of period $p > 1$, and we ask whether a stability length $L_{\mathrm{stab}}$ can exist.

    If $L_{\mathrm{stab}}$ were proportional to $p$, then
    \begin{equation*}
        \mathcal{A}^{(L_{\mathrm{stab}})} W = W \implies 
        \mathrm{Alg}(\mathcal{A}^{(L_{\mathrm{stab}})}) W = W.
    \end{equation*}
    However, 
    \begin{equation*}
        \mathrm{Alg}(\mathcal{A}^{(1)}) W = 
        \mathrm{span} \left( \bigcup_{j\geq 1} \mathcal{A}^{(j)} W \right) =
        \mathrm{span} \left( \bigcup_{j=0}^p W_j \right) \supsetneq W \ ,
    \end{equation*}
    which means that $\mathrm{Alg}(\mathcal{A}^{(1)}) \neq \mathrm{Alg}(\mathcal{A}^{(L_{\mathrm{stab}})})$ and thus contradicts the stability condition.

    Next, suppose that $L_{\mathrm{stab}}$ is \textit{not} proportional to the period of any periodic subspace of $\mathcal{A}^{(1)}$. Let $W$ be a $p$-periodic subspace of $\mathcal{A}^{(1)}$, and define 
    \begin{equation*}
        \tilde{W}_{i+1} := \mathcal{A}^{(L_\mathrm{stab})} \tilde{W}_i, \quad \tilde{W}_0 := W. 
    \end{equation*}
    Let $\tilde{p}$ be the smallest integer such that $\tilde{W}_{\tilde{p}} = \tilde{W}_0 = W$. Using that $\mathcal{A}^{(p)} W = W$, we have
    \begin{equation*}
        \tilde{W}_{\tilde{p}} = \mathcal{A}^{(\tilde{p} L_\mathrm{stab})} W = \mathcal{A}^{(r)} W = W_r,
    \end{equation*}
    where $r := \tilde{p}L_{\mathrm{stab}} - \lfloor \tilde{p}L_{\mathrm{stab}}/p \rfloor p$ is the remainder of dividing $\tilde{p} L_{\mathrm{stab}}$ by $p$, meaning that $0 \leq r < p$.

    For $\tilde{W}_{\tilde{p}} = W$ to hold, we must have $r = 0$, i.e. 
    \begin{equation*}
        \tilde{p} L_{\mathrm{stab}} = kp \quad \text{for some } k \in \mathbb{Z}_{\geq 1}.
    \end{equation*}
    Since we are assuming that $L_{\mathrm{stab}}$ is not proportional to $p$, the smallest such $\tilde{p}$ is $\tilde{p} = p$. Hence, $W$ remains a $p$-periodic subspace for $\mathcal{A}^{(L_{\mathrm{stab}})}$.

    Therefore,
    \begin{equation*}
        \mathrm{Alg}(\mathcal{A}^{(L_{\mathrm{stab}})}) W = 
        \mathrm{span} \left( \bigcup_{j\geq 1} \mathcal{A}^{(j L_{\mathrm{stab}})} W \right) =
        \mathrm{span} \left( \bigcup_{j=0}^p \tilde{W}_j \right)
        =
        \mathrm{span} \left( \bigcup_{j=0}^p W_j \right) ,
    \end{equation*}
    even though $\mathcal{A}^{(L_{\mathrm{stab}})} W = \tilde{W}_1 \subsetneq \mathrm{span} \left( \cup_{j=0}^p W_j \right)$, so $\mathcal{A}^{(L_{\mathrm{stab}})} \neq \mathrm{Alg}(\mathcal{A}^{(L_{\mathrm{stab}})})$, and $\{A^i\}$ cannot be stable.

    We conclude from the above that the existence of any periodic subspace implies non-stability. 

    Now assume that condition \textit{(i)} holds, i.e. there are no periodic subspaces, but condition \textit{(ii)} fails: at least two equivalent diagonal blocks of $\mathcal{A}^{(1)}$ are proportional with a proportionality constant $\lambda \neq 1$.

    From the decomposition in Eq. \eqref{eq:decomposition_span}, this means that for some $j$, $\Lambda_j \neq \mathds{1}_{d_j}$. Equivalently, there exist diagonal blocks $(m,m)$ and $(n,n)$ such that for all $\ell \geq 1$ and all $a \in \mathcal{A}^{(\ell)}$, it holds that $a_{nn} = \lambda^\ell a_{mm}$. Then we can choose
    \begin{align*}
        a \in \mathcal{A}^{(L_{\mathrm{stab}})}, 
            \quad &\text{with } a_{mm} = \mathds{1},\ a_{nn} = \lambda^{L_{\mathrm{stab}}}\mathds{1},\\
        b \in \mathcal{A}^{(2L_{\mathrm{stab}})}, 
            \quad &\text{with } b_{mm} = \mathds{1},\ b_{nn} = \lambda^{2 L_{\mathrm{stab}}} \mathds{1}.
    \end{align*}
    so that $(a-b) \in \mathrm{Alg}(\mathcal{A}^{(L_{\mathrm{stab}})})$, but 
    \begin{equation*}
        (a-b)_{mm} = 0, \quad (a-b)_{nn} = (\lambda^{L_{\mathrm{stab}}} - \lambda^{2L_{\mathrm{stab}}}) \mathds{1} \neq 0
    \end{equation*}
    where we used the assumption $\lambda \neq 0, 1$. Hence $a-b \notin \mathcal{A}^{(L_{\mathrm{stab}})}$, implying $\mathcal{A}^{(L_{\mathrm{stab}})} \neq \mathrm{Alg}(\mathcal{A}^{(L_{\mathrm{stab}})})$, so $\{A^i\}$ is non-stable. 
    
    This completes the proof of the $\Rightarrow$ direction.

    For the $\Leftarrow$ direction, assume that conditions \textit{(i)} and \textit{(ii)} hold. Then, for all $\ell \geq L_{BI}$, the decomposition in Eq. \eqref{eq:decomposition_span} reduces to
    \begin{equation*}
        \mathcal{A}^{(\ell)} = 
        \bigoplus_{j=1}^g \left( \mathds{1}_{d_j} \otimes \mathcal{M}_{D_j \times D_j} (\mathbb{C})
        \right) \oplus \mathbf{0}_{d_0}, \quad \forall \ell \geq L_{BI},
    \end{equation*}
    so $\mathcal{A}^{(L_{BI})} = \mathcal{A}^{(L_{BI}+s)}$ for all $s \geq 0$. Consequently, 
    \begin{equation} \label{eq:pbc_wielandt_aux1}
        \mathcal{A}^{(L_{BI} + s)} = \mathrm{Alg}(\mathcal{A}^{(L_{BI} + t)}) \ , \quad \forall s, t \geq 0.
    \end{equation}

    To identify this algebra with $\mathrm{Alg}(\mathcal{A}^{(1)})$, note that $\mathds{1} \oplus \mathbf{0}_{d_0} \in \mathcal{A}^{(\ell)}$ for all $\ell \geq L_{BI}$. Although this differs slightly from the assumption $\mathds{1} \in \mathcal{A}^{(\ell)}$ for all $\ell \geq L_{BI}$ in Lemma \ref{lemma:identity_contained_semisimple} (that we will prove later) due to the zero block, semisimplictity ensures that the proof of the lemma still holds. Thus,
    \begin{equation*}
        \mathrm{Alg}(\mathcal{A}^{(1)}) = \mathcal{A}^{(r_{\mathrm{alg}}+L_{BI}+s)}, \quad \forall s \geq 0,
    \end{equation*}
    where $r_{\mathrm{alg}} \leq D^2$. Using this together with Eq.~\eqref{eq:pbc_wielandt_aux1}, we conclude $\mathrm{Alg}(\mathcal{A}^{(1)}) = \mathrm{Alg}(\mathcal{A}^{(L_{BI})})$, as required for stability, so $L_{\mathrm{stab}} = L_{BI}$. 

    Finally, for any $\alpha \geq 1$, note that the inclusions 
    $\mathrm{Alg}(\mathcal{A}^{(\alpha L_{BI})}) \subseteq
        \mathrm{Alg}(\mathcal{A}^{(\alpha)}) \subseteq
        \mathrm{Alg}(\mathcal{A}^{(1)})
    $, together with $\mathrm{Alg}(\mathcal{A}^{(\alpha L_{BI})}) = \mathrm{Alg}(\mathcal{A}^{(L_{BI})}) = \mathrm{Alg}(\mathcal{A}^{(1)}) $, imply that $\mathrm{Alg}(\mathcal{A}^{(\alpha)}) = \mathrm{Alg}(\mathcal{A}^{(1)}) $ for each $\alpha \in \mathbb{Z}_{\geq 1}$.

    This concludes the proof of the claim.
\end{proof}

\subsection{Stability in the uniform MPS-X case} \label{app:sec_stability_assumptions_W1W2}

Let us now turn to the more general case of uniform MPS-X. A natural first question is whether the stability characterization established in Proposition \ref{prop:stability_semisimple} for semisimple sets of matrices still holds in the non-semisimple case. This turns out not to be true: the proof crucially relies on the block-diagonal (semisimple) structure, and fails once off-diagonal blocks are present.

To illustrate this, consider the following example (already presented in Sec. \ref{sec:algebra-span-thmbasis} of the main text). If we take a single matrix with a Jordan block, then
\begin{equation*}
    \mathcal{A}^{(1)} = \text{span} \left\{ {\scriptsize \begin{pmatrix}
        1 & 1 \\ 0 & 1
    \end{pmatrix}} \right\} \longrightarrow 
    \mathcal{A}^{(\ell)} = \text{span}\left\{ {\scriptsize \begin{pmatrix}
        1 & \ell \\ 0 & 1
    \end{pmatrix}} \right\}, \
    \text{Alg}(\mathcal{A}^{(\ell)}) = \left\{ {\scriptsize \begin{pmatrix}
        a & b \\ 0 & a
    \end{pmatrix}} \mid a,b \in \mathbb{C} \right\} .
\end{equation*}
Hence $\mathcal{A}^{(\ell)} \neq \mathrm{Alg}(\mathcal{A}^{(\ell)})$ for any $\ell$, so the span $\mathcal{A}^{(\ell)}$ is not stable for any $\ell$, even though it has no periodic subspaces and all proportionality constants on the diagonal blocks are equal to $1$. This shows that Proposition \ref{prop:stability_semisimple} does not extend directly to the non-semisimple case.

However, as we will prove in the rest of the appendix, the semisimple result \textit{can} be extended to non-semisimple sets as follows. Recall that $p$ is the least common multiple of the periods of all periodic subspaces of $\{A^i\}$, and $q$ is the additional amount of blocking that needs to be done such that all proportionality constants between diagonal blocks become simultaneously equal to 1 in $\mathcal{A}^{(pq)}$ (in the case that such $q$ does not exist, then $q = \infty$). Let $\mathds{1}_0$ be the identity matrix with zeros in the positions corresponding to vanishing diagonal blocks of the algebra. 
\genWielandtt*
This is our main result about the stability of matrices, which can be viewed as a generalization of the quantum Wielandt inequality \cite{Sanz2010, Michalek2018}, now applicable to arbitrary sets.

Note that blocking fewer sites than required in Proposition \ref{prop:stability_criterion} may already suffice to stabilize the span under additional assumptions. This occurs, for instance, in the semisimple case, where $\mathcal{A}^{(pq(L_{BI} + \alpha))} = \mathrm{Alg}(\mathcal{A}^{(pq(1+\beta))})$ for all $\alpha, \beta \geq 0$ (see Proposition \ref{prop:stability_semisimple}).

Before presenting the proof, we emphasize that two assumptions will be maintained throughout the next sections to ease the notation, until they are explicitly lifted in the last one. These are:
\begin{itemize}
    \item[\textit{(W1)}] $\mathcal{A}^{(1)}$ has no periodic subspaces and no proportionality constants between equivalent diagonal blocks other than $0$ or $1$ (i.e. $p = q = 1$). 
    
    This assumption is natural, as it guarantees the stability of the diagonal blocks of the matrices by Proposition \ref{prop:stability_semisimple}; otherwise, $\mathcal{A}^{(1)}$ would not be stable. It always holds upon blocking every $pq$ sites, when $q < \infty$.
    
    \item[\textit{(W2)}] $\mathcal{A}^{(1)}$ admits a basis with the properties described in Theorem \ref{prop:structure_span}, with a structure of $b$ blocks by $b$ blocks (i.e. $L_{\mathrm{span}} = 1$). This can always be achieved upon blocking every $L_{\mathrm{span}}$ sites.
\end{itemize}
Proposition \ref{prop:stability_semisimple} already provides necessary and sufficient conditions for the stability of the \textit{diagonal} part of the matrices. What remains is to understand how the \textit{off-diagonal} blocks affect stability.

The core of our proof will be to identify an integer $L_{\mathds{1}}$ such that, whenever $\mathds{1}_0$ eventually appears in the span, it must already be contained in $\mathcal{A}^{(L_{\mathds{1}})}$. Once this is established, the desired stability result will easily follow.

We begin by fixing some notation that will be used throughout the proofs:
\begin{itemize}
    \item Let $\mathds{1}_0$ be the identity matrix with zeros in the positions corresponding to vanishing diagonal blocks of the algebra, i.e. $\mathds{1}_0 := \sum_{\{0,s\} \in \Sigma_\infty} [\mathds{1}]_{\{0,s\}}$, which consists of the identity except for zeros in the diagonal entries belonging to the $[\{0,\varepsilon\}, \{0,\varepsilon\}]$ sector.

    \item $[A]^{\text{diag}}_{\{0,s\}}$ denotes the diagonal part of the basis element corresponding to symbol $\{0,s\} \in \Sigma_\infty$, with all off-diagonal blocks removed.

    \item $\mathcal{A}_{\text{nilp}}^{(\ell)}$ denotes the \textit{nilpotent part} of the span, i.e. the elements of $\mathcal{A}^{(\ell)}$ whose diagonal blocks are all zero. This can be written as
    \begin{equation*}
        \mathcal{A}_{\text{nilp}}^{(\ell)} :=
        \left\{
            a \in \mathcal{A}^{(\ell)} \mid a_{ii} = \mathbf{0}_{D_i \times D_i} , \ \forall i \in \{1, \dots, b\}
        \right\}
    \end{equation*}
    \item We use the notation
    \begin{equation*}
        \mathcal{A}_{\text{nilp}}^{(\ell), [\{0,i\},\{0,j\}]} 
        :=
        \left\{a \in \mathcal{A}_{\text{nilp}}^{(\ell)} \mid a_{mn} = 0 \text{ if } \{0,r_m\} \neq \{0,i\} \text{ or } \{0,r_n\} \neq \{0,j\} \right\} .
    \end{equation*}
    Similarly, we denote
    \begin{equation*}
        \mathcal{A}_{\text{nilp}}^{(\ell), \text{non-}\varepsilon} := \mathcal{A}_{\text{nilp}}^{(\ell)} \setminus \mathcal{A}_{\text{nilp}}^{(\ell), [\{0,\varepsilon\},\{0,\varepsilon\}]}.
    \end{equation*}
    As an example, given 
    \begin{equation*}
        \mathcal{A}^{(\ell)} = \left\{{\scriptsize \begin{pmatrix}
        0 & d & f \\  & c & e \\  &  & 0
    \end{pmatrix}} \mid c,d,e,f \in \mathbb{C}\right\},
    \end{equation*}
    with $\Sigma_\infty=\{\{0,1\}\}$ and $\Sigma_f = \{\{1\},\{2\},\{3\}\}$ we have that
    \begin{equation*}
        \mathcal{A}_{\text{nilp}}^{(\ell), [\{0,1\},\{0,\varepsilon\}]} =  
        \left\{{\scriptsize \begin{pmatrix}
            0 & 0 & 0 \\  & 0 & e \\  &  & 0
        \end{pmatrix}} \mid e \in \mathbb{C}\right\}
        , \quad 
        \mathcal{A}_{\text{nilp}}^{(\ell), \text{non-}\varepsilon} 
        = \left\{{\scriptsize \begin{pmatrix}
        0 & d & 0 \\  & 0 & e \\  &  & 0
    \end{pmatrix}} \mid d,e \in \mathbb{C}\right\} .
    \end{equation*}
\end{itemize}

One technical lemma we will use repeatedly is the ability to choose a convenient basis whenever $\mathds{1}_0$ lies in the relevant part of the span.
\begin{lemma} \label{lemma:choose_convenient_basis}
    Let $\ell \in \mathbb{N}$, and let $(i,j)$ be the first block (with respect to the $\preceq$-order) such that 
    \begin{equation*}
        \mathds{1}_0 \in \mathcal{A}_{\text{nilp}}^{(\ell)}\mid_{\prec (i,j)} \quad \text{but} \quad 
        \mathds{1}_0 \notin \mathcal{A}_{\text{nilp}}^{(\ell)}\mid_{\preceq (i,j)},
    \end{equation*}
    Then, we can choose a basis for $\mathcal{A}^{(n\ell)}$ for all $n \in \mathbb{N}$ in which there are no ``generalized Jordan blocks'' up to block $\prec (i,j)$. Concretely, for each $\{0,s\} \in \Sigma_\infty$ and every $n \in \mathbb{N}$, there exists $k_{ij;\{0,s\}}^{(\ell)} \in \mathbb{C}$ such that
    \begin{equation} \label{eq:lemma_choose_convenient_basis_aux1}
        [A]_{\{0,s\}}^{(n \ell)} \mid_{\preceq(i,j)} = 
        [A]^{\text{diag}}_{\{0,s\}} + [n k_{ij;\{0,s\}}^{(\ell)} A]_{ij},
    \end{equation}
    where $[B]_{ij}$ denotes a matrix that is zero everywhere except for the block located at $(i,j)$, which contains the matrix $B$. 
\end{lemma}

\begin{proof}
    Since we are assuming that $\mathcal{A}^{(1)}$ admits a basis with the convenient structure of Theorem \ref{prop:structure_span}, together with the condition $\mathds{1}_0 \in \mathcal{A}^{(\ell)}|_{\prec (i,j)}$, without loss of generality we can represent each basis element labeled by $\{0,s\} \in \Sigma_\infty$ as
    \begin{equation*}
        [A]^{(\ell)}_{\{0,s\}}\mid_{\preceq(i,j)} = [A]^{\text{diag}}_{\{0,s\}} + [k_{ij;\{0,s\}}^{(\ell)} A]_{ij}
    \end{equation*}
    for some $k_{ij;\{0,s\}}^{(\ell)} \in \mathbb{C}$ and any arbitrary $A$. Note that the diagonal part of the $\Sigma_\infty$ basis elements, $[A]^{(\ell),\text{diag}}_{\{0,s\}}$, is independent of $\ell$. This is because we are assuming that the constants corresponding to the diagonal blocks are either 0 or 1, so they remain unchanged under blocking.
    
    To obtain the form of the basis elements of $\mathcal{A}^{(n \ell)}$ for any $n \in \mathbb{N}$, we can just multiply $n$ times each of the $[A]^{(\ell)}_{\{0,s\}}$ elements, thus ensuring that the resulting elements still satisfy property (P1) for $\Gamma$ (defined in Appendix \ref{app:notation_and_basis}) and that Eq. \eqref{eq:lemma_choose_convenient_basis_aux1} holds.
\end{proof}

As an example, consider the following set of matrices,
\begin{equation*}
    \mathcal{A}^{(\ell)} = \left\{ {\scriptsize\begin{pmatrix}
    A & A+B & D \\ & A & -A+C \\ & & A
    \end{pmatrix}} \mid A, B, C, D \right\} \longrightarrow
    \mathcal{A}^{(n\ell)} = \left\{ {\scriptsize\begin{pmatrix}
    A & nA+B & D \\ & A & -nA+C \\ & & A
    \end{pmatrix}} \mid A, B, C, D \right\}, 
\end{equation*} 
Then, $\Sigma_\infty = \{\{0,1\}\}$ and $\Sigma_f = \{\{1\},\{2\},\{3\}\}$, and the basis elements can be defined as
\begin{equation*}
    [A]_{\{0,1\}}^{(n\ell)} := 
    {\scriptsize \begin{pmatrix}
    A & n A & 0 \\ & A & -n A \\ & & A
    \end{pmatrix}}, \ \ 
    [B]_{\{1\}}^{(n\ell)} := 
    {\scriptsize \begin{pmatrix}
    0 & B & 0 \\ & 0 & 0 \\ & & 0 
    \end{pmatrix}}, \ \ 
    [C]_{\{2\}}^{(n\ell)} := 
    {\scriptsize \begin{pmatrix}
    0 & 0 & 0 \\ & 0 & C \\ & & 0 
    \end{pmatrix}}, \ \
    [D]_{\{3\}}^{(n\ell)} := 
    {\scriptsize \begin{pmatrix}
    0 & 0 & D \\ & 0 & 0 \\ & & 0 
    \end{pmatrix}}.
\end{equation*}
Since $\mathds{1} \in \mathcal{A}^{(\ell)}$, we can pick a basis of $\mathcal{A}^{(\ell)}$ as prescribed by Lemma \ref{lemma:choose_convenient_basis}, where the $\Sigma_\infty$ basis elements are strictly block-diagonal. Indeed, since we can equivalently write
\begin{equation*}
    \mathcal{A}^{(n\ell)} = \left\{ {\scriptsize\begin{pmatrix}
    A & B & D \\ & A & C \\ & & A
    \end{pmatrix}} \mid A, B, C, D \right\} , \quad \forall n \in \mathbb{N},
\end{equation*} 
we can just take basis elements
\begin{equation*}
    [A]_{\{0,1\}}^{(n\ell)} := 
    {\scriptsize \begin{pmatrix}
    A & 0 & 0 \\ & A & 0 \\ & & A
    \end{pmatrix}}, \ \
    [B]_{\{1\}}^{(n\ell)} := 
    {\scriptsize \begin{pmatrix}
    0 & B & 0 \\ & 0 & 0 \\ & & 0 
    \end{pmatrix}}, \ \ 
    [C]_{\{2\}}^{(n\ell)} := 
    {\scriptsize \begin{pmatrix}
    0 & 0 & 0 \\ & 0 & C \\ & & 0 
    \end{pmatrix}}, \ \
    [D]_{\{3\}}^{(n\ell)} := 
    {\scriptsize \begin{pmatrix}
    0 & 0 & D \\ & 0 & 0 \\ & & 0 
    \end{pmatrix}}.
\end{equation*}

Another remark that we will use is encapsulated in the following lemma.
\begin{lemma} \label{lemma:non-epsilon_per_sectors}
    If $\mathcal{A}_{\text{nilp}}^{(\ell_1)} \mid_{\preceq n\text{-th diag}} = \mathcal{A}_{\text{nilp}}^{(\ell_2)} \mid_{\preceq n\text{-th diag}}$, then for each pair of $\{0,i\}, \{0,j\} \in \Sigma_\infty$,
    \begin{align*}
        \mathcal{A}_{\text{nilp}}^{(\ell_1),[\{0,i\},\{0,j\}]} \mid_{\preceq n\text{-th diag}} &= \mathcal{A}_{\text{nilp}}^{(\ell_2),[\{0,i\},\{0,j\}]} \mid_{\preceq n\text{-th diag}}, \\     \mathcal{A}_{\text{nilp}}^{(\ell_1),\text{non-}\varepsilon} \mid_{\preceq n\text{-th diag}} &= \mathcal{A}_{\text{nilp}}^{(\ell_2),\text{non-}\varepsilon} \mid_{\preceq n\text{-th diag}}.
    \end{align*}
\end{lemma}
\begin{proof}
    This is a direct consequence of the fact that blocks in different sectors of $\mathcal{A}^{(\ell)}$ are linearly independent from each other, under the assumption that $\mathcal{A}^{(\ell)}$ is already in the form described by Theorem \ref{prop:structure_span}.  
\end{proof}

\subsection{When does the nilpotent part of the span stabilize upon blocking?} \label{app:Wielandt:nilp_stabilize}

To tackle the central technical question of whether $\mathds{1}_0$ belongs to $\mathcal{A}^{(\ell)}$, we introduce two lemmas that characterize when the nilpotent part of the span remains invariant under a particular amount of blocking. The general statement is proved by induction in Lemma \ref{lemma:nilp_stability}, with the base case established first in Lemma \ref{lemma:nilp_1st_diag_stability}.
\begin{lemma} \label{lemma:nilp_1st_diag_stability}
    $\mathcal{A}_{\text{nilp}}^{(\ell)}\mid_{\preceq \text{1st diag}} = 
    \mathcal{A}_{\text{nilp}}^{(2)}\mid_{\preceq \text{1st diag}}$ for all $\ell \geq 2$.
\end{lemma}
\begin{proof}
    We will proceed by induction in $\ell$. For $\ell = 2$, we rewrite $\mathcal{A}^{(\ell)}_{\text{nilp}}\mid_{\preceq \text{1st diag}}$ in a more convenient way as follows. 
    \begin{equation} \label{eq:A_nilp_1stdiag}
        \mathcal{A}_{\text{nilp}}^{(2)} \mid_{\preceq \text{1st diag}} =
        \text{span}\left\{ [A]^{(1)}_{\{0,s\}} [B]^{(1)}_{\{t\}}, [A]^{(1)}_{\{s\}} [B]^{(1)}_{\{0,t\}},
        [A]^{(1)}_{\{s\}} [B]^{(1)}_{\{t\}}
        \right\}_{t,s,A,B}\mid_{\preceq \text{1st diag}}
    \end{equation}
    We note the following about the first two terms in Eq. \eqref{eq:A_nilp_1stdiag}:
    \begin{align*}
        [A]^{(1)}_{\{0,s\}} [B]^{(1)}_{\{t\}} \mid_{\preceq \text{1st diag}} 
        &= \left( [A]^{\text{diag}}_{\{0,s\}} + [A]^{(1), \text{off-diag}}_{\{0,s\}} \right) [B]^{(1)}_{\{t\}} \mid_{\preceq \text{1st diag}} \\ 
        &= [A]^{\text{diag}}_{\{0,s\}} [B]^{(1)}_{\{t\}} \mid_{\preceq \text{1st diag}} \\
        &=\begin{cases} [AB]_{\{t\}}^{(1)} \mid_{\preceq \text{1st diag}}, &\text{if } \{0,s\} = \{0,r_t^1\} \neq \{0,\varepsilon\} \\ 0, &\text{otherwise} \end{cases} \\
        [A]^{(1)}_{\{s\}} [B]^{(1)}_{\{0,t\}} \mid_{\preceq \text{1st diag}} 
        &= \begin{cases} [AB]_{\{s\}}^{(1)} \mid_{\preceq \text{1st diag}}, &\text{if } \{0,t\} = \{0,r_s^2\} \neq \{0,\varepsilon\} \\ 0, &\text{otherwise} \end{cases}
    \end{align*}
    where we have used that the diagonal part of $[A]_{\{0,s\}}^{(\ell)}$ is independent of $\ell$ since there are no constants different from 0 or 1 in the diagonal blocks. On the other hand, the 3rd term in Eq. \eqref{eq:A_nilp_1stdiag} is 0 when restricted to the 1st diagonal, since both elements $[A]^{(1)}_{\{s\}}$ and $[B]^{(1)}_{\{t\}}$ are nilpotent, and thus their product is supported on the 2nd or higher diagonals. Therefore, we get
    \begin{equation} \label{eq:A_nilp_2_1stdiag}
        \mathcal{A}_{\text{nilp}}^{(2)} \mid_{\preceq \text{1st diag}} =
        \text{span}\left\{ [AB]^{(1),\text{non-}\varepsilon}_{\{s\}}
        \right\}_{s,A,B}\mid_{\preceq \text{1st diag}}
        = \mathcal{A}_{\text{nilp}}^{(1),\text{non-}\varepsilon} \mid_{\preceq \text{1st diag}} 
    \end{equation}
    As a consequence, since $\mathcal{A}_{\text{nilp}}^{(1),\text{non-}\varepsilon} \mid_{\preceq \text{1st diag}}$ does not have any non-zero blocks in the $[\{0,\varepsilon\},\{0,\varepsilon\}]$ sector by definition, we have that
    \begin{equation} \label{eq:no_epsilon_in_A2}
        \mathcal{A}_{\text{nilp}}^{(2)} \mid_{\preceq \text{1st diag}}
        =
        \mathcal{A}_{\text{nilp}}^{(2),\text{non-}\varepsilon} \mid_{\preceq \text{1st diag}} \, .
    \end{equation}

    Now we assume that the claim is true for $\ell$, and we want to prove it for $\ell+1$. By Lemma \ref{lemma:non-epsilon_per_sectors}, the inductive hypothesis for $\ell$ implies that $\mathcal{A}_{\text{nilp}}^{(\ell),\text{non-}\varepsilon} \mid_{\preceq \text{1st diag}} = \mathcal{A}_{\text{nilp}}^{(2),\text{non-}\varepsilon} \mid_{\preceq \text{1st diag}}$. Similarly as before, we can write
    \begin{align*}
        \mathcal{A}_{\text{nilp}}^{(\ell+1)} \mid_{\preceq \text{1st diag}} &=
        \text{span}\left\{ [A]^{(1)}_{\{0,s\}} [B]^{(\ell)}_{\{t\}}, [A]^{(1)}_{\{s\}} [B]^{(\ell)}_{\{0,t\}},
        [A]^{(1)}_{\{s\}} [B]^{(\ell)}_{\{t\}}
        \right\}_{t,s,A,B}\mid_{\preceq \text{1st diag}} \\
        &= \text{span}\left\{ \mathcal{A}_{\text{nilp}}^{(1),\text{non-}\varepsilon}, \mathcal{A}_{\text{nilp}}^{(\ell),\text{non-}\varepsilon}\right\} \mid_{\preceq \text{1st diag}} \\
        &= \mathcal{A}_{\text{nilp}}^{(1),\text{non-}\varepsilon}\mid_{\preceq \text{1st diag}} \\
        &= \mathcal{A}_{\text{nilp}}^{(2)}\mid_{\preceq \text{1st diag}}
    \end{align*}
    where we have used the induction hypothesis on $\ell$, as well as Eq. \eqref{eq:A_nilp_2_1stdiag} and \eqref{eq:no_epsilon_in_A2}.
\end{proof}

The above statement can be generalized to hold for higher diagonals as follows.
\begin{lemma} \label{lemma:nilp_stability}
    Given $p \geq 1$, and assuming that $\mathds{1}_0 \in \mathcal{A}^{(2^{p-1-\ell})}\mid_{\preceq (p-1-\ell) \text{-th diag}}$ for all $\ell \in \{0, \dots, p-2\}$, then it necessarily holds that
    \begin{equation*}
        \mathcal{A}_{\text{nilp}}^{(2^{p} + \alpha 2^{p-1})}\mid_{\preceq p\text{-th diag}} = \mathcal{A}_{\text{nilp}}^{(2^{p})}\mid_{\preceq p\text{-th diag}}\ , \quad  \forall \alpha \geq 1. 
    \end{equation*}
\end{lemma}
\begin{proof}
    We will proceed by induction in $p$. In turn, for each fixed value of $p$ we will also do induction on $\alpha$. The base case with $p = 1$ is proven by Lemma \ref{lemma:nilp_1st_diag_stability}.
    
    {}

    Let us now prove that the claim holds for $p>1$ when it is assumed to be true for $p-1$. To show the statement for $p$, we proceed by induction on $\alpha$ again. Due to the assumption $\mathds{1}_0 \in \mathcal{A}^{(2^{p-1})}\mid_{\preceq (p-1) \text{-th diag}}$, we can use Lemma \ref{lemma:choose_convenient_basis} to choose \textit{wlog} a basis of $\mathcal{A}^{(2^{p-1})}$ such that the $\Sigma_\infty$ basis elements, denoted as $[A]^{(2^{p-1})}_{\{0,s\}}$, are zero in the first $p-1$ diagonals.
    
    First, we find a convenient rewriting for the $\alpha = 0$ case that will be useful to prove the general case with arbitrary $\alpha$. For this purpose, we rewrite $\mathcal{A}_{\text{nilp}}^{(2^{p})}\mid_{\preceq \text{$p$-th diag}}$ as
    \begin{equation} \label{eq:A_nilp_p-thdiag}
        \mathcal{A}_{\text{nilp}}^{(2^{p})} \mid_{\preceq \text{$p$-th diag}} =
        \text{span}\left\{ [A]^{(2^{p-1})}_{\{0,s\}} [B]^{(2^{p-1})}_{\{t\}}, [A]^{(2^{p-1})}_{\{s\}} [B]^{(2^{p-1})}_{\{0,t\}},
        [A]^{(2^{p-1})}_{\{s\}} [B]^{(2^{p-1})}_{\{t\}}
        \right\}\mid_{\preceq \text{$p$-th diag}}
    \end{equation}
    The first two terms in Eq. \eqref{eq:A_nilp_p-thdiag} can be rewritten using the fact that 
    \begin{align*}
        [A]^{(2^{p-1})}_{\{0,s\}} [B]^{(2^{p-1})}_{\{t\}} \mid_{\preceq \text{$p$-th diag}}
        &= \left( [A]^{\text{diag}}_{\{0,s\}} + [A]^{(2^{p-1}), p\text{-th diag}}_{\{0,s\}} \right) [B]^{(2^{p-1})}_{\{t\}} \mid_{\preceq \text{$p$-th diag}} \\
        &= [A]^{\text{diag}}_{\{0,s\}} [B]^{(2^{p-1})}_{\{t\}} \mid_{\preceq \text{$p$-th diag}} \\
        &= \begin{cases} [AB]_{\{t\}}^{(2^{p-1})} \mid_{\preceq p\text{-th diag}}, &\text{if } \{0,s\} = \{0,r^1_t\} \neq \{0,\varepsilon\} \\ 0, &\text{otherwise} \end{cases}
    \end{align*}
    and similarly $[A]^{(2^{p-1})}_{\{s\}} [B]^{(2^{p-1})}_{\{0,t\}}\mid_{\preceq \text{$p$-th diag}} = [AB]_{\{s\}}^{(2^{p-1})}$ if $\{0,t\} = \{0,r_s^2\} \neq \{0,\varepsilon\}$, or $0$ otherwise. Regarding the third term in Eq. \eqref{eq:A_nilp_p-thdiag}, since both elements in the product are nilpotent, we get that
    \begin{align*}
        [A]^{(2^{p-1})}_{\{s\}} [B]^{(2^{p-1})}_{\{t\}} \mid_{\preceq \text{$p$-th diag}} = 
        \left( [A]^{(2^{p-1})}_{\{s\}} \mid_{\preceq \text{$(p-1)$-th diag}} [B]^{(2^{p-1})}_{\{t\}} \mid_{\preceq \text{$(p-1)$-th diag}} \right) \mid_{\preceq \text{$p$-th diag}}
    \end{align*}
    Putting everything together into Eq. \eqref{eq:A_nilp_p-thdiag}, we have that
    \begin{equation} \label{eq:expression_Anilp}
        \mathcal{A}_{\text{nilp}}^{(2^{p})} \mid_{\preceq \text{$p$-th diag}} =
        \text{span}\left\{ \mathcal{A}_{\text{nilp}}^{(2^{p-1}),\text{non-}\varepsilon}, \ 
        \mathcal{A}_{\text{nilp}}^{(2^{p-1})}\mid_{\preceq \text{$(p-1)$-th diag}} \cdot \mathcal{A}_{\text{nilp}}^{(2^{p-1})}\mid_{\preceq \text{$(p-1)$-th diag}}
        \right\}\mid_{\preceq \text{$p$-th diag}} .
    \end{equation}
    
    Now we prove that the claim holds by doing induction on $\alpha$. That is, assume that the statement is true for $\alpha$ (i.e. $\mathcal{A}_{\text{nilp}}^{(2^{p} + \alpha 2^{p-1})} \mid_{\preceq \text{$p$-th diag}} = \mathcal{A}_{\text{nilp}}^{(2^{p})} \mid_{\preceq \text{$p$-th diag}}$), and we want to show it is also true for $\alpha + 1$. Similarly as for the $p = 2$ case, we have
    \begin{align}
        &\mathcal{A}_{\text{nilp}}^{(2^{p} + (\alpha+1)2^{p-1})} \mid_{\preceq \text{$p$-th diag}} \nonumber\\
        & =
        \text{span}\left\{ [A]^{(2^{p-1})}_{\{0,s\}} [B]^{(2^{p} + \alpha 2^{p-1})}_{\{t\}}, [A]^{(2^{p-1})}_{\{s\}} [B]^{(2^{p} + \alpha 2^{p-1})}_{\{0,t\}},
        [A]^{(2^{p-1})}_{\{s\}} [B]^{(2^{p} + \alpha 2^{p-1})}_{\{t\}}
        \right\}\mid_{\preceq \text{$p$-th diag}} \nonumber \\
        & = \text{span}\left\{ 
        \mathcal{A}_{\text{nilp}}^{(2^{p} + \alpha 2^{p-1}),\text{non-}\varepsilon}, \ 
        \mathcal{A}_{\text{nilp}}^{(2^{p-1}),\text{non-}\varepsilon}, \
        \mathcal{A}_{\text{nilp}}^{(2^{p-1})}\mid_{\preceq \text{$(p-1)$-th diag}} \cdot \mathcal{A}_{\text{nilp}}^{(2^{p} + \alpha 2^{p-1})}\mid_{\preceq \text{$(p-1)$-th diag}}
        \right\} \mid_{\preceq \text{$p$-th diag}}  \label{eq:nilp_p-th_0} \\
        &= \text{span}\left\{ 
        \mathcal{A}_{\text{nilp}}^{(2^{p}),\text{non-}\varepsilon}, \ 
        \mathcal{A}_{\text{nilp}}^{(2^{p-1}),\text{non-}\varepsilon}, \
        \mathcal{A}_{\text{nilp}}^{(2^{p-1})}\mid_{\preceq \text{$(p-1)$-th diag}} \cdot \mathcal{A}_{\text{nilp}}^{(2^{p} + \alpha 2^{p-1})}\mid_{\preceq \text{$(p-1)$-th diag}}
        \right\} \mid_{\preceq \text{$p$-th diag}} \label{eq:nilp_p-th_1} \\
        &= \text{span}\left\{ 
        \mathcal{A}_{\text{nilp}}^{(2^{p}),\text{non-}\varepsilon}, \ 
        \mathcal{A}_{\text{nilp}}^{(2^{p-1}),\text{non-}\varepsilon}, \
        \mathcal{A}_{\text{nilp}}^{(2^{p-1})} \mid_{\preceq \text{$(p-1)$-th diag}} \cdot \mathcal{A}_{\text{nilp}}^{(2^{p-1})}\mid_{\preceq \text{$(p-1)$-th diag}}
        \right\} \mid_{\preceq \text{$p$-th diag}} \label{eq:nilp_p-th_2} \\
        &= \mathcal{A}_{\text{nilp}}^{(2^{p})} \mid_{\preceq \text{$p$-th diag}}, \label{eq:nilp_p-th_3}
    \end{align}
    where we used the following at each step:
    \begin{itemize}
        \item To obtain Eq. \eqref{eq:nilp_p-th_0}, reason analogously to the case of $\alpha = 0$. 
        
        \item To obtain Eq. \eqref{eq:nilp_p-th_1}, use the inductive hypothesis on $\alpha$, i.e. $\mathcal{A}_{\text{nilp}}^{(2^{p} + \alpha 2^{p-1})} \mid_{\preceq \text{$p$-th diag}} = \mathcal{A}_{\text{nilp}}^{(2^{p})} \mid_{\preceq \text{$p$-th diag}}$, which by Lemma \ref{lemma:non-epsilon_per_sectors} implies that $\mathcal{A}_{\text{nilp}}^{(2^{p} + \alpha 2^{p-1}), \text{non-}\varepsilon} \mid_{\preceq \text{$p$-th diag}} = \mathcal{A}_{\text{nilp}}^{(2^{p}), \text{non-}\varepsilon} \mid_{\preceq \text{$p$-th diag}}$. 
        
        \item To obtain Eq. \eqref{eq:nilp_p-th_2}, use the inductive hypothesis on $p-1$: due to the assumption $\mathds{1}_0 \in \mathcal{A}^{(2^{p-1-l})}\mid_{\preceq (p-1-l) \text{-th diag}}$ for all $l \in \{0, \dots, p-2\}$, we know that
        \begin{equation*}
            \mathcal{A}_{\text{nilp}}^{(2^{p} + \alpha 2^{p-1})}\mid_{\preceq \text{$(p-1)$-th diag}} = 
            \mathcal{A}_{\text{nilp}}^{(2^{p-1} + (2+2\alpha) 2^{p-2})}\mid_{\preceq \text{$(p-1)$-th diag}}
            = \mathcal{A}_{\text{nilp}}^{(2^{p-1})}\mid_{\preceq \text{$(p-1)$-th diag}} .
        \end{equation*}
        \item Finally, to obtain the desired conclusion in Eq. \eqref{eq:nilp_p-th_3}, use Eq. \eqref{eq:expression_Anilp} together with the fact that
        \begin{equation*}
            \mathcal{A}_{\text{nilp}}^{(2^p),\text{non-}\varepsilon} = 
            \text{span}\left\{ 
            \mathcal{A}_{\text{nilp}}^{(2^{p-1}),\text{non-}\varepsilon}, \left(\mathcal{A}_{\text{nilp}}^{(2^{p-1})}\mid_{\preceq \text{$(p-1)$-th diag}} \cdot \mathcal{A}_{\text{nilp}}^{(2^{p-1})}\mid_{\preceq \text{$(p-1)$-th diag}} \right)^{\text{non-}\varepsilon}
            \right\},
        \end{equation*}
        where the second term is contained inside of $\mathcal{A}_{\text{nilp}}^{(2^{p-1})}\mid_{\preceq \text{$(p-1)$-th diag}} \cdot \mathcal{A}_{\text{nilp}}^{(2^{p-1})}\mid_{\preceq \text{$(p-1)$-th diag}}$ appearing in Eq. \eqref{eq:expression_Anilp}.
    \end{itemize}
    The proof is now complete.
\end{proof}

\subsection{When is the identity contained in the span?}

We now turn to the problem of upper bounding the blocking length required for the element $\mathds{1}_0$ to appear in the span. This question is subtle: starting from $\mathcal{A}^{(1)}$ in gCF, it is in general \emph{not} true that $\mathds{1}_0 \notin \mathcal{A}^{(1)}$ implies that $\mathds{1}_0 \notin \mathcal{A}^{(\ell)}$ for all $\ell$. For instance, consider 
\begin{equation} \label{eq:example_jordan_disappears}
    \mathcal{A}^{(1)} = \left\{{\scriptsize\begin{pmatrix}
    A & B & A \\ & A & B \\ & & A
    \end{pmatrix}} \mid A, B \right\} \longrightarrow \mathcal{A}^{(2)} = \left\{{\scriptsize\begin{pmatrix}
        A & B & 2A+C \\ & A & B \\ & & A
    \end{pmatrix}} \mid A, B, C \right\}
\end{equation} 
Here, $\mathds{1}_0 \notin \mathcal{A}^{(1)}$ but $\mathds{1}_0 \in \mathcal{A}^{(2)}$. In this subsection, we determine an explicit value of $L_{\mathds{1}}$ such that, whenever $\mathds{1}_0 \in \mathcal{A}^{(\ell)}$ for some $\ell \geq 1$, then it is guaranteed to be contained in $\mathcal{A}^{(L_{\mathds{1}})}$; that is, $\mathds{1}_0 \in \mathcal{A}^{(L_{\mathds{1}})}$. We will still work under assumptions \textit{(W1)} and \textit{(W2)} stated in \ref{app:sec_stability_assumptions_W1W2}.

To address this, we first prove a partial result for the $p$-th diagonal and blocking lengths that are multiples of $2^p$ (Proposition \ref{prop:key2_Wielandt}). Building on this, we then lift the argument to the general setting, which yields the desired explicit upper bound on $L_{\mathds{1}}$ (Corollary \ref{cor:Wielandt}).

\begin{proposition} \label{prop:key2_Wielandt}
    $\mathds{1}_0 \in \mathcal{A}^{(\alpha 2^{p})}\mid_{\preceq p\text{-th diag}}$ for some $\alpha \in \mathbb{Z}_{\geq 1} \implies \mathds{1}_0 \in \mathcal{A}^{(2^{p})}\mid_{\preceq p\text{-th diag}}$.
\end{proposition}
\begin{proof}
    We prove this by induction. We start with $p = 1$, so we want to show that $\mathds{1}_0 \in \mathcal{A}^{(2\alpha)}\mid_{\preceq 1\text{st diag}}$ for any $\alpha \in \mathbb{Z}_{\geq 1}$ implies that $\mathds{1}_0 \in \mathcal{A}^{(2)}\mid_{\preceq 1\text{st diag}}$. For each $j \in \Sigma_\infty$, in gCF we know that there exist constants $k_{m,m+1;\{0,j\}}^{(2)}\in \mathbb{C}$ such that the corresponding basis elements can be written as
    \begin{align*}
        [A]_{\{0,j\}}^{(2)} \mid_{\preceq 1\text{st diag}} &= [A]^{\text{diag}}_{\{0,j\}} + \sum_m [k_{m,m+1;\{0,j\}}^{(2)}A]_{(m,m+1)} \implies \\
        \implies [A]_{\{0,j\}}^{(2\alpha)} \mid_{\preceq 1\text{st diag}} &= [A]^{\text{diag}}_{\{0,j\}} + \sum_m [\alpha k_{m,m+1;\{0,j\}}^{(2)}A]_{(m,m+1)}.
    \end{align*}
    Due to the assumption that $\mathds{1}_0 \in \mathcal{A}^{(2\alpha)}\mid_{\preceq 1\text{st diag}}$, we know that there must exist $b \in \mathcal{A}_{\text{nilp}}^{(2\alpha)}$ such that
    \begin{equation} \label{eq:induction_p=1}
        \left( [\mathds{1}]_{\{0,j\}}^{(2\alpha)}  + b \right) \mid_{\preceq 1\text{st diag}} = [\mathds{1}]_{\{0,j\}}^{\text{diag}}\implies b\mid_{\preceq 1\text{st diag}} = \sum_m [-\alpha k_{m,m+1;\{0,j\}}^{(2)} \mathds{1}]_{(m,m+1)}. 
    \end{equation}

    Lemma \ref{lemma:nilp_1st_diag_stability} implies that $\mathcal{A}_{\mathrm{nilp}}^{(2)}\mid_{\preceq 1\text{st diag}} = \mathcal{A}_{\mathrm{nilp}}^{(2\alpha)}\mid_{\preceq 1\text{st diag}}$, so there exists $\tilde{b} \in \mathcal{A}_{\mathrm{nilp}}^{(2)}$ such that $\tilde{b} \mid_{\preceq 1\text{st diag}} = b \mid_{\preceq 1\text{st diag}}$, which means that
    \begin{align*}
        \left( \frac{1}{\alpha} \tilde{b} + [\mathds{1}]_{\{0,j\}}^{(2)} \right) \mid_{\preceq 1\text{st diag}} &= 
        [\mathds{1}]_{\{0,j\}}^{\text{diag}} + 
        \sum_m \underbrace{\left[ \frac{1}{\alpha} \underbrace{\tilde{b}_{m,m+1}}_{=b_{m,m+1}} + k_{m,m+1;\{0,j\}}^{(2)} \mathds{1} \right]_{(m,m+1)}}_{=0, \ \forall m \  \text{ (by Eq. \eqref{eq:induction_p=1})}} \\
        &= [\mathds{1}]_{\{0,j\}}^{\text{diag}} \in \mathcal{A}^{(2)}\mid_{\preceq 1\text{st diag}}.
    \end{align*}
    Since we can obtain the same conclusion for each $j \in \Sigma_\infty$, we have that
    \begin{equation*}
        \mathds{1}_0 = \sum_{j\in\Sigma_\infty} [\mathds{1}]_{\{0,j\}}^{\text{diag}} \in \mathcal{A}^{(2)} \mid_{\preceq 1\text{st diag}},
    \end{equation*}
    which is what we aimed for. 

    Now we show that the claim is true for any $p \geq 2$, assuming that it holds for all $1, 2, \dots, p-1$. Our inductive hypothesis can be rewritten as follows: for each $l \in \{0, \dots, p-2\}$, it holds that
    \begin{equation} \label{eq:Wielandt_IH}
        \mathds{1}_0 \in \mathcal{A}^{(\beta_l 2^{p-1-l})}\mid_{\preceq (p-1-l)\text{-th diag}} \text{ for some } \beta_l \in \mathbb{Z}_{\geq 1} \implies \mathds{1}_0 \in \mathcal{A}^{(2^{p-1-l})}\mid_{\preceq (p-1-l)\text{-th diag}} .
    \end{equation}
    To prove the claim for $p$, we start from the assumption that $\mathds{1}_0 \in \mathcal{A}^{(\alpha 2^{p})}\mid_{\preceq p\text{-th diag}}$ for some $\alpha \geq 1$. In particular, we can rewrite this assumption such that for each $l \in \{0, \dots, p-2\}$,
    \begin{equation*}
        \mathds{1}_0 \in \mathcal{A}^{(\alpha 2^{p})}\mid_{\preceq (p-1-l)\text{-th diag}} = 
        \mathcal{A}^{((\alpha 2^{1+l}) 2^{p-1-l})}\mid_{\preceq (p-1-l)\text{-th diag}}.
    \end{equation*}
    This assumption is precisely what allows us to invoke the inductive hypothesis in Eq. \eqref{eq:Wielandt_IH}, now applied with $\beta_l = \alpha 2^{1+l}$. Therefore, for each $l \in \{0, \dots, p-2\}$, we obtain
    \begin{equation}
        \mathds{1}_0 \in \mathcal{A}^{(2^{p-1-l})}\mid_{\preceq (p-1-l)\text{-th diag}} .
    \end{equation}
    The key observation is that this condition is exactly what is required to apply Lemma \ref{lemma:nilp_stability}, which in turn implies $$\mathcal{A}_{\mathrm{nilp}}^{(2^p+\gamma 2^{p-1})}\mid_{\preceq p\text{-th diag}} = \mathcal{A}_{\mathrm{nilp}}^{(2^p)}\mid_{\preceq p\text{-th diag}}, \quad \forall \gamma \geq 1.$$ In particular, choosing $\gamma = 2(\alpha - 1)$ yields the desired relation:
    \begin{equation} \label{eq:nilp_assumption_required}
        \mathcal{A}_{\text{nilp}}^{(\alpha 2^{p})}\mid_{\preceq p\text{-th diag}} = \mathcal{A}_{\text{nilp}}^{(2^{p})}\mid_{\preceq p\text{-th diag}}, \quad \forall \alpha \geq 1 .
    \end{equation}
    
    Due to the fact that $\mathds{1}_0 \in \mathcal{A}^{(2^{p-1})}\mid_{\preceq (p-1)\text{-th diag}}$ (and hence $\mathds{1}_0 \in \mathcal{A}^{(2^{p})}\mid_{\preceq (p-1)\text{-th diag}}$), we can use Lemma \ref{lemma:choose_convenient_basis} to choose \textit{wlog} a basis of $\mathcal{A}^{(2^{p})}$ such that the $\Sigma_\infty$ basis elements are zero in the first $p-1$ diagonals. This means that, for each $j \in \Sigma_\infty$, in gCF there exist constants $k_{m,m+p;\{0,j\}}^{(2^{p})}\in \mathbb{C}$ such that the corresponding basis elements can be written as
    \begin{align*}
        [A]_{\{0,j\}}^{(2^{p})} \mid_{\preceq p\text{-th diag}} = [A]^{\text{diag}}_{\{0,j\}} + \sum_m [k_{m,m+p;\{0,j\}}^{(2^{p})}A]_{(m,m+p)}
    \end{align*}
    
    On the other hand, since $\mathds{1}_0 \in \mathcal{A}^{(\alpha 2^{p})} \mid_{\preceq p\text{-th diag}}$, for each $j \in \Sigma_\infty$ there must exist $b \in \mathcal{A}_{\text{nilp}}^{(\alpha 2^{p})}$ such that 
    \begin{equation} \label{eq:induction_p=p}
        \left( [\mathds{1}]_{\{0,j\}}^{(\alpha 2^{p})} + b \right) \mid_{\preceq p\text{-th diag}} = [\mathds{1}]_{\{0,j\}}^{\text{diag}}\implies b\mid_{\preceq p\text{-th diag}} = \sum_m [-\alpha k_{m,m+p;\{0,j\}}^{(2^{p})}\mathds{1}]_{(m,m+p)}. 
    \end{equation}
    In turn, from Eq. \eqref{eq:nilp_assumption_required}, it follows that 
    \begin{equation*}
        \exists \ \tilde{b} \in \mathcal{A}^{(2^{p})}_{\text{nilp}} \quad \text{s.t.} \quad \tilde{b}\mid_{\preceq p\text{-th diag}} = b\mid_{\preceq p\text{-th diag}}. 
    \end{equation*}
    If this is the case, then $[\mathds{1}]_{\{0,j\}}^{\text{diag}} \in \mathcal{A}^{(2^{p})} \mid_{\preceq p\text{-th diag}}$, because $[\mathds{1}]_{\{0,j\}}^{\text{diag}}$ can be obtained by adding two elements in $\mathcal{A}^{(2^{p})} \mid_{\preceq p\text{-th diag}}$, as the following equation shows:
    \begin{equation*}
        \left( \frac{1}{\alpha} \tilde{b} + [\mathds{1}]_{\{0,j\}}^{(2^{p})} \right) \mid_{\preceq p\text{-th diag}} = 
        [\mathds{1}]_{\{0,j\}}^{\text{diag}} + 
        \sum_m \underbrace{\left[ \frac{1}{\alpha} \underbrace{\tilde{b}_{m,m+p}}_{=b_{m,m+p}} + k_{m,m+p;\{0,j\}}^{(2^{p})} \mathds{1}\right]_{(m,m+p)}}_{=0, \ \forall m \  \text{ (Eq. \eqref{eq:induction_p=p})}} = [\mathds{1}]_{\{0,j\}}^{\text{diag}}.
    \end{equation*}
Since we can reach the same conclusion for each $j \in \Sigma_\infty$, we have that $\mathds{1}_0 \in \mathcal{A}^{(2^{p})}\mid_{\preceq p\text{-th diag}}$, and the claim is proven.
\end{proof}

We are now ready to address the question posed at the beginning of this subsection: what is a value of $L_{\mathds{1}}$ such that, if $\mathds{1}_0 \notin \mathcal{A}^{(L_{\mathds{1}})}$, then $\mathds{1}_0 \notin \mathcal{A}^{(\ell)}$ for any $\ell\geq 1$? The following corollary, which plays a central role in establishing our main stability result, shows that under assumptions \textit{(W1)} and \textit{(W2)}, one can take $L_{\mathds{1}} = 2^b$.
\begin{corollary} \label{cor:Wielandt}
    $\mathds{1}_0 \in \mathcal{A}^{(\ell)}$ for some $\ell \geq 1 \iff \mathds{1}_0 \in \mathcal{A}^{(2^b)}$.  
\end{corollary}
\begin{proof}
    First, observe that whenever $\mathds{1}_0 \in \mathcal{A}^{(n)}$ for some $n$, it follows that $\mathds{1}_0 \in \mathcal{A}^{(\alpha n)}$ for all $\alpha \in \mathbb{Z}_{\geq 1}$, since multiplying $\mathds{1}_0$ by itself any number of times leaves it unchanged. Hence,
    \begin{equation*}
        \mathds{1}_0 \in \mathcal{A}^{(\ell)} \implies \mathds{1}_0 \in \mathcal{A}^{(\ell 2^{b})}.
    \end{equation*}
    Since the matrices have a structure of $b$ blocks by $b$ blocks, the condition $\mathds{1}_0 \in \mathcal{A}^{(\ell 2^{b})}$ can equivalently be written as
    \begin{equation}
        \mathds{1}_0 \in \mathcal{A}^{(\ell 2^{b})}\mid_{\preceq b\text{-th diag}} .
    \end{equation}
    We can therefore use Proposition \ref{prop:key2_Wielandt} to conclude that 
    \begin{equation}
        \mathds{1}_0 \in \mathcal{A}^{(2^{b})}\mid_{\preceq b\text{-th diag}} = \mathcal{A}^{(2^{b})}.
    \end{equation}
    The converse direction of the claim is straightforward by choosing $\ell = 2^b$, so this completes the proof.
\end{proof}

\subsection{The span becomes an algebra upon blocking once it contains $\mathds{1}_0$}

As a preliminary step, we establish a simple upper bound on the length of an algebra. This is typically assumed without proof in the literature, but is included here for completeness.
\begin{lemma} \label{lemma:algebra_length}
    Given a set of matrices in $\mathcal{M}_{D}(\mathbb{C})$, there exists $r_{alg} \leq D^2$ such that
    \begin{equation*}
        \bigcup_{k=1}^{r_{alg}} \mathcal{A}^{(k)} = \text{Alg}(\mathcal{A}^{(1)}).
    \end{equation*}
\end{lemma}
\begin{proof}
    Let $\mathcal{A}^{(\leq n)} := \cup_{k=1}^{n} \mathcal{A}^{(k)}$. By construction, $\mathcal{A}^{(\leq n)} \subseteq \mathcal{A}^{(\leq m)}$ for any $n \leq m$. We claim that $r_{alg}$ is the smallest integer such that $\mathcal{A}^{(\leq r_{alg})} = \mathcal{A}^{(\leq r_{alg}+1)}$. If this is the case, then
    \begin{equation}
        \mathcal{A}^{(r_{alg}+2)} \subseteq \text{span}(\mathcal{A}^{(\leq r_{alg}+1)} \cdot \mathcal{A}^{(1)}) = \text{span}(\mathcal{A}^{(\leq r_{alg})} \cdot \mathcal{A}^{(1)}) \subseteq \mathcal{A}^{(\leq r_{alg}+1)},
    \end{equation}
    so $\mathcal{A}^{(\leq r_{alg}+2)} = \mathcal{A}^{(\leq r_{alg}+1)} \cup \mathcal{A}^{(r_{alg}+2)} = \mathcal{A}^{(\leq r_{alg}+1)} = \mathcal{A}^{(\leq r_{alg})}$. Similarly, $\mathcal{A}^{(\leq r_{alg}+k)} = \mathcal{A}^{(\leq r_{alg})}$ for any $k \in \mathbb{Z}_{\geq 1}$. Since the dimension of $\mathcal{A}^{(\leq n)} \leq D^2$ for any $n$, and in the worst case $\dim\mathcal{A}^{(\leq n)} = \dim \mathcal{A}^{(\leq n-1)} + 1$ for all $n$ with $2 \leq n \leq r_{alg}$, it follows that $r_{alg} \leq D^2$.
\end{proof}

Now we are going to see that, if $\mathds{1} \in \mathcal{A}^{(\ell)}$ for certain blocking lengths $\ell$ (i.e. assuming $\mathds{1}_0$ has no zeros in the diagonal), the span eventually becomes an algebra. 
\begin{lemma}\label{lemma:identity_contained_semisimple_v0}
    If $\mathds{1} \in \mathcal{A}^{(1)}$ (i.e. $L_{\mathds{1}} = 1$), then
    \begin{equation*}
        \mathcal{A}^{(r_{alg} + s)} = \text{Alg}(\mathcal{A}^{(1+t)}), \quad \forall s,t \geq 0.
    \end{equation*}
    In particular, $\mathcal{A}^{(1)}$ is stable with $L_{\mathrm{stab}} \leq r_{\mathrm{alg}}$.
\end{lemma}
\begin{proof}
    Since $\mathds{1} \in \mathcal{A}^{(1)}$, the family $\{\mathcal{A}^{(n)}\}_{n \geq 1}$ is increasing, i.e. $\mathcal{A}^{(m)} \subseteq \mathcal{A}^{(n)}$ whenever $m \leq n$. Thus,
    \begin{equation*}
        \text{Alg}(\mathcal{A}^{(1)}) = \cup_{n=1}^{r_{alg}} \mathcal{A}^{(n)} = \mathcal{A}^{(r_{alg})}.
    \end{equation*} 
    Due to the facts that $\mathcal{A}^{(r_{\mathrm{alg}})} \subseteq \mathcal{A}^{(r_{\mathrm{alg}}+s)}$ and $\mathcal{A}^{(r_{\mathrm{alg}}+s)} \subseteq \mathrm{Alg}(\mathcal{A}^{(1)})$ for every $s \geq 0$, it follows that $\text{Alg}(\mathcal{A}^{(1)}) = \mathcal{A}^{(r_{\mathrm{alg}}+s)}$.

    Now, for any $t \geq 0$, we have that
    \begin{equation*}
        \text{Alg}(\mathcal{A}^{(1+t)}) \subseteq
        \text{Alg}(\mathcal{A}^{(1)}) = \mathcal{A}^{(r_{\mathrm{alg}}+s)}, \quad \forall s \geq 0, 
    \end{equation*}
    and conversely 
    \begin{equation*}
        \text{Alg}(\mathcal{A}^{(1+t)}) \supseteq 
        \mathcal{A}^{((1+t)r_{\mathrm{alg}})} = \mathcal{A}^{(r_{\mathrm{alg}}+s)}. 
    \end{equation*}
    Therefore, $\text{Alg}(\mathcal{A}^{(1+t)}) = \mathcal{A}^{(r_{\mathrm{alg}}+s)}$ for every $s, t \geq 0$, and $\mathcal{A}^{(1)}$ is stable with $L_{\mathrm{stab}} \leq r_{\mathrm{alg}}$. 
\end{proof} 

The next lemma refines the estimate from Lemma \ref{lemma:identity_contained_semisimple_v0}, demonstrating that both the stabilizing length and the required blocking may be smaller when additional structural information is known. That is, whenever $\mathds{1} \in \mathcal{A}^{(L)}$, Lemma \ref{lemma:identity_contained_semisimple_v0} asserts that $\mathcal{A}^{(L(r_{\mathrm{alg}}+s))} = \mathrm{Alg}(\mathcal{A}^{(L(1+t))})$ for all $s,t \ge 0$, but the following lemma shows that blocking less is enough under a stronger assumption.

\begin{lemma} \label{lemma:identity_contained_semisimple}
    If there exists some integer $L$ such that $\mathds{1} \in \mathcal{A}^{(\ell)}$ for all $\ell \geq L$, then
    \begin{equation*}
         \mathcal{A}^{(r_{alg} + L + s)} = \mathrm{Alg}(\mathcal{A}^{(1+t)}), \ \forall s,t \in \mathbb{Z}_{\geq 0}.
    \end{equation*}
    In particular, $\mathcal{A}^{(1)}$ is stable with $L_{\mathrm{stab}} \leq r_{\mathrm{alg}} + L$.
\end{lemma}
\begin{proof}
    Since $\mathds{1} \in \mathcal{A}^{(L + s)}$ for all $s \geq 0$, then we know that
    \begin{equation*}
        \mathcal{A}^{(n)} \subseteq \mathcal{A}^{(n + L + s)}, \ \forall n, s \in \mathbb{Z}_{\geq 0}.
    \end{equation*}
    This means that
    \begin{equation*}
        \mathcal{A}^{(1)}, \dots, \mathcal{A}^{(r_{alg})} \subseteq \mathcal{A}^{(r_{alg} + L + s)}, \ \forall s \in \mathbb{Z}_{\geq 0}.
    \end{equation*}
    Therefore, we can use this to conclude that
    \begin{align*}
        \text{Alg}(\mathcal{A}^{(1)}) = \cup_{n=1}^{r_{alg}} \mathcal{A}^{(n)} = \mathcal{A}^{(r_{alg} + L + s)} , \ \forall s \in \mathbb{Z}_{\geq 0},
    \end{align*}
    which in turn implies that $\mathcal{A}^{(r_{\mathrm{alg}}+L)} = \mathcal{A}^{(r_{\mathrm{alg}}+L+s)}$ for all $s \geq 0$. 
    
    Now, for any $t \geq 0$, we have that 
    \begin{equation*}
        \text{Alg}(\mathcal{A}^{(1+t)}) \subseteq
        \text{Alg}(\mathcal{A}^{(1)}) = \mathcal{A}^{(r_{\mathrm{alg}}+L+s)}, \quad \forall s \geq 0, 
    \end{equation*}
    and conversely 
    \begin{equation*}
        \text{Alg}(\mathcal{A}^{(1+t)}) \supseteq 
        \mathcal{A}^{((1+t)(r_{\mathrm{alg}}+L))} = \mathcal{A}^{(r_{\mathrm{alg}}+L+s)}. 
    \end{equation*}
    Therefore, $\text{Alg}(\mathcal{A}^{(1+t)}) = \mathcal{A}^{(r_{\mathrm{alg}}+L+s)}$ for every $s, t \geq 0$, and $\mathcal{A}^{(1)}$ is stable with $L_{\mathrm{stab}} \leq r_{\mathrm{alg}}+L$. 
\end{proof}

In the general case, however, $\mathds{1}_0$ might have zeros in the diagonal. When this is the case, even if $\mathds{1}_0 \in \mathcal{A}^{(L)}$ for some $L$, it no longer necessarily holds that $\mathcal{A}^{(n)} \subseteq \mathcal{A}^{(n+L+s)}$ for any $s \geq 0$, which were the starting points of both Lemma \ref{lemma:identity_contained_semisimple_v0} and Lemma \ref{lemma:identity_contained_semisimple}. Therefore, the proof needs to be refined, since there might be blocks in the $[\{0,\varepsilon\}, \{0,\varepsilon\}]$-sector that disappear upon blocking, while others do not. Indeed, consider the example
\begin{equation*}
    \mathcal{A}^{(1)} = \left\{ {\scriptsize \begin{pmatrix}
    0 & B & C \\
    & A & B \\
    & & 0 \\
\end{pmatrix}} \mid A, B, C \right\}.
\end{equation*}
Block $(1,3)$ is in sector $[\{0,\varepsilon\}, \{0,\varepsilon\}]$ but it does not disappear upon blocking. Instead, given 
\begin{equation*}
    \mathcal{A}^{(1)} = \left\{ {\scriptsize\begin{pmatrix}
        0 & B \\
        & 0
    \end{pmatrix}} \mid B \right\},
\end{equation*}
then block $(1,2)$ is in sector $[\{0,\varepsilon\}, \{0,\varepsilon\}]$ and it disappears upon blocking.

\begin{lemma} \label{lemma:este_es_wielandt_with_zeros}
    If $\mathds{1}_0 \in \mathcal{A}^{(1)}$, then the span \emph{stabilizes} upon further blocking, meaning that 
    \begin{equation}
        \mathcal{A}^{(r_{alg} b + s)} = \text{Alg}(\mathcal{A}^{(b(1+t))}), \ \forall s,t \in \mathbb{Z}_{\geq 0} .
    \end{equation}
    In particular, $\mathcal{A}^{(b)}$ is stable with $L_{\mathrm{stab}} \leq r_{\mathrm{alg}}$.
\end{lemma}
\begin{proof}
    Due to assumption \textit{(W2)}, $\mathcal{A}^{(1)}$ admits a basis as described in Theorem \ref{prop:structure_span}, and similarly for $\mathcal{A}^{(n)}, \ \forall n \geq 1$. First, note that for any $a \in \mathcal{A}^{(b)}$, there exists a set of constants $c_{e_1 e_2 \dots e_b} \in \mathbb{C}$ and matrices $A^{[n]}_x$ for $n \in \{1, \dots, b\}$ and $x \in \Sigma$ such that
    \begin{equation} \label{eq:product_aux1}
        a = \sum_{e_1, \dots, e_b \in \Sigma} c_{e_1 e_2 \dots e_b} [A^{[1]}_{e_1}]_{e_1}^{(1)} [A^{[2]}_{e_2}]_{e_2}^{(1)} \dots [A^{[b]}_{e_b}]_{e_b}^{(1)}
    \end{equation}
    Note that any non-zero terms in the sum above must necessarily have at least one term $e_i \notin [\{0,\varepsilon\}, \{0, \varepsilon\}]$. If this was not the case, since all $[A]_{e_i}^{(1)}$ would be strictly nilpotent elements of block-nilpotency order $\leq b$, the product of a number $b$ of them would necessarily vanish, i.e.
    \begin{equation*}
        [A^{[1]}_{e_1}]_{e_1}^{(1)} [A^{[2]}_{e_2}]_{e_2}^{(1)} \dots [A^{[b]}_{e_b}]_{e_b}^{(1)} = 0 \quad \text{if } e_i \in [\{0,\varepsilon\}, \{0, \varepsilon\}], \ \forall i. 
    \end{equation*}
    Thus, for each of the non-zero terms in the sum of Eq. \eqref{eq:product_aux1}, there exists at least one $e_i \notin [\{0,\varepsilon\}, \{0,\varepsilon\}]$. Consequently:
    \begin{itemize}
        \item If $e_i \in [\{0,t\}, \ast]$ with $t \neq \varepsilon$, we can insert as many $\mathds{1}_0$ as we want in front of $[A_{e_i}^{[i]}]_{e_i}^{(1)}$ without changing the result, i.e.
        \begin{align*}
            [A^{[1]}_{e_1}]_{e_1}^{(1)} [A^{[2]}_{e_2}]_{e_2}^{(1)} \dots [A^{[b]}_{e_b}]_{e_b}^{(1)} &= 
            [A^{[1]}_{e_1}]_{e_1}^{(1)} [A^{[2]}_{e_2}]_{e_2}^{(1)} \dots \mathds{1}_0 \cdot \mathds{1}_0 \dots \mathds{1}_0 \cdot [A^{[i]}_{e_i}]_{e_i}^{(1)} \dots [A^{[b]}_{e_b}]_{e_b}^{(1)} \\
            &\in \mathcal{A}^{(b + s)}, \ \forall s \in \mathbb{Z}_{\geq 0}.
        \end{align*}

        \item Similarly, if $e_i \in [\ast, \{0,t\}]$ with $t \neq \varepsilon$, we can insert as many $\mathds{1}_0$ as we want after $[A_{e_i}^{[i]}]_{e_i}^{(1)}$ without changing the result, i.e.
        \begin{align*}
            [A^{[1]}_{e_1}]_{e_1}^{(1)} [A^{[2]}_{e_2}]_{e_2}^{(1)} \dots [A^{[b]}_{e_b}]_{e_b}^{(1)} &= 
            [A^{[1]}_{e_1}]_{e_1}^{(1)} [A^{[2]}_{e_2}]_{e_2}^{(1)} \dots [A^{[i]}_{e_i}]_{e_i}^{(1)}\cdot \mathds{1}_0 \cdot \mathds{1}_0 \dots \mathds{1}_0 \dots [A^{[b]}_{e_b}]_{e_b}^{(1)} \\
            &\in \mathcal{A}^{(b + s)}, \ \forall s \in \mathbb{Z}_{\geq 0}.
        \end{align*}
    \end{itemize}
    This implies that, if $\mathds{1}_0 \in \mathcal{A}^{(1)}$, then $\mathcal{A}^{(b)} \subseteq \mathcal{A}^{(b+s)}$ for all $s \geq 0$. In fact, $\mathcal{A}^{(\ell_1)} \subseteq \mathcal{A}^{(\ell_2)}$ for all $\ell_1, \ell_2$ such that $b \leq \ell_1 \leq \ell_2$. Therefore, 
    \begin{equation*}
        \text{Alg}(\mathcal{A}^{(b)}) = 
        \cup_{s=1}^{r_{alg}} \mathcal{A}^{(s b)}
        = \mathcal{A}^{(r_{alg} b)} = \mathcal{A}^{(r_{alg} b + s)}, \ \forall s \in \mathbb{Z}_{\geq 0}\ .
    \end{equation*}

    Now, for any $t \geq 0$, we have that
    \begin{equation*}
        \text{Alg}(\mathcal{A}^{(b(1+t))}) \subseteq
        \text{Alg}(\mathcal{A}^{(b)}) = \mathcal{A}^{(r_{\mathrm{alg}}b+s)}, \quad \forall s \geq 0, 
    \end{equation*}
    and conversely 
    \begin{equation*}
        \text{Alg}(\mathcal{A}^{(b(1+t))}) \supseteq 
        \mathcal{A}^{(b(1+t)r_{\mathrm{alg}})} = \mathcal{A}^{(r_{\mathrm{alg}}+s)}. 
    \end{equation*}
    Therefore, $\text{Alg}(\mathcal{A}^{(b(1+t))}) = \mathcal{A}^{(r_{\mathrm{alg}}b+s)}$ for every $s, t \geq 0$, and $\mathcal{A}^{(b)}$ is stable with $L_{\mathrm{stab}} \leq r_{\mathrm{alg}}$.
\end{proof}

\subsection{Proving the main result about the stability of matrices}

We now have all the ingredients to prove our main result, that we restate below for convenience. 

\genWielandtt*
\begin{proof}
    We begin with the $\Rightarrow$ direction.
    Assume that the set of matrices is stable under blocking; in particular, that $\mathcal{A}^{(pqL_{\mathrm{span}} b 2^b r_{\mathrm{alg}})}$ forms an algebra. Since it is an algebra, it necessarily contains the element $\mathds{1}_0$.
    We can therefore apply Corollary \ref{cor:Wielandt} to conclude that $\mathds{1}_0 \in \mathcal{A}^{(pqL_{\mathrm{span}} 2^b)}$.
    Note that the factor ``$pqL_{\mathrm{span}}$'' appears to ensure that the assumptions \textit{(W1)} and \textit{(W2)} required by the corollary are satisfied.
    
    For the $\Leftarrow$ direction, suppose now that $\mathds{1}_0 \in \mathcal{A}^{(pqL_{\mathrm{span}} 2^b)}$.
    Then Lemma \ref{lemma:este_es_wielandt_with_zeros} directly yields Eq. \eqref{eq:main_stability_result}.
    Setting $s = 0$, $t = r_{\mathrm{alg}} - 1$ and $0$, we obtain
    \begin{equation*}
    \mathcal{A}^{(pqL_{BI} b 2^b r_{\mathrm{alg}})}
    = \mathrm{Alg}\big(\mathcal{A}^{(pqL_{\mathrm{span}} b 2^b r_{\mathrm{alg}})}\big)
    = \mathrm{Alg}\big(\mathcal{A}^{(pqL_{\mathrm{span}} b 2^b)}\big).
    \end{equation*}
    Hence, the set of matrices $\{A^i\}$ is stable under blocking every $pqL_{BI} b 2^b$ sites, with $L_{\mathrm{stab}} = r_{\mathrm{alg}}$.
    
    Now, assume that either $q = \infty$ or $\mathds{1}_0 \notin \mathcal{A}^{(pqL_{BI} 2^b)}$, so that the set $\{A^i\}$ is not stable upon any blocking. By Corollary \ref{cor:Wielandt}, this implies that $\mathds{1}_0 \notin \mathcal{A}^{(\ell)}$ for all $\ell$. We will now prove the last part of the claim by showing, via contradiction, that this entails $\mathcal{A}^{(n)} \nsubseteq \mathcal{A}^{(m+n)}$ for any values of $n,m$. 
    
    For this purpose, assume that there exist some values of $m, n \in \mathbb{Z}_{\geq 1}$ such that $\mathcal{A}^{(n)} \subseteq \mathcal{A}^{(n+m)}$. Then, 
    \begin{equation*}
        \mathcal{A}^{(n+\beta_1 m)} \subseteq \mathcal{A}^{(n+\beta_2 m)}, \quad \forall \beta_1, \beta_2 \in \mathbb{Z}_{\geq 0} \ \text{s.t.} \ \beta_1 \leq \beta_2 . 
    \end{equation*}
    In particular,
    \begin{equation*}
        \mathcal{A}^{(mn)} = \text{span}\{\mathcal{A}^{((m-1)n)} \cdot \mathcal{A}^{(n)}\} \subseteq \text{span}\{\mathcal{A}^{((m-1)n)} \cdot \mathcal{A}^{(n+\beta m)}\} = \mathcal{A}^{(m(n+\beta))}, \ \forall \beta \in \mathbb{Z}_{\geq 0}
    \end{equation*}
    This implies that
    \begin{equation*}
        \mathcal{A}^{(mnk)} \subseteq \mathcal{A}^{(mnl)}, \ \forall k \leq l
    \end{equation*}
    Therefore, we have
    \begin{equation*}
        \bigcup_{k=1}^{r_{alg}} \mathcal{A}^{(mnk)} = \mathcal{A}^{(mnr_{alg})} = \text{Alg}(\mathcal{A}^{(mn)})
    \end{equation*}
    which means that $\mathds{1}_0 \in \mathcal{A}^{(mnr_{alg})}$ and hence we reach a contradiction. The claim is thus proven.
\end{proof}

Thanks to the general characterization of the span structure provided by Theorem \ref{prop:structure_span}, we can also describe precisely the form of \textit{non-stable} sets of matrices. That is, according to Theorem \ref{prop:structure_span}, the element $\mathds{1}_0$ is not contained in $\mathcal{A}^{(\ell)}$ for any $\ell$ only if we are in one of the following two scenarios:
\begin{itemize}
    \item $q = \infty$: In this case, it is impossible to block in such a way that all proportionality constants between diagonal blocks become simultaneously equal to 1. Consequently, the algebra will always differ from the span: for any blocking length, at least one pair of blocks that are proportional in the span are independent from each other in the algebra.

    \item $q < \infty$ with \textit{persistent} Jordan-type structures: This case occurs when there are off-diagonal blocks that are proportional to a diagonal block. An example of this behavior was already presented in the main text in Eq. \eqref{eq:ex_differences_alg-span_2}, which we reproduce below for convenience:
    \begin{equation*} 
        \mathcal{A}^{(1)} = \text{span} \left\{ {\scriptsize \begin{pmatrix}
            1 & 1 \\ 0 & 1
        \end{pmatrix}} \right\} \implies
        \mathcal{A}^{(\ell)} = \text{span}\left\{ {\scriptsize \begin{pmatrix}
            1 & \ell \\ 0 & 1
        \end{pmatrix}} \right\}, \
        \text{Alg}(\mathcal{A}^{(\ell)}) = \left\{ {\scriptsize \begin{pmatrix}
            a & b \\ 0 & a
        \end{pmatrix}} \mid a,b \in \mathbb{C} \right\} .
    \end{equation*}
    Nevertheless, not every span with a Jordan-type structure is necessarily non-stable: in some cases these components may vanish upon blocking, as illustrated in Eq. \eqref{eq:example_jordan_disappears}.
\end{itemize}

\newpage

{}

\section{Key technical lemmas}

This section presents the key technical lemmas that are used in the proofs of the rest of the appendices.

\subsection{On the dependence on the diagonal free blocks}

The lemmas in this subsection are central to establishing both the algebra structure of Proposition \ref{prop:structure_subalgebra} and the span structure of Theorem \ref{prop:structure_span}. They provide an explicit construction of the change-of-basis matrix for structures of 2 blocks by 2 blocks, which allows one to control the dependence of the off-diagonal blocks on the diagonal ones. This construction guarantees the existence of a basis in which the off-diagonal block takes the desired form: either $0$ or free in the algebra case, and either $0$, proportional to the diagonal block, or free in the span case. Each of the individual lemmas below address the different possible relations between the diagonal blocks.

\begin{lemma} \label{lemma:technical1}
    Given a set of matrices $\mathcal{B} \subset \mathcal{M}_{D_A+D_B}(\mathbb{C})$ of the form
    \begin{equation}
        {\small \mathcal{B} = \left\{  {\footnotesize \begin{pmatrix}
            A & f(A) + g(B) \\ & B
        \end{pmatrix} \mid {\scriptsize A \in \mathcal{M}_{D_A}(\mathbb{C}), B \in \mathcal{M}_{D_B}(\mathbb{C}) } }
        \right\}, 
        \text{ with }
        \begin{cases}
            \text{(i)}& Af(B) = \hat{f}(AB) \\
            \text{(ii)}& g(A)B = \hat{g}(AB) \\
            \text{(iii)}& Ag(B)+f(A)B = 0
        \end{cases}}, \ \forall A, B
    \end{equation}
    for some linear functions $f, \hat{f}: \mathcal{M}_{D_A}(\mathbb{C}) \to \mathcal{M}_{D_A \times D_B}(\mathbb{C})$ and $g, \hat{g}: \mathcal{M}_{D_B}(\mathbb{C}) \to \mathcal{M}_{D_A \times D_B}(\mathbb{C})$, then 
    \begin{equation}
        \exists P : \ P \mathcal{B} P^{-1} = \left\{ {\footnotesize \begin{pmatrix} A & 0 \\ & B \end{pmatrix} } \mid {\footnotesize A \in \mathcal{M}_{D_A}(\mathbb{C}), B \in \mathcal{M}_{D_B}(\mathbb{C})} \right\}.
    \end{equation}
    An explicit construction for $P$ is provided in the proof.
\end{lemma}
\begin{proof}
    Let $f^{ij} := f(\dyad{i}{j})$ and $g^{ij} := g(\dyad{i}{j})$. Property \textit{(i)} implies that $\dyad{i}{i} f^{jk} = 0$ for any $i,j,k$ with $i \neq j$, meaning that $f^{jk}$ is all 0s except for row $j$. Additionally, $\hat{f}(\dyad{i}{l}) = \dyad{i}{j} f^{jl} = \dyad{i}{k} f^{kl}$ for any $j,k$, so the elements of $f^{ij}$ are independent of $i$. Similarly, from \textit{(ii)} follows that $g^{ij}$ is all 0s except for column $j$, and its elements are independent of $j$. Therefore, we can write 
    \begin{equation}
        f^{ij} = \sum_{n=1}^{D_B} f^{\ast j}_{\ast n} \dyad{i}{n}, \quad g^{ij} = \sum_{m=1}^{D_A} g^{i\ast}_{m\ast} \dyad{m}{j}.
    \end{equation}
    Here and in the following, the symbol ``$\ast$'' denotes a dummy index: for example, $f^{\ast j}_{\ast n}$ refers to the coefficient of $f^{ij}$ in column $n$ (nonzero only in row $i$, with a value that is independent of $i$ for any $i$, as argued above), and $g^{i\ast}_{m\ast}$ refers to the coefficient of $g^{ij}$ in row $m$ (nonzero only in column $j$, with a value that is independent of $j$ for any $j$).
    From property \textit{(iii)} follows that $\dyad{i}{j}g^{kl} + f^{ij}\dyad{k}{l} = 0$, which implies that $g^{k\ast}_{j\ast} = -f^{\ast j}_{\ast k}$, $\forall j,k$.

    Using these structural properties of $f, g$, we proceed to construct the desired $P$. First, denote
    \begin{equation}
        F_{ij} := {\footnotesize \begin{pmatrix} \dyad{i}{j} & f^{ij} \\ & 0 \end{pmatrix}}, \quad G_{ij} := {\footnotesize \begin{pmatrix} 0 & g^{ij} \\ & \dyad{i}{j} \end{pmatrix}}. 
    \end{equation}
    Note that $\mathcal{B} = \text{span}\{F_{ij}, G_{ij} \mid i,j \}$, and $F_{ij}$ and $G_{ij}$ act on the standard basis elements as
    \begin{align} 
        &{\small\begin{cases}
            F_{ij} \ket{m} = \delta_{jm} \ket{i}, \quad m \in \{1, \dots, D_A\} \\
            F_{ij} \ket{D_A + n} = f^{\ast j}_{\ast n} \ket{i}, \quad n \in \{1, \dots, D_B\}
        \end{cases}} \\
        & {\small \begin{cases}
            G_{ij} \ket{m} = 0, \quad m \in \{1, \dots, D_A\} \\
            G_{ij} \ket{D_A + n} = \delta_{jn} \left( \ket{D_A + i} + \sum_{l=1}^{D_A} g^{i\ast}_{l\ast} \ket{l} \right), \quad n \in \{1, \dots, D_B\}.
        \end{cases}}
    \end{align}
    Let us define new vectors
    \begin{equation}
        {\small \begin{cases}
            \ket{v_m} := \ket{m}, \quad m \in \{1, \dots, D_A\} \\
            \ket{v_{D_A + n}} := \ket{D_A + n} + \sum_{l=1}^{D_A} g^{n\ast}_{l\ast} \ket{l} = 
            \ket{D_A + n} - \sum_{l=1}^{D_A} f^{\ast l}_{\ast n} \ket{l}, \quad n \in \{1, \dots, D_B\}.
        \end{cases}}
    \end{equation}
    In the basis $\{\ket{v_i}\}$, the action of $F_{ij}$ can be expressed as
    \begin{align}
        &F_{ij} \ket{v_m} = \delta_{jm} \ket{i} = \delta_{jm} \ket{v_i} , \\
        &F_{ij} \ket{v_{{D_A}+n}} = F_{ij} \ket{D_A + n} - \sum_{l=1}^{D_A} f^{\ast l}_{\ast n} \underbrace{F_{ij} \ket{l}}_{= \delta_{jl} \ket{i}} = 0 .
    \end{align}
    which means that $P F_{ij} P^{-1} = {\scriptsize \begin{pmatrix} \dyad{i}{j} &  \\ & 0 \end{pmatrix}}$ where $P$ is the change of basis matrix from $\{\ket{i}\}$ to $\{\ket{v_i}\}$ such that $\ket{v_\alpha} = P \ket{\alpha}$. On the other hand, 
    \begin{align}
        &G_{ij} \ket{v_m} = 0 , \\
        &G_{ij} \ket{v_{{D_A}+n}} = G_{ij} \ket{D_A + n} + \sum_{n=1}^{D_A} g^{n\ast}_{l\ast} \underbrace{G_{ij} \ket{l}}_{= 0} = \delta_{jn} \left( \ket{D_A + i} + \sum_{l=1}^{D_A} g^{i\ast}_{l\ast} \ket{l} \right) = \delta_{jn} \ket{v_{D_A+i}},
    \end{align}
    so $P G_{ij} P^{-1} = {\scriptsize \begin{pmatrix} 0 &  \\ & \dyad{i}{j} \end{pmatrix}}$. In conclusion,
    \begin{equation}
        P \mathcal{B} P^{-1} = \left\{ {\footnotesize \begin{pmatrix} A & 0 \\ & B \end{pmatrix} } \mid A,B \right\}, \quad 
        P := {\footnotesize \begin{pmatrix} \mathds{1} & \tilde{P} \\ & \mathds{1} \end{pmatrix} } \text{ with } 
        \tilde{P} := {\scriptsize \begin{pmatrix} 
        f_{\ast 1}^{\ast 1} & \dots & f_{\ast N}^{\ast 1} \\
        \vdots & \ddots & \vdots \\
        f_{\ast 1}^{\ast N} & \dots & f_{\ast N}^{\ast N}
    \end{pmatrix} }.
    \end{equation}
    Note that $P^{-1} = {\scriptsize \begin{pmatrix} \mathds{1} & -\tilde{P} \\ & \mathds{1} \end{pmatrix} }$, and therefore it also holds that
    \begin{equation*}
        \begin{cases}
            f(A) - A\tilde{P} = 0 \\
            g(B) + \tilde{P}B = 0
        \end{cases}\ , 
        \quad \forall A, B. 
    \end{equation*}
\end{proof}

\begin{corollary} \label{cor:technical1}
    Given a set of matrices $\mathcal{B} \subset \mathcal{M}_{D_A+D_B}(\mathbb{C})$ of the form
    \begin{equation}
        {\small \mathcal{B} = \left\{  {\footnotesize \begin{pmatrix}
            A & f(A) \\ & 0
        \end{pmatrix} \mid {\scriptsize A \in \mathcal{M}_{D_A}(\mathbb{C}) } }
        \right\}, }
        \text{ with }
        Af(B) = \hat{f}(AB), \ \forall A, B
    \end{equation}
    for some linear functions $f, \hat{f}: \mathcal{M}_{D_A}(\mathbb{C}) \to \mathcal{M}_{D_A \times D_B}(\mathbb{C})$, then 
    \begin{equation}
        \exists P : \ P \mathcal{B} P^{-1} = \left\{ {\footnotesize \begin{pmatrix} A & 0 \\ & 0 \end{pmatrix} } \mid {\footnotesize A \in \mathcal{M}_{D_A}(\mathbb{C})} \right\},
    \end{equation}
    where
    \begin{equation}
        P = {\footnotesize \begin{pmatrix} \mathds{1} & \tilde{P} \\ & \mathds{1} \end{pmatrix} } \text{ with } 
        \tilde{P} := {\scriptsize \begin{pmatrix} 
        f_{\ast 1}^{\ast 1} & \dots & f_{\ast D_B}^{\ast 1} \\
        \vdots & \ddots & \vdots \\
        f_{\ast 1}^{\ast D_A} & \dots & f_{\ast D_B}^{\ast D_A}
    \end{pmatrix} }.
    \end{equation}
\end{corollary}

\begin{corollary} \label{cor:technical2}
    Given a set of matrices $\mathcal{B} \subset \mathcal{M}_{D_A+D_B}(\mathbb{C})$ of the form
    \begin{equation}
        {\small \mathcal{B} = \left\{  {\footnotesize \begin{pmatrix}
            0 & g(A) \\ & A
        \end{pmatrix} \mid {\scriptsize A \in \mathcal{M}_{D_A}(\mathbb{C}) } }
        \right\}, }
        \text{ with }
        g(A)B = \hat{g}(AB), \ \forall A, B
    \end{equation}
    for some linear functions $g, \hat{g}: \mathcal{M}_{D_B}(\mathbb{C}) \to \mathcal{M}_{D_A \times D_B}(\mathbb{C})$, then 
    \begin{equation}
        \exists P : \ P \mathcal{B} P^{-1} = \left\{ {\footnotesize \begin{pmatrix} 0 & 0 \\ & A \end{pmatrix} } \mid {\footnotesize A \in \mathcal{M}_{D_A}(\mathbb{C})} \right\},
    \end{equation}
    where
    \begin{equation}
        P = {\footnotesize \begin{pmatrix} \mathds{1} & \tilde{P} \\ & \mathds{1} \end{pmatrix} } \text{ with } 
        \tilde{P} := {\scriptsize \begin{pmatrix} 
        -g_{1\ast}^{1\ast} & \dots & -g_{1 \ast}^{D_B \ast} \\
        \vdots & \ddots & \vdots \\
        -g_{D_A\ast}^{1\ast} & \dots & -g_{D_A\ast}^{D_B\ast}
    \end{pmatrix} }.
    \end{equation}
\end{corollary}

\begin{lemma} \label{lemma:technical2}
    Given a set of matrices $\mathcal{B} \subset \mathcal{M}_{2D}(\mathbb{C})$ of the form
    \begin{equation}
        {\small \mathcal{B} = \left\{  {\footnotesize \begin{pmatrix}
            A & f(A) \\ & \omega A
        \end{pmatrix} \mid {\scriptsize A \in \mathcal{M}_{D}(\mathbb{C}) } }
        \right\}, 
        \text{ with }
        \begin{cases}
            \text{(i)}& A\tilde{f}(B) + \mu\tilde{f}(A)B + \delta AB = f(AB) \\
            \text{(ii)}& A f(B) + \lambda f(A) B + \gamma AB = \hat{f}(AB)
        \end{cases}}, \ \forall A, B
    \end{equation}
    for some linear functions $f, \hat{f}, \tilde{f}: \mathcal{M}_{D}(\mathbb{C}) \to \mathcal{M}_{D}(\mathbb{C})$, and constants $\lambda, \mu, \delta, \gamma, \omega \in \mathbb{C}$, $\lambda, \mu, \omega \neq 0$, where we use the notation $f^{ij} := f(\dyad{i}{j}) = \sum_{m,n} f^{ij}_{mn} \dyad{m}{n}$ and similarly for $\hat{f}, \tilde{f}$. Then, 
    \begin{itemize}
        \item If the quantity $k_i:= (\hat{f}^{i1}_{i1} - f^{i1}_{i1} - \gamma)/\lambda$ is independent of $i$, then
        \begin{equation}
            \exists P : \ P \mathcal{B} P^{-1} = \left\{ {\footnotesize \begin{pmatrix} A & k A \\ & \omega A \end{pmatrix} } \mid {\footnotesize A \in \mathcal{M}_{D}(\mathbb{C}) } \right\}.
        \end{equation}
        An explicit construction for $P$ is provided in the proof.
        \item Otherwise, if $k_i$ is not independent of $i$, then there is an element of the form ${\tiny \begin{pmatrix} 0 & \highg{\neq 0} \\ & 0 \end{pmatrix} }$ in $\mathcal{B}_2 := \{a b \mid a, b \in \mathcal{B}\}$.
    \end{itemize}
\end{lemma}
\begin{proof}
    From Property \textit{(ii)}, we have
    \begin{align}
        \hat{f}^{ij} &:= {\scriptsize \begin{pmatrix}
            \hat{f}_{11}^{ij} & \dots & \hat{f}_{1D}^{ij} \\
            \vdots & & \vdots \\
            \hat{f}_{D1}^{ij} & \dots & \hat{f}_{DD}^{ij}
        \end{pmatrix}} = \dyad{i}{k} f^{kj} + \lambda f^{ik} \dyad{k}{j} + \gamma \dyad{i}{j} = \\
        &= {\scriptsize \begin{blockarray}{cccc} \begin{block}{(ccc)c} 
          0 & \dots & 0 &  \\
          \vdots & & \vdots & \\
          f_{k1}^{kj} & \dots & f_{kD}^{kj} & \leftarrow i \\
          \vdots &  & \vdots & \\
          0 & \dots & 0 & \\
        \end{block} \end{blockarray} + 
        \begin{blockarray}{ccccc}
         & & j & & & \\
        \begin{block}{(ccccc)} 
          0 & \dots & \lambda f_{1k}^{ik} & \dots & 0 \\
          \vdots & & \vdots & & \vdots \\
          0 & \dots & \lambda f_{Dk}^{ik} & \dots & 0 \\
        \end{block} \end{blockarray} }
        + \gamma \dyad{i}{j},  \ 
        \forall i,j,k ,
    \end{align}
    from which follows that $f^{mk}_{nk} = f^{mj}_{nj} =: f^{m\ast}_{n\ast}$ and $f^{km}_{kn} = f^{jm}_{jn} =: f^{\ast m}_{\ast n}$ for all $j,k$, given $m \neq n$, and also $\hat{f}^{ij}_{ij} = \lambda f^{ik}_{ik} + f^{kj}_{kj} + \gamma$, for all $i,j,k$.

    In a similar way, using Property \textit{(i)}, we can obtain more information about the structure of $f$. That is, due to the fact that
    \begin{equation}
        f^{ij} := {\scriptsize \begin{pmatrix}
            f_{11}^{ij} & \dots & f_{1D}^{ij} \\
            \vdots & & \vdots \\
            f_{D1}^{ij} & \dots & f_{DD}^{ij}
        \end{pmatrix}} = \dyad{i}{k} \tilde{f}^{kj} + \mu \tilde{f}^{ik} \dyad{k}{j} + \delta \dyad{i}{j}, \ \forall i,j,k ,
    \end{equation}
    then function $f^{ij}$ can only have non-zero elements only on the $i$-th row and the $j$-th column, i.e.
    \begin{equation} \label{eq:f_structure}
        f^{ij} = f^{ij}_{ij} \dyad{i}{j} + \sum_{\alpha \neq j} f^{\ast j}_{\ast \alpha} \dyad{i}{\alpha} + \sum_{\beta \neq i} f^{i\ast}_{\beta\ast} \dyad{\beta}{j}.
    \end{equation}
    Moreover, from Property \textit{(ii)} again, we have $\dyad{i}{j} f^{kl} + \lambda f^{ij} \dyad{k}{l} = 0$ for any $j \neq k$, which implies that $f^{\ast j}_{\ast k} = - \frac{1}{\lambda} f^{k\ast}_{j\ast}$ for any $j \neq k$.

    We proceed now to construct the desired $P$. First, denote
    \begin{equation}
        F_{ij} := \begin{pmatrix} \dyad{i}{j} & f^{ij} \\ & \omega \dyad{i}{j} \end{pmatrix}.
    \end{equation}
    Note that $\mathcal{B} = \text{span}\{F_{ij} \mid i,j \}$, and $F_{ij}$ acts on the standard basis elements as
    \begin{equation} 
        {\small\begin{cases}
            F_{ij} \ket{m} = \delta_{jm} \ket{i}, \\
            F_{ij} \ket{D + n} = 
            {\footnotesize \begin{cases}
                \omega \ket{D+i} + f^{ij}_{ij} \ket{i} + \sum_{\beta \neq i, \beta=1}^D f^{i\ast}_{\beta\ast} \ket{\beta}, \quad j = n, \\
                f^{\ast j}_{\ast n} \ket{i}, \quad j \neq n,
            \end{cases}}
        \end{cases} }
    \end{equation}
    for $m, n \in \{1, \dots, D\}$. Let us define new vectors
    \begin{equation}
        {\small \begin{cases}
            \ket{w_m} := \ket{m}, \\
            \ket{w_{D + n}} := \ket{D + n} + \frac{f^{n1}_{n1}}{\lambda} \ket{n} + \sum_{\alpha\neq n, \alpha=1}^{D} \frac{f^{n\ast}_{\alpha\ast}}{\lambda} \ket{\alpha}.
        \end{cases}}
    \end{equation}
    In the basis $\{\ket{w_i}\}$, and using the structural properties of $f$ that were derived above, the action of $F_{ij}$ can be expressed as
    \begin{align}
        F_{ij} \ket{w_m} &= \delta_{jm} \ket{i} , \\
        F_{ij} \ket{w_{D+n}} &= F_{ij} \ket{D + n} + 
        \frac{f^{n1}_{n1}}{\lambda} F_{ij} \ket{n} +
        \sum_{\alpha \neq n} \frac{f^{n \ast}_{\alpha \ast}}{\lambda} F_{ij} \ket{\alpha} = \\
        &= \begin{cases}
            \omega \ket{D+i} + \underbrace{\left( f^{in}_{in} + \frac{f^{n1}_{n1}}{\lambda} \right)}_{=\frac{1}{\lambda} (\hat{f}^{i1}_{i1} - \gamma)}  \ket{i} + \sum_{\beta\neq i} \frac{f^{i\ast}_{\beta\ast}}{\lambda} \ket{\beta} =
        \ket{w_{D+i}} + \underbrace{\frac{1}{\lambda} ( \hat{f}^{i1}_{i1} - f^{i1}_{i1} - \gamma)}_{=k_i} \ket{w_i}, \text{ if } j = n, \\
     (f^{\ast j}_{\ast n} + \frac{f^{n\ast}_{j\ast}}{\lambda}) \ket{i} = 0, \text{ if } j \neq n,
        \end{cases}
    \end{align}
    which means that $P F_{ij} P^{-1} = {\scriptsize \begin{pmatrix} \dyad{i}{j} & k_i\dyad{i}{j} \\ & \omega \dyad{i}{j} \end{pmatrix}}$. Then,
    \begin{itemize}
        \item If $k_i$ is independent of $i$, that we denote as $k_i =: \alpha$, it means that 
        \begin{equation}
            P \mathcal{B} P^{-1} = \left\{ {\footnotesize \begin{pmatrix} A & \alpha A \\ & \omega A \end{pmatrix} } \mid A \right\}, \quad 
            P := {\footnotesize \begin{pmatrix} \mathds{1} & \tilde{P} \\ & \mathds{1} \end{pmatrix} } \text{ with } 
            \tilde{P} := {\scriptsize \begin{pmatrix} 
            -f_{11}^{11} & -f^{2\ast}_{1\ast} & \dots & -f^{D\ast}_{1\ast} \\
             -f^{1\ast}_{2\ast} & -f^{21}_{21} & \dots & -f^{D\ast}_{2\ast} \\
            \vdots & \vdots & \ddots & \vdots \\
            -f^{1\ast}_{D\ast} & -f^{2\ast}_{D\ast} & \dots & -f^{D1}_{D1}
        \end{pmatrix} }.
        \end{equation}
        Note that $P^{-1} = {\scriptsize \begin{pmatrix} \mathds{1} & -\tilde{P} \\ & \mathds{1} \end{pmatrix} }$, and $f(A) = kA + A\tilde{P} - \omega \tilde{P}A$ for all $A$.
        \item Otherwise, if $k_i$ is not independent of $i$, there exists a pair $i,j$ such that $k_i \neq k_j$, and
        \begin{align}
            \begin{pmatrix} \dyad{i}{i} & k_i \dyad{i}{i} \\ 0 & \omega \dyad{i}{i} \end{pmatrix} \begin{pmatrix} \dyad{i}{i} & k_i \dyad{i}{i} \\ 0 & \omega \dyad{i}{i} \end{pmatrix} 
            &= \begin{pmatrix} \dyad{i}{i} & k_i(1+\omega) \dyad{i}{i} \\ 0 & \omega^2 \dyad{i}{i} \end{pmatrix} \in \mathcal{B}_{2} \\
            \begin{pmatrix} \dyad{i}{j} & k_i \dyad{i}{j} \\ 0 & \omega \dyad{i}{j} \end{pmatrix} \begin{pmatrix} \dyad{j}{i} & k_j \dyad{j}{i} \\ 0 & \omega \dyad{j}{i} \end{pmatrix} 
            &= \begin{pmatrix} \dyad{i}{i} & (\omega k_i + k_j) \dyad{i}{i} \\ 0 & \omega^2 \dyad{i}{i} \end{pmatrix} \in \mathcal{B}_{2},
        \end{align}
        where $\mathcal{B}_2 := \{ab \mid a, b \in \mathcal{B}\}$. Subtracting these two elements, we have ${\scriptsize \begin{pmatrix} 0 & (k_i - k_j) \dyad{i}{i} \\ & 0 \end{pmatrix}} \in \mathcal{B}_{2}$,
        and thus an element of the form ${\tiny \begin{pmatrix} 0 & \highg{\neq 0} \\ & 0 \end{pmatrix} }$ is in $\mathcal{B}_2$.
    \end{itemize}
    
\end{proof}

\subsection{On the dependence on the off-diagonal free blocks}

The following result is used in both the algebra and span cases to show that off-diagonal blocks can depend on the off-diagonal free blocks labeled by $\Sigma_f$ only through linear combinations.

\begin{lemma} \label{lemma:technical3}
    Let $f, g_1, g_2: \mathcal{M}_{D_1 \times D_2} (\mathbb{C}) \to \mathcal{M}_{D_1 \times D_2} (\mathbb{C})$ be linear maps and let $\alpha, \beta \in \mathbb{C}$. Suppose that for all $A, B$,
    \begin{equation}
        \begin{cases}
            \text{(a)} \ Af(B) + \alpha AB = g_1(AB) \ , \\
            \text{(b)} \ f(A)B + \beta AB = g_2(AB) \ .
        \end{cases}
    \end{equation}
    Then there exists $k \in \mathbb{C}$ such that
    \begin{equation}
        \begin{cases}
            f(A) = kA \\
            g_1(A) = (k+\alpha)A \\
            g_2(A) = (k+\beta)A
        \end{cases}.
    \end{equation}
\end{lemma}
\begin{proof}
    Let $f^{ij} := f(\dyad{i}{j})$. Property \textit{(a)} with $A = \dyad{i}{j}$ and $B = \dyad{k}{l}$ with $j \neq k$ implies that $0 = \dyad{i}{j} f^{kl}$, meaning that $f^{kl}$ is all 0s except for the $k$-th row. Similarly, Property \textit{(b)} implies that $f^{ij}$ is all 0s except for the $j$-th column. Therefore, $f^{ij} = f^{ij}_{ij} \dyad{i}{j}$. 

    Property \textit{(a)} also implies that, for any $i$, $k$, we have
    \begin{align}
        g_1^{ik} &= \dyad{i}{j} f^{jk} + \alpha \dyad{i}{k} \\
        &= (f^{jk}_{jk} + \alpha) \dyad{i}{k} \\
        &= (f^{j'k}_{j'k} + \alpha) \dyad{i}{k}, \ \forall j \neq j'.
    \end{align}
    Therefore, $f^{jk}_{jk} = f^{j'k}_{j'k}$ for any $j, j'$. Similarly, Property \textit{(b)} also implies that, for any $j, j'$, $f^{ij}_{ij} = f^{ij'}_{ij'}$. This means that $f_{ij}^{ij}$ is independent both of $i,j$, and therefore $f^{ij} = k \dyad{i}{j}$ and $f(A) = kA$ for some constant $k \in \mathbb{C}$. 
    Properties \textit{(a)} and \textit{(b)} finally imply that $g_1(A) = (k+\alpha)A$ and $g_2(A) = (k+\beta)A$, and the proof is complete.
\end{proof}

\subsection{On the persistence of the structure under blocking}

The lemmas in this subsection aim to show that the span structure is preserved under blocking. In particular, Lemma \ref{lemma:nonzero_elements} shows that isolatable blocks remain isolatable. Lemmas \ref{lemma:same_form_AB_2x2} and \ref{lemma:same_form_AA_2x2} establish the base case for structures of 2 blocks by 2 blocks, demonstrating that generalized Jordan blocks or zeros persist for any blocking length. Finally, Lemma \ref{lemma:same_form_AA_2x2_extension} extends this result to structures of $b$ blocks by $b$ blocks, which is then used in the inductive step of the proof of Theorem \ref{prop:structure_span} on the span structure.

\begin{lemma} \label{lemma:nonzero_elements}
    Let $\mathcal{A}^{(\ell)}$ have a block structure with $b$ blocks by $b$ blocks, and suppose that block $(i,j)$ is isolatable in $\mathcal{A}^{(\ell)}$ with $(i,j) \notin [\{0,\varepsilon\}, \{0,\varepsilon\}]$. Then, for every $m \geq \ell$, the block $(i,j)$ remains isolatable in $\mathcal{A}^{(m)}$ and becomes a full free block in $\mathcal{A}^{(2L_0^{\mathrm{diag}}+m)}$.
\end{lemma}
\begin{proof}
    First, assume that $\mathcal{A}^{(\ell)}$ has a structure of 2 blocks by 2 blocks, and consider the off-diagonal block, i.e. $(i,j) = (1,2)$. The hypothesis implies the existence of ${\scriptsize \begin{pmatrix} 0 & Z \\ & 0 \end{pmatrix}} \in \mathcal{A}^{(\ell)}$ for some $Z \neq 0$. 
    
    Consider multiplying this element on the right by any of the MPS matrices. We obtain:
    \begin{equation} \small
        \begin{pmatrix} 0 & Z \\ & 0 \end{pmatrix} \begin{pmatrix} A_1^i & B^i \\ & A_2^i \end{pmatrix} = \begin{pmatrix} 0 & ZA_2^i \\ & 0 \end{pmatrix} \in \mathcal{A}^{(\ell+1)}, \quad \forall i.
    \end{equation}
    Assume that $Z A_2^i = 0$ for all $i$. Then, $Z \mathcal{C}^{(n)} = 0$ for all $n$, where $\mathcal{C}^{(n)} := \text{span}\{A_2^{i_1} \dots A_2^{i_n}\mid i_1, \dots, i_n\}$. Due to the fact that $\mathcal{C}^{(n)} = \mathcal{M}_{D_2}(\mathbb{C})$ for all $n \geq L_0^{\text{diag}}$, this would necessarily imply that $Z = 0$, which contradicts the initial assumption.
     
    If instead $\{0,r_2\} = \{0,\varepsilon\}$, then $\{0,r_1\} \neq \{0,\varepsilon\}$, and we multiply the isolatable element by any of the MPS matrices on the left, obtaining:
    \begin{equation*} {\small
        \begin{pmatrix} A_1^i & B^i \\ & A_2^i \end{pmatrix} 
        \begin{pmatrix} 0 & Z \\ & 0 \end{pmatrix} 
        = 
        \begin{pmatrix} 0 & A_1^i Z \\ & 0 \end{pmatrix} \in \mathcal{A}^{(\ell+1)}, \quad \forall i.}
    \end{equation*}
    An analogous argument can be applied to conclude the proof for the first part of the claim.

    For the second claim, we now show that block $(1,2)$ becomes a full free block in $\mathcal{A}^{(2L_0^{\text{diag}}+m)}$, for any $m \geq \ell$. By rescaling if necessary, assume $Z = \dyad{j}{k} + \sum_{(\alpha, \beta)\neq (j,k)} z_{\alpha \beta} \dyad{\alpha}{\beta}$. 
    
    In the case that $\{0,r_1\}, \{0,r_2\} \neq \{0,\varepsilon\}$, the diagonal blocks span the full matrix algebras upon blocking every $L_0^{\text{diag}}$ sites or more. Therefore, we have that for all $p \in \mathbb{Z}_{\geq 0}$, 
    \begin{equation}
        {\small \underbrace{\begin{pmatrix} \dyad{i}{j} & \ast \\ & \ast \end{pmatrix}}_{\in \mathcal{A}^{(L_0^{\text{diag}})}} 
        \underbrace{\begin{pmatrix} 0 & \dyad{j}{k} + \dots \\ & 0 \end{pmatrix}}_{\in \mathcal{A}^{(\ell)}}
        \underbrace{\begin{pmatrix} \ast & \ast \\ & \dyad{k}{l} \end{pmatrix}}_{\in \mathcal{A}^{(L_0^{\text{diag}}+p)}} = 
        \begin{pmatrix} 0 & \dyad{i}{l} \\ & 0 \end{pmatrix}} \in \mathcal{A}^{(2L_0^{\text{diag}}+\ell+p)} \quad \forall i,l,
    \end{equation}
    meaning that block $(1,2)$ becomes a free block.
    
    Note that, if $\{0,r_1\}$ or $\{0,r_2\}$ equals $\{0,\varepsilon\}$, then even less blocking is required. For example, if $\{0,r_2\} = \{0,\varepsilon\}$, then $Z = \dyad{j}{1} + \dots$ and
    \begin{equation*}
        \underbrace{{\small \begin{pmatrix} \dyad{i}{j} & \ast \\ & 0 \end{pmatrix}}}_{\in \mathcal{A}^{(L_0^{\text{diag}}+p)}}
        \underbrace{\small{\begin{pmatrix} 0 & \dyad{j}{1} + \dots \\ & 0 \end{pmatrix}}}_{\in \mathcal{A}^{(\ell)}} = 
        \begin{pmatrix} 0 & \dyad{i}{1} \\ & 0 \end{pmatrix} \in \mathcal{A}^{(L_0^{\text{diag}}+\ell+p)}, \quad \forall p \in \mathbb{Z}_{\geq 0},
    \end{equation*}
    An analogous argument holds if $\{0,r_1\} = \{0,\varepsilon\}$. 

    Finally, the argument extends directly to the general case of a $b \times b$ block structure and any off-diagonal block $(i,j)$. Indeed, one can always select an isolatable element in $\mathcal{A}^{(\ell)}$ that vanishes on all blocks $\prec (i,j)$ and is nonzero in block $(i,j)$, say with entry $Z \neq 0$. In this setting, the multiplication steps described above remain unchanged when restricted to the submatrix
    \begin{equation*}
        {\scriptsize \begin{pmatrix}
            \mathcal{A}^{(\ell)}_{ii} & \mathcal{A}^{(\ell)}_{ij} \\
            0 & \mathcal{A}^{(\ell)}_{jj}
        \end{pmatrix}}
    \end{equation*}
    Therefore, without loss of generality, the entire argument extends to arbitrary $(i,j)$, completing the proof.
\end{proof}

The following lemma formalizes the fact that if, for some blocking length, the span of a 2 blocks by 2 blocks structure has zero as the off-diagonal part, then the same holds for all larger blocking lengths.

\begin{lemma} \label{lemma:same_form_AB_2x2}
    Given that $\mathcal{A}^{(L)}$ consists of 2 blocks by 2 blocks matrices of type AB, of the form
    \begin{equation} \label{eq:nice_form_1}
        \mathcal{A}^{(L)} =
        \left\{  {\scriptsize
        \begin{pmatrix}
            A & 0 \\ & B
        \end{pmatrix}} \mid A, B \right\},
    \end{equation}
    then $\mathcal{A}^{(\ell)}$ has the same form for all $\ell \geq L$ (and thus $m_{12} = \infty)$. The same holds for type A0 and 0A structures.
\end{lemma}
\begin{proof}
    The assumption implies that $\mathcal{A}^{(mL)}$ has the same form as Eq. \eqref{eq:nice_form_1}. This means that no element of the form ${\scriptsize \begin{pmatrix} 0 & \neq 0 \\ & 0 \end{pmatrix}}$ exists in $\mathcal{A}^{(\ell)}$ for any $\ell \geq 1$, as it would lead to a contradiction by Lemma \ref{lemma:nonzero_elements}. Therefore, for any $\ell \geq L$, 
    \begin{equation} \label{eq:form_nonzero_element_1}
        \mathcal{A}^{(\ell)} = \left\{  {\footnotesize
        \begin{pmatrix}
            A & f_{\{0,r_1\}}^{(\ell)}(A) + f_{\{0,r_2\}}^{(\ell)}(B) \\ & B
        \end{pmatrix}} \mid A, B \right\}, 
    \end{equation}
    for some linear functions $f_{\{0,r_1\}}^{(\ell)}$, $f_{\{0,r_2\}}^{(\ell)}$. Then, for any $s \in \mathbb{N}$ and arbitrary $A, B$, 
    \begin{enumerate}
        \item ${\footnotesize 
            \begin{pmatrix} A & f_{\{0,1\}}^{(L+s)}(A) \\ & 0 \end{pmatrix}}
            {\footnotesize\begin{pmatrix} 0 & 0 \\ & B \end{pmatrix}} =
            {\footnotesize\begin{pmatrix} 0 & f_{\{0,r_1\}}^{(L+s)}(A) B \\ & 0 \end{pmatrix}} \in \mathcal{A}^{(2L+s)} \implies 
            f_{\{0,r_1\}}^{(L+s)} = 0$.
        \item ${\footnotesize 
            \begin{pmatrix} A & 0 \\ & 0 \end{pmatrix}}
            {\footnotesize\begin{pmatrix} 0 & f_{\{0,r_2\}}^{(L+s)}(B) \\ & B \end{pmatrix}} =
            {\footnotesize\begin{pmatrix} 0 & A f_{\{0,r_2\}}^{(L+s)}(B) \\ & 0 \end{pmatrix}} \in \mathcal{A}^{(2L+s)} \implies 
            f_{\{0,r_2\}}^{(L+s)} = 0$.
    \end{enumerate}
    Therefore, $\mathcal{A}^{(\ell)}$ is of the form of Eq. \eqref{eq:nice_form_1} for all $\ell \geq L$. Note that the same argument applies if $\mathcal{A}^{(L)}$ is originally of type A0 or 0A.
\end{proof}

The following is an auxiliary result that we will use to prove the subsequent Lemma \ref{lemma:same_form_AA_2x2}.

\begin{lemma}\label{lemma:span_A1_tambien_Jordan}
    If there exists a linear function $g$ and constants $\alpha, \beta \in \mathbb{C}$ with $\alpha \neq 0$ that satisfy
    \begin{equation}
        Ag(B) + \alpha g(A)B = \beta AB, \quad \forall A, B,
    \end{equation}
    then $g(A) = \frac{\beta}{1+\alpha} A$ for all $A$.
\end{lemma}
\begin{proof}
    First, we take $A = \dyad{i}{k}, B = \dyad{k}{j}$ for any $i, j, k$. Then, we have that
    \begin{align}
        \beta \dyad{i}{j} &= \dyad{i}{k} g^{kj} + \alpha g^{ik} \dyad{k}{j} = \\
        &= {\scriptsize \begin{blockarray}{cccc} \begin{block}{(ccc)c} 
          0 & \dots & 0 &  \\
          \vdots & & \vdots & \\
          g_{k1}^{kj} & \dots & \alpha g_{kD}^{kj} & \leftarrow i \\
          \vdots &  & \vdots & \\
          0 & \dots & 0 & \\
        \end{block} \end{blockarray} + 
        \begin{blockarray}{ccccc}
         & & j & & & \\
        \begin{block}{(ccccc)} 
          0 & \dots & \alpha g_{1k}^{ik} & \dots & 0 \\
          \vdots & & \vdots & & \vdots \\
          0 & \dots & \alpha g_{Dk}^{ik} & \dots & 0 \\
        \end{block} \end{blockarray},}  \ 
        \forall i,j,k ,
    \end{align}
    from which follows that \textit{(i)} $g^{mk}_{nk} = g^{km}_{kn} = 0$ for all $m,n,k$ with $m \neq n$, and \textit{(ii)} $\beta = \alpha g_{ik}^{ik} + g^{kj}_{kj}$, for all $i,j,k$. Choosing $i = j = k$, \textit{(ii)} implies that $g_{ii}^{ii} = \frac{\beta}{1+\alpha}$ for any $i$. Then again, this together with \textit{(ii)} choosing any $i,j,k$ with $i = k$, implies that
    \begin{equation}
        g_{ij}^{ij} = \frac{\beta}{1+\alpha}.
    \end{equation}
    Finally, let us choose $A = \dyad{i}{j}$ and $B = \dyad{k}{l}$ for any $i, j, k, l$ with $j \neq k$. Then, we have that
    \begin{align}
        0 &= \dyad{i}{j} g^{kl} + \alpha g^{ij} \dyad{k}{l} = \\
        &= \sum_{n \neq l} g^{kl}_{jn} \dyad{i}{n} + \sum_{m \neq i} g^{ij}_{mk} \dyad{m}{l}.
    \end{align}
    This implies that $g^{ij}_{mn} = 0$ for all $i, j, m, n$ with $i \neq m$ and $j \neq n$.

    Putting everything together, we obtain that $g^{ij} = \frac{\beta}{1+\alpha} \dyad{i}{j}$ and therefore,
    \begin{equation}
        g(A) = \frac{\beta}{1 + \alpha} A, \quad \forall A.
    \end{equation}
\end{proof}

Now we proceed to prove the lemma ensuring that for the case of 2 blocks by 2 blocks structures, if the span has a structure with a generalized Jordan block for a certain specific blocking length, then the span also has this same structure in $\mathcal{A}^{(\ell)}$ for all sufficiently large $\ell$.

\begin{lemma} \label{lemma:same_form_AA_2x2}
    Given that $\mathcal{A}^{(2L)}$ consists of 2 blocks by 2 blocks matrices of type AA, of the form
    \begin{equation} \label{eq:nice_form_2}
        \mathcal{A}^{(2L)} =
        \left\{  {\footnotesize
        \begin{pmatrix}
            A & k_{12;\{0,r_1\}}^{(2L)} A \\ & k_{22;\{0,r_1\}}^{(2L)}A
        \end{pmatrix}} \mid A \right\},
    \end{equation}
    then $\mathcal{A}^{(\ell)}$ has the same form for all $\ell \geq L$ (and thus $m_{12} = \infty$).
\end{lemma}
\begin{proof}
    The assumption implies that $\mathcal{A}^{(m2L)}$ has the same form as Eq.~\eqref{eq:nice_form_2}, with updated constants
    \begin{equation*}
        k_{12;\{0,1\}}^{(m2L)} = m k_{12;\{0,1\}}^{(2L)} \quad \text{and} \quad k_{22;\{0,1\}}^{(m2L)} = (k_{22;\{0,1\}}^{(2L)})^m .
    \end{equation*}
    This implies that no element of the form ${\scriptsize \begin{pmatrix} 0 & \neq 0 \\ & 0 \end{pmatrix}}$ exists in $\mathcal{A}^{(\ell)}$ for any $\ell \geq 1$, as this would lead to a contradiction by Lemma \ref{lemma:nonzero_elements}. Therefore, for any $\ell \geq L$, 
    \begin{equation} \label{eq:form_nonzero_element_2_aux}
        \mathcal{A}^{(\ell)} = \left\{  {\footnotesize
        \begin{pmatrix}
            A & f_{\{0,r_1\}}^{(\ell)}(A)\\ & k_{22;\{0,r_1\}}^{(\ell)} A
        \end{pmatrix}} \mid A \right\}.
    \end{equation}
    for some linear functions $f_{\{0,r_1\}}^{(\ell)}$.
    
    The initial assumption also implies that $\mathcal{A}^{(L)}$ has the same form as $\mathcal{A}^{(2L)}$ in Eq. \eqref{eq:nice_form_2}, since for arbitrary $A, B$, it must hold that 
    \begin{align*}
        &{\footnotesize 
        \begin{pmatrix} A & f_{\{0,r_1\}}^{(L)}(A) \\ & k_{22;\{0,r_1\}}^{(L)} A \end{pmatrix}}
        {\footnotesize\begin{pmatrix} B & f_{\{0,r_1\}}^{(L)}(B) \\ & k_{22;\{0,r_1\}}^{(L)} B \end{pmatrix}} =
        {\footnotesize\begin{pmatrix} AB & A f_{\{0,r_1\}}^{(L)}(B) + k_{22;\{0,r_1\}}^{(L)} f_{\{0,r_1\}}^{(L)}(A) B \\
        0 & (k_{22;\{0,r_1\}}^{(L)})^2 AB \end{pmatrix}} \in P \mathcal{A}^{(2L)} P^{-1} \\
        &\quad \implies \implies A f_{\{0,r_1\}}^{(L)}(B) + k_{22;\{0,r_1\}}^{(L)} f_{\{0,r_1\}}^{(L)}(A) B = k_{12;\{0,r_1\}}^{(2L)} AB
    \end{align*}
    This corresponds to the condition of Lemma \ref{lemma:span_A1_tambien_Jordan}, $A g(B) + \alpha g(A) B = \beta AB$. Therefore, one can conclude that 
    {\small\begin{equation}
        f_{\{0,r_1\}}^{(L)}(A) = \frac{k_{12;\{0,r_1\}}^{(2L)}}{1+k_{22;\{0,r_1\}}^{(L)}} A =: k_{12;\{0,r_1\}}^{(L)} A , \quad \forall A .
    \end{equation}}

    Then, for any $s \in \mathbb{N}$ and arbitrary $A, B$,
    \begin{enumerate}
        \item 
            ${\footnotesize \begin{pmatrix} 
            A & f_{\{0,r_1\}}^{(L+s)}(A) \\ 
            & k_{22;\{0,r_1\}}^{(L+s)} A 
            \end{pmatrix}}
            {\footnotesize\begin{pmatrix} 
            B & k_{12;\{0,r_1\}}^{(L)} B \\ 
            & k_{22;\{0,r_1\}}^{(L)} B 
            \end{pmatrix}} =
            {\footnotesize\begin{pmatrix} 
            AB & k_{12;\{0,r_1\}}^{(L)} AB + k_{22;\{0,r_1\}}^{(L)} f_{\{0,r_1\}}^{(L+s)}(A) B \\
            0 & k_{22;\{0,r_1\}}^{(L+s)} k_{22;\{0,r_1\}}^{(L)} AB 
            \end{pmatrix}} \in \mathcal{A}^{(2L+s)}$
            
        \item $ {\footnotesize\begin{pmatrix} 
            A & k_{12;\{0,r_1\}}^{(L)} A \\ 
            & k_{22;\{0,r_1\}}^{(L)} A 
            \end{pmatrix}}
            {\footnotesize \begin{pmatrix} 
            B & f_{\{0,r_1\}}^{(L+s)}(B) \\ 
            & k_{22;\{0,r_1\}}^{(L+s)} B 
            \end{pmatrix}}
            =
            {\footnotesize\begin{pmatrix} 
            AB & A f_{\{0,r_1\}}^{(L+s)}(B) + 
            k_{12;\{0,r_1\}}^{(L)} k_{22;\{0,r_1\}}^{(L+s)} AB \\
            0 & k_{22;\{0,r_1\}}^{(L+s)} k_{22;\{0,r_1\}}^{(L)} AB 
            \end{pmatrix}} \in \mathcal{A}^{(2L+s)}$
    \end{enumerate}
    These lead to the following conditions on $f_{\{0,r_1\}}^{(2L+s)}$,
    \begin{equation}
        \begin{cases}
            k_{22;\{0,r_1\}}^{(L)} f_{\{0,r_1\}}^{(L+s)}(A) B + k_{12;\{0,r_1\}}^{(L)} AB = f_{\{0,r_1\}}^{(2L+s)} (AB) \\
            A f_{\{0,r_1\}}^{(L+s)}(B) + 
            k_{12;\{0,r_1\}}^{(L)} k_{22;\{0,r_1\}}^{(L+s)} AB = f_{\{0,r_1\}}^{(2L+s)} (AB)
        \end{cases},
    \end{equation}
    coinciding with those of Lemma \ref{lemma:technical3}, which tells us that $f_{\{0,r_1\}}^{(\ell)}(A) = k_{12;\{0,r_1\}}^{(\ell)} A$ and thus $\mathcal{A}^{(\ell)}$ is of the form of Eq. \eqref{eq:nice_form_2} for all $\ell \geq L$.
\end{proof}

Finally, we generalize the lemma above so that we can use it in the proof of the span structure of Theorem \ref{prop:structure_span} in the general case of matrices with $b$ blocks by $b$ blocks.

\begin{lemma} [Generalization of Lemma \ref{lemma:same_form_AA_2x2}]\label{lemma:same_form_AA_2x2_extension}
    Given that block $(i,j)$ of $\mathcal{A}^{(2L)}$ is of type AA with $m_{ij} > 4L$, and assuming that for any $a \in \mathcal{A}^{(2L)}$ of the form $a \mid_{\prec (i,j)} = \sum_{e \in \Sigma} [A_e]_e^{(2L)}$, its $(i,j)$-th block has the form
    \begin{equation} \label{eq:lemma_gen_AA_aux1}
        a_{ij} = k_{ij;\{0,r_i\}}^{(2L)} A_{\{0,r_i\}} + \sum_{e \neq \{0,r_i\}} f_e^{(2L)}(A_e),
    \end{equation}
    then this form persists under blocking, meaning that for all $\ell$ with $2L \leq \ell < m_{ij}$, block $(i,j)$ in $\mathcal{A}^{(\ell)}$ has the same form of Eq. \eqref{eq:lemma_gen_AA_aux1} with updated constants $k_{ij;\{0,r_i\}}^{(\ell)}$ and functions $f_e^{(\ell)}$.
\end{lemma}
\begin{proof}
    From the definition of $m_{ij}$, we know that for all $\ell$ with $\ell < m_{ij}$, given $a \mid_{\prec (i,j)} = \sum_{e \in \Sigma} [A_e]_e^{(\ell)}$, there exist linear functions $f_e^{(\ell)}$ such that
    \begin{equation*}
        a_{ij} = \sum_{e \in \Sigma} f^{(\ell)}_{e} (A_e) .
    \end{equation*} 
    Taking $a = [A]^{(2L)}_{\{0,r_i\}} \in \mathcal{A}^{(2L)}$ and $b = [B]^{(2L+s)}_{\{0,r_i\}} \in \mathcal{A}^{(2L+s)}$ for arbitrary $A, B$ and $0 \leq s < m_{ij} - 4L$, then because of property (P1) of the structure constants tensor $\Gamma^{(2L, 2L+s)}$, we have on the one hand that
    \begin{equation*}
        (ab)_{ij} = \left([AB]_{\{0,r_i\}}^{(4L+s)}\right)_{ij} = f_{\{0,r_i\}}^{(4L+s)} (AB).
    \end{equation*}
    On the other hand, using the assumption that $f_{\{0,r_i\}}^{(2L)}(A) = k_{ij;\{0,r_i\}}^{(2L)} A$,
    \begin{equation*}
        (ab)_{ij} = 
        k_{ii;\{0,r_i\}}^{(2L)} A f_{\{0,r_i\}}^{(2L+s)}(B)
        + k_{jj;\{0,r_i\}}^{(2L+s)} k_{ij;\{0,r_i\}}^{(2L)} A B .
    \end{equation*}
    Taking $a = [A]^{(2L+s)}_{\{0,r_i\}} \in \mathcal{A}^{(2L+s)}$ and $b = [B]^{(2L)}_{\{0,r_i\}} \in \mathcal{A}^{(2L)}$, we obtain a similar equation. Then, we have
    \begin{equation*} \begin{cases}
        k_{ii;\{0,r_i\}}^{(2L)} A f_{\{0,r_i\}}^{(2L+s)}(B)
        + k_{jj;\{0,r_i\}}^{(2L+s)} k_{ij;\{0,r_i\}}^{(2L)} A B = f_{\{0,r_i\}}^{(4L+s)} (AB) , \\
        k_{ii;\{0,r_i\}}^{(2L+s)} k_{ij;\{0,r_i\}}^{(2L)} A B
        + k_{jj;\{0,r_i\}}^{(2L)} f_{\{0,r_i\}}^{(2L+s)}(A) B = f_{\{0,r_i\}}^{(4L+s)} (AB) ,
    \end{cases} \end{equation*}
    which coincide with the conditions of Lemma \ref{lemma:technical3}. This tells us that $f_{\{0,r_i\}}^{(\ell)}(A) = k^{(\ell)}_{ij;\{0,r_i\}} A$ and thus $\mathcal{A}^{(\ell)}$ is of the form of Eq. \eqref{eq:lemma_gen_AA_aux1} for all $\ell$ with $2L \leq \ell < m_{ij}$.
\end{proof}

\newpage

\end{document}